\documentclass[review]{elsarticle}

\usepackage[colorlinks,citecolor=blue,linktoc=all,linkcolor=cyan]{hyperref}
\usepackage{graphicx}
\usepackage{bm} 
\usepackage{mathbbol}
\usepackage[T1]{fontenc}
\usepackage{dsfont}             
\usepackage{mathrsfs}     
\usepackage{slashed}              
\usepackage{amsmath}
\usepackage{amssymb}
\usepackage{amsbsy}
\usepackage{amsfonts}
\usepackage{bbm}
\usepackage[dvipsnames]{xcolor}
\usepackage{pifont}

\usepackage[compat=1.1.0]{tikz-feynman}

\numberwithin{equation}{section}
\numberwithin{table}{section}
\numberwithin{figure}{section}

\DeclareMathOperator*{\SumInt}{%
\mathchoice%
  {\ooalign{$\displaystyle\sum$\cr\hidewidth$\displaystyle\int$\hidewidth\cr}}
  {\ooalign{\raisebox{.14\height}{\scalebox{.7}{$\textstyle\sum$}}\cr\hidewidth$\textstyle\int$\hidewidth\cr}}
  {\ooalign{\raisebox{.2\height}{\scalebox{.6}{$\scriptstyle\sum$}}\cr$\scriptstyle\int$\cr}}
  {\ooalign{\raisebox{.2\height}{\scalebox{.6}{$\scriptstyle\sum$}}\cr$\scriptstyle\int$\cr}}
}

\def\d{\mathrm{d}}
\def\D{\mathcal{D}}

\def\L{\mathcal{L}}

\def\C{\mathcal{C}}
\def\O{\mathcal{O}}
\def\N{\mathcal{N}}
\def\vec{\bm}
\def\i{\mathrm{i}}

\def\e{\mathrm{e}}

\def\M{\mathcal{M}}

\def\vecx{\vec{x}}

\def\tr{\mathrm{Tr}}
\def\hc{{\rm h.c.}}

\journal{Progress in Particle and Nuclear Physics}

\usepackage{titlesec}
\usepackage{sectsty}
\titleformat{\section}{\normalfont\Large\bfseries}{\thesection}{1em}{}
\titleformat{\subsection}{\normalfont\large\bfseries}{\thesubsection}{1em}{}
\titleformat{\subsubsection}{\normalfont\normalsize\bfseries}{\thesubsubsection}{1em}{}

\begin{document}

\makeatletter
\def\ps@pprintTitle{%
   \let\@oddhead\@empty
   \let\@evenhead\@empty
   \let\@oddfoot\@empty
   \let\@evenfoot\@oddfoot}
\makeatother
    
\begin{frontmatter}
		
\title{Status of the strong $CP$ problem}

\author[address1,address2]{Wen-Yuan Ai\corref{mycorrespondingauthor}}
\cortext[mycorrespondingauthor]{wenyuanai@sjtu.edu.cn}

\address[address1]{State Key Laboratory of Dark Matter Physics,\\ Tsung-Dao Lee Institute and School of Physics and Astronomy,\\ Shanghai Jiao Tong University, Shanghai 201210, China}
\address[address2]{Key Laboratory for Particle Astrophysics and Cosmology (MOE),\\ and Shanghai Key Laboratory for Particle Physics and Cosmology,\\ Shanghai Jiao Tong University, Shanghai 201210, China }
		
\begin{abstract}
The strong $CP$ problem remains one of the major open questions in the Standard Model of particle physics. Its best-known solution is the Peccei--Quinn mechanism, which predicts a new pseudoscalar boson, the QCD axion. In recent years, alternative solutions and new perspectives on the foundations of the strong $CP$ problem have also received renewed attention. In this review, we provide a pedagogical account of the underlying physics and summarise the current theoretical and phenomenological status of the problem.
\end{abstract}
		
\end{frontmatter}

\newpage
\thispagestyle{empty}
\tableofcontents
	

\newpage

\section{Introduction: what is the strong \texorpdfstring{$CP$}{CP} problem}

Quantum Chromodynamics (QCD) provides a remarkably successful description of the strong interactions across a wide range of energy scales. At the fundamental level, the most general renormalisable Lagrangian consistent with gauge invariance and Lorentz symmetry reads
\begin{align}
    \mathcal{L}_{\rm M}
    =
    -\frac{1}{4}G^a_{\mu\nu}G^{a\mu\nu}
    +
    \bar{\psi}_i\left(\i\slashed{D}-M_{ij}\right)\psi_j
    +\frac{\theta g_s^2}{32\pi^2}\, G^a_{\mu\nu}\widetilde{G}^{a\mu\nu},
    \label{eq:QCD-Lagrangian}
\end{align}
where $\psi$ denotes the quark multiplet in flavour space. The notation will be introduced more carefully below. In this review, we mainly focus on the three lightest flavours, $\psi=(u,d,s)^T$. The mass matrix $M$ is generally complex. In the diagonalised basis, it can be written as $M={\rm diag}(\e^{\i\alpha_1\gamma^5} m_u, \e^{\i\alpha_2\gamma^5} m_d, \e^{\i\alpha_3\gamma^5}m_s)$. Then the mass term can be written as 
\begin{align}
    \bar\psi M\psi = \bar\psi_L\M \psi_R +\bar\psi_R\M^\dagger\psi_L,
    \quad 
    \M = {\rm diag}(\e^{\i\alpha_1}m_u, \e^{\i\alpha_2}m_d,\e^{\i\alpha_3}m_s).
    \label{eq:Dirac-mass}
\end{align}
Although the last term in Eq.~\eqref{eq:QCD-Lagrangian} is a total derivative at the classical level, it has physical consequences due to the nontrivial topological structure of non-Abelian gauge theory~\cite{Belavin:1975fg,tHooft:1976snw}.

It is important to emphasise that neither the $\theta$ parameter nor the chiral phases in the quark mass matrix are independently physical. Under chiral field redefinitions
\begin{align}
    \psi_i \rightarrow \mathrm{e}^{\i\beta_i\gamma^5}\psi_i,
    \qquad
    \bar{\psi}_i \rightarrow \bar{\psi}_i\,\mathrm{e}^{\i\beta_i\gamma^5},
    \label{eq:chiralrotations}
\end{align}
we have~\cite{Fujikawa:1979ay,Fujikawa:1980eg}
\begin{align}
    \alpha_i \rightarrow \alpha_i+2\beta_i,
    \qquad
    \theta \rightarrow \theta-2\sum_i\beta_i.
\end{align}
Therefore, only the combination
\begin{equation}
    \bar{\theta} = \theta + \arg\det \M
\end{equation}
is invariant under chiral redefinitions and thus represents the physical strong $CP$ phase.

A nonzero $\bar{\theta}$ is widely believed to induce $CP$-violating effects in the strong interactions. In particular, it is expected to generate a neutron electric dipole moment (nEDM), as well as CP-odd pion-nucleon couplings and other hadronic observables. Experimentally, however, the neutron EDM is extremely constrained. The current bound~\cite{Baker:2006ts,Abel:2020pzs}
\begin{equation}
    |d_n| \lesssim 10^{-26}\,e\,\mathrm{cm},
\end{equation}
implies
\begin{equation}
    |\bar{\theta}| \lesssim 10^{-10},
\end{equation}
which is many orders of magnitude below the ``natural'' expectation $\bar{\theta}\sim\mathcal{O}(1)$ in the absence of a protective mechanism. This hierarchy constitutes the \emph{strong $CP$ problem}: why is the parameter $\bar{\theta}$ so small?

The puzzle is sharpened by the existence of $CP$ violation in the weak interactions through the Cabibbo--Kobayashi--Maskawa (CKM) phase~\cite{Kobayashi:1973fv}. Within the Standard Model, there is therefore no symmetry reason that enforces an approximately $CP$-conserving strong sector.

From a theoretical perspective, the meaning of $\bar{\theta}$ is deeply tied to the structure of the QCD vacuum, the role of topological sectors, and the interplay between chiral symmetry and axial anomalies. A proper understanding therefore requires a careful treatment of instantons, large gauge transformations, and the emergence of the $\theta$-vacuum, which will be developed in the following sections.

\medskip

Conventionally, proposed resolutions of the strong $CP$ problem fall into three broad classes:
\begin{itemize}
\item[(1)] \textbf{A vanishing light-quark mass}: in particular, the possibility that the up-quark mass is zero, in which case $\bar\theta$ can be rotated away and becomes unphysical.

\item[(2)] \textbf{$CP$- and parity-based solutions}: $CP$ or parity is imposed as a fundamental symmetry in the ultraviolet theory and subsequently broken in such a way that the strong $CP$ phase remains sufficiently small while allowing the observed weak-interaction $CP$ violation.

\item[(3)] \textbf{Dynamical relaxation}: the strong $CP$ phase is promoted to a dynamical field whose potential drives it to a $CP$-conserving minimum. The most prominent realisation is the Peccei--Quinn (PQ) mechanism~\cite{Peccei:1977hh}, which predicts the axion as a pseudo-Nambu--Goldstone boson~\cite{Weinberg:1977ma,Wilczek:1977pj}.
\end{itemize}

A logically distinct question is whether the conventional connection between a nonzero $\bar\theta$ and observable strong-interaction $CP$ violation is itself unavoidable. Alternative viewpoints on this issue have been explored in recent years~\cite{Ai:2020ptm,Nakamura:2021meh,Ai:2022htq,Ai:2024vfa,Schierholz:2024var,Ai:2024cnp,Kaplan:2024ezz,Williams:2026cec}\footnote{See also Refs.~\cite{Torrieri:2020nin,Yamanaka:2022vdt,Yamanaka:2022bfj} for some  other perspectives.} and have prompted an active debate~\cite{Ai:2020ptm,Albandea:2024fui,Ai:2024vfa,Ai:2024cnp,Benabou:2025viy,Ai:2025quf,Khoze:2025auv,Bhattacharya:2025qsk,Ringwald:2026apz,Aghaie:2026pkf,Sannino:2026wgx}. The purpose of this review is to provide a coherent and pedagogical account of the current status of the strong $CP$ problem, from its conceptual foundations to proposed solutions and phenomenological implications, while presenting the assumptions and arguments underlying these alternative viewpoints in a neutral manner.

A substantial body of review literature has developed around the strong $CP$ problem and axion physics. Pedagogical introductions to the strong $CP$ problem can be found in the classic review by Cheng~\cite{Cheng:1987gp}, the lectures by Peccei~\cite{Peccei:2006as} and Hook~\cite{Hook:2018dlk}, and the recent accounts by Bonanno, Bonati and D'Elia~\cite{Bonanno:2025wcv} and Sannino~\cite{Sannino:2026wgx}. Classic and modern reviews of axion theory and phenomenology include Kim~\cite{Kim:1986ax}, Kim and Carosi~\cite{Kim:2008hd}, Marsh~\cite{Marsh:2015xka}, Di Luzio, Giannotti, Nardi and Visinelli~\cite{DiLuzio:2020wdo}, Choi, Im and Shin~\cite{Choi:2020rgn}, and Chadha-Day, Ellis and Marsh~\cite{Chadha-Day:2021szb}. From the perspective of low-energy hadronic physics and EDM phenomenology, detailed reviews are provided by Pospelov and Ritz~\cite{Pospelov:2005pr,Pospelov:2025vzj} and by Engel, Ramsey-Musolf and van Kolck~\cite{Engel:2013lsa}. Axion astrophysics, cosmology and experimental searches are reviewed from complementary perspectives in Refs.~\cite{Raffelt:2006cw,Sikivie:2006ni,Graham:2015ouw,Irastorza:2018dyq,Caputo:2024oqc,OHare:2024nmr,Carenza:2024ehj,Baryakhtar:2025jwh,Cicoli:2026fqp}. For QCD topology and $\theta$ dependence, the classic review by Vicari and Panagopoulos~\cite{Vicari:2008jw} and the recent accounts in Refs.~\cite{Bonanno:2025wcv,Bonanno:2026zzf} provide complementary perspectives. Together, these works cover the theoretical foundations, phenomenology, cosmology and experimental implications relevant to the strong $CP$ problem and QCD axion.

The structure of this review is as follows. We begin in Part~\ref{part:origin} by reviewing the theoretical origin of the strong $CP$ problem, from the anomalous and topological structure of QCD to the emergence of the physical phase $\bar\theta$ and its low-energy $CP$-violating consequences, with particular emphasis on electric dipole moments (EDMs). In Part~\ref{part:solutions}, we discuss what constitutes a viable solution to the strong $CP$ problem and review the main conventional possibilities: a vanishing light-quark mass, $CP$- and parity-based constructions, and dynamical relaxation through the PQ mechanism. Part~\ref{part:QCD-axions} develops the QCD axion in greater detail, covering its benchmark realisations and low-energy properties, cosmology, astrophysical constraints and experimental searches, together with the axion quality problem and possible ultraviolet resolutions. Finally, in Part~\ref{part:reassessment}, we revisit the theoretical foundations of the strong $CP$ problem itself. We review recent proposals that question the conventional interpretation of $\bar\theta$, as well as the ensuing debate, and summarise the present status of these discussions.

\section{Notation}
\label{sec:notation}

When discussing the vacuum structure of Yang--Mills theory or QCD, one often encounters instantons, which are classical solutions of the Euclidean field equations.\footnote{Instantons typically describe quantum tunnelling. For a pedagogical introduction, see Ref.~\cite{Coleman:1985rnk}.} It is therefore useful to first establish our conventions for Euclidean quantities. Following Ref.~\cite{Vainshtein:1981wh}, we place a hat on Euclidean quantities whenever necessary to distinguish them from their Minkowskian counterparts. Lorentz indices $\mu,\nu,\cdots$ take values $0,\cdots,3$ in Minkowski spacetime and $1,\cdots,4$ in Euclidean spacetime. In Euclidean spacetime, there is no distinction between upper and lower indices, and we use lower indices throughout, except for $\gamma^5$.

\subsection{QCD in Minkowski spacetime}

The action in Minkowski spacetime reads (without the topological term for the moment)
\begin{align}
    S_{\rm M} = \int\d^4x\, \left[-\frac{1}{4}G^a_{\mu\nu}G^{a\mu\nu }+\bar{\psi}_i\left(\i\slashed{D}-M_{ij}\right)\psi_j\right],
    \label{eq:SM1}
\end{align}
where $\slashed{D}=\gamma^\mu D_\mu$, $G^a_{\mu\nu}=\partial_\mu A^a_\nu-\partial_\nu A_\mu^a+g_s f^{abc}A^b_\mu A^c_\nu$. The structure constants $f^{abc}$ satisfy $[T^a,T^b]=\i f^{abc}T^c$ with
$T^a$ being the generators of the gauge group. QCD has gauge group $SU(3)_c$. The topological structure relevant here is already present in an $SU(2)$ subgroup. For simplicity, we therefore restrict the explicit instanton calculations to $SU(2)$, for which $f^{abc}=\varepsilon^{abc}$. The Dirac matrices $\gamma^\mu$ satisfy $\{\gamma^\mu,\gamma^\nu\}=2\eta^{\mu\nu}$. The covariant derivative takes the form
\begin{align}
    D_\mu\psi_i = \left(\partial_\mu-\i g_s A_\mu^a T^a\right)\psi_i
\end{align}
if $\psi_i$ lives in the fundamental representation of the gauge group, and 
\begin{align}
D_\mu\psi_i=\partial_\mu\psi_i-\i g_s A_\mu^a[T^a,\psi_i]
\end{align}
if $\psi_i$ lives in the adjoint representation. The subscript $i$ on the fermions is the flavour index and throughout the paper, we suppress the colour indices on the quark fields.  In Eq.~\eqref{eq:SM1} we have not included the topological term yet. We will see that this term will emerge from the nontrivial vacuum structure of non-Abelian gauge fields.

Since chiral fermions lie at the centre of the discussion of the Adler-Bell-Jackiw (ABJ) anomaly~\cite{Bell:1969ts,Adler:1969gk} and the strong $CP$ problem, it is convenient to employ the Weyl basis for the Dirac matrices:
\begin{align}
    \gamma^\mu
    = 
    \begin{pmatrix}
    0 & \sigma^\mu \\
    \bar{\sigma}^\mu & 0
    \end{pmatrix},
    \qquad 
    \gamma^5=\i\gamma^0\gamma^1\gamma^2\gamma^3
    =\begin{pmatrix}
    -\mathbb{1}_2 & 0\\
    0 & \mathbb{1}_2
    \end{pmatrix},
\end{align}
where $\sigma^\mu=\left(\mathbb{1}_2,\vec{\sigma}\right)$ and $\bar{\sigma}^\mu=(\mathbb{1}_2,-\vec{\sigma})$ with $\mathbb{1}_2$ being the unit $2\times 2$ matrix and $\vec{\sigma}^i$ the Pauli matrices. The gamma matrices have the properties $\gamma^{0\dagger}=\gamma^0$, $\gamma^{i\dagger}=-\gamma^i$, $\{\gamma^5,\gamma^\mu\}=0$. The left-handed and right-handed fermions are $\psi_{\rm L/R}=\frac{1}{2}(1\mp \gamma^5)\psi$, respectively. They satisfy $\gamma^5\psi_{\rm L/R}=\mp \psi_{\rm L/R}$.

\subsection{QCD in Euclidean spacetime}

When going from Minkowski spacetime to Euclidean spacetime, one performs the Wick rotation $t\rightarrow-\i\tau$. For theories where the metric structure does not play an essential role, this procedure is often sufficient. In Euclidean QCD, however, it is convenient to further redefine the vector fields and gamma matrices in order to obtain a manifestly Euclidean formulation with a positive-definite metric.

For the vector field, we define $\hat{A}_{\mu}$ such that
\begin{align}
    \eta^{\mu\nu}A_{\mu}A_\nu=-\delta^{\mu\nu}\hat{A}_{\mu}\hat{A}_{\nu}.
\end{align}
Then we have~\cite{Vainshtein:1981wh} 
\begin{align}
    \hat{A}_i = A_i,
    \qquad 
    \hat{A}_4 = -\i A_0.
\end{align}
Correspondingly, the covariant derivative in Euclidean space takes the form
\begin{align}
    \hat{D}_\mu=\partial_\mu-\i g_s \hat{A}_\mu^a T^a,
\end{align}
such that from Minkowski to Euclidean space, we have $D_0\rightarrow \i\hat{D}_4$ and $D_i\rightarrow \hat{D}_i$. For the field strength, the substitution is $G^a_{ij}\rightarrow \hat{G}^a_{ij}$ and $G^a_{0i}\rightarrow \i\hat{G}^a_{4i}$. With these redefined covariant tensors, one can obtain the contravariant tensors by raising the indices via the Euclidean metric tensor $\delta^{\mu\nu}$.

For the gamma matrices, we have $\hat{\gamma}_4=\gamma_0$ and $\hat{\gamma}_i=\i\gamma_i$. These matrices satisfy the properties $\{\hat{\gamma}_\mu,\hat{\gamma}_\nu\}=2\delta_{\mu\nu}$, $\hat{\gamma}_{\mu}^{\dagger}=\hat{\gamma}_\mu$. Note that $\gamma^5=\i\gamma^0\gamma^1\gamma^2\gamma^3=-\hat{\gamma}_1\hat{\gamma}_2\hat{\gamma}_3\hat{\gamma}_4$. Finally, the fermions in the Euclidean space are defined as the same as their Minkowskian counterparts
\begin{align}
    \hat{\psi} = \psi,
    \qquad 
    \hat{\bar{\psi}} = \bar{\psi}.
\end{align}
Note that while $\bar{\psi}$ transforms as $\psi^\dagger\gamma^0$ in Minkowski space, $\hat{\bar{\psi}}$ transforms as $\hat{\psi}^\dagger$ in Euclidean space.  To see it, recall that under Lorentz transformations we have 
\begin{subequations}
\begin{align}
    \psi'(x') &= \e^{-\frac{\i}{4}\omega^{\mu\nu}\sigma_{\mu\nu}}\psi(x),
    \label{eq:psi-transf}\\
    \bar{\psi}'(x') &= \psi'^\dagger(x')\gamma^0=\bar{\psi}(x)\e^{\frac{\i}{4}\omega^{\mu\nu}\sigma_{\mu\nu}},
    \label{eq:barpsi-transf}
\end{align}
\label{eq:transf}
\end{subequations}
where $\sigma_{\mu\nu}=\frac{\i}{2} [\gamma_\mu,\gamma_\nu]$ and we have used $\gamma^0\left(\sigma_{\mu\nu}\right)^\dagger\gamma^0=\sigma_{\mu\nu}$. When going to Euclidean space, we have $\omega^{ij}\rightarrow \hat{\omega}_{ij}$, $\omega^{0i}\rightarrow -\i\hat{\omega}_{4i}$ (as the field strength tensor does), $\sigma_{ij}\rightarrow -\hat{\sigma}_{ij}$ and $\sigma_{0i}\rightarrow -\i\hat{\sigma}_{4i}$. From Eq.~\eqref{eq:transf}, one obtains
\begin{subequations}
\begin{align}
    \hat{\psi}'(x') &= \e^{\frac{\i}{4}\hat{\omega}_{\mu\nu}\hat{\sigma}_{\mu\nu}}\hat{\psi}(x),
    \label{eq:psi-transf-Euc}\\
    \hat{\bar{\psi}}'(x') &= \hat{\bar{\psi}}(x)\e^{-\frac{\i}{4}\hat{\omega}_{\mu\nu}\hat{\sigma}_{\mu\nu}}.
    \label{eq:barpsi-transf-Euc}
\end{align}
\end{subequations}
Taking the Hermitian conjugate of Eq.~\eqref{eq:psi-transf-Euc} and using $\left(\hat{\sigma}_{\mu\nu}\right)^\dagger=\hat{\sigma}_{\mu\nu}$, one obtains the same transformation as Eq.~\eqref{eq:barpsi-transf-Euc}.

The Euclidean action $S_{\rm E}$ is obtained from the Minkowskian action $S_{\rm M}$ through a Wick rotation $t\rightarrow -\i\tau$ together with $\i S_{\rm M}\rightarrow -S_{\rm E}$. In terms of the redefined fields, we have 
\begin{align}
    S_{\rm E} = \int\d^4 x\left[\frac{1}{4}\hat{G}^a_{\mu\nu}\hat{G}^a_{\mu\nu} + \hat{\bar{\psi}}_i\left(\hat{\slashed{D}} + M_{ij}\right)\hat{\psi}_j\right],
\end{align}
where $\hat{\slashed{D}}\equiv \hat{\gamma}_\mu \hat{D}_\mu$. The relation between $\hat{\slashed{D}}$ and $\slashed{D}$ is $\hat{\slashed{D}}\leftrightarrow -\i\slashed{D}$.

\part{The origin of the strong \texorpdfstring{$CP$}{CP} problem}
\label{part:origin}

\section{Chiral anomaly and QCD vacuum structure}

\subsection{The \texorpdfstring{$\eta'$}{eta-prime} puzzle and chiral anomaly}

We begin our discussion with the chiral symmetry breaking of QCD. Consider the idealised limit $M\to 0$
for $N_f$ quark flavours. The QCD Lagrangian then has the classical global symmetry
\begin{equation}
    U(N_f)_L \times U(N_f)_R \simeq SU(N_f)_L \times SU(N_f)_R \times U(1)_V \times U(1)_A,
\end{equation}
up to discrete identifications. The vector $U(1)_V$ factor is baryon number. The axial symmetries are spontaneously broken by a
nonvanishing quark condensate $\langle \bar{\psi}\psi\rangle\neq 0$, leading to
\begin{equation}
    SU(N_f)_L \times SU(N_f)_R \times U(1)_V \times U(1)_A \rightarrow SU(N_f)_V \times U(1)_V.
    \label{eq:chiral-symmetry-breaking-without-anomaly}
\end{equation}
The number of Goldstone bosons equals the number of the generators of the broken symmetries. Since the quarks have masses, the chiral symmetries are only approximate, and one then has pseudo-Goldstone bosons instead of strictly massless Goldstone bosons. The breaking of the approximate non-Abelian axial symmetries $SU(3)_A$ accounts for the pseudoscalar octet $(\pi,K,\eta)$ as pseudo-Nambu--Goldstone bosons.

If the singlet axial symmetry $U(1)_A$ were also an exact global symmetry of the quantum theory in the limit of $M\to 0$, one would expect a ninth pseudoscalar Nambu--Goldstone boson. This expectation is not realised in the hadron spectrum: the $\eta'$ is much heavier than the $(\pi,K,\eta)$ octet states. This mismatch is the traditional $U(1)_A$ or $\eta'$ puzzle~\cite{Weinberg:1975ui}. The mass of $\eta'$ is incompatible with treating the singlet axial symmetry as an ordinary approximate chiral symmetry broken only by light quark masses.

In terms of left- and right-handed fields, a singlet axial rotation acts as
\begin{equation}
    \psi_L \rightarrow \mathrm{e}^{-\mathrm{i}\beta}\psi_L,
    \qquad
    \psi_R \rightarrow \mathrm{e}^{\mathrm{i}\beta}\psi_R,
\end{equation}
whereas a non-singlet transformation inserts a traceless flavour generator. Classically, in the massless theory, both transformations appear to give conserved currents. Quantum mechanically, however, only the non-singlet axial currents remain conserved in the chiral limit. The singlet current is anomalous. This is the mechanism that removes the unwanted ninth Goldstone boson from the low-energy spectrum and, at the same time, prepares the ground for the physical meaning of the QCD vacuum angle~\cite{tHooft:1976rip,tHooft:1986ooh,Witten:1979vv,Veneziano:1979ec}.

The flavour-singlet axial current reads $J^{5\mu}=\bar{\psi}\gamma^\mu\gamma^5\psi$, with the sum over the $N_f$ light flavours understood. At the classical level one obtains
\begin{equation}
    \partial_\mu J^{5\mu} = 2\mathrm{i}\,\bar{\psi}M\gamma^5\psi,
\end{equation}
so the current is conserved when $M=0$. This conclusion is modified by quantisation. With the conventions of Eq.~\eqref{eq:QCD-Lagrangian}, the anomalous Ward identity is\footnote{Note that Ref.~\cite{Peskin:1995ev} uses a different convention $\epsilon^{0123}=+1$, which leads to an extra minus sign in the second term in Eq.~\eqref{eq:singlet-anomaly}.}
\begin{align}
    \partial_\mu J_5^\mu
    &= 2\mathrm{i}\,\bar{\psi}M\gamma^5\psi
    +\frac{N_f g_s^2}{16\pi^2}\,
    G^a_{\mu\nu}\widetilde{G}^{a\mu\nu},
    \label{eq:singlet-anomaly}
\end{align}
where $\widetilde{G}^{a\mu\nu} = \frac{1}{2}\epsilon^{\mu\nu\rho\sigma}G^a_{\rho\sigma}$. We adopt the convention 
\begin{align}
    \epsilon^{0123} = -1.
\end{align} The coefficient of the last term is fixed by the Adler--Bell--Jackiw anomaly and is not renormalised by strong dynamics~\cite{Adler:1969gk,Bell:1969ts}. In a path-integral derivation, the same term arises from the non-invariance of the fermion measure under a chiral change of variables~\cite{Fujikawa:1979ay,Fujikawa:1980eg}. For a general axial current $\bar{\psi}\gamma^\mu\gamma^5 T\psi$, the gluonic anomaly is proportional to $\tr T$; hence it vanishes for traceless generators of $SU(N_f)_A$ but survives for the singlet generator.

Equation~\eqref{eq:singlet-anomaly} is often described as an explicit breaking of $U(1)_A$, and provides the basic explanation for the absence of the ninth Goldstone boson in the hadron spectrum. However, it does not by itself provide a microscopic picture of how the axial charge is violated. This is the role played by instantons in gauge theory~\cite{tHooft:1976snw,tHooft:1976rip}, which we discuss next.

\subsection{Topological sectors and instantons}

The essential additional ingredient is the topology of non-Abelian gauge fields. This structure is most transparent in the Euclidean path integral,
\begin{align}
    Z[\eta,\bar \eta] = \int{\cal D}\hat{\bar\psi}\,{\cal D}\hat{\psi}\,{\cal D}\hat{A}\, {\rm e}^{-\int_\Omega {\rm d}^4x\,\left({\cal L}_{\rm E}-\bar \eta\hat{\psi}-\hat{\bar\psi} \eta\right)}.
    \label{partition:function}
\end{align}
Here $\Omega$ denotes the spacetime volume, which is infinite for $\mathbb{R}^4$. We include external fermionic sources $\eta(x)$ and $\bar\eta(x)$ for later use in deriving quark correlation functions.

There are two complementary ways to understand the physical meaning of Euclidean field theory. First, the Euclidean path integral can be regarded as a formulation of transition amplitudes in Euclidean time, which can be used to extract quantities such as energy eigenvalues. Alternatively, it can be viewed as an analytic continuation of the Minkowski theory, where relations derived in Euclidean spacetime are analytically continued back to Minkowski spacetime. In this review, the second viewpoint is mostly used~\cite{tHooft:1976rip}. In both viewpoints, instantons are classical solutions of the Euclidean equations of motion and provide the starting point for the semiclassical expansion of the Euclidean path integral.\footnote{Typically, after analytic continuation back to Minkowski spacetime, instantons do not correspond to real classical solutions of the equations of motion. Nevertheless, the resulting complex saddles can still play an important role in the Minkowskian path integral, as understood within Picard--Lefschetz theory~\cite{Cherman:2014sba,Ai:2019fri}.}

In the path integral, we are concerned with finite-action gauge configurations,
\begin{equation}
    S_{\rm E}^{\rm YM} = \frac{1}{4}\int_\Omega \d^4x\,\hat{G}^a_{\mu\nu}\hat{G}^a_{\mu\nu} < \infty.
    \label{eq:finite-action}
\end{equation}
For infinite spacetime volume, $\Omega\rightarrow\infty$, the condition~\eqref{eq:finite-action} implies that the field strength vanishes sufficiently rapidly at infinity. Hence, at large Euclidean radius, the gauge potential approaches a pure gauge,
\begin{equation}
    \hat{A}_\mu(x) \rightarrow \frac{\i}{g_s}\,\omega(x)\,\partial_\mu\omega^{-1}(x)
    \quad
    \text{at } \partial\mathbb{R}^4\simeq S^3.
    \label{eq:pure-gauge}
\end{equation}
The asymptotic gauge function $\omega$ defines a map $S^3\rightarrow SU(N_c)$. For $N_c\geq2$, such maps are classified by the third homotopy gropu $\pi_3(SU(N_c))=\mathbb{Z}$.
The corresponding integer labels disconnected classes of gauge-field configurations. In the continuum, it is represented by the topological charge (or winding number)
\begin{equation}
    Q = \int \d^4x\,q(x),
    \qquad 
    q(x)=\frac{g_s^2}{32\pi^2} \hat{G}^a_{\mu\nu} \widetilde{\hat{G}}^{a\mu\nu}.
    \label{eq:topological-charge}
\end{equation}
With the boundary condition~\eqref{eq:pure-gauge}, $Q$ is an integer. This gives rise to the {\it topological quantisation} of gauge-field configurations. For the later discussion, we emphasise that this topological quantisation in the spacetime $\mathbb R^4$ relies on the limit $\Omega\rightarrow\infty$.

The topological charge labels the sectors but does not determine the configurations within them. When computing observables through the path integral, it is useful to perform the semiclassical expansion. In the latter, the minimum-action configurations in sectors with nonzero $Q$ are (anti-)self-dual instanton solutions~\cite{tHooft:1976rip,tHooft:1976snw,Coleman:1985rnk,Diakonov:2002fq}. For $Q=\pm1$, these are the Belavin--Polyakov--Schwartz--Tyupkin (BPST) (anti-)instantons~\cite{Belavin:1975fg}, with classical Yang--Mills action
\begin{align}
    S_{\rm BPST} = \frac{8\pi^2}{g_s^2}.
\end{align}
Explicitly, the BPST instanton in the regular gauge reads
\begin{align}
    \hat A^a_\mu = \frac{2\eta_{a\mu\nu}}{g_s}\frac{(x-x_0)_\nu}{(x-x_0)^2+\rho^2},
    \label{eq:BPSTinst}
\end{align}
where $x_0$ and $\rho$ denote its centre and size, respectively, and are called collective coordinates. Here $\eta_{a\mu\nu}$ are the 't Hooft symbols~\cite{tHooft:1976snw},
\begin{align}
    \eta_{a\mu\nu} = 
    \begin{cases} 
    \varepsilon_{a\mu\nu}, & \mu,\nu=1,2,3,\\
    -\delta_{a\nu}, & \mu=4,\\
    \delta_{a\mu}, & \nu=4,\\
    0, & \mu=\nu=4.
    \end{cases}
\end{align}
The anti-instanton is obtained by replacing $\eta_{a\mu\nu}$ with the corresponding anti-self-dual symbol $\bar{\eta}_{a\mu\nu}$.

For $|Q|>1$, the minimum-action (anti-)self-dual configurations are described by the Atiyah--Drinfeld--Hitchin--Manin (ADHM) construction~\cite{Atiyah:1978ri}. A configuration with $Q>0$ consists of a multi-instanton solution, while one with $Q<0$ consists of the corresponding multi-anti-instanton solution. More generally, in the dilute instanton gas approximation (DIGA), a sector of fixed topological charge $Q$ may contain both instantons and anti-instantons. Denoting their numbers by $n$ and $\bar n$, respectively, one has $n-\bar n=Q$. The (anti-)self-dual minimum-action configurations correspond to $\bar n=0$ for $Q>0$ and $n=0$ for $Q<0$, whereas generic dilute instanton configurations may contain additional instanton--anti-instanton pairs. Each constituent carries collective coordinates describing its position, size, and gauge orientation.

\subsection{Chiral charge violation in a fixed topological sector}
\label{sec:chiral-charge-violation}

The connection between gauge-field topology and chiral symmetry violation becomes sharp once fermions are included. In a fixed Euclidean gauge background, the change of the chiral charge is encoded in the eigenequation of the massless Dirac operator,
\begin{align}
\label{eq:eigenequation-Dirac}
\hat{\slashed{D}}\psi_\lambda=\lambda\,\psi_\lambda.
\end{align}
Note that the Euclidean massless Dirac operator is anti-Hermitian and so its eigenfunctions can be readily assumed to be orthonormal and its eigenvalues are purely imaginary $\lambda^*=-\lambda$.

Since $\{\hat{\slashed{D}},\gamma^5\}=0$, if $\psi_\lambda$ is an eigenfunction of the massless Dirac operator with eigenvalue $\lambda$, then $\gamma^5\psi_\lambda$ is also an eigenfunction, with eigenvalue $-\lambda$:
\begin{align}
    \hat{\slashed{D}}(\gamma^5\psi_{\lambda}) = -\lambda(\gamma^5\psi_{\lambda}).
\end{align}
Thus, all nonzero eigenvalues occur in pairs of opposite sign. For any zero mode $\psi_0$, we may decompose it as
\begin{align}
    \psi_0=P_L\psi_0+P_R\psi_0 \equiv \psi_{0L}+\psi_{0R}.
\end{align}
Provided that $\psi_{0L}$ or $\psi_{0R}$ is nonvanishing, it is itself a zero mode. Therefore, the zero modes can always be chosen to be eigenfunctions of $\gamma^5$,
\begin{align}
    \gamma^5\psi_0 = \pm\psi_0.
\end{align}
We denote the numbers of right- and left-handed zero modes by $n_R$ and $n_L$, respectively. The index theorem~\cite{Atiyah:1963zz} states that
\begin{align}
    n_{R,i}-n_{L,i} = 2T(R)\,Q,
    \qquad 
    \text{for each flavour $i$},
    \label{eq:index-theorem}
\end{align}
where $T(R)$ is the Dynkin index of the gauge-group representation $R$. For the fundamental representation, $T(R)=1/2$. In particular, in a BPST instanton background there is one right-handed zero mode and no left-handed zero mode.\footnote{Note that some references, such as Ref.~\cite{Coleman:1985rnk}, adopt a definition of $\gamma^5$ that differs from ours by an overall minus sign. Consequently, their chirality assignments are opposite to those used here.
} Conversely, an anti-instanton background supports one left-handed zero mode and no right-handed zero mode.

Below, we derive the Euclidean anomalous Ward identity for the flavour-singlet axial current. For simplicity, let us consider $N_f=1$ and assume a real mass $m$. We compute
\begin{align}
    \partial_\mu \hat{J}_\mu^5 
    = 
    \partial_\mu \langle \hat{\bar\psi}\hat{\gamma}_\mu \gamma^5 \hat\psi\rangle_Q
    = 
    -\partial_\mu \langle \hat\psi\hat{\bar\psi}\hat{\gamma}_\mu \gamma^5\rangle_Q.
    \label{eq:partial-mu-J5-1}
\end{align}
The fermionic Green's function $S_Q\equiv \langle \hat\psi(x) \hat{\bar\psi}(x')\rangle_Q$ satisfies
\begin{align}
   \left(\hat{\slashed{D}}_x+m\right) S_Q(x,y) = \delta^4(x-y).
\end{align}
Inserting the spectral sum for the Green's function,
\begin{align}
    S_Q(x,x')= \SumInt_\lambda\frac{\psi_\lambda(x)\psi_\lambda^\dagger(x') }{\lambda+m},
\end{align}
into Eq.~\eqref{eq:partial-mu-J5-1} and using the eigenequation~\eqref{eq:eigenequation-Dirac}, one obtains
\begin{align}
    \partial_\mu \hat{J}_\mu^5 = 2m \langle \hat{\bar\psi}\gamma^5 \hat\psi\rangle_Q + 2 \SumInt_\lambda \psi_\lambda^\dagger \gamma^5 \psi_\lambda,
\end{align}
where we have used $\lambda^*=-\lambda$. The first term on the right-hand side originates from the nonvanishing mass. Since for non-zero modes, $\gamma^5\psi_n$ is orthogonal to $\psi_n$, the second term only receives contribution from the zero modes when integrating the equation over spacetime. Therefore, we have 
\begin{align}
    \int \d^4 x\,\partial_\mu \hat{J}^5_\mu(x) = \int\d^4 x\, 2m\langle \hat{\bar{\psi}}(x) \gamma^5 \hat{\psi}(x)\rangle_Q + 2(n_R-n_L)=\int\d^4 x\, 2m\langle \hat{\bar{\psi}}(x) \gamma^5 \hat{\psi}(x)\rangle_Q + 2\int \d^4x\, q(x).
    \label{eq:integrated-anomaly}
\end{align}
Adding the factor of $N_f$, one may infer the anomalous Ward identity in Euclidean spacetime,
\begin{align}
    \partial_\mu\hat{J}^5_\mu = 2\bar\psi M\gamma^5 \psi +\frac{N_f g_s^2}{16\pi^2} \hat{G}^a_{\mu\nu}\widetilde{\hat{G}}^{a\mu\nu}.
    \label{eq:anomalous-Ward-Euclidean}
\end{align}
Analytically continuing this equation to Minkowski spacetime, one recovers Eq.~\eqref{eq:singlet-anomaly}. For general complex mass, $m=|m|\e^{\i\alpha\gamma^5}$, the derivation is more involved, see Ref.~\cite{Ai:2020ptm}. There is actually a gap between Eqs.~\eqref{eq:integrated-anomaly} and~\eqref{eq:anomalous-Ward-Euclidean}.  A rigorous derivation of the anomalous Ward identity is reviewed in Appendix~\ref{app:Fujikawa-anomaly}.

\subsection{Large gauge transformations, the pre-vacuum, and the first caveat}
\label{sec:pre-vacuum}

We have seen how the chiral charge is anomalously violated in a Euclidean gauge-field background with definite topological charge. Among the inequivalent configurations in a sector with charge $Q$, instantons provide the semiclassical saddle points and thus a simple picture of the associated chiral-charge violation. Instantons are also interpreted as describing tunnelling processes, which naturally raises the question of the corresponding initial and final states. To address this, we consider gauge-field configurations on a spatial hypersurface at fixed time, namely the three-dimensional space $\mathbb{R}^3$.

For definiteness, we work in the temporal gauge $A_0=0$. To identify the vacuum states, we first seek the classically minimal-energy, i.e. zero-energy, configurations. These are pure gauges,
\begin{equation}
    A_i(\vecx) = \frac{\i}{g_s}\,\omega(\vecx)\partial_i\omega^{-1}(\vecx),
    \label{eq:pure-gauge-vacua}
\end{equation}
where $\omega(\vecx)$ is a gauge transformation matrix. If we require $\omega(\vecx)$ to approach a direction-independent limit at spatial infinity~\cite{Callan:1976je},
\begin{align}
    \omega(\vecx)\to {\rm const} \quad \text{for} \quad |\vecx|\to\infty,
    \label{eq:constant-assumption}
\end{align}
then $\mathbb{R}^3$ can be compactified to $S^3$, and $\omega(\vecx)$ defines a map $S^3\to SU(N_c)$. The zero-energy gauge configurations consequently fall into different topological classes characterised by an integer, the Chern--Simons (CS) number,
\begin{equation}
    n_{\rm CS}[A] = \frac{g_s^2}{4\pi^2} \epsilon_{ijk}\int \d^3x\,\tr\!\left[\frac{1}{2}A_i\partial_j A_k - \frac{\i g_s}{3}A_iA_jA_k\right],
    \label{eq:winding-number-large-gauge}
\end{equation}
where $A_i=A_i^aT^a$ and $\tr(T^aT^b)=\delta^{ab}/2$. The pure-gauge configurations in Eq.~\eqref{eq:pure-gauge-vacua} have $n_{\rm CS}\in\mathbb{Z}$. Note that for a general field configuration that is not pure gauge, the CS number is not necessarily an integer.

It is important to note that Eq.~\eqref{eq:constant-assumption} imposes a condition not only on the admissible pure-gauge configurations but also on the admissible gauge transformations. Because of this condition, the group of admissible gauge transformations is disconnected. In the conventional construction, two configurations related by a gauge transformation continuously connected to the identity are regarded as gauge equivalent. Such transformations are called \emph{small gauge transformations} and form the identity component $\mathcal G_0$. Pure-gauge configurations with the same CS number can be related by small gauge transformations. We may therefore choose a representative configuration in each topological class and denote the corresponding state by $|n_{\rm CS}\rangle$, where $n_{\rm CS}$ labels its CS number. These states are called \emph{pre-vacua}.

Gauge transformations that are not connected to the identity are called \emph{large gauge transformations}. Denoting the full group of admissible gauge transformations by $\mathcal G$, one has $\mathcal G/\mathcal G_0\simeq\mathbb Z$, where the integer $n\in\mathbb Z$ the homotopy classes. A large gauge transformation $\mathcal U_m$ belonging to the component labelled by $m$ maps
\begin{equation}
    \mathcal U_m|n_{\rm CS}\rangle
    =
    |n_{\rm CS}+m\rangle.
\end{equation}
The Euclidean instanton is then interpreted semiclassically as a tunnelling trajectory between adjacent pre-vacua. More generally, a field history interpolating between $|n^i_{\rm CS}\rangle$ and $|n^f_{\rm CS}\rangle$ carries topological charge\footnote{We distinguish the winding number, defined in the four-dimensional Euclidean spacetime, from the CS number, defined on a three-dimensional spatial hypersurface. Unless specified otherwise, the topological charge refers to the winding number throughout this review.}
\begin{equation}
    Q = n^f_{\rm CS}-n_{\rm CS}^i.
\end{equation}

\paragraph{The gauge-redundancy caveat.}
There is, however, a conceptual caveat that should be kept explicit. As noted in Refs.~\cite{Jackiw:1979ur,itzykson2006quantum}, the condition~\eqref{eq:constant-assumption} does not have a first-principles derivation and should be regarded as an assumption. For example, in Ref.~\cite{Jackiw:1979ur}, Jackiw wrote [above Eq.~(36b)]
\begin{quote}
``{\it We shall make a very important hypothesis concerning the physically admissible finite transformations. While some plausible arguments can be given in support of this hypothesis (see below) in the end we must recognize it as an assumption, without which subsequent development cannot be made.}''
\end{quote}
Gauge transformations connected to the identity implement the Gauss-law constraint and must act trivially on physical states. Reference~\cite{Ai:2024vfa} argues that, if the constraint~\eqref{eq:constant-assumption} is removed, the space of gauge transformations becomes connected and there is only one pre-vacuum;\footnote{Even if there is only one pre-vacuum, there still can be different winding numbers $Q$.} within this construction, gauge invariance then enforces CP symmetry in Yang--Mills theory. For the purposes of the present part of the review, we first follow the conventional treatment. Later, in Part~\ref{part:reassessment}, we return to the question of whether relaxing the constraint~\eqref{eq:constant-assumption} may lead to different physical conclusions.

\subsection{From the pre-vacuum to the \texorpdfstring{$\theta$}{theta}-vacuum}

The pre-vacuum state $|n_{\rm CS}\rangle$ with a definite CS number cannot represent a physical vacuum state for two reasons. First, under a large gauge transformation it is mapped to a different state, rather than merely acquiring an overall phase. Second, a state with fixed $n_{\rm CS}$ does not satisfy the principle of cluster decomposition, which requires that sufficiently separated experiments become statistically independent in a vacuum state~\cite{streater2000pct,Weinberg:1995mt}. These considerations motivate the introduction of the $\theta$-vacuum.

Since the Hamiltonian commutes with large gauge transformations, it can be diagonalised simultaneously with them. In the basis of pre-vacua, the conventional eigenstates are
\begin{equation}
    |\theta\rangle = \sum_{n_{\rm CS}\in\mathbb{Z}}\e^{-\i n_{\rm CS}\theta}|n_{\rm CS}\rangle,
    \qquad 
    0\leq\theta<2\pi,
    \label{eq:theta-vacuum}
\end{equation}
up to an overall normalisation. With the convention above,
\begin{equation}
    \mathcal{U}_1|\theta\rangle = \e^{\i\theta}|\theta\rangle .
\end{equation}
$|\theta\rangle$ is called the $\theta$-vacuum. Different values of $\theta$ label superselection sectors in the idealised infinite-volume theory: local gauge-invariant operators cannot change the eigenvalue of $\mathcal{U}_1$. In this sense, the classical Yang--Mills Lagrangian does not select a particular value of $\theta$; rather, $\theta$ specifies an additional parameter of the quantum theory.

The equivalence between Eq.~\eqref{eq:theta-vacuum} and the $\theta$ term in the Lagrangian can be understood by inserting complete sets of pre-vacua into the Euclidean transition amplitude. If $\langle n^f_{\rm CS}|\e^{-HT}|n^i_{\rm CS}\rangle$ receives contributions from field histories with $Q=n^f_{\rm CS}-n^i_{\rm CS}$, then
\begin{align}
    \lim_{T\to\infty}\langle\theta|\e^{-HT}|\theta\rangle 
    &=\N \sum_{Q\in\mathbb{Z}}\e^{\i\theta Q}Z_Q 
    = \int \D \hat{\bar\psi}\D\hat\psi \D \hat{A} \,\e^{-S_{\rm E}[\hat{A},\hat{\bar\psi},\hat\psi]+\i\theta Q[\hat{A}]},
\end{align}
where $\N$ is a common normalisation factor and can be absorbed into the path integral measure. This corresponds to adding a topological term to the Lagrangian,
\begin{align}
    S_{\rm E} = \int\d^4x\left[\frac{1}{4}\hat{G}^a_{\mu\nu}\hat{G}^a_{\mu\nu} 
    + \hat{\bar{\psi}}_i\left(\hat{\gamma}_\mu\hat{D}_\mu+M_{ij}\right)\hat{\psi}_j 
    -\frac{\i\theta g_s^2}{32\pi^2} \hat{G}^a_{\mu\nu}\widetilde{\hat{G}}^a_{\mu\nu}\right].
    \label{eq:Euclidean-action-full}
\end{align}
After analytic continuation back to Minkowski space, we have
\begin{align}
    S_{\rm M} = \int\d^4x\, \left[-\frac{1}{4}G^a_{\mu\nu}G^{a\mu\nu } + \bar{\psi}_i\left(\i\slashed{D}-M_{ij}\right)\psi_j +\frac{\theta g_s^2}{32\pi^2}\, G^a_{\mu\nu}\widetilde{G}^{a\mu\nu}\right],
    \label{eq:SM2}
\end{align}
Thus, the canonical $\theta$-vacuum and the Lagrangian $\theta G\widetilde{G}$ term are two equivalent ways of encoding the same $\theta$ dependence in the conventional formulation of Yang--Mills theory.\footnote{In the canonical quantisation of Yang--Mills theory, additional angular parameters may be introduced~\cite{Ai:2024vfa}; see Sec.~\ref{sec:reassessment-physical-Hilbert-space}. In that formulation, an invariant combination of these angular parameters corresponds to the parameter appearing in the path integral.} Since there are different conventions used for the topological term in the literature, we summarise them in Table~\ref{tab:different_conventions}.

\begin{table}[ht]
    \centering
    \begin{tabular}{|c||c|c|c}
        \hline
        \textit{} & $\epsilon^{0123}=-1$ & $\epsilon^{0123}=+1$
        \\
        \hline\hline
         $|\theta\rangle=\sum_{n_{\rm CS}}  \e^{-\i n_{\rm CS}\theta} |n_{\rm CS}\rangle$ &  $\begin{gathered}
            \L^\theta_{\rm E}=-\frac{\i\theta g_s^2}{32\pi^2}\hat{G}^a_{\mu\nu}\widetilde{\hat{G}}^a_{\mu\nu}, \\ \L_{\rm M}= +\frac{\theta g_s^2}{32\pi^2}\, G^a_{\mu\nu}\widetilde{G}^{a\mu\nu}\\ \text{(our convention)}
         \end{gathered}$ & $\begin{gathered}
        \L^\theta_{\rm E}=-\frac{\i\theta g_s^2}{32\pi^2}\hat{G}^a_{\mu\nu}\widetilde{\hat{G}}^a_{\mu\nu}, \\ \L_{\rm M}= -\frac{\theta g_s^2}{32\pi^2}\, G^a_{\mu\nu}\widetilde{G}^{a\mu\nu} 
         \end{gathered}$ \\
         \hline
         $|\theta\rangle=\sum_{n_{\rm CS}} \e^{+\i n_{\rm CS}\theta} |n_{\rm CS}\rangle$  & $\begin{gathered}
            \L^\theta_{\rm E}=+\frac{\i\theta g_s^2}{32\pi^2}\hat{G}^a_{\mu\nu}\widetilde{\hat{G}}^a_{\mu\nu}, \\ \L_{\rm M}= -\frac{\theta g_s^2}{32\pi^2}\, G^a_{\mu\nu}\widetilde{G}^{a\mu\nu}
         \end{gathered}$ & $\begin{gathered}
            \L^\theta_{\rm E}=+\frac{\i\theta g_s^2}{32\pi^2}\hat{G}^a_{\mu\nu}\widetilde{\hat{G}}^a_{\mu\nu}, \\ \L_{\rm M}= +\frac{\theta g_s^2}{32\pi^2}\, G^a_{\mu\nu}\widetilde{G}^{a\mu\nu}
         \end{gathered}$ \\
         \hline
    \end{tabular}
    \caption{Different conventions used for the topological term. Here we have fixed $Q=\frac{g_s^2}{32\pi^2} \int\d^4 x\, 
\hat{G}^a_{\mu\nu}
\widetilde{\hat{G}}^{a\mu\nu} = n^f_{\rm CS}-n^i_{\rm CS} $ and $\epsilon_{1234}=+1$.}
\label{tab:different_conventions}
\end{table}

\section{The 't Hooft vertex and fermionic correlation functions}

\subsection{Instanton effects represented by an effective operator}
\label{sec:thooft_vertex}

We have seen that, in a fixed gauge-field background with nonzero topological charge, the anomalous Ward identity implies a violation of axial charge. In this section, we show how this effect can be represented by an effective fermionic operator.

First, let us consider the vacuum transition amplitude in a BPST instanton background ($n=1,\bar n=0$),
\begin{align}
    Z_{1,0} = \int\mathcal{D}\hat{\bar\psi} \mathcal{D}\hat\psi\mathcal{D}\hat A_{1,0} \,\e^{-\int_\Omega \d^4x\left[\hat{\bar\psi}\hat{\slashed{D}} \hat\psi+\frac{1}{4}\hat G_{\mu\nu}^a\hat G_{\mu\nu}^a\right]}.\label{eq:one_instanton_amplitude}
\end{align}
where the integration over $\hat A_{1,0}$ is restricted to the one-instanton sector, including fluctuations and moduli about the classical solution. For each flavour, the fermion fields may be expanded in a complete set of eigenfunctions of the massless Dirac operator,
\begin{equation}
    \hat\psi_i(x) = a_{i0}\psi_0(x)+\sum_{n\ne0}a_{in}\psi_n(x),\qquad 
    \hat{\bar\psi}_i(x) = \bar b_{i0}\psi_0^\dagger(x)+\sum_{n\ne0}\bar b_{in}\psi_n^\dagger(x), \label{eq:thooft_mode_expansion}
\end{equation}
where $\hat{\slashed{D}}_{\rm inst}\psi_n=\lambda_n\psi_n$ and $\lambda_0=0$. In the Euclidean path integral, $\hat\psi$ and $\hat{\bar\psi}$ are independent Grassmann variables, and hence $a_{in}$ and $\bar b_{in}$ are also independent. Correspondingly, the fermionic measure becomes
\begin{equation}
    \int\mathcal{D}\hat{\bar\psi}\mathcal{D}\hat\psi
    \rightarrow 
    \int\prod_{i=1}^{N_f}\left[\left(\prod_n \d\bar b_{in}\right) \left(\prod_n \d a_{in}\right)\right].\label{eq:thooft_fermion_measure}
\end{equation}
Substituting Eq.~\eqref{eq:thooft_mode_expansion} into the fermionic action gives
\begin{equation}
    S_{{\rm E}}^{\rm fermion} = \sum_{i=1}^{N_f}\sum_{n\ne0}\lambda_n\bar b_{in}a_{in}.
\label{eq:thooft_fermion_action}
\end{equation}
The zero-mode coefficients $a_{i0}$ and $\bar b_{i0}$ are absent from the exponent. Consequently, their Grassmann integrals vanish,
\begin{equation}
    \int\d\bar b_{i0}\d a_{i0}=0.
    \label{eq:thooft_zero_mode_integral}
\end{equation}
The vacuum-to-vacuum transition amplitude therefore vanishes in the massless limit. This result reflects the axial-charge selection rule associated with the instanton background~\cite{tHooft:1976snw,tHooft:1976rip}: for a single massless flavour, the transition changes the axial charge by two units, whereas the states $|n\rangle$ and $|n+1\rangle$ contain no fermionic excitations that could supply this charge.

A nonvanishing amplitude can instead be obtained by inserting fermion fields that saturate the Grassmann zero modes. With our convention, the zero mode in the $Q=1$ instanton background is right-handed, so that $a_{i0}\psi_0\subset\hat\psi_{iR}$ and $\bar b_{i0}\psi_0^\dagger\subset\hat{\bar\psi}_{iL}$. We must therefore insert one factor of $\hat{\bar\psi}_{iL}(x)\hat\psi_{iR}(y)$ for each flavour in order to saturate the zero-mode measure $\prod_i\d\bar b_{i0}\d a_{i0}$.

More generally, one may introduce a local scalar source $J_{ij}(x)$ in flavour space and consider
\begin{align}
    Z_{1,0}[J] = \int\mathcal{D}\hat{\bar\psi}\mathcal{D}\hat\psi\mathcal{D}\hat A_{1,0}\,
    \e^{-\int_\Omega \d^4x\left[\hat{\bar\psi}\hat{\slashed{D}}\hat\psi+\frac{1}{4}\hat G_{\mu\nu}^a\hat G_{\mu\nu}^a+\hat{\bar\psi}_i J_{ij}\hat\psi_j\right]}.\label{eq:one_instanton_amplitude_source}
\end{align}
The source is taken to be a gauge singlet. Integrating over the fermion fields gives the functional determinant
\begin{equation}
\det\left(\hat{\slashed{D}}_{\rm inst}+J\right).
\end{equation}
To leading order in the source, it is sufficient to retain its effect in the zero-mode subspace and neglect its effect on the nonzero-mode determinant. The zero eigenvalues are then lifted, and their contribution becomes
\begin{equation}
\left.\det\left(\hat{\slashed{D}}_{\rm inst}+J\right)\right|_{\rm zero\ modes}\approx\det_{ij}\left[\int \d^4x\,\psi_0^\dagger(x)J_{ij}(x)\psi_0(x)\right].
\label{eq:thooft_source_determinant}
\end{equation}

We now seek to reproduce the instanton-induced effect by an effective interaction evaluated in the topologically trivial sector. Schematically,
\begin{align}
    Z_{1,0}[J] 
    \overset{\rm eff}{=} 
    \int\mathcal{D}\hat{\bar\psi} \mathcal{D}\hat\psi\mathcal{D}\hat A_{0,0}
    \,\left(\int\d^4 x\,\O_{I}(x)\right)
    \e^{-\int_\Omega \d^4x \left[\hat{\bar\psi}\hat{\slashed{D}}_0 \hat\psi+\frac{1}{4}\hat G_{\mu\nu}^a\hat G_{\mu\nu}^a +\hat{\bar\psi}_i J_{ij}\hat\psi_j\right]}.\label{eq:thooft_effective_matching}
\end{align}
Here the integration over $\hat A_{0,0}$ is restricted to the topologically trivial sector. After integrating over the instanton collective coordinates and taking the low-momentum limit, the leading instanton contribution can be represented by~\cite{tHooft:1976rip}\footnote{When expanding $Z_{1,0}[J]$ in powers of $J$, the leading nonvanishing contribution occurs at order $\O(J^{N_f})$ and arises from $N_f$ insertions of $\int\d^4x\,\hat{\bar\psi}_{iL}J_{ij}\hat\psi_{jR}$, which saturate the fermionic zero modes and generate the flavour determinant in Eq.~\eqref{eq:thooft_source_determinant}. Similarly, in the topologically trivial sector with the insertion $\int\d^4x\,\O_I(x)$, the leading source dependence arises from the same $N_f$ insertions. The sign of $\Gamma$ in Eq.~\eqref{eq:OI} may depend on the value of $N_f$. }
\begin{equation}
    \O_{I}(x) = -\Gamma\det_{ij}\left(\hat{\bar\psi}_{iR}(x)\hat\psi_{jL}(x) \right), 
    \label{eq:OI}
\end{equation}
where $\Gamma$ may be computed from one-loop corrections (see the next section for more details).
The anti-instanton induces the Hermitian-conjugate operator, $\O_{\bar I}=\O_{I}^\dagger$.

The transformation properties of $\O_{I/\bar I}$ exhibit its connection with the axial anomaly. Under $SU(N_f)_L\times SU(N_f)_R$ and the vector symmetry $U(1)_V$, $\O_{I/\bar I}(x)$ is invariant. Under the axial transformation $\hat\psi\rightarrow\e^{\i\beta\gamma^5}\hat\psi$, however,
\begin{equation}
    \det_{ij}\left(\hat{\bar\psi}_{iR}\hat\psi_{jL} \right) \rightarrow \e^{-2\i N_f\beta}\det_{ij}\left(\hat{\bar\psi}_{iR}\hat\psi_{jL}\right).
    \label{eq:thooft_U1A_violation}
\end{equation}
Therefore, $\O_{I/\bar{I}}$ violates the continuous $U(1)_A$ symmetry while preserving its discrete $\mathbb{Z}_{2N_f}$ subgroup. Equivalently, a gauge configuration with $|Q|=1$ violates the axial charge by $2N_f$ units in magnitude, with instantons and anti-instantons producing violations of opposite signs.

Now we can generalise the analysis to the $\theta$-vacuum,
\begin{align}
    Z[J] 
    &= \int\mathcal{D}\hat{\bar\psi} \mathcal{D}\hat\psi\mathcal{D}\hat A\,\e^{-\int_\Omega \d^4x\left[\hat{\bar\psi}\hat{\slashed{D}}\hat\psi+\frac{1}{4}\hat G_{\mu\nu}^a\hat G_{\mu\nu}^a+\hat{\bar\psi}_i J_{ij}\hat\psi_j -\frac{\i\theta g_s^2}{32\pi^2} \hat{G}^a_{\mu\nu}\widetilde{\hat{G}}^a_{\mu\nu}\right]}\notag\\
    &= \mathcal{N} \sum_{\Delta n=-\infty}^{\infty}\e^{\i\Delta n\theta}Z_{\Delta n}=\sum_{\Delta n=-\infty}^\infty \int \mathcal{D}\hat{\bar\psi} \mathcal{D}\hat\psi \mathcal{D}\hat A_{\Delta n} \,\e^{\i \Delta n\theta}\e^{-\int_\Omega \d^4x \left[\hat{\bar\psi}\hat{\slashed{D}}\hat\psi+\frac{1}{4} \hat G_{\mu\nu}^a\hat G_{\mu\nu}^a +\hat{\bar\psi}_i J_{ij}\hat\psi_j\right]} \notag\\ 
    &\simeq\sum_{n=0}^\infty \sum_{\bar n=0}^{\infty}\frac{1}{n!\bar{n}!}\int \mathcal{D}\hat{\bar\psi} \mathcal{D}\hat\psi \mathcal{D}\hat A_{n,\bar{n}} \,\e^{\i (n-\bar n)\theta}\e^{-\int_\Omega \d^4x \left[\hat{\bar\psi}\hat{\slashed{D}} \hat\psi+\frac{1}{4} \hat G_{\mu\nu}^a\hat G_{\mu\nu}^a +\hat{\bar\psi}_i J_{ij}\hat\psi_j\right]},
\end{align}
where the subscript $\Delta n$ indicates that the gauge field configurations are restricted to the $\Delta n$ topological sector. In the last equality, we have used the dilute instanton gas approximation (DIGA) as an illustrative approach. We will discuss it more in the next section. Again, we seek to reproduce the instanton-induced effect by an effective interaction in the topologically trivial sector,
\begin{align}
    Z[J]\overset{\rm eff}{=}\int\mathcal{D}\hat{\bar\psi}\mathcal{D}\hat\psi\mathcal{D}\hat A_{0,0}\,\times\hat{\mathcal{O}}
    \times 
    \e^{-\int_\Omega \d^4x\left[\hat{\bar\psi} \hat{\slashed{D}}_0\hat\psi+\frac{1}{4} \hat G_{\mu\nu}^a\hat G_{\mu\nu}^a +\hat{\bar\psi}_i J_{ij}\hat\psi_j \right]}. \label{eq:thooft_effective_matching-2}
\end{align}
Within the DIGA, one may have~\cite{tHooft:1986ooh}
\begin{align}
   \hat{\mathcal{O}} = \sum_{n=0}^\infty\sum_{\bar{n}=0}^\infty \frac{1}{n!{\bar n}!} \e^{\i(n-\bar n)\theta} \left[\int\d^4 x\,\O_{I}(x)\right]^{n}  \left[\int\d^4 x\,\O_{\bar I}(x)\right]^{\bar n}=\e^{-\int\d^4 x\,\hat{\O}^{(\theta)}_{\rm 't\; Hooft}(x) },
\end{align}
where
\begin{align}
    \hat{\O}^{(\theta)}_{\rm 't\; Hooft}(x) = \Gamma\,\e^{\i\theta} \det_{ij}\left(\hat{\bar{\psi}}_{iR}(x)\hat{\psi}_{jL}(x)\right)  + \hc .
    \label{eq:Hooft-operator1}
\end{align}
Analytically continued to Minkowski spacetime, one then has an additional term in the Lagrangian
\begin{align}
    \Delta\L_{\rm M} = -\Gamma\,\e^{\i\theta}\det_{ij}\left({\bar{\psi}}_{iR}(x) {\psi}_{jL}(x)\right)  + \hc .
    \label{eq:Delta-LM1}
\end{align}
Comparing to the Dirac mass term~\eqref{eq:Dirac-mass}, the chiral phases in the 't Hooft vertex and the Dirac mass term cannot be removed by field redefinitions unless $\theta+\bar\alpha=0$. Therefore, a nonvanishing $\bar\theta$ would signal $CP$-violating effects, which we will discuss in detail in Sec.~\ref{sec:theta-CP-effects}.

In the above derivation, there is, however, an order-of-limits subtlety. Reference~\cite{Ai:2020ptm} argues that carefully taking care of this subtlety would replace the phase $\exp(\i\theta)$ multiplied with $\det_{ij}\left(\hat{\bar\psi}_{iR}\hat\psi_{jL}\right)$ with $\exp(-\i\bar\alpha)$. In this case, one would have 
\begin{align}
    \Delta\L_{\rm M} = -\Gamma\,\e^{-\i\bar\alpha} \det_{ij}\left({\bar{\psi}}_{iR}(x){\psi}_{jL}(x)\right)  + \hc .
    \label{eq:Delta-LM2}
\end{align}
One might think that the phase $\exp(\i\theta)$ in Eq.~\eqref{eq:Delta-LM1} is enforced by the selection rule; the theory should be invariant under a chiral transformation supplemented with (spurious) changes in $\alpha_i$, $\theta$ going as follows:
\begin{align}
    {\psi}_i\to \e^{\i\beta\gamma^5}{\psi}_i,\qquad \alpha_i\to \alpha_i-2\beta,\qquad \theta\to \theta+2 N_f\beta.
\end{align}
However, with Eq.~\eqref{eq:Delta-LM2}, the above spurious symmetry is also observed. Since we are ultimately interested in fermion correlation functions, we shall discuss the order-of-limits subtlety directly in the computation of them.

\subsection{Fermionic correlation functions and the second caveat}
\label{sec:diga}

As an illustrative semiclassical approach, the DIGA approximates the relevant gauge-field configurations by dilute ensembles of $n$ instantons and $\bar n$ anti-instantons, labelled by the pair $(n,\bar n)$. In a weakly coupled theory, or at short distances where asymptotic freedom renders the gauge coupling small, the semiclassical expansion around individual instantons provides a controlled framework for understanding how different topological sectors contribute to the path integral~\cite{tHooft:1976snw,Vainshtein:1981wh}. In low-energy QCD at zero temperature, however, large instantons are not parametrically suppressed, and the DIGA should therefore not be regarded as a quantitatively reliable description of the physical QCD vacuum in the infrared. What is robust is not the DIGA itself, but the existence of topologically nontrivial sectors and the possibility that the path integral receives contributions from all integer values of $Q$~\cite{Callan:1976je,Bonanno:2025wcv,Bonanno:2026zzf}. In the derivation below, we adopt the simplest version of the DIGA, in which instantons and anti-instantons are treated as independent. Extensions incorporating interactions between instantons and anti-instantons have been studied extensively; see, for example, Refs.~\cite{Schafer:1996wv,Diakonov:2002fq}.

For simplicity, we consider $N_f=1$. We allow a general chiral phase associated with the mass, $M= m \exp(\i\alpha\gamma^5)$ with $m>0$. The relevant quark correlation function is given by
\begin{align}
    \langle\hat\psi(x)\hat{\bar\psi}(x^\prime)\rangle 
    =
    -\left.\frac{\delta^2 \log Z[\eta,\bar{\eta}]}{\delta \bar \eta(x) \delta \eta(x^\prime)}\right|_{\eta=\bar{\eta}=0}
    =
    \frac{\int{\cal D}\hat{\bar\psi}\,{\cal D}\hat\psi\,{\cal D}\hat{A}\,\hat{\psi}(x)\hat{\bar\psi}(x^\prime) {\rm e}^{-\int_\Omega \d^4x\,\L_{\rm E}}}{\int{\cal D}\hat{\bar\psi}\,{\cal D}\hat{\psi}\,{\cal D}\hat{A}\, {\rm e}^{-\int_\Omega \d^4x\,\L_{\rm E}}}, 
    \label{quark:correlation}
\end{align}
where now the Euclidean Lagrangian includes the topological term, see Eq.~\eqref{eq:Euclidean-action-full}.
To proceed, we approximate the Green's function of the quarks in the background of one anti-instanton
\begin{align}
    S(x,x^\prime) \approx \frac{\hat\psi_{0\rm L}(x)\hat\psi_{0{\rm L}}^{\dagger}(x^{\prime})}{m\, {\rm e}^{-{\rm i}\alpha}}+\SumInt\limits_{\lambda\not=0} \frac{\hat\psi_\lambda(x)\hat\psi^{\dagger}_\lambda(x^{\prime})}{\lambda}.
\end{align}
Note that the first term in the approximate expression aligns with the $CP$ phase $\alpha$ pertaining to the quark mass. For real masses, this approximation has been used in Refs.~\cite{Shifman:1979uw,Diakonov:1985eg}.

Now consider a quasistationary background given by $(n,\bar{n})$ in the DIGA, the Green's function for the quarks should be well approximated by~\cite{Diakonov:1985eg}
\begin{align}
    S_{n,\bar n}(x,x^{\prime})\approx S_{0}(x,x^{\prime})+\sum_{\nu=1}^n\frac{\hat\psi_{0{\rm R}}(x-x_{0,\nu}){\hat\psi^\dagger_{0{\rm R}}}(x^{\prime}-x_{0,\nu})}{m\, {\rm e}^{{\rm i}\alpha}}{+}\sum_{\bar\nu=1}^{\bar n}\frac{\hat\psi_{0{\rm L}}(x-x_{0,\bar\nu}){\hat\psi^\dagger_{0{\rm L}}}(x^{\prime}-x_{0,\bar \nu})}{m\, {\rm e}^{-{\rm i}\alpha}}. 
    \label{Greens:functions}
\end{align}
Here $x_{0,\nu}$ and $x_{0,\bar\nu}$ are the locations of instantons and anti-instantons, respectively, $S_{0}$ is the Green's function of a Dirac fermion with mass $m \exp({\rm i}\alpha\gamma^5)$ in a translation-invariant (i.e., void of instantons) background. This approximation neglects contributions from overlapping instantons, which are more suppressed as the instanton gas becomes more dilute. While the Green's function close to the individual instantons and anti-instantons is dominated and therefore approximated by the `t~Hooft zero modes, sufficiently far away from the points $x_{0,\nu}$ and $x_{0,\bar\nu}$ the Green's function is given by the form in the background without instantons, i.e. 
\begin{align}
    S_{0}(x,x^{\prime}) = (-\gamma_{\mu}\hat{\partial}_\mu + m\, {\rm e}^{-{\rm i}\alpha \gamma^5})\int\frac{{\rm d}^4p}{(2\pi)^4}\, {\rm e}^{-{\rm i}p(x-x^{\prime})}\frac{1}{p^2+m^2} .\label{eq:S0}
\end{align}
In Eq.~(\ref{Greens:functions}), we note the alignment of the instanton-induced breaking of chiral symmetry with the quark masses so that there is no indication of $CP$ violation at this level but also note that $\theta$ has not yet entered into the calculation.

Using the Green's functions in Eq.~\eqref{Greens:functions}, we can evaluate the fermion two-point function in the sector $(n,\bar n)$:
\begin{align}
    \langle \hat\psi(x)\hat{\bar\psi}(x^\prime)\rangle_{n,\bar n} 
    &= 
    \int\mathcal{D}\hat A_{n,\bar n}\,\mathcal{D}\hat{\bar\psi}\,\mathcal{D}\hat{\psi}\,\hat\psi(x)\hat{\bar\psi}(x^\prime)\e^{-S_{\rm E}[\hat{A},\hat{\bar\psi},\hat{\psi}]}\notag\\
    &= 
    \left(\prod_{\nu=1}^{n}\int_{\Omega}\d^4x_{0,\nu}\,\d\Sigma_\nu\, J_\nu\right)\left(\prod_{\bar\nu=1}^{\bar n} \int_{\Omega}\d^4x_{0,\bar\nu}\,\d\Sigma_{\bar\nu}\,J_{\bar\nu} \right)S_{n,\bar n}(x,x^\prime) \e^{-S_{\rm BPST}(n+\bar n)} \e^{\i(n-\bar n)(\alpha+\theta)} (\Theta\varpi)^{n+\bar n}. \label{eq:psibarpsi-nbarn}
\end{align}
The integrations over collective coordinates other than the positions of the instantons and anti-instantons are denoted by $\d\Sigma_{\nu}$ and $\d\Sigma_{\bar\nu}$, respectively. The corresponding Jacobians arising from the replacement of the bosonic zero-mode integrations by integrations over collective coordinates are denoted by $J_\nu$ and $J_{\bar\nu}$. The one-loop gauge-field determinant around a single instanton or anti-instanton, denoted collectively by $\bar A$, with the zero modes omitted and normalised to the determinant in the trivial background $A=0$, is given by
\begin{align}
    \varpi\equiv\frac{1}{\sqrt{\det_{\bar A}^{\prime}/\det_{A=0}}}. \label{eq:varpidef}
\end{align}
Here the prime indicates that the zero eigenvalues are omitted. In an instanton background with $\eta=+1$, or an anti-instanton background with $\eta=-1$, one has~\cite{Ai:2020ptm}
\begin{align}
    \frac{\det\left( \hat{\slashed{D}}+m\,\e^{\i\alpha\gamma^5}\right)}{\det\left(\hat{\slashed{\partial}}+ m\,\e^{\i\alpha\gamma^5}\right)} 
    =
    \e^{\i\eta\alpha}\left|\frac{\det\left(\hat{\slashed{D}} +m\,\e^{\i\alpha\gamma^5}\right)}{\det\left(\hat{\slashed{\partial}} +m\,\e^{\i\alpha\gamma^5} \right)}\right|\equiv\e^{\i\eta\alpha}\Theta,
    \label{eq:fermion-determinant-ratio}
\end{align}
where the positive quantity $\Theta$ is defined by the final equality. In Refs.~\cite{Ai:2020ptm,Ai:2024cnp}, the fermionic path integral was instead taken to yield $\det(-\hat{\slashed{D}}-m\,\e^{\i\alpha\gamma^5})$. The overall sign inside the functional determinant may reflect a different convention for the ordering of the Euclidean Grassmann integration measure. It was then claimed that
\begin{align}
    \frac{\det\left(-\hat{\slashed{D}}-m\,\e^{\i\alpha\gamma^5}\right)}{\det\left(-\hat{\slashed{\partial}} -m\,\e^{\i\alpha\gamma^5} \right)}\overset{?}{=}-\e^{\i\eta\alpha}\left|\frac{\det\left( -\hat{\slashed{D}}-m\,\e^{\i\alpha\gamma^5}\right)}{\det\left(-\hat{\slashed{\partial}}-m\,\e^{\i\alpha\gamma^5}\right)}\right|. 
    \label{eq:fermion-determinant-ratio-incorrect}
\end{align}
We argue that the minus sign on the right-hand side should be absent. With a common regulator and a consistent choice of orientation for the Grassmann measure, the instanton background does not change the total number of regulated fermionic modes. Any overall sign generated by replacing the Dirac operator by its negative therefore appears identically in the numerator and denominator and cancels completely in their ratio. Finally, we note that both $\varpi$ and $\Theta$ have the same value for an instanton and an anti-instanton.

Next, we turn to the collective coordinates and integrate out the location of a single anti-instanton as
\begin{align}
\begin{aligned}
    \int_{\Omega}{\rm d}^4 x_{0,\bar\nu}\ S(x,x^\prime) &\approx  \int_{\Omega}{\rm d}^4 x_{0,\bar\nu} \left[S_{0}(x,x^\prime) {+}\frac{\hat\psi_{0{\rm L}}(x-x_{0,\bar\nu})\hat\psi_{0{\rm L}}^\dagger(x^\prime-x_{0,\bar\nu})}{m\, {\rm e}^{-{\rm i}\alpha}}+\cdots\right]\\
   &= \Omega\,(S_{0}(x,x^\prime)+\cdots){ +} m^{-1} {\rm e}^{{\rm i}\alpha} h(x,x^\prime)P_{\rm L}.
\end{aligned}
\end{align}
The dots above represent contributions from the zero modes of the (anti)-instantons whose centres were not integrated over. This expression defines the overlap function $h(x,x^\prime)$---a rank-two tensor in spinor space:
\begin{align}
    h(x,x^\prime)P_{\rm L} &= \int_{\Omega}{\rm d}^4 x_{0,\bar\nu}\,\hat\psi_{0{\rm L}}(x-x_{0,\bar\nu}) \hat\psi_{0 {\rm L}}^\dagger(x^\prime-x_{0,\bar\nu}),\\ 
    h(x,x^\prime)P_{\rm R} &= \int_{\Omega}{\rm d}^4 x_{0,\nu}\,\hat\psi_{0{\rm R}}(x-x_{0,\nu}) \hat\psi_{0{\rm R}}^\dagger(x^\prime-x_{0,\nu}).
\end{align}
Further, we integrate over the remaining collective coordinates as
\begin{align}
    \bar h(x,x^\prime)\equiv \frac{\int \d\Sigma \, J\times  h(x,x^\prime)}{\int \d\Sigma\, J}.
\end{align}
Notice that we ignore here the fact that for the classical instanton, the integral over the dilatational mode is divergent. A consistent treatment of the large-instanton region requires infrared physics beyond the DIGA.

Summing over all possible $(n,\bar n)$ sectors and integrating now over all locations of instantons and anti-instantons, we obtain 
\begin{align}
    &\sum\limits_{\bar n,n\geq 0}\frac{1}{n!\bar n!} \Big[ {\,\bar h(x,x^\prime)}(\bar n\, m^{-1} {\rm e}^{{\rm i}\alpha} P_{\rm L}+n\, m^{-1} {\rm e}^{-{\rm i}\alpha} P_{\rm R}) \Omega^{\bar n+n -1} \!\!+ S_{0}(x,x^\prime) \Omega^{\bar n+n}\! \Big] \kappa^{\bar n +n} {\rm e}^{{\rm i}\Delta n(\alpha + \theta)} \notag\\
    &\quad=\left[\left({\rm e}^{-\i\theta} P_{\rm L}+{\rm e}^{{\rm i}\theta}  P_{\rm R}\right)\frac{  \kappa}{m}\bar h(x,x^\prime) + S_{0}(x,x^\prime)\right]{\rm e}^{2\kappa \Omega\cos(\alpha+\theta)}, 
    \label{eq:fermion-two-point-incorrect-order}
\end{align}
where $\kappa={\textstyle\int}\!{\rm d}\Sigma\,J\,
\,\Theta\,\varpi\,{\rm e}^{-S_{\rm BPST}}
$ is the instanton density per spacetime volume.
Proceeding as for the fermion correlation, we obtain the partition function
\begin{align}
    Z = \sum_{n,\bar n}\frac{1}{n!\bar n!}(\kappa \Omega)^{\bar n+n}{ \rm e}^{{\rm i}  (n-\bar n)(\alpha+\theta)}={\rm e}^{2\kappa \Omega \cos(\alpha+\theta)}. 
    \label{eq:vacuum-parti-incorrect-order}
\end{align}
Taking the ratio, the overall exponential factors cancel, we obtain the final result for the fermion correlation function
\begin{align}
    \langle\hat{\psi}(x)\hat{\bar\psi}(x')\rangle=\left(\frac{\kappa}{m}\right)\e^{\i\theta\gamma^5}\,\bar h(x,x^\prime) + S_{0}(x,x^\prime). 
    \label{eq:correlation-incorrect-order}
\end{align}
The first term represents the instanton-induced contribution. Recalling that $S_0(x,x')$ contains the chiral phase $\e^{-\i\alpha\gamma^5}$, as shown in Eq.~\eqref{eq:S0}, the correlation function above would correspond to Eq.~\eqref{eq:Delta-LM1}, which leads to $CP$ violation.

From Eq.~\eqref{eq:vacuum-parti-incorrect-order}, one may infer the energy density of the $\theta$-vacuum,
\begin{align}
    \rho(\theta) = -2\kappa \cos(\alpha+\theta),
\end{align}
which has a minimum at $\alpha+\theta=0$. Since $\alpha+\theta$ is a theory parameter instead of a dynamical field, there is no reason for it to choose a value that minimises the energy.

The derivations of Eqs.~\eqref{eq:vacuum-parti-incorrect-order} and~\eqref{eq:correlation-incorrect-order} can be straightforwardly generalised to $N_f$ flavours, see Ref.~\cite{Ai:2020ptm}. Note that, since now there is no minus sign in Eq.~\eqref{eq:fermion-determinant-ratio}, for $N_f$ flavours one would simply get $\rho(\theta)=-2\kappa_{N_f}\cos(\bar\alpha+\theta)$, while the incorrect Eq.~\eqref{eq:fermion-determinant-ratio-incorrect} would give $\rho(\theta)=-2\kappa_{N_f}\cos(\bar\alpha+\theta+N_f\pi)$.

\paragraph{The order-of-limits caveat.}
In the calculation above, we have kept $\Omega$ finite throughout and taken the thermodynamic limit $\Omega\to\infty$ only at the end. However, as emphasised below Eq.~\eqref{eq:topological-charge}, topological quantisation itself arises only in the infinite-spacetime-volume limit for $\mathbb{R}^4$. This observation motivates an alternative prescription in which one first takes $\Omega\to\infty$ within each topological sector and only subsequently sums over the sectors to infinity. Reference~\cite{Ai:2020ptm} has demonstrated the non-commutativity of the order of taking these two limits and shown that the alternative order of limits can lead to a form for the fermion correlation functions where the $CP$-odd phases can be removed, corresponding to Eq.~\eqref{eq:Delta-LM2}.  Other authors have defended the conventional approach~\cite{Benabou:2025viy,Bhattacharya:2025qsk,Khoze:2025auv,Aghaie:2026pkf}. The present review will not assume that this issue is resolved by fiat. Instead, we will separate the standard framework, used throughout most phenomenology, from the later conceptual reassessment.

\subsection{Chiral perturbation theory}
\label{sec:chiral-perturbation-theory}

The DIGA discussed above provides an intuitive microscopic picture of the anomalous breaking of $U(1)_A$ and of the dependence on the vacuum angle. At zero temperature and low energies, however, the dilute instanton gas is not quantitatively controlled. A more systematic description of the infrared dynamics is provided by chiral perturbation theory (ChPT)~\cite{Pich:1995bw,Scherer:2002tk}, which is based on the pattern of spontaneous chiral symmetry breaking and organises low-energy observables as an expansion in a generic soft momentum, conventionally denoted by $p$, and in the light-quark masses~\cite{Gasser:1983yg,Gasser:1984gg}. In the conventional formulation of QCD, ChPT provides a controlled description of the dependence of the vacuum energy and low-energy observables on $\bar\theta$ without assuming a dilute ensemble of instantons~\cite{Crewther:1979pi,Leutwyler:1992yt,GrilliDiCortona:2015jxo,DiLuzio:2020wdo}. As we shall see, however, it is useful to distinguish the constraints imposed by chiral symmetry and spurion analysis from the matching of the effective theory to the underlying QCD generating functional.

We first formulate the theory for a general number $N_f$ of light quark flavours. Temporarily neglecting the $U(1)_A$ anomaly, the approximate symmetry $U(N_f)_L\times U(N_f)_R$ is spontaneously broken to $U(N_f)_V$, up to discrete identifications. The corresponding pseudo-Nambu--Goldstone fields can be collected into a matrix $U(x)\in U(N_f)$. Ignoring the anomalous singlet interaction for the moment, the leading-order chiral Lagrangian is~\cite{Gasser:1983yg,Gasser:1984gg}\footnote{In our convention, $\M$ is defined through $\bar\psi_L\M\psi_R+\hc$, while some references instead define the mass matrix through $\bar\psi_R\M\psi_L+\hc$. With the latter convention, the mass term in Eq.~\eqref{eq:chiral-Lagrangian-U-Nf} is correspondingly written as $\tr(\M^\dagger U+U^\dagger\M)$.}
\begin{align}
    \mathcal{L}^{(2)}_{\chi} = \frac{F_0^2}{4} \tr\left( \partial_\mu U\,\partial^\mu U^\dagger \right) 
    + 
    \frac{F_0^2B_0}{2} \tr\left( \M U+U^\dagger\M^\dagger \right),\label{eq:chiral-Lagrangian-U-Nf}
\end{align}
where $F_0$ denotes the pseudoscalar decay constant in the chiral limit, while $B_0$ is related to the magnitude of the quark condensate through
$\langle \bar\psi_{jL}\psi_{iR}\rangle = -(F_0^2B_0/2)\langle U_{ij}\rangle$.\footnote{This relation can be understood by treating the quark mass matrix $\M$ as an external source. Taking the functional derivative of the QCD generating functional, in which the quarks are the dynamical fields, and of the corresponding low-energy generating functional, in which $U$ is the dynamical field, with respect to $\M$, and then matching the two results yields the stated relation.} The mesonic field carries the same chiral quantum numbers as $U_{ij}\sim\bar\psi_{jL}\psi_{iR}$ and therefore transforms as $U\to RUL^\dagger$ under $U(N_f)_L\times U(N_f)_R$. Correspondingly, the mass matrix is treated as a spurion transforming as $\M\to L\M R^\dagger$, so that $\tr(\M U)$ is invariant.

The anomalous breaking of the singlet axial symmetry must be incorporated separately. A particularly transparent way of encoding its symmetry structure is to introduce a determinantal interaction, motivated by the Kobayashi--Maskawa--'t Hooft interaction~\cite{Kobayashi:1970ji,Kobayashi:1971qz,tHooft:1976snw},
\begin{align}
    \Delta\mathcal{L}_{\rm anom}
    =
    |\lambda|F_0^4 \left(\e^{-\i\xi}\det U + \e^{\i\xi}\det U^\dagger \right),
    \label{eq:chiral-determinant-spurion}
\end{align}
where $\xi$ denotes the phase carried by the corresponding low-energy coupling. This interaction preserves $SU(N_f)_L\times SU(N_f)_R\times U(1)_V$ while breaking the continuous $U(1)_A$ symmetry, and provides a simple way of keeping track of the anomalous singlet phase.

The relevant spurionic symmetry can now be made explicit. Under an axial transformation $\psi_i\to\e^{\i\beta\gamma^5}\psi_i$, the effective theory is invariant if the fields and spurions are simultaneously transformed according to
\begin{align}
    \M \to \e^{-2\i\beta}\M,
    \qquad
    U \to \e^{2\i\beta}U,
    \qquad 
    \theta\to\theta+2N_f\beta.
    \label{eq:chiral-spurion-transformation}
\end{align}
Since $\bar\alpha\equiv\arg\det\M$, the first transformation implies $\bar\alpha\to\bar\alpha-2N_f\beta$. Invariance of Eq.~\eqref{eq:chiral-determinant-spurion} further requires $\xi\to\xi+2N_f\beta$. This transformation property is satisfied by both of the assignments~\cite{Ai:2020ptm,Ai:2024cnp,Ai:2025quf}
\begin{align}
    \xi= 
    \begin{cases}
    \theta, &\text{(conventional matching)},\\
    -\bar\alpha,&\text{(alternative matching)}.
    \end{cases}
    \label{eq:options}
\end{align}
Thus, the spurion transformation properties by themselves do not distinguish between these two possibilities. The value of $\xi$ must ultimately be fixed by matching the low-energy effective theory to the underlying QCD generating functional or, equivalently, to an appropriate set of QCD correlation functions.

It is sometimes argued that ChPT by itself is sufficient to establish the presence of $CP$ violation. Such an argument, however, generally involves an implicit assumption about the vacuum alignment of the chiral condensate. Let the quark mass matrix be diagonal,
$\M={\rm diag}(m_1\e^{\i\alpha_1},\ldots,m_{N_f}\e^{\i\alpha_{N_f}})$,
and parameterise the vacuum orientation of the chiral condensate as
\begin{align}
    \langle U\rangle = {\rm diag} \left( \e^{\i\phi_1}, \ldots, \e^{\i\phi_{N_f}} \right).
\end{align}
In the conventional treatment, it is usually assumed that $\sum_i\phi_i=\theta$, or after rotating the angle $\theta$ into the quark mass matrix, $\sum_i\phi_i=0$. However, as we show below, this assumption is ultimately tied to the choice $\xi=\theta$ and should therefore not be imposed independently of the matching of the anomalous interaction.

The leading potential following from Eqs.~\eqref{eq:chiral-Lagrangian-U-Nf} and~\eqref{eq:chiral-determinant-spurion} is
\begin{align}
    V(\phi_1,\ldots,\phi_{N_f}) = -F_0^2B_0 \sum_{i=1}^{N_f} m_i\cos(\alpha_i+\phi_i) - 2|\lambda|F_0^4 \cos\left(\xi-\sum_i\phi_i \right).
    \label{eq:general-chiral-potential}
\end{align}
The vacuum phases $\phi_i$ are determined by the vacuum-alignment conditions $\partial V/\partial\phi_i=0$, which give
\begin{align}
    m_i\sin(\alpha_i+\phi_i) = \frac{2|\lambda|F_0^2}{B_0} \sin\left(\xi-\sum_j\phi_j \right).
    \label{eq:general-vacuum-condition}
\end{align}
When the anomalous singlet interaction dominates over the explicit quark-mass terms, as is appropriate when the singlet mode is parametrically heavier than the pseudo-Nambu--Goldstone bosons, Eq.~\eqref{eq:general-vacuum-condition} implies $\sum_i\phi_i\approx\xi$, up to corrections suppressed by the light-quark masses relative to the anomalous singlet scale. Consequently, the residual phase controlling $CP$ violation in the low-energy theory is $\sum_i(\alpha_i+\phi_i)\approx\bar\alpha+\xi$. For the conventional choice $\xi=\theta$, this reduces to $\bar\theta=\theta+\bar\alpha$, whereas for $\xi=-\bar\alpha$ it vanishes. Correspondingly, $CP$-odd observables are controlled by this residual phase and, sufficiently close to the $CP$-conserving point, are odd functions of it and begin linearly in its expansion.

To make this point more transparent, we can perform an anomalous singlet axial rotation to transfer the topological angle into the quark mass matrix, followed by non-singlet axial rotations that distribute the resulting total phase uniformly among the flavours. In this basis, one has
\begin{align}
    \M_{\bar\theta} = \e^{\i\bar\theta/N_f} {\rm diag}(m_1,\ldots,m_{N_f}),
    \qquad
    \xi\to\xi' =
    \begin{cases}
    0, & \text{(conventional matching)},\\
    -\bar\theta, & \text{(alternative matching)}.
    \end{cases}
    \label{eq:chiral-mass-spurion-theta}
\end{align}
In the limit in which the anomalous singlet mode is much heavier than the non-singlet pseudoscalars, its vacuum alignment gives $\sum_i\phi_i\simeq\xi'$. At energies well below the $\eta'$ mass, the singlet fluctuation can therefore be integrated out, and the chiral field may be written as
\begin{align}
    U = \e^{\i\xi'/N_f}\widetilde U,
    \qquad
    \widetilde U\in SU(N_f). 
    \label{eq:singlet-integrated-U}
\end{align}
The phase $\e^{\i\xi'/N_f}$ represents the vacuum orientation of the heavy singlet mode and should not be discarded when the singlet fluctuation is integrated out. The leading-order chiral Lagrangian for the remaining light fields then becomes
\begin{align} 
    \mathcal{L}^{(2)}_{\chi} 
    = 
    \frac{F_0^2}{4} \tr\left( \partial_\mu\widetilde U\, \partial^\mu\widetilde U^\dagger \right) 
    + 
    \frac{F_0^2B_0}{2} \tr\left[ \e^{\i\xi'/N_f}\M_{\bar\theta}\widetilde U + \widetilde U^\dagger \M_{\bar\theta}^\dagger \e^{-\i\xi'/N_f} \right].\label{eq:chiral-Lagrangian-theta-2}
\end{align}
For the conventional matching, $\xi'=0$, and we reproduce the standard form. For the alternative matching, however, $\xi'=-\bar\theta$, and the phase in $\M_{\bar\theta}$ is cancelled by the additional phase $\xi'$. The remaining $SU(N_f)$ theory therefore contains no $CP$-odd phase.

The same issue can be expressed in the large-$N_c$ formulation in which an auxiliary field $q(x)$ is introduced to represent the topological charge density~\cite{DiVecchia:1980yfw},
\begin{align}
    \L^{\rm EFT} 
    &\supset
    \frac{F_0^2}{4} {\rm Tr}(\partial_\mu U)\partial^\mu U^\dagger
    + \frac{F_0^2B_0}{2} {\rm Tr}(\M U+U^\dagger\M^\dagger) - \frac{\i}{2}q(x) {\rm Tr}\left[ \log U-\log U^\dagger \right] + \frac{N_c}{aF_0^2}q^2(x) - \xi q(x), 
    \label{eq:large-Nc-chiral-Lagrangian}
\end{align}
where $q(x)$ is an auxiliary field and $a$ is a parameter that does not scale with $N_c$ at leading order. In the conventional matching, $\xi=\theta$, and the last term has the familiar form as $-\theta q(x)$.\footnote{Upon identifying $q(x)$ with the topological charge density, $q(x)=\frac{g_s^2}{32\pi^2} G(x)^{\mu\nu}\widetilde{G}(x)_{\mu\nu}$, it may appear natural to identify the coefficient of $q(x)$ in the effective theory directly with the fundamental parameter $\theta$. Such an identification, however, does not follow from the correspondence between the operators alone. More generally, parameters multiplying operators in an effective theory are matching coefficients: after degrees of freedom are integrated out, they can encode quantum effects from the underlying theory and are fixed by requiring the effective theory to reproduce appropriate correlation functions or observables of the fundamental theory. Consequently, the coefficient multiplying the effective operator $q(x)$ must ultimately be determined by matching, rather than inferred solely from the appearance of the corresponding term in the ultraviolet Lagrangian. Furthermore, the EFT action is minimised with respect to variations of $q(x)$, so that $\int_\Omega \d^4x\,q(x)$ is not subject to a global topological constraint, unlike in the underlying QCD theory.} Reference~\cite{Benabou:2025viy} argues that this choice is required in matching the chiral theory to QCD, whereas Ref.~\cite{Ai:2025quf} emphasises that the spurion analysis itself allows the second choice in Eq.~\eqref{eq:options} and argues that the phase of the effective coupling must instead be fixed by matching physical correlation functions. With the order of limits advocated in Refs.~\cite{Ai:2020ptm,Ai:2025quf}, this matching leads to $\xi=-\bar\alpha$. The point of disagreement is therefore not the chiral operator content itself, but the matching of the anomalous singlet phase to the underlying QCD generating functional.

For the remainder of this subsection, and in Sec.~\ref{sec:theta-CP-effects}, where we review the standard phenomenology of strong $CP$ violation, we adopt the conventional matching $\xi=\theta$ (or equivalently $\xi'=0$). We therefore take Eq.~\eqref{eq:chiral-Lagrangian-theta-2} with $\xi'=0$ as our starting point. An important step in extracting its physical consequences is the alignment of the non-singlet chiral vacuum. In general, for $\bar\theta\neq0$, the minimum of the potential does not occur at $\langle\widetilde U\rangle=\mathbb{1}_{N_f}$. Writing
\begin{align}
   \langle \widetilde{U}\rangle 
   &= 
   {\rm diag} 
   \left( \e^{\i\varphi_1}, \ldots, \e^{\i\varphi_{N_f}} \right), \qquad \sum_i\varphi_i=0,
\end{align}
the leading-order potential is
\begin{align}
    V(\langle \widetilde{U}\rangle,\bar\theta) = -F_0^2B_0 \sum_{i=1}^{N_f} m_i \cos\left( \frac{\bar\theta}{N_f}+\varphi_i \right). \label{eq:chiral-theta-potential}
\end{align}
Minimisation subject to $\sum_i\varphi_i=0$ gives
\begin{align}
    m_i\sin\left( \frac{\bar\theta}{N_f}+\varphi_i \right) = m_j\sin\left( \frac{\bar\theta}{N_f}+\varphi_j \right) \label{eq:chiral-vacuum-alignment}
\end{align}
for any $i$ and $j$. Vacuum alignment is essential: expanding directly around $\widetilde U=\mathbb{1}_{N_f}$ at nonzero $\bar\theta$ would in general generate tadpoles and obscure the organisation of the low-energy expansion~\cite{Crewther:1979pi,Leutwyler:1992yt}.

For sufficiently small $\bar\theta$, the minimisation can be carried out analytically. Defining $m_*^{-1}\equiv\sum_i m_i^{-1}$, one finds
\begin{align}
    \frac{\bar\theta}{N_f}+\varphi_i
    =
    \frac{m_*}{m_i}\bar\theta
    +
    \mathcal{O}(\bar\theta^3).
    \label{eq:vacuum-alignment-small-theta}
\end{align}
Substituting this solution into Eq.~\eqref{eq:chiral-theta-potential}, the vacuum energy becomes
\begin{align}
    E(\bar\theta) = E(0) + \frac{1}{2}F_0^2B_0m_*\bar\theta^2 + \mathcal{O}(\bar\theta^4), 
    \label{eq:chiral-vacuum-energy-small-theta}
\end{align}
and the leading-order topological susceptibility is therefore
\begin{align}
    \chi \equiv \left. \frac{\partial^2E(\bar\theta)} {\partial\bar\theta^2}\right|_{\bar\theta=0}
    =
    F_0^2B_0m_* + \mathcal{O}(m_q^2).
    \label{eq:chiral-topological-susceptibility}
\end{align}
For three light flavours, $m_*=(m_u^{-1}+m_d^{-1}+m_s^{-1})^{-1}$. The dependence on $\bar\theta$ is therefore suppressed by the light-quark masses. In particular, if any quark were exactly massless, $m_*$ would vanish and the vacuum energy would become independent of $\bar\theta$, consistently with the fact that the vacuum angle can then be removed by an exact chiral rotation~\cite{Crewther:1979pi,Leutwyler:1992yt}.

The structure is especially transparent for $N_f=2$, for which the vacuum alignment can be performed exactly at leading order. Writing
$\langle\widetilde U\rangle={\rm diag}
\left(\e^{\i\varphi},\e^{-\i\varphi}\right)$,
Eq.~\eqref{eq:chiral-theta-potential} becomes
\begin{align}
    V(\varphi,\bar\theta)
    = -F^2_0 B_0 \left[m_u \cos\left(\frac{\bar\theta}{2}+\varphi \right) + m_d\cos\left(\frac{\bar\theta}{2}-\varphi \right) \right].
    \label{eq:two-flavour-potential}
\end{align}
At leading order, $F_0$ may be identified with the pion decay constant $F_\pi$, while $B_0$ is related to the pion mass at $\bar\theta=0$ through $m_\pi^2=B_0(m_u+m_d)$. The stationary condition gives
\begin{align}
    \tan\varphi = \frac{m_d-m_u}{m_u+m_d} \tan\frac{\bar\theta}{2}.
    \label{eq:two-flavour-vacuum-angle}
\end{align}
For fixed $\bar\theta$, Eq.~\eqref{eq:two-flavour-vacuum-angle} has two solutions differing by $\pi$. These generally correspond to the minimum and the maximum of the potential, respectively, except at the special point $m_u=m_d$ and $\bar\theta=\pi$ discussed below. Their stationary values are
\begin{align}
    V_{\rm min,max}(\bar\theta) 
    =
    \mp F_0^2B_0 \sqrt{ (m_u+m_d)^2 \cos^2\left(\frac{\bar\theta}{2}\right) + (m_d-m_u)^2 \sin^2\left(\frac{\bar\theta}{2}\right) }.
\end{align}
Thus the physical vacuum is unique at this order unless the square root vanishes. Taking the lower stationary value and using the leading-order relation for $m_\pi$ yields the familiar result~\cite{DiVecchia:1980yfw,GrilliDiCortona:2015jxo}
\begin{align}
    E(\bar\theta) = -m_\pi^2F_\pi^2
    \sqrt{ 1- \frac{4m_um_d}{(m_u+m_d)^2} \sin^2\left(\frac{\bar\theta}{2}\right)}.
    \label{eq:chiral-vacuum-energy-two-flavour}
\end{align}
Expanding around $\bar\theta=0$ gives
\begin{align}
    \chi_{\rm LO} = m_\pi^2F_\pi^2 \frac{m_um_d}{(m_u+m_d)^2},
    \label{eq:two-flavour-susceptibility}
\end{align}
in agreement with Eq.~\eqref{eq:chiral-topological-susceptibility}. In contrast to the simple cosine dependence obtained in the DIGA, Eq.~\eqref{eq:chiral-vacuum-energy-two-flavour} exhibits the nontrivial light-quark-mass dependence dictated by chiral symmetry, illustrating why the dilute instanton gas is not a quantitative description of the zero-temperature QCD vacuum.

A noteworthy special case occurs at $\bar\theta=\pi$. Since $CP$ sends $\bar\theta\to-\bar\theta$ and the theory is $2\pi$ periodic in $\bar\theta$, this is itself a $CP$-symmetric point. At $\bar\theta=\pi$, Eq.~\eqref{eq:two-flavour-potential} reduces to
\begin{align}
    V(\varphi,\pi) = -F_0^2B_0(m_d-m_u)\sin\varphi .
\end{align}
For the physical ordering $m_d>m_u$, the leading-order vacuum therefore lies at $\varphi=\pi/2$. Naively, $CP$ acts as $\varphi\to-\varphi$ together with $\bar\theta\to-\bar\theta$. However, identifying $-\pi$ with $\pi$ in the parametrisation of the mass term $\M_{\bar\theta}\widetilde U+\hc$ is accompanied by the shift $\varphi\to\varphi+\pi$. Thus, within the theory at $\bar\theta=\pi$, the effective action of $CP$ is
$\varphi\to\pi-\varphi$, under which the vacuum at $\varphi=\pi/2$ is invariant. The vacuum therefore preserves $CP$ for nondegenerate physical quark masses. In the isospin-symmetric limit $m_u=m_d=m$, by contrast, the leading-order potential becomes flat in $\varphi$ at $\bar\theta=\pi$ and cannot by itself determine the vacuum structure. Higher-order chiral terms, or a formulation retaining the anomalous singlet dynamics explicitly, lift this accidental degeneracy and can yield two isolated, degenerate vacua related by $CP$. Choosing either of them then {\it spontaneously} breaks $CP$, giving rise to the Dashen phenomenon~\cite{Dashen:1970et,Leutwyler:1992yt,Smilga:1998dh}. More complete analyses show that spontaneous $CP$ breaking at $\bar\theta=\pi$ can persist over a finite range of nondegenerate quark masses around the isospin-symmetric limit~\cite{Creutz:2003xu}, although the physical QCD quark masses lie outside this region.

ChPT is systematically improvable. Higher-order corrections arise from loops generated by Eq.~\eqref{eq:chiral-Lagrangian-theta-2} and from higher-order operators in the chiral expansion accompanied by additional low-energy constants~\cite{Gasser:1983yg,Gasser:1984gg}. Such corrections are important for precision calculations of the topological susceptibility, the QCD axion mass, and other quantities that depend on the $\bar\theta$-dependent vacuum energy~\cite{GrilliDiCortona:2015jxo,DiLuzio:2020wdo}. In the $SU(2)$ theory, the strange quark is integrated out and its effects are absorbed into the low-energy constants, while the $SU(3)$ theory treats the kaons and the $\eta$ as explicit degrees of freedom. The two descriptions are useful in complementary regimes.

It is worth stressing the logical status of this analysis in the context of the present review. Equations~\eqref{eq:chiral-vacuum-energy-small-theta} and~\eqref{eq:chiral-vacuum-energy-two-flavour} are the standard ChPT results obtained after matching the low-energy theory to the conventional $\bar\theta$-dependent QCD generating functional. Chiral symmetry and spurion analysis determine the allowed operator structures and their transformation properties, but do not by themselves determine all matching coefficients and phases. The recent debate concerns whether the conventional matching $\xi=\theta$ is uniquely implied by QCD, or whether a different prescription for defining the infinite-volume path integral leads to the alternative matching $\xi=-\bar\alpha$ while respecting the same spurionic symmetries~\cite{Ai:2020ptm,Ai:2024cnp,Benabou:2025viy,Ai:2025quf}. We return to this underlying conceptual question in Part~\ref{part:reassessment}.

The same vacuum alignment that determines the $\bar\theta$-dependent vacuum energy also fixes the $CP$-odd interactions among the low-energy hadronic degrees of freedom. In the conventional framework, expanding the chiral effective theory around the aligned vacuum generates $CP$-odd mesonic and baryonic interactions proportional to $\bar\theta$ and to the light-quark masses. We turn to these physical consequences in Sec.~\ref{sec:theta-CP-effects}.

\section{\texorpdfstring{$CP$}{CP}-violating effects from a nonzero \texorpdfstring{$\bar\theta$}{theta-bar}}
\label{sec:theta-CP-effects}

\subsection{\texorpdfstring{$CP$}{CP}-odd hadronic interactions in chiral effective theory}
\label{sec:CP-odd-hadronic-interactions}

Throughout this section, we adopt the conventional matching discussed in Sec.~\ref{sec:chiral-perturbation-theory}, for which the low-energy theory depends on the invariant angle $\bar\theta$. A nonzero $\bar\theta$ then gives rise to $CP$-odd hadronic interactions. These effects were first analysed using current algebra and early chiral methods~\cite{Baluni:1978rf,Crewther:1979pi}, and were subsequently formulated more systematically within chiral effective theory~\cite{Pich:1991fq,Mereghetti:2010tp,deVries:2015una}; see also Refs.~\cite{Pospelov:2005pr,Engel:2013lsa,Pospelov:2025vzj,Sannino:2026wgx} for reviews and recent discussions.

The vacuum alignment derived in the previous section is important not only for determining the vacuum energy, but also for identifying the $CP$-odd interactions. After expanding around the aligned vacuum, the quark-mass matrix contains the phases $(\bar\theta/N_f+\phi_i)$, so that its pseudoscalar component for each flavour is proportional to $m_i\sin(\bar\theta/N_f+\phi_i)$. For sufficiently small $\bar\theta$, Eq.~\eqref{eq:vacuum-alignment-small-theta} gives
$m_i\sin(\bar\theta/N_f+\phi_i)=m_*\bar\theta+\mathcal{O}(\bar\theta^3)$,
independently of the flavour index at leading order. At the quark level, the resulting $CP$-odd mass term can therefore be written, up to an overall sign convention, as a flavour-singlet pseudoscalar operator proportional to
$m_*\bar\theta\sum_i\bar\psi_i\,\i\gamma^5\psi_i$~\cite{Baluni:1978rf,Crewther:1979pi,Mereghetti:2010tp}. Since this term originates from the same quark-mass spurion as the ordinary scalar mass term, chiral symmetry relates several $CP$-odd low-energy constants to $CP$-even quantities governing the quark-mass dependence of hadronic observables.

Before turning to the baryon sector, it is useful to illustrate explicitly how a $CP$-odd hadronic interaction emerges after vacuum alignment. For simplicity, consider two light flavours and temporarily retain the $U(1)_A$ singlet pseudoscalar rather than integrating it out. We write
\begin{align}
    U=\langle U\rangle\,\e^{\frac{\i\Phi}{F_\pi}},
    \qquad
    \Phi= \eta_0\,\mathbb{1}_2+\pi^a\tau^a
    =
    \begin{pmatrix}
        \eta_0+\pi^0 & \sqrt{2}\pi^+\\
        \sqrt{2}\pi^- & \eta_0-\pi^0
    \end{pmatrix},
    \label{eq:U2-meson-expansion}
\end{align}
where $\eta_0$ denotes the flavour-singlet pseudoscalar in the two-flavour theory.\footnote{The field $\eta_0$ in the two-flavour truncation should not be identified quantitatively with the physical $\eta'$, for which the strange-quark component and $\eta$-$\eta'$ mixing must be included.} Denoting the aligned phases of the two mass terms by $A_u$ and $A_d$, the vacuum-alignment conditions imply
\begin{align}
    m_u\sin A_u = m_d\sin A_d\equiv K,
    \qquad
    A_u+A_d=\bar\theta ,
\end{align}
where
\begin{align}
    K = \frac{m_um_d\sin\bar\theta}
    {\sqrt{m_u^2+m_d^2+2m_um_d\cos\bar\theta}}.
    \label{eq:K-theta}
\end{align}
The $CP$-odd part of the mass interaction is proportional to
$\i K\,\tr(\langle U\rangle^\dagger U-U^\dagger\langle U\rangle)$, and its expansion to cubic order gives
\begin{align}
    \Delta\mathcal{L}^{CP}_{\eta_0\pi\pi} 
    = \frac{B_0K}{F_\pi}\, \eta_0 \left[(\pi^0)^2+2\pi^+\pi^-
    \right] +\cdots.
    \label{eq:eta0-pipi-CP}
\end{align}
For $|\bar\theta|\ll1$,
\begin{align}
    \Delta\mathcal{L}^{CP}_{\eta_0\pi\pi}
    =
    \frac{B_0m_um_d}{F_\pi(m_u+m_d)} \bar\theta\, \eta_0 \left[
    (\pi^0)^2+2\pi^+\pi^- \right] +\mathcal{O}(\bar\theta^3).
    \label{eq:eta0-pipi-CP-small-theta}
\end{align}
The anomalous singlet interaction determines the singlet mass and participates in the vacuum alignment, but does not itself directly generate the $\eta_0\pi\pi$ vertex. Equation~\eqref{eq:eta0-pipi-CP-small-theta} thus provides a simple illustration of how the vacuum angle is converted into a physical $CP$-odd hadronic interaction after vacuum alignment. The corresponding $CP$-violating $\eta^{(\prime)}\rightarrow\pi\pi$ amplitudes have been studied within chiral effective theory in Ref.~\cite{Pich:1991fq}.

For EDM phenomenology, the most important low-energy interactions involve baryons. ChPT can be extended to include the nucleon doublet $N=(p,n)^T$; at momenta well below the nucleon mass, it is particularly convenient to employ heavy-baryon ChPT~\cite{Jenkins:1990jv,Bernard:1995dp}. The leading nonderivative $CP$-odd pion-nucleon interactions can be decomposed into isoscalar, isovector, and isotensor structures~\cite{Mereghetti:2010tp,deVries:2012ab,Bsaisou:2014oka,Engel:2013lsa},
\begin{align}
    \mathcal{L}^{CP}_{\pi N} 
    &= 
    \bar g_0\,\bar N\bm{\tau}\cdot\bm{\pi}N + \bar g_1\,\bar N N\,\pi^0 + \bar g_2\,\bar N \left( 3\tau_3\pi^0-\bm{\tau}\cdot\bm{\pi} \right)N +\cdots ,
    \label{eq:CP-odd-pion-nucleon}
\end{align}
where $\bar g_0$, $\bar g_1$, and $\bar g_2$ multiply isospin tensors of rank $0$, $1$, and $2$, respectively, conventionally referred to as the $\Delta I=0,1,2$ interactions.\footnote{Conventions for the $CP$-odd pion-nucleon couplings differ in the literature, in particular by factors of $F_\pi$ and in the normalisation of the isotensor coupling. Equation~\eqref{eq:CP-odd-pion-nucleon} defines the convention used throughout this review.} These couplings connect the underlying $CP$-violating quark interaction to both single-nucleon and nuclear observables. They generate long-distance contributions to nucleon electric dipole form factors through pion loops and, in nuclei, long-range $CP$-odd forces through one-pion exchange.

For the $\bar\theta$ term, the leading $CP$-odd quark-mass interaction is an isoscalar under $SU(2)_V$. It can therefore generate $\bar g_0$ without additional isospin breaking. By contrast, $\bar g_1$ requires an insertion proportional to $m_d-m_u$ and is suppressed in the isospin limit, while the isotensor coupling $\bar g_2$ arises only at still higher order in isospin breaking and is generally negligible for the $\bar\theta$ source at the accuracy relevant for present phenomenology~\cite{Mereghetti:2010tp,deVries:2015una,Bsaisou:2014oka}. Consequently, $\bar g_0$ is the dominant $CP$-odd pion-nucleon coupling induced by $\bar\theta$, although $\bar g_1$ can be important in observables for which the $\bar g_0$ contribution is suppressed by spin-isospin selection rules.

A particularly useful consequence of chiral symmetry is the relation between $\bar g_0$ and the strong neutron-proton mass splitting. The $CP$-odd quark-mass interaction induced by $\bar\theta$ and the isospin-breaking mass term proportional to $m_d-m_u$ arise from the same chiral-symmetry-breaking spurion, so that their leading hadronic realisations involve the same low-energy constant. Defining $\bar m=(m_u+m_d)/2$, $\epsilon=(m_d-m_u)/(m_u+m_d)$, and $\delta m_N^{\rm str}=(m_n-m_p)_{\rm QCD}$, one obtains~\cite{Crewther:1979pi,Mereghetti:2010tp,deVries:2015una}
\begin{align}
    \bar g_0 = -\frac{\delta m_N^{\rm str}}{2F_\pi} \frac{m_*}{\bar m\,\epsilon}\, \bar\theta +\cdots ,
    \label{eq:g0-mass-splitting}
\end{align}
in the convention of Eq.~\eqref{eq:CP-odd-pion-nucleon}. In the two-flavour limit, $m_*=\bar m(1-\epsilon^2)/2$, so that
$\bar g_0=-[\delta m_N^{\rm str}/(4F_\pi)][(1-\epsilon^2)/\epsilon]\bar\theta+\cdots$.
This relation replaces the direct determination of a $CP$-odd hadronic matrix element by the $CP$-even strong neutron-proton mass splitting, which is accessible to lattice QCD. In $SU(3)$ ChPT the relation is preserved by loop corrections through next-to-next-to-leading order, while the local terms that violate it are expected to induce only few-percent corrections~\cite{deVries:2015una}. Using updated lattice inputs, a recent analysis finds~\cite{deVries:2015una,FlavourLatticeAveragingGroupFLAG:2024oxs,Mulder:2025esr}
\begin{align}
    \bar g_0 \simeq -17.2(2.0)\times10^{-3}\,\bar\theta .
    \label{eq:g0-numerical}
\end{align}

The coupling $\bar g_1$ is less precisely known. Although it can be related at leading order to isospin-breaking quantities, higher-order contributions include an independent short-distance low-energy constant, substantially weakening the predictive power of the relation. The commonly used two-flavour estimate, including an estimate of this short-distance contribution, is~\cite{Bsaisou:2012rg,Mulder:2025esr}
\begin{align}
    \bar g_1 \simeq  3.4(1.5)\times10^{-3}\,\bar\theta ,
    \label{eq:g1-numerical}
\end{align}
with an uncertainty of roughly $50\%$.

The pion-nucleon couplings do not, however, exhaust the leading low-energy information required for nucleon EDMs. Chiral effective theory also contains local $CP$-odd photon-nucleon operators that encode unresolved short-distance dynamics. In a relativistic notation convenient for the discussion below, they may be written as~\cite{Hockings:2005cn,Mereghetti:2010tp,Mereghetti:2010kp}
\begin{align}
    \mathcal{L}_{N\gamma}^{CP}
    =
    -\frac{\i}{2}\,
    \bar N
    \left[ \bar d_0^{\,r}(\mu) + \bar d_1^{\,r}(\mu)\tau_3
    \right] \sigma^{\mu\nu}\gamma^5N\,F_{\mu\nu}
    +\cdots ,
    \label{eq:CP-odd-Ngamma}
\end{align}
where $\mu$ is the renormalisation scale, and $\bar d_0^{\,r}$ and $\bar d_1^{\,r}$ are the renormalised isoscalar and isovector low-energy constants, respectively. These coefficients encode the response to the $\bar\theta$ source arising from QCD dynamics at scales not resolved explicitly in the chiral theory. They therefore do not represent independent sources of $CP$ violation: in the present framework, where $\bar\theta$ is the only $CP$-violating parameter, $\bar d_{0,1}^{\,r}$ must vanish for $\bar\theta=0$ and are odd functions of $\bar\theta$. Consequently, for $|\bar\theta|\ll1$, they are linear in $\bar\theta$. Unlike $\bar g_0$, these short-distance coefficients are not fixed by a corresponding relation to a single $CP$-even hadronic observable, because electromagnetic interactions allow additional independent chiral structures~\cite{Mereghetti:2010tp}.
Although the local operators occur at a higher chiral index than the leading $\bar g_0$ interaction, their tree-level contributions to the nucleon EDM are of the same chiral order as the leading pion-loop contribution generated by $\bar g_0$~\cite{Hockings:2005cn,Mereghetti:2010kp}. The coefficients $\bar d_{0,1}^{\,r}$ must therefore be treated, together with the $CP$-odd pion-nucleon couplings, as leading hadronic inputs for the nucleon EDM. Their interplay with the long-distance pion contribution will be discussed in Sec.~\ref{sec:nucleon-EDMs}.

For completeness, three-flavour $SU(3)$ ChPT also generates a $CP$-odd $\eta$-nucleon interaction. With the convention
\begin{align}
\mathcal{L}^{CP}_{\eta N}
=
\bar g_{0\eta}\,\bar N N\,\eta+\cdots ,
\end{align}
the leading coupling can be related to the light- and strange-quark nucleon sigma terms~\cite{deVries:2015una},
\begin{align}
    \bar g_{0\eta} = \frac{2m_*\bar\theta}{\sqrt{3}F_\eta} \left( \frac{\sigma_{Ns}}{m_s} - \frac{\sigma_{Nl}}{2\bar m} \right) +\cdots ,
    \label{eq:g0eta}
\end{align}
where $\sigma_{Nq}=m_q\langle N|\bar q q|N\rangle$ and $\sigma_{Nl}=\sigma_{Nu}+\sigma_{Nd}$. This relation is less precise than the corresponding one for $\bar g_0$, with higher-order $SU(3)$-breaking effects estimated at roughly the $30\%$ level~\cite{deVries:2015una}. The coupling $\bar g_{0\eta}$ contributes to hadronic observables such as the nucleon EDM. In nuclear systems, however, the typical momentum transfer is well below $m_\eta$, so that $\eta$ exchange is effectively short-ranged. Upon matching onto an $SU(2)$ nuclear effective field theory (EFT) in which the $\eta$ is integrated out, its contribution is absorbed into $CP$-odd nucleon-nucleon contact interactions. 

The interactions displayed above are the most relevant ones for the discussion that follows, but they do not exhaust the $CP$-odd chiral Lagrangian. Higher-order terms include derivative pion-nucleon interactions, purely pionic operators, short-range $CP$-odd nucleon-nucleon interactions, and additional electromagnetic operators~\cite{Mereghetti:2010tp,deVries:2012ab,Bsaisou:2014oka}. Their relative importance depends on the chiral and isospin transformation properties of the underlying source. In particular, higher-dimensional $CP$-violating quark and gluon operators generally produce a hierarchy of hadronic low-energy constants different from that induced by the QCD $\bar\theta$ term~\cite{deVries:2012ab,Bsaisou:2014oka,Pospelov:2025vzj}.

\subsection{Nucleon electric dipole moments}
\label{sec:nucleon-EDMs}

Among the hadronic observables induced by a nonzero $\bar\theta$, the EDMs of the neutron and proton provide particularly direct probes of $CP$ violation. The neutron EDM has played a central role in quantifying the strong $CP$ problem since the earliest analyses~\cite{Baluni:1978rf,Crewther:1979pi}. The relation between the underlying $\bar\theta$ term and the nucleon EDMs has subsequently been studied using chiral effective theory~\cite{Pich:1991fq,Borasoy:2000pq,Hockings:2005cn,Ottnad:2009jw,Mereghetti:2010kp,Guo:2012vf}, QCD sum rules~\cite{Narison:2008jp}, and lattice QCD~\cite{Dragos:2019oxn,Liang:2023jfj,Blum:2026hul}; see also Refs.~\cite{Pospelov:2005pr,Engel:2013lsa,Liu:2024kqy,Pospelov:2025vzj} for reviews.

For an on-shell spin-$1/2$ nucleon, the $CP$-odd coupling to a single photon is parametrized by the electric dipole form factor (EDFF), $F_{3,N}(q^2)$~\cite{Hockings:2005cn,Mereghetti:2010kp,Guo:2012vf}. A convenient decomposition of the electromagnetic-current matrix element is
\begin{align}
    \langle N(p')|J_{\rm em}^{\mu}|N(p)\rangle 
    = 
    e\,\bar u_N(p')
    \left[\gamma^\mu F_{1,N}(q^2) + \frac{\i\sigma^{\mu\nu}q_\nu}{2m_N}F_{2,N}(q^2)
    - \frac{\sigma^{\mu\nu}q_\nu\gamma^5}{2m_N}F_{3,N}(q^2) +\cdots \right] u_N(p),
    \label{eq:nucleon-EDFF}
\end{align}
where $q=p'-p$, $e>0$ denotes the magnitude of the elementary electric charge, and $F_{1,N}$ and $F_{2,N}$ are the usual Dirac and Pauli form factors. With the normalisation adopted in Eq.~\eqref{eq:nucleon-EDFF}, $F_{3,N}$ is dimensionless and its zero-momentum limit determines the nucleon EDM,
\begin{align}
    d_N = \frac{e}{2m_N}F_{3,N}(0).
    \label{eq:nucleon-EDM-F3}
\end{align}
At sufficiently low momentum, the physical EDM can equivalently be represented by
\begin{align}
    \mathcal{L}_{\rm EDM} = -\frac{\i}{2}d_N\, \bar N\sigma^{\mu\nu}\gamma^5N\,F_{\mu\nu}, \label{eq:nucleon-EDM-operator}
\end{align}
which reduces in the nonrelativistic limit to $\mathcal{L}_{\rm EDM}=d_N N^\dagger\bm{\sigma}\cdot\bm{E}\,N$, or equivalently to $H_{\rm EDM}=-d_N\bm{\sigma}\cdot\bm{E}$ for a single nucleon. Here $d_N$ denotes the full physical EDM and should not be confused with the short-distance low-energy constants $\bar d_{0,1}^{\,r}$ introduced in Eq.~\eqref{eq:CP-odd-Ngamma}. Different conventions for the overall sign of $F_{3,N}$, and for whether the electric charge $e$ is included in the definition of the current or of the form factor, are common in the literature.

\begin{figure}[ht!]
\centering
\begin{tikzpicture}[baseline={-0.025cm*height("$=$")}]
  \begin{feynman}
    \vertex (a) at (-3,0);
    \vertex (i1) at (-1.5,0);
    \vertex [dot] (i2) at (1.5,0);
    \vertex (b) at (3,0);
    \vertex [dot] (i3) at (0,1.5);
    \vertex (c) at (0,3);

    \diagram* {
      (a) -- [fermion, edge label'=\(n\)] (i1)
          -- [fermion, edge label'=\(p\)] (i2)
          -- [fermion, edge label'=\(n\)] (b),
      (i1) -- [charged scalar, quarter left, edge label'=\(\pi^-\)] (i3)
           -- [charged scalar, quarter left, edge label'=\(\pi^-\)] (i2),
      (i3) -- [photon, edge label'=\(\gamma\)] (c),
    };

    \node at (i1) {\(\times\)};
  \end{feynman}
\end{tikzpicture}
\qquad + \qquad
\begin{tikzpicture}[baseline={-0.025cm*height("$=$")}]
  \begin{feynman}
    \vertex (a) at (-3,0);
    \vertex [dot] (i1) at (-1.5,0);
    \vertex (i2) at (1.5,0);
    \vertex (b) at (3,0);
    \vertex [dot] (i3) at (0,1.5);
    \vertex (c) at (0,3);

    \diagram* {
      (a) -- [fermion, edge label'=\(n\)] (i1)
          -- [fermion, edge label'=\(p\)] (i2)
          -- [fermion, edge label'=\(n\)] (b),
      (i1) -- [charged scalar, quarter left, edge label'=\(\pi^-\)] (i3)
           -- [charged scalar, quarter left, edge label'=\(\pi^-\)] (i2),
      (i3) -- [photon, edge label'=\(\gamma\)] (c),
    };

    \node at (i2) {\(\times\)};
  \end{feynman}
\end{tikzpicture}
\caption{Representative leading pion-loop contributions to the neutron electric dipole form factor. The crossed vertex denotes the $CP$-odd pion-nucleon interaction proportional to $\bar g_0$, while the uncrossed pion-nucleon vertex denotes the ordinary $CP$-even axial coupling proportional to $g_A$. }
\label{fig:edm-pion-loop}
\end{figure}
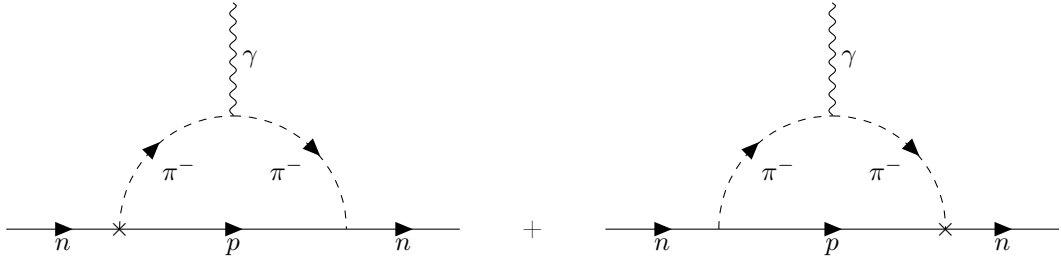

The chiral EFT ingredients entering the nucleon EDM were introduced in Sec.~\ref{sec:CP-odd-hadronic-interactions}. The local operators proportional to $\bar d_{0,1}^{\,r}$ contribute directly at tree level, while long-distance contributions arise from pion loops involving one $CP$-odd pion-nucleon vertex and one ordinary $CP$-even pion-nucleon vertex, as illustrated for the neutron in Fig.~\ref{fig:edm-pion-loop}. In heavy-baryon ChPT, the leading $CP$-even one-pion interaction is
\begin{align}
    \mathcal{L}_{\pi N}^{CP\text{-even}}
    =
    -\frac{g_A}{F_\pi}\,
    \bar N
    S^\mu
    \bm{\tau}\cdot\partial_\mu\bm{\pi}\,
    N
    +\cdots,
    \label{eq:CP-even-pion-nucleon}
\end{align}
where $g_A\simeq1.27$ is the nucleon isovector axial charge and $S^\mu$ is the covariant heavy-baryon spin vector, with $S^\mu=(0,\bm{\sigma}/2)$ in the nucleon rest frame~\cite{Jenkins:1990jv,Bernard:1995dp,Hockings:2005cn}. Together with the $CP$-odd coupling $\bar g_0$ in Eq.~\eqref{eq:CP-odd-pion-nucleon}, this interaction generates the leading nonanalytic pion-loop contribution to the nucleon EDM.

It is useful to introduce the physical isoscalar and isovector combinations
$d_0=(d_p+d_n)/2$ and $d_1=(d_p-d_n)/2$. They can be decomposed as
\begin{align}
    d_0 = \bar d_0^{\,r}(\mu) + d_0^{\rm loop}(\mu),
    \qquad
    d_1 = \bar d_1^{\,r}(\mu) + d_1^{\rm loop}(\mu),
    \label{eq:nucleon-edm-decomposition}
\end{align}
The separation between local and loop contributions is renormalisation-scheme and scale dependent, whereas the physical combinations $d_0$ and $d_1$ are independent of $\mu$. The significance of the pion loops is that they contain nonanalytic dependence on the light-quark masses and momentum that cannot be mimicked by local operators and therefore represents genuine long-distance information~\cite{Crewther:1979pi,Hockings:2005cn,Mereghetti:2010kp}.

At leading one-loop order and in the isospin-symmetric limit, the nonanalytic pion contribution is purely isovector. It arises from one insertion of the axial coupling $g_A$ and one insertion of the $CP$-odd coupling $\bar g_0$~\cite{Crewther:1979pi,Hockings:2005cn,Mereghetti:2010kp}. In the convention of Eq.~\eqref{eq:CP-odd-pion-nucleon}, the renormalised loop contribution contains the leading nonanalytic term
\begin{align}
    d_1^{\rm loop}(\mu)
    \supset
    -\frac{e\,g_A\bar g_0}{8\pi^2F_\pi}
    \log\frac{\mu^2}{m_\pi^2}.
    \label{eq:edm-chiral-log}
\end{align}
The coefficient of the logarithmic dependence on $m_\pi$ is fixed by chiral symmetry, while the $\mu$ dependence is cancelled by that of the renormalised short-distance coefficient $\bar d_1^{\,r}(\mu)$. At subleading order, recoil and isospin-breaking corrections modify the leading isovector result, and the EDFF develops additional nonanalytic structure, including an isoscalar contribution~\cite{Ottnad:2009jw,Mereghetti:2010kp}. Thus, ChPT predicts the characteristic long-distance quark-mass and momentum dependence of the nucleon EDFF, but not the complete static EDMs: the latter also depend on the independent short-distance coefficients $\bar d_{0,1}^{\,r}$ and therefore require additional nonperturbative input.

As an illustration, a covariant baryon-ChPT analysis used lattice-QCD data to constrain the relevant short-distance low-energy constants and obtained nucleon EDMs of order $10^{-16}\bar\theta\,e\,{\rm cm}$ at the physical point~\cite{Guo:2012vf}. Ultimately, however, a first-principles determination of the complete nucleon EDM requires nonperturbative QCD.

Direct lattice simulations at a real nonzero $\bar\theta$ are complicated by the complex Euclidean weight. In the phenomenologically relevant regime $|\bar\theta|\ll1$, one can instead expand around the $CP$-conserving theory,
\begin{align}
    \langle\mathcal{O}\rangle_{\bar\theta}
    =
    \langle\mathcal{O}\rangle_{0}
    +
    \i\bar\theta
    \langle\mathcal{O}Q\rangle_{0}
    +
    \mathcal{O}(\bar\theta^2),
    \label{eq:lattice-small-theta}
\end{align}
where $\langle Q\rangle_0=0$ has been used. The EDM may then be extracted from the $CP$-odd form factor $F_3(q^2)$ or, alternatively, from the spin-dependent energy shift in a background electric field. Important systematic uncertainties include topological sampling, excited-state contamination, finite-volume and discretisation effects, and the extrapolations to the physical quark masses and, for form-factor calculations, to $q^2=0$~\cite{Liu:2024kqy}.

Representative determinations of the $\bar\theta$-induced nucleon EDMs are summarised in Table~\ref{tab:nucleon-edm-theory}. For ease of comparison, all results are expressed in units of $10^{-3}\bar\theta\,e\,{\rm fm}$. The first two direct lattice determinations in Table~\ref{tab:nucleon-edm-theory} (from Refs.~\cite{Dragos:2019oxn,Liang:2023jfj}) give very similar neutron central values, whereas the more recent background-field result is substantially larger in magnitude. In Ref.~\cite{Blum:2026hul}, the quoted systematic uncertainty includes excited-state and background-field effects, while discretisation, finite-volume, and chiral-extrapolation uncertainties have not yet been fully quantified. A precision first-principles determination of $d_n/\bar\theta$ therefore remains an active problem.

\begin{table}[t]
\centering
\begin{tabular}{lccc}
\hline
Method & $d_n$ & $d_p$ & Reference\\
\hline\hline
Covariant baryon ChPT + lattice input & $-2.9\pm0.9$ & $1.1\pm1.1$ & \cite{Guo:2012vf} \\
Lattice QCD, gradient-flow topological charge & $-1.52(71)$ & $1.1(1.0)$ & \cite{Dragos:2019oxn} \\
Lattice QCD, overlap fermions + CDER & $-1.48(14)(31)$ & $3.8(11)(8)$ & \cite{Liang:2023jfj} \\
Lattice QCD, background electric field & $-5.0(0.4)(0.8)$ & $---$ & \cite{Blum:2026hul} \\
\hline
\end{tabular}
\caption{Representative determinations of the nucleon EDMs induced by the QCD $\bar\theta$ term. The entries for $d_n$ and $d_p$ are given in units of $10^{-3}\bar\theta\,e\,{\rm fm}$. }
\label{tab:nucleon-edm-theory}
\end{table}

Experimentally, the strongest direct nucleon-EDM constraint is provided by the neutron. The current measurement gives~\cite{Abel:2020pzs}
\begin{align}
    d_n = \left( 0.0\pm1.1_{\rm stat}\pm0.2_{\rm sys}
    \right) \times10^{-26}\,e\,{\rm cm},
\end{align}
corresponding to
\begin{align}
    |d_n| < 1.8\times10^{-26}\,e\,{\rm cm}
    \qquad
    (90\%~{\rm C.L.}).
    \label{eq:neutron-edm-bound}
\end{align}
Using, for example, the result of Ref.~\cite{Liang:2023jfj} gives the representative constraint~\cite{Mulder:2025esr}
\begin{align}
    |\bar\theta|
    \lesssim
    1.2\times10^{-10}
    \qquad
    (90\%~{\rm C.L.}),
    \label{eq:theta-bound-neutron}
\end{align}
although the precise conversion depends on the nonperturbative determination of $d_n/\bar\theta$. It is therefore appropriate at present to summarise the neutron-EDM implication as the order-of-magnitude constraint $|\bar\theta|\lesssim10^{-10}$.

The nucleon EDMs also constitute important one-body contributions to the EDMs of composite nuclei, while the $CP$-odd pion-nucleon couplings introduced in Eq.~\eqref{eq:CP-odd-pion-nucleon} generate long-range $CP$-odd nuclear forces. Their interplay is discussed in the next subsection.

\subsection{Nuclear, atomic, and molecular electric dipole moments}
\label{sec:nuclear-atomic-molecular-EDMs}

The effects of a nonzero $\bar\theta$ are not restricted to individual nucleons. The hadronic interactions discussed in Secs.~\ref{sec:CP-odd-hadronic-interactions} and~\ref{sec:nucleon-EDMs}, in particular the intrinsic nucleon EDMs and the $CP$-odd pion-nucleon couplings, induce $CP$-odd nuclear forces and moments, which in turn generate observable effects in atomic and molecular systems. EDM measurements across these different systems therefore provide complementary probes of hadronic $CP$ violation~\cite{Engel:2013lsa,Yamanaka:2016umw,Chupp:2017rkp,Safronova:2017xyt,Engel:2025uci,deVries:2026quu}.

We begin with nuclear EDMs. At low energies, the relevant hadronic ingredients are the intrinsic nucleon EDMs, the $CP$-odd pion-nucleon couplings $\bar g_{0,1,2}$ introduced in Eq.~\eqref{eq:CP-odd-pion-nucleon}, and short-range $CP$-odd nucleon-nucleon interactions. In chiral EFT, the leading long-range $CP$-odd nuclear force is generated by one-pion exchange involving one ordinary $CP$-even pion-nucleon vertex and one $CP$-odd vertex~\cite{Mereghetti:2010tp,deVries:2012ab,Bsaisou:2014oka}. Short-range effects are encoded in independent nucleon-nucleon contact operators. For the QCD $\bar\theta$ term, these contact interactions enter at higher chiral order and are expected to be subleading to one-pion exchange, whereas their relative importance can be substantially different for other microscopic sources of $CP$ violation~\cite{deVries:2012ab,Bsaisou:2014oka,deVries:2015una}.

It is useful to separate the nuclear EDM schematically into
\begin{align}
    d_A = d_A^{(N)} + d_A^{\rm pol} + d_A^{(2{\rm b})} +\cdots , 
    \label{eq:nuclear-EDM-decomposition}
\end{align}
where $d_A^{(N)}$ denotes the one-body contribution from the intrinsic proton and neutron EDMs, $d_A^{\rm pol}$ arises from the $CP$-odd polarization of the nuclear wave function, and $d_A^{(2{\rm b})}$ collects irreducible two-body electromagnetic-current contributions and other subleading corrections~\cite{Bsaisou:2014oka,Yamanaka:2016umw}. For a nucleus polarised along the $z$ axis, with an unperturbed ground state $|\Psi_0\rangle$ of definite parity, the leading one-body and polarization contributions may be written schematically as
\begin{align}
    d_A^{(N)}
    &=
    \left\langle\Psi_0\left| \sum_{i=1}^{A} \left(d_0+d_1\tau_i^3\right) \sigma_i^3 \right|\Psi_0\right\rangle ,
    \nonumber\\
    d_A^{\rm pol}
    &=
    2\,\operatorname{Re} \sum_{n\neq0} \frac{\langle\Psi_0|D_z|\Psi_n\rangle\langle\Psi_n|V_{\not P\not T}|\Psi_0\rangle}{E_0-E_n},
    \label{eq:nuclear-EDM-polarization}
\end{align}
where $d_0$ and $d_1$ are the physical isoscalar and isovector nucleon EDMs defined in Eq.~\eqref{eq:nucleon-edm-decomposition}, $D_z$ is the ordinary nuclear electric-dipole operator, and $V_{\not P\not T}$ denotes the $P$- and $T$-violating nuclear potential. The second line makes the origin of the polarization contribution transparent: $V_{\not P\not T}$ admixes opposite-parity excited states into the nuclear ground state, allowing the ordinary electric-dipole operator to acquire a nonvanishing expectation value.

Light nuclei are theoretically attractive because their few-body structure permits controlled calculations within chiral EFT and related ab initio approaches~\cite{Dekens:2014jka,Bsaisou:2014oka,Yamanaka:2016umw}. The deuteron provides a particularly instructive example. Up to small nuclear-structure corrections to the one-body contribution, its EDM may be written schematically as
\begin{align}
    d_D = d_n+d_p+d_D^{\rm pol}+\cdots .
    \label{eq:deuteron-EDM}
\end{align}
For the $\bar\theta$ source, the leading one-pion-exchange contribution proportional to the isoscalar coupling $\bar g_0$ vanishes because of the spin-isospin selection rules of the deuteron. The leading polarization contribution is therefore proportional to the isospin-breaking coupling $\bar g_1$, while higher-order effects generate only a small dependence on $\bar g_0$~\cite{Bsaisou:2012rg,Dekens:2014jka,Bsaisou:2014oka}. Consequently, the $\bar\theta$-induced deuteron EDM is expected to remain relatively close to the one-body contribution $d_n+d_p$, with a smaller but complementary contribution from the $CP$-odd nuclear force.

The three-nucleon systems $^3{\rm He}$ and $^3{\rm H}$ exhibit a different pattern. Their EDMs receive contributions from both the intrinsic nucleon EDMs and $CP$-odd nuclear forces proportional to $\bar g_0$ and $\bar g_1$. For the $\bar\theta$ source, chiral-EFT calculations find that $\bar g_0$ provides the dominant nuclear contribution, with a smaller contribution from $\bar g_1$~\cite{Dekens:2014jka,Bsaisou:2014oka}. The one-body contribution to the helion EDM is predominantly sensitive to $d_n$, whereas that of the triton is predominantly sensitive to $d_p$, reflecting the approximate pairing of the two proton spins in $^3{\rm He}$ and of the two neutron spins in $^3{\rm H}$. Their nuclear two-body contributions need not be small compared with the corresponding one-body terms~\cite{Dekens:2014jka,Bsaisou:2014oka}. Measurements of several light-nuclear EDMs would therefore probe complementary combinations of $d_n$, $d_p$, $\bar g_0$, $\bar g_1$, and short-range interactions, and could help distinguish the QCD $\bar\theta$ term from other microscopic sources of $CP$ violation~\cite{Dekens:2014jka,Bsaisou:2014oka,Yamanaka:2016umw}.

For neutral atoms, the relation between a nuclear EDM and the observable atomic EDM is more subtle. In the idealised limit of pointlike, nonrelativistic constituents interacting only through electrostatic forces, Schiff's theorem implies that the nuclear EDM is completely screened by the rearrangement of the atomic electrons~\cite{Engel:2013lsa,Chupp:2017rkp,Engel:2025uci}. The finite spatial extent of the nucleus spoils this exact cancellation, leaving a residual $P$- and $T$-odd electrostatic interaction with the electrons whose leading contribution is parametrized by the nuclear Schiff moment. Treating the protons as point charges and neglecting intrinsic nucleon EDMs, the corresponding charge-distribution part of the Schiff operator is~\cite{Engel:2025uci}
\begin{align}
    \bm S^{\rm ch}
    = \frac{e}{10} \sum_{p=1}^{Z}
    \left(r_p^2-\frac{5}{3}\langle r^2\rangle_{\rm ch}
    \right) \bm r_p ,
    \label{eq:Schiff-moment-schematic}
\end{align}
where $\bm r_p$ denotes the position of the $p$th proton relative to the nuclear centre of mass and $\langle r^2\rangle_{\rm ch}$ is the mean-square nuclear charge radius. The subtraction term proportional to $\langle r^2\rangle_{\rm ch}$ is a direct consequence of Schiff screening. The complete Schiff operator also receives contributions from the intrinsic nucleon EDMs, together with smaller quadrupole and relativistic corrections~\cite{Engel:2025uci}.

The nuclear Schiff moment is induced by the same hadronic $CP$-odd interactions that contribute to nuclear EDMs. For a polarised nucleus, denoting by $S_A$ the component of the Schiff moment along the nuclear-spin quantisation axis, its dependence on the leading hadronic couplings may be written schematically as~\cite{Engel:2025uci}
\begin{align}
    S_A = a_0^{(A)}\bar g_0 + a_1^{(A)}\bar g_1 + a_2^{(A)}\bar g_2 + a_n^{(A)}d_n + a_p^{(A)}d_p +\cdots , 
    \label{eq:Schiff-hadronic-matching}
\end{align}
where the coefficients $a_i^{(A)}$ encode the nuclear many-body response in the convention for the $CP$-odd pion-nucleon couplings adopted in Eq.~\eqref{eq:CP-odd-pion-nucleon}. The ellipsis includes contributions from short-range $CP$-odd nucleon-nucleon interactions and higher-order operators. Once $S_A$ is determined, its contribution to the EDM of a given atom can be written as
\begin{align}
    d_{\rm atom} = k_S^{(A)}\,S_A +\cdots , 
    \label{eq:atomic-EDM-Schiff}
\end{align}
where $k_S^{(A)}$ is an atom-dependent coefficient determined by the electronic structure~\cite{Engel:2013lsa,Chupp:2017rkp}. Thus, relating an atomic EDM measurement to $\bar\theta$ involves a sequence of matching steps: the underlying QCD source determines the hadronic low-energy couplings, nuclear many-body dynamics converts these couplings into $S_A$, and atomic structure determines the response to the resulting Schiff moment.

The most stringent current EDM limit from a diamagnetic atom is provided by $^{199}{\rm Hg}$~\cite{Pospelov:2025vzj},
\begin{align}
    |d_{\rm Hg}| < 7.4\times10^{-30}\,e\,{\rm cm}
    \qquad
    (95\%~{\rm C.L.}),
    \label{eq:Hg-EDM-bound}
\end{align}
as obtained by Ref.~\cite{Graner:2016ses}. Translating this remarkable experimental sensitivity into a constraint on $\bar\theta$ requires both the atomic response to the nuclear Schiff moment and, more importantly, a nuclear-structure calculation of the $^{199}{\rm Hg}$ Schiff moment. The latter remains a major source of theoretical uncertainty: existing many-body calculations exhibit a substantial spread in the relevant Schiff-moment coefficients~\cite{Engel:2025uci,Pospelov:2025vzj}. Consequently, the mercury measurement can yield a constraint on $\bar\theta$ potentially comparable to that from the neutron EDM, but its interpretation is presently subject to significantly larger nuclear-structure uncertainties~\cite{Pospelov:2025vzj}.

Other important diamagnetic atoms include $^{129}{\rm Xe}$~\cite{Sachdeva:2019rkt,Allmendinger:2019jrk} and radioactive species such as $^{225}{\rm Ra}$~\cite{Parker:2015yka,Bishof:2016uqx}. The latter is particularly promising because its octupole-deformed nucleus can possess a large collective Schiff moment in the intrinsic frame. Moreover, low-lying nuclear states of opposite parity can be mixed efficiently by $P$- and $T$-violating interactions, allowing the intrinsic Schiff moment to appear in the laboratory frame. The small energy splitting between these parity partners provides an additional enhancement. As a result, the Schiff moment of octupole-deformed nuclei such as $^{225}{\rm Ra}$ can be substantially larger than that of approximately spherical nuclei such as $^{199}{\rm Hg}$~\cite{Engel:2013lsa,Engel:2025uci}.

Molecules provide additional opportunities because their closely spaced opposite-parity rotational levels allow them to be strongly polarised by relatively modest laboratory electric fields, thereby enhancing sensitivity to $P$- and $T$-violating interactions~\cite{Chupp:2017rkp,Safronova:2017xyt}. Molecules containing heavy nuclei with enhanced Schiff moments are therefore promising probes of hadronic $CP$ violation, particularly when octupole-deformed radioactive nuclei are involved. Another important hadronic $P$- and $T$-odd observable is the nuclear magnetic quadrupole moment (MQM), which can occur for nuclei with spin $I\geq1$ and can be probed efficiently in suitable molecules with nonzero electronic angular momentum~\cite{Chupp:2017rkp,deVries:2026quu}.

Traditionally, paramagnetic atoms and molecules are regarded primarily as probes of the electron EDM and $CP$-odd electron-nucleon interactions, whereas diamagnetic systems are more directly sensitive to hadronic $CP$ violation through nuclear moments~\cite{Chupp:2017rkp,Safronova:2017xyt}. This distinction, however, is not absolute. Hadronic $CP$ violation can induce scalar-pseudoscalar electron-nucleon interactions through hadronic and electromagnetic loops. A recent heavy-baryon ChPT analysis computed these interactions for the QCD $\bar\theta$ term and, using the ${\rm HfF}^{+}$ measurement, obtained the constraint~\cite{Mulder:2025esr}
\begin{align}
    |\bar\theta| < 1.5\times10^{-8}
    \qquad
    (90\%~{\rm C.L.}).
\end{align}
Although this bound is currently weaker than that inferred from the neutron EDM, it demonstrates that paramagnetic molecules provide an independent probe of hadronic $CP$ violation. A complementary EFT framework has recently been developed to relate paramagnetic-molecule observables to the intrinsic neutron and proton EDMs, with the required nuclear matrix elements calculated explicitly for BaF~\cite{Dekens:2025skl}. These developments broaden the range of atomic and molecular systems sensitive to hadronic $CP$ violation and further blur the traditional separation between paramagnetic and diamagnetic EDM searches; see Ref.~\cite{deVries:2026quu} for a recent overview.

No single EDM measurement provides a model-independent determination of $\bar\theta$. Different systems probe different combinations of the intrinsic nucleon EDMs, $CP$-odd pion-nucleon couplings, short-range nuclear interactions, and the corresponding nuclear, atomic, or molecular response coefficients. Light nuclei are particularly attractive because their nuclear structure can be treated comparatively systematically, while atomic and molecular systems can provide substantial experimental enhancement at the cost of additional matching and, in many cases, nuclear-structure uncertainties. This complementarity is important not only for improving constraints on $\bar\theta$, but also for identifying the microscopic source of $CP$ violation should a nonzero EDM eventually be observed. The associated theoretical uncertainties and experimental prospects are discussed in the following subsection.

\subsection{Theory uncertainties and experimental prospects}
\label{sec:EDM-uncertainties-prospects}

Within the conventional framework adopted throughout this section, interpreting EDM measurements in terms of $\bar\theta$ requires a sequence of theoretical matching steps,
\begin{align}
    \bar\theta 
    \longrightarrow
    \left\{d_n,d_p,\bar g_0,\bar g_1,\ldots\right\}
    \longrightarrow
    \left\{d_A,S_A,\ldots\right\}
    \longrightarrow
    d_{\rm atom/mol}.
    \label{eq:EDM-matching-chain}
\end{align}
These steps involve, respectively, nonperturbative QCD, nuclear many-body dynamics, and atomic or molecular structure, and the dominant theoretical uncertainty depends strongly on the system under consideration~\cite{Engel:2013lsa,Chupp:2017rkp,Pospelov:2025vzj,deVries:2026quu}.

At the hadronic level, the situation is uneven. The leading $CP$-odd pion-nucleon coupling $\bar g_0$ is comparatively well constrained through its chiral relation to the strong neutron-proton mass splitting~\cite{Mereghetti:2010tp,deVries:2015una}, whereas the isospin-breaking coupling $\bar g_1$ receives larger higher-order and short-distance uncertainties~\cite{Bsaisou:2012rg,Mulder:2025esr}. The nucleon EDMs remain a more substantial challenge. As discussed in Sec.~\ref{sec:nucleon-EDMs}, chiral symmetry determines their characteristic nonanalytic dependence on the light-quark masses, but the short-distance coefficients $\bar d_{0,1}^{\,r}$ are not fixed by symmetry. A first-principles determination therefore requires lattice QCD, for which the spread among current calculations shows that a precision determination of $d_{n,p}/\bar\theta$ has not yet been reached~\cite{Liu:2024kqy,Blum:2026hul}.

The subsequent nuclear and atomic matching introduces a different hierarchy of uncertainties. For light nuclei such as the deuteron, helion, and triton, chiral EFT combined with few-body and ab initio methods provides a comparatively controlled treatment of nuclear structure~\cite{Dekens:2014jka,Bsaisou:2014oka,Yamanaka:2016umw}. Their interpretation is therefore especially sensitive to improved determinations of the underlying hadronic couplings $d_{n,p}$ and $\bar g_{0,1}$. In heavy diamagnetic atoms, by contrast, the nuclear Schiff moment can constitute the dominant theoretical uncertainty. This is particularly important for $^{199}{\rm Hg}$, for which existing many-body calculations show a substantial spread, while recent beyond-mean-field studies continue to improve the treatment of nuclear correlations~\cite{Engel:2025uci,Pospelov:2025vzj,Zhou:2025jfi}. The corresponding atomic and molecular response coefficients are generally better controlled than the nuclear Schiff moments, although their precision remains system dependent~\cite{Safronova:2017xyt,Pospelov:2025vzj}.

Experimentally, substantial improvements are being pursued in several complementary directions. Next-generation neutron searches such as n2EDM at PSI and TUCAN at TRIUMF target sensitivities around $10^{-27}\,e\,{\rm cm}$, approximately an order of magnitude beyond the present limit~\cite{n2EDM:2021yah,TUCAN:2025rjm}. Storage rings offer direct access to charged hadrons: the proposed proton EDM experiment targets a sensitivity near $10^{-29}\,e\,{\rm cm}$~\cite{pEDM:2022ytu}, while the recent first direct limit on the deuteron EDM,
$|d_D|<2.5\times10^{-17}\,e\,{\rm cm}$ at $95\%$ confidence level, demonstrates the feasibility of the storage-ring approach~\cite{JEDI:2026dwx}. Atomic and molecular searches provide complementary opportunities, particularly through enhanced Schiff moments in octupole-deformed nuclei and radioactive molecules~\cite{Engel:2025uci,Arrowsmith-Kron:2023hcr}. As discussed in Sec.~\ref{sec:nuclear-atomic-molecular-EDMs}, paramagnetic molecules have also emerged as probes of hadronic $CP$ violation through induced electron-nucleon interactions and intrinsic nucleon EDMs~\cite{Mulder:2025esr,Dekens:2025skl,deVries:2026quu}.

This experimental diversity is important because a nonzero EDM in a single system would establish $P$ and $T$ violation but would not, by itself, identify the microscopic source. The QCD $\bar\theta$ term predicts a correlated pattern among the nucleon EDMs, $CP$-odd pion-nucleon couplings, nuclear EDMs, and Schiff moments, while higher-dimensional sources of $CP$ violation generally lead to different hierarchies of hadronic interactions~\cite{deVries:2012ab,Bsaisou:2014oka,Pospelov:2025vzj}. Measurements across several systems, combined with sufficiently precise hadronic and nuclear theory, are therefore essential both for strengthening constraints on $\bar\theta$ and for diagnosing the origin of any future EDM signal.

At present, the neutron EDM implies the familiar order-of-magnitude constraint $|\bar\theta|\lesssim10^{-10}$ within the conventional framework~\cite{Abel:2020pzs}. Further progress requires advances in experiment and theory in parallel: improved EDM sensitivities must be accompanied by more precise lattice-QCD determinations of nucleon observables, controlled calculations of light-nuclear EDMs and heavy-nuclear Schiff moments, and accurate atomic and molecular response coefficients. As throughout Sec.~\ref{sec:theta-CP-effects}, this phenomenological interpretation assumes the conventional $\bar\theta$-dependent matching discussed in Sec.~\ref{sec:chiral-perturbation-theory}; the underlying conceptual assumptions will be revisited in Part~\ref{part:reassessment}.

\part{Solutions of the strong \texorpdfstring{$CP$}{CP} problem}
\label{part:solutions}

\section{What counts as a solution?}
\label{sec:what-counts-as-solution}

\subsection{Eliminating, enforcing, or dynamically relaxing \texorpdfstring{$\bar\theta$}{theta-bar}}
\label{sec:mechanisms-strong-CP-solutions}

In this part of the review, we use the term ``solution'' in its conventional sense: a mechanism that explains why the parameter controlling strong $CP$ violation is absent or sufficiently small, rather than simply taking $|\bar\theta|\lesssim10^{-10}$ as an unexplained input. The possibility that the conventional interpretation of $\bar\theta$ itself may require reassessment will be considered separately in Part~\ref{part:reassessment}. Within the conventional framework, proposed solutions can be organised according to what happens to the physical strong $CP$ phase. There are three logically distinct possibilities: $\bar\theta$ may cease to be a physical parameter, it may be constrained to vanish or be very small by a symmetry of the underlying theory, or it may be promoted to a dynamical quantity whose vacuum value relaxes to the $CP$-conserving point~\cite{Cheng:1987gp,Peccei:2006as,Hook:2018dlk}.

The first possibility is to eliminate the physical parameter altogether. If one quark is exactly massless, an anomalous chiral rotation of that quark can shift the QCD vacuum angle without introducing a compensating phase into its mass. Different values of $\theta$ are then related by a field redefinition, and the vacuum angle is unphysical~\cite{Crewther:1979pi,Leutwyler:1992yt}.\footnote{The massless-quark mechanism is not the only way to render the strong $CP$ phase unphysical. Enlarged gauge structures or additional anomalous symmetries can provide alternative realisations in which the QCD vacuum angle, or an appropriate combination of vacuum angles, becomes unphysical~\cite{Aldazabal:2002py,Hsu:2004mf,Hook:2014cda}.} This is conceptually stronger than explaining why a physical parameter happens to be small: there is no distinguished value of $\bar\theta$ that needs to be selected. In phenomenological discussions this possibility is usually associated with a massless up quark. Whether QCD permits such a possibility is a quantitative question about the light-quark masses, which we discuss in Sec.~\ref{sec:massless-quark-possibility}.

A second strategy is to enforce strong $CP$ conservation through a symmetry imposed on the ultraviolet theory. The most prominent examples invoke $CP$ or parity. Since $G^a_{\mu\nu}\widetilde G^{a\mu\nu}$ is odd under both $P$ and $CP$, an exact $CP$ or parity symmetry forbids the corresponding $\theta$ term. The quark-mass sector must simultaneously be constrained such that no unacceptable phase is generated in $\arg\det\M$. The challenge is that $CP$ violation is observed in the weak interactions, and the ultraviolet symmetry must therefore be broken while preserving the required suppression of $\bar\theta$. In the Nelson--Barr construction, $CP$ is an exact symmetry of the underlying theory and is broken spontaneously; the quark-mass matrix is arranged such that an unsuppressed CKM phase can be generated while its determinant remains real at tree level~\cite{Nelson:1983zb,Barr:1984qx}. Parity-based constructions instead use a parity symmetry, often realised in a left-right-symmetric or mirror framework, to constrain both the fundamental $\theta$ parameter and the structure of the quark mass matrices~\cite{Mohapatra:1978fy}. In both cases, the smallness of strong $CP$ violation is therefore traced to a symmetry of the ultraviolet theory rather than to a small parameter chosen directly in low-energy QCD.

The third possibility is dynamical relaxation. The paradigmatic example is the PQ mechanism~\cite{Peccei:1977hh}, in which an anomalous global $U(1)_{\rm PQ}$ symmetry is spontaneously broken. The associated pseudo-Nambu--Goldstone boson, the axion~\cite{Weinberg:1977ma,Wilczek:1977pj}, couples to the QCD topological charge density so that the effective vacuum angle becomes field dependent. In a normalisation where the QCD anomaly coefficient is absorbed into the axion decay constant, one may write $\bar\theta_{\rm eff}(x)=\bar\theta+a(x)/f_a$. QCD dynamics then generates an axion potential $V(a)=E(\bar\theta_{\rm eff})$, where $E(\bar\theta)$ is the QCD vacuum energy discussed in Sec.~\ref{sec:chiral-perturbation-theory}. Within the conventional $\bar\theta$-dependent description of QCD, this potential is minimised at $\langle\bar\theta_{\rm eff}\rangle=0$ modulo $2\pi$. The PQ mechanism therefore does not require the original parameter $\bar\theta$ to be small: instead, the physical effective angle is driven dynamically towards the $CP$-conserving vacuum.

None of these mechanisms, however, is sufficient by itself to establish a realistic solution. In the massless-quark scenario, the vanishing quark mass must remain exact if the vacuum angle is to remain unphysical; a $CP$- or parity-based construction must ensure that symmetry breaking and quantum corrections do not regenerate an unacceptably large $\bar\theta$; and the PQ mechanism must be protected against additional sources of explicit PQ breaking that could appreciably shift the axion minimum away from the $CP$-conserving point. Thus, beyond identifying a mechanism that eliminates, suppresses, or dynamically relaxes $\bar\theta$ at leading order, one must ask whether its effect remains stable once quantum corrections are included. We turn to this question next.

\subsection{Radiative stability}
\label{sec:radiative-stability}

Explaining why $\bar\theta$ vanishes or is small at tree level is not, by itself, sufficient. Since the experimental constraint requires $|\bar\theta|\lesssim10^{-10}$, quantum corrections generated after the symmetry responsible for suppressing $\bar\theta$ is broken must remain extraordinarily small. Radiative stability is therefore a central criterion for any proposed solution of the strong $CP$ problem~\cite{Cheng:1987gp,Hook:2018dlk}.

The origin of radiative corrections to the strong $CP$ phase can be seen directly from the definition of $\bar\theta$. For a nonsingular quark mass matrix $\M$, a small correction $\M\to\M+\delta\M$, accompanied in general by $\theta\to\theta+\delta\theta$, gives
\begin{align}
    \delta\bar\theta = \delta\theta + {\rm Im}\,\tr\left(\M^{-1}\delta\M \right) +\mathcal{O}(\delta\M^2),
    \label{eq:radiative-theta}
\end{align}
where we have used $\delta\log\det\M=\tr(\M^{-1}\delta\M)$. Thus, even if a symmetry enforces $\bar\theta=0$ at some scale, complex radiative corrections to the quark mass matrix can regenerate a nonzero strong $CP$ phase once that symmetry is broken. The same observation applies to threshold corrections generated when heavy degrees of freedom are integrated out. Equation~\eqref{eq:radiative-theta} therefore provides a useful diagnostic for the symmetry-based solutions discussed below.

It is useful here to distinguish radiative stability from an explanation of the smallness of $\bar\theta$. Within the Standard Model itself, once a small value of $\bar\theta$ is chosen as a boundary condition, the observed CKM phase does not generically drive it back to an order-one value. The weak contribution to $\bar\theta$ is strongly suppressed by the flavour structure of the Standard Model and arises only at high perturbative order; classic estimates find effects many orders of magnitude below the present bound~\cite{Ellis:1978hq}. The strong $CP$ problem is therefore not that a small $\bar\theta$ is destabilised by large Standard-Model radiative corrections. Rather, the Standard Model provides no known reason for the renormalised value of $\bar\theta$ to be as small as $10^{-10}$ in the first place. A viable solution should explain the small boundary value while preserving it against corrections from whatever new physics is introduced to provide that explanation.

For the massless-quark possibility, radiative stability follows directly if the vanishing quark mass is enforced by an exact chiral symmetry. Perturbative corrections cannot generate an operator forbidden by the symmetry, and the chiral rotation that removes the vacuum angle remains available. The principal difficulty of this proposal is therefore not radiative instability but whether an exactly vanishing light-quark mass is compatible with QCD phenomenology. We return to this question in Sec.~\ref{sec:massless-quark-possibility}.

The issue is more restrictive for solutions based on spontaneously broken $CP$ or parity. Above the symmetry-breaking scale, the symmetry can enforce $\bar\theta=0$. Below that scale, however, the same breaking must ultimately produce the observed weak $CP$ violation, and the complex parameters responsible for the CKM phase can feed back into the strong $CP$ phase through radiative corrections. The success of such a construction therefore depends not merely on obtaining $\bar\theta=0$ at tree level, but on arranging its flavour and symmetry structure so that the combination in Eq.~\eqref{eq:radiative-theta} vanishes or is sufficiently suppressed at the first several perturbative orders. This is a central aspect of Nelson--Barr constructions~\cite{Nelson:1983zb,Barr:1984qx}, in which special structures of the enlarged quark mass matrix make its determinant real at tree level but radiative corrections generally generate a nonzero value. The size and loop order of these corrections are model dependent and have played an important role in assessing realistic Nelson--Barr models. Similar considerations apply to parity-based solutions. In particular constructions, the leading contribution to $\bar\theta$ can be postponed to two loops and remain below the experimental limit, illustrating how the symmetry and particle content can provide additional suppression.

The PQ mechanism has a somewhat different notion of radiative stability. Because the effective angle is a dynamical field, corrections that respect the PQ symmetry do not simply generate an arbitrary constant $\bar\theta$; the axion continues to relax in the full effective potential. The dangerous effects are instead interactions that explicitly break the PQ symmetry independently of the QCD anomaly. If such interactions generate an additional axion potential whose minimum is not aligned with the QCD minimum, the vacuum value of $\bar\theta_{\rm eff}$ is shifted away from zero. The requirement that this shift remain below approximately $10^{-10}$ is the essence of the axion quality problem~\cite{DiLuzio:2020wdo}, which will be discussed in detail in Sec.~\ref{sec:axion-quality-problem}.

\subsection{UV robustness and phenomenological viability}
\label{sec:UV-robustness-viability}

Radiative stability addresses whether a proposed solution survives quantum corrections within a given theory. A logically separate question is whether the structure responsible for suppressing $\bar\theta$ remains robust once the theory is regarded as an EFT embedded in a more complete ultraviolet description. Unless forbidden by an exact symmetry, all operators compatible with the microscopic gauge symmetries are expected to appear in the effective Lagrangian, suppressed by appropriate powers of the ultraviolet scale. A convincing solution should therefore not rely on the unexplained absence or extreme suppression of operators that would generically regenerate an unacceptable strong $CP$ phase.

This consideration is particularly transparent for mechanisms based on global symmetries. The PQ symmetry must be violated by the QCD anomaly, but any additional explicit breaking generically contributes to the axion potential. For example, if a PQ-charged field $\Phi$ acquires a vacuum expectation value (VEV) of order $f_a$, a higher-dimensional interaction of the schematic form
\begin{align}
    \Delta\L_{\rm UV} = \frac{c_d}{\Lambda_{\rm UV}^{\,d-4}}\, \Phi^d+\hc
    \label{eq:PQ-breaking-UV-operator}
\end{align}
can generate an additional contribution to the axion potential with a phase unrelated to the QCD contribution. Even when $\Lambda_{\rm UV}$ is very large, the required bound on the residual effective angle, $|\bar\theta_{\rm eff}|\lesssim10^{-10}$, makes the PQ mechanism unusually sensitive to such operators~\cite{Barr:1992qq,Holman:1992us}. This observation motivates constructions in which the PQ symmetry arises accidentally from gauge or discrete symmetries, so that the leading PQ-violating operators occur only at sufficiently high dimension. We postpone a quantitative discussion of this axion quality problem to Sec.~\ref{sec:axion-quality-problem}.

Solutions based on $CP$ or parity face a related, although structurally different, issue. If $CP$ or parity is an exact symmetry of the ultraviolet theory, operators generated above the symmetry-breaking scale must respect it. Once the symmetry is spontaneously broken, however, higher-dimensional interactions connecting the symmetry-breaking sector to the Standard Model can transmit new complex phases to the quark mass matrices. A tree-level texture that gives $\arg\det(\M_u\M_d)=0$ may therefore cease to do so after additional heavy states are integrated out. The relevant question is not merely whether a particular renormalisable Lagrangian possesses the desired mass-matrix structure, but whether this structure follows sufficiently robustly from the symmetries of the ultraviolet theory. This requirement is especially important in Nelson--Barr models, where generic interactions between the $CP$-breaking sector and the Standard Model can spoil the special determinant structure unless they are forbidden or sufficiently suppressed~\cite{Dine:2015jga}. Analogous considerations apply to parity-based solutions, where the pattern of parity breaking and the heavy sector determine how strongly the low-energy theory departs from the parity-symmetric boundary conditions.

The ultraviolet question is therefore closely related to naturalness, but it is useful to formulate the criterion more concretely. A proposed solution should not merely replace the small parameter $\bar\theta$ by another comparably severe and unexplained small parameter. For example, if obtaining $|\bar\theta|\lesssim10^{-10}$ requires a dimensionless coupling in the ultraviolet theory to be tuned to one part in $10^{10}$ without a symmetry or dynamical reason, the original puzzle has largely been transferred rather than resolved. Some degree of hierarchy among scales or couplings may of course arise naturally from symmetry breaking, dimensional transmutation, approximate symmetries, or other dynamical mechanisms. The relevant question is whether the small quantities required by the construction are themselves technically and structurally explained.

A successful solution must also be phenomenologically viable. The mechanisms discussed in the following sections necessarily introduce additional assumptions or degrees of freedom beyond ordinary low-energy QCD. A massless-up-quark solution makes a definite statement about the light-quark mass spectrum; Nelson--Barr models typically introduce additional quarks and a sector responsible for spontaneous $CP$ breaking; parity solutions require an enlarged gauge, fermion, or scalar sector associated with the restoration of parity at high energies; and the PQ mechanism predicts the axion together with the dynamics responsible for PQ breaking. These ingredients are constrained, depending on the construction, by flavour observables, collider searches, precision measurements, astrophysics, and cosmology. Phenomenological consistency is therefore part of the assessment of a solution rather than an optional consideration added after the strong $CP$ problem has been addressed.

Taken together, these considerations provide a set of complementary criteria for assessing proposed solutions to the strong $CP$ problem: the mechanism should explain why $\bar\theta$ is absent or sufficiently small, this explanation should remain stable under radiative and ultraviolet corrections, and the resulting theory should be phenomenologically viable without introducing an equivalent unexplained tuning elsewhere. With these criteria in mind, we now turn to the principal proposed solutions, beginning with the conceptually simplest possibility that one of the light quarks is exactly massless, in which case the strong $CP$ phase becomes unphysical.

\section{The massless-quark possibility}
\label{sec:massless-quark-possibility}

\subsection{The up-quark mass and the Kaplan--Manohar ambiguity}
\label{sec:Kaplan-Manohar-ambiguity}

The phenomenological question underlying the massless-quark solution is whether the up-quark mass can in fact vanish. Quark masses are parameters of the QCD Lagrangian rather than directly observable quantities, and historically their values were inferred primarily from the pattern of explicit chiral-symmetry breaking in the pseudoscalar-meson spectrum. At leading order in three-flavour ChPT, taking the quark mass matrix to be real and diagonal, $\M={\rm diag}(m_u,m_d,m_s)$, the pion and kaon masses satisfy~\cite{Gasser:1984gg}
\begin{align}
    m_{\pi^{\pm}}^2 = B_0(m_u+m_d),\qquad
    m_{K^{\pm}}^2 = B_0(m_u+m_s),\qquad
    m_{K^0}^2 = B_0(m_d+m_s),
    \label{eq:LO-meson-masses-quark-masses}
\end{align}
where electromagnetic effects and higher-order chiral corrections have been omitted. After accounting for electromagnetic mass splittings, these relations lead to Weinberg's leading-order estimates $m_u/m_d\simeq0.56$ and $m_s/m_d\simeq20$~\cite{Weinberg:1977hb}. Taken literally, the first of these would appear to exclude a massless up quark.

The difficulty is that the extraction of individual quark-mass ratios becomes less direct beyond leading order. In the standard chiral power counting, a quark-mass insertion counts as two powers of the soft scale, $B_0\M=\mathcal{O}(p^2)$. The next-to-leading-order chiral Lagrangian, of $\mathcal{O}(p^4)$, therefore contains operators quadratic in the quark masses, whose coefficients include the low-energy constants $L_6$, $L_7$, and $L_8$~\cite{Gasser:1984gg}. Kaplan and Manohar observed that, for three flavours, there exists a reparametrisation of the quark mass matrix that cannot be distinguished by the pseudoscalar-meson observables at this order~\cite{Kaplan:1986ru}. In our convention for the mass matrix, it may be written as
\begin{align}
    \M \rightarrow \M' = \M + \lambda_{\rm KM}\,{\rm adj}\!\left(\M^\dagger\right),
    \label{eq:Kaplan-Manohar-transformation}
\end{align}
where ${\rm adj}(\M^\dagger)$ denotes the adjugate matrix. For a nonsingular matrix, ${\rm adj}(\M^\dagger)=(\det\M^\dagger)(\M^\dagger)^{-1}$. Since $L,R\in SU(3)$ have unit determinant, ${\rm adj}(\M^\dagger)$ transforms in the same way as $\M$ under $SU(3)_L\times SU(3)_R$. Moreover, it is quadratic in the quark masses and therefore contributes at $\mathcal{O}(p^4)$.

For real diagonal quark masses, Eq.~\eqref{eq:Kaplan-Manohar-transformation} takes the particularly transparent form
\begin{align}
    m_u\rightarrow m_u'=m_u+\lambda_{\rm KM}m_dm_s,
    \quad
    m_d\rightarrow m_d'=m_d+\lambda_{\rm KM}m_um_s,
    \quad
    m_s\rightarrow m_s'=m_s+\lambda_{\rm KM}m_um_d. \label{eq:Kaplan-Manohar-diagonal}
\end{align}
The change induced in the leading-order mass term is of $\mathcal{O}(p^4)$ and can be compensated by corresponding shifts of the next-to-leading-order low-energy constants $L_6$, $L_7$, and $L_8$. Consequently, pseudoscalar masses, decay constants, and related low-energy observables computed through this order cannot separately determine $\lambda_{\rm KM}$ and the quark masses. This is known as the Kaplan--Manohar ambiguity.

The relevance to the strong $CP$ problem is immediate. By an appropriate choice of $\lambda_{\rm KM}$, a parameter choice with $m_u=0$ can be mapped, within the effective theory, onto another parametrisation with $m_u'\neq0$, while leaving the relevant low-energy observables unchanged through the order considered. Kaplan and Manohar showed explicitly that a massless up quark could be accommodated by the pseudoscalar spectrum if sufficiently large next-to-leading-order chiral corrections were allowed~\cite{Kaplan:1986ru}. Thus, the apparent exclusion of $m_u=0$ from the leading-order mass relations was not, by itself, conclusive.

A useful combination that is particularly well constrained by the low-energy meson sector is~\cite{Leutwyler:1996qg}
\begin{align}
    Q^2 \equiv \frac{m_s^2-\bar m^2}{m_d^2-m_u^2},
    \qquad
    \bar m=\frac{m_u+m_d}{2}.
    \label{eq:Leutwyler-Q}
\end{align}
Although $Q$ is not strictly invariant under the Kaplan--Manohar reparametrisation, the corresponding variation is of second order in isospin breaking and is numerically very small. One may instead define a slightly modified ratio $\widetilde Q$ that is reparametrisation invariant to the order relevant here and differs from $Q$ only by terms proportional to $(m_d-m_u)^2$~\cite{Leutwyler:2023ldz}. To this accuracy, a fixed value of $Q$ therefore constrains the quark-mass ratios to the well-known Leutwyler ellipse in the $(m_u/m_d,m_s/m_d)$ plane~\cite{Leutwyler:1996qg}.

\begin{figure}[t]
    \centering
    \includegraphics[width=0.5\linewidth]{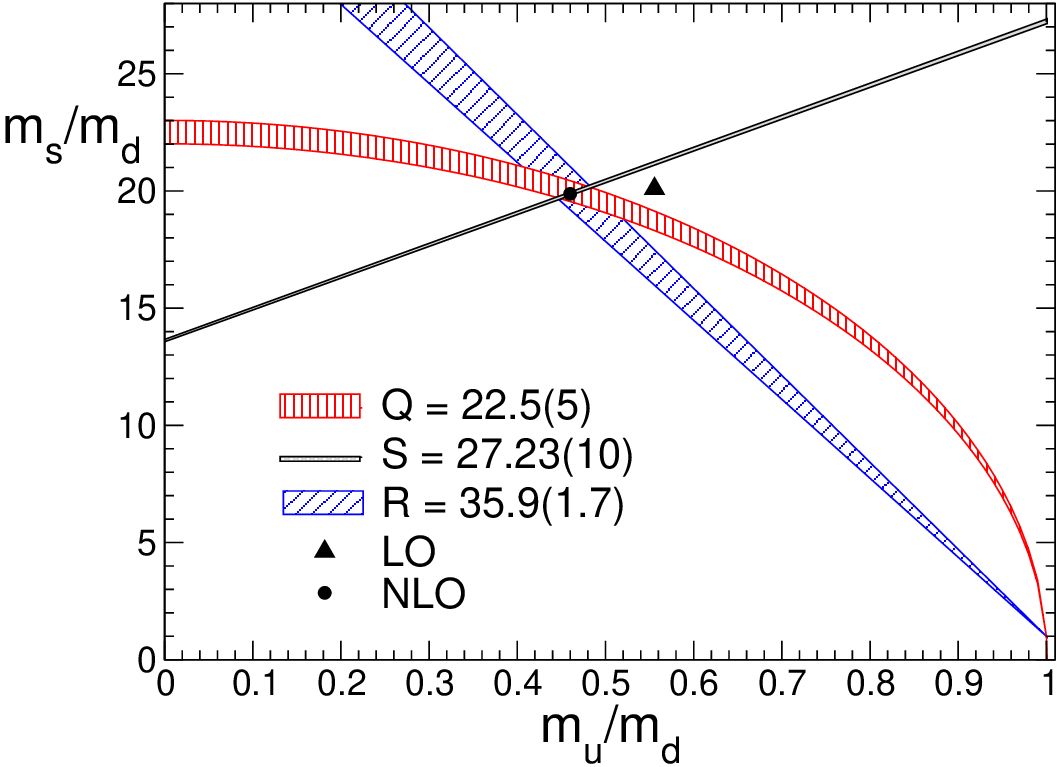}
    \caption{Constraints on the light-quark mass ratios in the $(m_u/m_d,m_s/m_d)$ plane. The red, black, and blue bands correspond to the ratios $Q$, $S$, and $R$, respectively, using the $N_f=2+1+1$ lattice averages quoted in Ref.~\cite{Leutwyler:2023ldz}. The triangle labelled LO denotes Weinberg's leading-order estimate, while the black dot labelled NLO denotes the determination obtained using the NLO chiral low-energy theorem discussed in Ref.~\cite{Leutwyler:2023ldz}. Reproduced from Fig.~2 of the arXiv version of Ref.~\cite{Leutwyler:2023ldz} under the CC BY 4.0 licence.}
    \label{fig:ellipse}
\end{figure}

Figure~\ref{fig:ellipse} illustrates the present constraints on the light-quark mass ratios. In addition to $Q$, it also displays $S\equiv m_s/\bar m$, $R\equiv (m_s-\bar m)/(m_d-m_u)$,
which measure, respectively, the hierarchy between the strange- and average light-quark masses and the relative strengths of $SU(3)$-flavour and isospin breaking. These ratios obey the exact relation $Q^2=R(S+1)/2$.
The red, black, and blue bands in Fig.~\ref{fig:ellipse} show the constraints from $Q$, $S$, and $R$, respectively. The particularly narrow $S$ band reflects the high precision with which $m_s/\bar m$ is determined on the lattice, whereas $R$ is less precisely known because it depends on the small isospin-breaking difference $m_d-m_u$ and is correspondingly more sensitive to electromagnetic effects~\cite{Leutwyler:2023ldz}. The triangle denotes the classic leading-order estimate obtained by Weinberg from the pseudoscalar-meson masses, using Dashen's theorem for the electromagnetic self-energies~\cite{Weinberg:1977hb,Dashen:1969eg}, while the black dot shows the corresponding NLO determination discussed in Ref.~\cite{Leutwyler:2023ldz}. The fact that the three modern constraints intersect in a small region illustrates how additional information beyond finite-order ChPT fixes the position along the Leutwyler ellipse and removes the freedom associated with the Kaplan--Manohar ambiguity.

The ordinary $SU(3)$ chiral theory does not retain the anomalous singlet degree of freedom explicitly. When the effective theory is extended to include the $U(1)_A$ sector together with the large-$N_c$ expansion, additional constraints inherited from QCD restrict the allowed dependence on the quark mass matrix and the vacuum angle. Kaiser and Leutwyler showed that the Kaplan--Manohar transformation is incompatible with the large-$N_c$ counting rules and is absent order by order in the combined chiral and $1/N_c$ expansion~\cite{Kaiser:2000gs}. Together with phenomenological information beyond the pseudoscalar spectrum, these considerations already strongly disfavoured $m_u=0$~\cite{Leutwyler:1996qg}. A more direct determination, however, is provided by lattice QCD, where the quark masses are parameters of the discretised microscopic theory and are tuned through hadronic observables to the physical point. The Kaplan--Manohar reparametrisation of the low-energy effective theory therefore does not constitute an ambiguity in the lattice determination of the renormalised quark masses. We turn next to the lattice-QCD status of $m_u$.

\subsection{Lattice-QCD status}
\label{sec:massless-up-lattice}

Lattice QCD provides a more direct way of addressing the massless-up-quark possibility. The Kaplan--Manohar ambiguity arises because, at a fixed order in ChPT, different choices of the quark-mass parameters and low-energy constants can reproduce the same set of low-energy observables. On the lattice, by contrast, the quark masses enter as parameters of the microscopic discretised QCD action, and the dependence of hadronic observables on these parameters can be computed nonperturbatively. By matching a suitable set of lattice observables to their physical values, while controlling the continuum, finite-volume, and other systematic limits, one can determine the corresponding renormalised quark masses in a specified renormalisation scheme and at a specified scale. The Kaplan--Manohar transformation is not a symmetry of the underlying QCD theory and therefore does not represent an ambiguity in such a microscopic determination.

Early lattice studies of the massless-up-quark possibility focused on determining the low-energy constants governing the next-to-leading-order chiral corrections. Calculations with dynamical quarks already found the relevant combinations of low-energy constants to lie outside the range required to accommodate $m_u=0$~\cite{Irving:2001vy}. A subsequent calculation with three dynamical light flavours obtained $m_u/m_d=0.410(36)$, again disfavouring a vanishing up-quark mass~\cite{Nelson:2001bhq}. The MILC Collaboration later performed a more extensive three-flavour analysis using partially quenched staggered ChPT for pseudoscalar masses and decay constants, finding
$m_u/m_d=0.43(0)(1)(8)$, well outside the region compatible with $m_u=0$~\cite{MILC:2004qnl}. Although these early calculations were still subject to appreciable uncertainties from chiral extrapolation, discretisation, renormalisation, and electromagnetic effects, they already provided strong evidence against the massless-up-quark possibility; see Ref.~\cite{FlavourLatticeAveragingGroupFLAG:2024oxs} for a review of the subsequent developments.

Modern lattice calculations determine the light-quark masses with substantially improved control over these systematic effects. In isospin-symmetric QCD, where $m_u=m_d\equiv m_{ud}$, lattice simulations can determine the average light-quark mass $m_{ud}=(m_u+m_d)/2$ with high precision. Separating $m_u$ and $m_d$ additionally requires control over strong and electromagnetic isospin breaking, since physical mass splittings such as $m_{K^0}-m_{K^+}$ receive contributions from both $m_d-m_u$ and QED. These effects may be incorporated directly in QCD+QED simulations or treated perturbatively around the isospin-symmetric theory. The latter strategy is implemented systematically in the RM123 approach~\cite{deDivitiis:2013xla}, in which observables are expanded in the small parameters $(m_d-m_u)/\Lambda_{\rm QCD}$ and $\alpha_{\rm em}$~\cite{deDivitiis:2013xla,Giusti:2017dmp}; see also Ref.~\cite{FlavourLatticeAveragingGroupFLAG:2024oxs} for a review of lattice treatments of strong and electromagnetic isospin breaking.

A representative determination by the Budapest--Marseille--Wuppertal Collaboration~\cite{Fodor:2016bgu}, based on $N_f=2+1$ QCD supplemented by quenched QED, obtained
\begin{align}
    m_u^{\overline{\rm MS}}(2\,{\rm GeV})
    =2.27(6)(5)(4)\ {\rm MeV},
    \quad
    m_d^{\overline{\rm MS}}(2\,{\rm GeV})
    =4.67(6)(5)(4)\ {\rm MeV},
    \quad
    \frac{m_u}{m_d}
    =0.485(11)(8)(14),
    \label{eq:BMW-light-quark-masses}
\end{align}
where the first uncertainty is statistical, the second accounts for lattice systematic effects, and the third estimates the uncertainty associated with quenching QED. This determination excludes $m_u=0$ by more than $24$ standard deviations. Independent calculations using different lattice formulations and analysis strategies have reached compatible conclusions. In particular, the $N_f=2+1+1$ RM123 analysis of isospin breaking~\cite{Giusti:2017dmp} and the Fermilab Lattice/MILC/TUMQCD calculation with four dynamical HISQ flavours~\cite{FermilabLattice:2018est} provide important inputs and cross-checks for modern determinations of the individual light-quark masses~\cite{FlavourLatticeAveragingGroupFLAG:2024oxs}.

The current lattice status is conveniently summarised by the FLAG Review 2024~\cite{FlavourLatticeAveragingGroupFLAG:2024oxs}. For $N_f=2+1+1$, FLAG quotes the averages
\begin{align}
    m_u^{\overline{\rm MS}}(2\,{\rm GeV})=2.14(8)\ {\rm MeV},
    \quad
    m_d^{\overline{\rm MS}}(2\,{\rm GeV})=4.70(5)\ {\rm MeV},
    \quad 
    \frac{m_u}{m_d}=0.465(24),
    \label{eq:FLAG-light-quark-masses}
\end{align}
which incorporate, in particular, the $N_f=2+1+1$ determinations of Refs.~\cite{Giusti:2017dmp,FermilabLattice:2018est}; see Ref.~\cite{FlavourLatticeAveragingGroupFLAG:2024oxs} for the detailed averaging procedure and the treatment of correlations among the inputs. The corresponding $N_f=2+1$ averages,
$m_u=2.27(9)\ {\rm MeV}$, $m_d=4.67(9)\ {\rm MeV}$, and $m_u/m_d=0.485(19)$, are fully compatible. Although the quoted uncertainties combine statistical and systematic information from several analyses and should not be interpreted as the significance of a single measurement, the resulting value of $m_u$ is manifestly far from zero.

Independent lattice determinations using different fermion discretisations provide further cross-checks of this conclusion. The Extended Twisted Mass Collaboration has determined the average light- and strange-quark masses using $N_f=2+1+1$ Wilson-clover twisted-mass ensembles at three lattice spacings, including ensembles at the physical pion mass, together with nonperturbative RI$'$-MOM renormalisation~\cite{ExtendedTwistedMass:2021gbo}. More recently, the CLQCD Collaboration~\cite{CLQCD:2023sdb} used $N_f=2+1$ tadpole-improved clover ensembles at three lattice spacings and pion masses extending to the physical point, obtaining
$m_u=2.45(22)(20)\ {\rm MeV}$ and $m_d=4.74(11)(9)\ {\rm MeV}$ in the $\overline{\rm MS}$ scheme at $2\,{\rm GeV}$. These determinations use lattice formulations and renormalisation strategies different from those underlying the most precise averages and nevertheless give mutually consistent light-quark masses. The lattice evidence for a nonzero up-quark mass is therefore robust against the choice of discretisation and analysis strategy.

There remains a precision subtlety associated with electromagnetism. Quark masses are renormalisation-scheme and scale dependent, which is why Eq.~\eqref{eq:FLAG-light-quark-masses} quotes them in the $\overline{\rm MS}$ scheme at $2\,{\rm GeV}$. Once QED is included, the separation of hadronic quantities into ``pure-QCD'' and electromagnetic contributions is itself convention dependent, and the up- and down-quark masses run differently because of their different electric charges. In addition, some calculations entering the present FLAG averages for the individual light-quark masses employ electro-quenched QED~\cite{FlavourLatticeAveragingGroupFLAG:2024oxs}. These issues matter for percent-level precision, but are far too small to affect the conclusion that $m_u$ is nonzero.

A separate caveat concerns possible nonperturbative contributions with the same low-energy quantum numbers as an up-quark mass. Because the axial symmetry of a single quark is anomalous, instanton-induced interactions involving the other massive flavours can generate such terms. It has therefore been suggested that the phenomenologically inferred $m_u$ might contain a substantial nonperturbative ``additive'' or topological contribution even if the microscopic up-quark mass vanished. This possibility is conceptually distinct from the Kaplan--Manohar ambiguity, which is a reparametrisation freedom of the low-energy effective theory. This proposed loophole has also been tested directly on the lattice. Reference~\cite{Alexandrou:2020bkd} studied the dependence of the pion mass on the dynamical strange-quark mass in order to isolate the nonperturbative contribution that could mimic an up-quark mass. The effect was found to be much too small to account for the physical value of $m_u$. Together with the direct determinations summarised in Eq.~\eqref{eq:FLAG-light-quark-masses}, this strongly excludes the possibility that the observed up-quark mass is generated predominantly by such a topological contribution.

The lattice evidence therefore places the massless-up-quark possibility on a much firmer footing. Already at leading order in $SU(3)$ ChPT, the pseudoscalar-meson mass relations favour a nonzero up-quark mass, while phenomenological analyses incorporating higher-order chiral information strongly disfavoured $m_u=0$. The Kaplan--Manohar ambiguity qualified the simplest extraction by showing that, at a fixed order in $SU(3)$ ChPT, pseudoscalar observables alone do not uniquely determine the individual quark-mass ratios without additional information on the low-energy constants. Modern lattice calculations bypass this limitation by probing the quark-mass dependence of QCD directly and consistently find $m_u\simeq2\,{\rm MeV}$. Within the Standard Model, the possibility of removing the physical strong $CP$ phase through an exactly massless up quark is therefore regarded as excluded. 

The exclusion of $m_u=0$ within the Standard Model does not, however, rule out more general ultraviolet constructions that exploit related chiral-symmetry ideas while generating the observed light-quark masses through additional dynamics. One example relates the small $CP$-violating component of the up-quark mass to Dirac neutrino masses, while the dominant real contribution to the physical up-quark mass is generated by instantons of an additional QCD-like interaction~\cite{Carena:2019nnd}. More recently, gauged quark flavour and non-invertible PQ symmetries arising from fractional instantons have been used to construct an ultraviolet massless-down-quark solution, with an $SU(9)$ colour-flavour unified completion that generates realistic quark masses, mixing, and an order-one CKM phase~\cite{Cordova:2024ypu}. These constructions should be distinguished from the minimal Standard-Model proposal excluded by the lattice determinations above: the vanishing microscopic quark mass is protected in an enlarged theory, while additional dynamics are responsible for reproducing the observed low-energy quark spectrum.

\section{\texorpdfstring{$CP$}{CP}- and parity-based solutions}
\label{sec:CP&P}

\subsection{Spontaneous \texorpdfstring{$CP$}{CP} breaking and the Nelson--Barr mechanism}
\label{sec:Nelson-Barr}

A conceptually attractive possibility is that $CP$ is an exact symmetry of the ultraviolet theory. If $CP$ is exact, there exists a $CP$ basis in which the QCD vacuum angle vanishes and all parameters of the Lagrangian are real. In discussing the full three-generation quark sector, we denote the up- and down-type quark mass matrices by $\M_u$ and $\M_d$, respectively, defined through
\begin{align}
    \mathcal{L}_{\rm mass} = -\bar u_L\M_u u_R -\bar d_L\M_d d_R +\hc ,
    \label{eq:quark-mass-matrices-CP-solutions}
\end{align}
where $u=(u,c,t)^T$ and $d=(d,s,b)^T$. These matrices act in generation space and should be distinguished from the light-quark mass matrix $\M$ in Eq.~\eqref{eq:Dirac-mass}. In this notation,
\begin{align}
    \bar\theta = \theta+\arg\det(\M_u\M_d).
    \label{eq:bartheta-full-quark-sector}
\end{align}
Since $\M_u$ and $\M_d$ are real before $CP$ breaking, their determinants are real. Exact $CP$ therefore restricts the quark-mass contribution to $\bar\theta$ to $0$ or $\pi$, rather than selecting $0$ uniquely. Throughout this subsection we restrict to the phenomenologically relevant branch in which the product of the real coloured-fermion mass determinants is positive, so that $\theta=0$ corresponds to $\bar\theta=0$. For definiteness, before introducing additional coloured fermions, we take $\det\M_u>0$ and $\det\M_d>0$.

The observed weak interactions, however, do violate $CP$ through the CKM phase. Therefore, $CP$ must be broken spontaneously. The central challenge is to generate an order-one weak $CP$ phase without simultaneously inducing an unacceptably large $\bar\theta$. To see why this is nontrivial, consider a scalar field $S$ whose potential is $CP$-invariant but whose VEV is complex,
\begin{align}
    \langle S\rangle
    =
    v_S\e^{\i\delta_S},
    \qquad
    \delta_S\neq0,\pi.
    \label{eq:spontaneous-CP-vev}
\end{align}
Under $CP$, $\langle S\rangle\to\langle S\rangle^*$, so the vacuum in Eq.~\eqref{eq:spontaneous-CP-vev} spontaneously breaks $CP$. Couplings of $S$ to the quark sector can then transmit this phase to the quark mass matrices and generate weak $CP$ violation. If this transmission is generic, however, the same phase will also enter the determinants of the quark mass matrices and induce an unacceptably large $\bar\theta$. Spontaneous $CP$ breaking by itself therefore does not solve the strong $CP$ problem. What is required is a special structure that transmits the $CP$-violating phase to the flavour rotations responsible for the CKM phase while preventing it from entering the quark-mass contribution to $\bar\theta$.

The Nelson--Barr mechanism provides a systematic way of realising this separation~\cite{Nelson:1983zb,Barr:1984qx,Barr:1984fh}; see also Refs.~\cite{Cheng:1987gp,Dine:2015jga,Vecchi:2014hpa,Valenti:2021xjp} for later discussions. Barr formulated the underlying structure in terms of a simple set of sufficient conditions~\cite{Barr:1984qx}. Let $F$ denote the fermions carrying the gauge quantum numbers of the ordinary chiral families, and let $R$ denote an additional set of fermions forming a real set of representations under the Standard-Model gauge group, in the sense that every complex representation appearing in $R$ is accompanied by its conjugate. The latter can therefore contain heavy states with gauge-invariant masses and is commonly realised by vector-like fermions. Barr's sufficient conditions are:
\begin{enumerate}
    \item[(1)] vacuum expectation values (VEVs) that break $SU(2)_L\times U(1)_Y$ appear only in $F$-$F$ Yukawa interactions, and not in $F$-$R$ or $R$-$R$ terms;
    \item[(2)] VEVs that break $CP$ appear only in interactions connecting the $F$ and $R$ sectors, and not in $F$-$F$ or $R$-$R$ terms.
\end{enumerate}
Together with the gauge quantum numbers of the fermions, these conditions lead to a pattern of allowed and forbidden mass terms in which the spontaneous $CP$-violating phases enter the mixing between the ordinary and additional fermions but do not contribute to the determinants of the relevant coloured-fermion mass matrices. The same mixing nevertheless changes the flavour composition of the light fermion eigenstates. After the heavy states are integrated out, the light quarks are generally complex linear combinations of the original $F$ and $R$ fields, so that their effective Yukawa matrices can contain a physical CKM phase even though the quark-mass contribution to $\bar\theta$ vanishes at tree level~\cite{Nelson:1983zb,Barr:1984qx}.

A particularly transparent low-dimensional illustration is provided by the model of Bento, Branco and Parada (BBP)~\cite{Bento:1991ez}, which contains one vector-like down-type quark $D_{L,R}$ and a complex Standard-Model-singlet scalar $S$. In the language of Barr's theorem, the ordinary Standard-Model quarks belong to the $F$ sector, whereas the vector-like pair $D_{L,R}$ belongs to the additional $R$ sector.\footnote{More precisely, when all fermions are written as left-handed Weyl fields, the vector-like pair corresponds to $D_L$ and $(D_R)^c$, which transform in conjugate gauge representations and therefore form a real set in the sense of Barr's theorem. The scalar fields $H$ and $S$ are not classified as $F$ or $R$; the $F/R$ distinction refers only to fermions.} A discrete $\mathbb{Z}_2$ symmetry is imposed under which $D_L$, $D_R$, and $S$ are odd while all Standard-Model fields are even. Its purpose is to realise Barr's separation between electroweak and $CP$ breaking at the renormalisable level.

At the renormalisable level, the relevant interactions can be written as
\begin{align}
    -\mathcal{L}_{\rm NB} \supset  y^d_{ij}\, \bar Q_{iL}H d_{jR} 
    + \bar D_L\,M_DD_R
    + \left(f_j S+f_j' S^\dagger \right)\bar D_Ld_{jR}
    +\hc , 
    \label{eq:NB-BBP-Lagrangian}
\end{align}
where $i,j=1,2,3$, $Q_{iL}=(u_{iL},d_{iL})^T$, and the parameters $y^d_{ij}$, $f_j$, $f_j'$, and $M_D$ are real as a consequence of the fundamental $CP$ symmetry. The role of the $\mathbb{Z}_2$ symmetry in implementing Barr's sufficient conditions is summarised in Table~\ref{tab:BBP-Barr-conditions}. It forbids $\bar Q_{iL}HD_R$, preventing the electroweak-breaking VEV from entering an $F$-$R$ mass term, and forbids $S\bar D_LD_R$, preventing the $CP$-violating vacuum expectation value from entering the $R$-$R$ mass. At the same time, the interaction $S\bar D_Ld_{jR}$ is allowed, so that spontaneous $CP$ violation is transmitted through mixing between the ordinary and vector-like fermions, as required by Barr's second condition. The $\mathbb{Z}_2$ symmetry is somewhat more restrictive than Barr's general sufficient conditions: for example, it also forbids the $CP$-conserving bare mixing $\bar D_Ld_{jR}$, although such a term would not by itself spoil the determinant argument.

\begin{table}[t]
    \centering
    \small
    \renewcommand{\arraystretch}{1.25}
    \begin{tabular}{|p{2.8cm}||p{1.6cm}|p{0.8cm}|p{2.6cm}|p{4.5cm}|}
        \hline
        Interaction & $F/R$ sector & $\mathbb{Z}_2$ & Relevant breaking & Barr's conditions \\
        \hline\hline
        $\bar Q_{iL}H d_{jR}$ & $F$-$F$ & even & EW &
        EW-breaking VEV in $F$-$F$ (\ding{51}) \\
        \hline
        $\bar Q_{iL}H D_R$ & $F$-$R$ & odd & EW &
        EW-breaking VEV in $F$-$R$ (\ding{55}) \\
        \hline
        $S\,\bar D_Ld_{jR}$, $S^\dagger\bar D_Ld_{jR}$ & $R$-$F$ & even & $CP$ &
        $CP$-violating VEV in $F$-$R$ (\ding{51}) \\
        \hline
        $S\,\bar D_LD_R$, $S^\dagger\bar D_LD_R$ & $R$-$R$ & odd & $CP$ &
        $CP$-violating VEV in $R$-$R$ (\ding{55}) \\
        \hline
        $\bar D_LM_DD_R$ & $R$-$R$ & even & none &
        $CP$-conserving $R$-$R$ mass (\ding{51}) \\
        \hline
    \end{tabular}
    \caption{Role of the $\mathbb{Z}_2$ symmetry in the BBP realisation of the Nelson--Barr mechanism. The $F/R$ labels refer to the fermion sectors in Barr's theorem. The scalar $H$ carries the electroweak-breaking VEV, whereas the singlet $S$ carries the spontaneous $CP$-violating phase.}
    \label{tab:BBP-Barr-conditions}
\end{table}

After electroweak and $CP$ symmetry breaking, Eq.~\eqref{eq:NB-BBP-Lagrangian} gives
\begin{align}
    -\mathcal{L}_{{\rm mass},d}
    =
    \begin{pmatrix}
        \bar d_{iL} & \bar D_L
    \end{pmatrix}
    \mathbb{M}_d
    \begin{pmatrix}
        d_{jR}\\
        D_R
    \end{pmatrix}
    +\hc ,
    \qquad
    \mathbb{M}_d
    =
    \begin{pmatrix}
        m_d & 0\\
        B & M_D
    \end{pmatrix},
    \label{eq:NB-mass-matrix}
\end{align}
where
\begin{align}
    (m_d)_{ij} = y^d_{ij}v,
    \qquad
    B_j = v_S \left(f_j\e^{\i\delta_S}
    + f_j'\e^{-\i\delta_S} \right),
    \label{eq:NB-B-mixing}
\end{align}
and $v$ denotes the neutral Higgs VEV in this normalisation. The $3\times3$ block $m_d$ is real and arises from the ordinary $F$-$F$ Yukawa interaction, while $B$ carries the spontaneous $CP$-violating phases and mixes $D_L$ with the three ordinary right-handed down-type quarks. The upper-right block vanishes because $\bar Q_{iL}HD_R$ is forbidden, while $M_D$ is a real vector-like mass. Equation~\eqref{eq:NB-mass-matrix} is therefore block triangular, and
\begin{align}
    \det\mathbb{M}_d = M_D\det m_d,
    \label{eq:NB-real-determinant}
\end{align}
which is completely independent of the complex mixing parameters $B_j$. If, in addition, $\det(\mathbb{M}_d)>0$, one then obtains
\begin{align}
    \bar\theta_{\rm tree} = \theta + \arg\det\M_u + \arg\det\mathbb{M}_d = 0.
    \label{eq:NB-tree-theta}
\end{align}

The fact that the complex parameters $B_j$ drop out of Eq.~\eqref{eq:NB-real-determinant} does not mean that they are physically irrelevant. The CKM matrix is determined by the flavour rotations of the left-handed quarks, which follow from the Hermitian mass-squared matrices rather than from their determinants. In the down-type sector,
\begin{align}
    \mathbb{M}_d\mathbb{M}_d^\dagger
    =
    \begin{pmatrix}
        m_dm_d^T & m_dB^\dagger \\
        Bm_d^T & M_D^2+BB^\dagger
    \end{pmatrix}.
    \label{eq:NB-MMdagger}
\end{align}
Here $BB^\dagger=\sum_i|B_i|^2$ is a real positive number, whereas the off-diagonal blocks are generally complex. To see how the heavy state modifies the light sector, consider a light eigenstate with components $\ell$ and $h$ in the ordinary and vector-like sectors,
\begin{align}
    \mathbb{M}_d\mathbb{M}_d^\dagger
    \begin{pmatrix}
        \ell\\
        h
    \end{pmatrix}
    =
    \lambda
    \begin{pmatrix}
        \ell\\
        h
    \end{pmatrix}.
\end{align}
For $\lambda\ll M_D^2+BB^\dagger$, the heavy component is
\begin{align}
    h\approx -\frac{1}{M_D^2+BB^\dagger}\,
    Bm_d^T\,\ell .
    \label{eq:NB-heavy-component}
\end{align}
Substituting this relation back into the light-component equation gives, to leading order, the effective Hermitian mass-squared matrix governing the three light left-handed down-type quarks,
\begin{align}
    \left(m_{d,L}^2\right)_{\rm eff}
    \approx
    m_dm_d^T-\frac{m_dB^\dagger Bm_d^T}{M_D^2+BB^\dagger}.
    \label{eq:NB-effective-Hd}
\end{align}
The physical light-quark masses and the corresponding left-handed flavour rotation are then determined by
\begin{align}
    V_{dL}^\dagger\left(m_{d,L}^2\right)_{\rm eff}V_{dL}
    =
    {\rm diag}(m_d^2,m_s^2,m_b^2).
    \label{eq:NB-diagonal-light-masses}
\end{align}

The origin of the weak $CP$ phase is now explicit. The matrix $B^\dagger B$ has elements
$(B^\dagger B)_{ij}=B_i^*B_j$. Writing $B_i=|B_i|\e^{\i\phi_i}$, its off-diagonal entries depend on the relative phases $\phi_j-\phi_i$. If all components share a common phase, $B=\e^{\i\phi}b$ with $b$ real, that phase cancels from $B^\dagger B$ and can be removed by rephasing the vector-like quark. A nontrivial complex structure therefore requires at least two nonzero components of $B$ with a relative phase different from $0$ or $\pi$.

Even then, a complex $V_{dL}$ by itself does not imply physical $CP$ violation, since phases can also be removed by rephasing the light quarks. The physical weak phase resides in the relative left-handed rotation
\begin{align}
    V_{\rm CKM} = V_{uL}^\dagger V_{dL}.
    \label{eq:NB-CKM}
\end{align}
In the present construction the up-type mass matrix is real, so $V_{uL}$ may be chosen real. For three generations, nontrivial relative phases in $B$, together with a generic flavour orientation of the real matrix $m_d$, can then leave a rephasing-invariant phase in $V_{\rm CKM}$ and generate weak $CP$ violation. The effect must also be sufficiently large: when $|B|\ll M_D$, the complex correction in Eq.~\eqref{eq:NB-effective-Hd} is suppressed by $|B|^2/M_D^2$. An order-one CKM phase therefore typically requires both nontrivial relative phases and appreciable heavy-light mixing~\cite{Dine:2015jga,Valenti:2021xjp}. The latter also induces small deviations from unitarity in the effective $3\times3$ CKM matrix, which disappear in the decoupling limit.

The BBP model thus makes the characteristic Nelson--Barr separation manifest: the spontaneous $CP$-violating phases drop out of the determinant but can survive in the flavour rotations that determine weak $CP$ violation. More general constructions may contain several vector-like quarks, place the additional states in the up-type sector, or employ different gauge representations while retaining the same basic separation~\cite{Barr:1984qx,Barr:1984fh,Vecchi:2014hpa,Valenti:2021xjp}. The tree-level result is nevertheless only the first requirement of a viable solution. Loop corrections and higher-dimensional operators can perturb the zero and reality structures after $CP$ breaking and regenerate $\bar\theta$~\cite{Bento:1991ez,Dine:2015jga,Vecchi:2014hpa}. We return to radiative stability and ultraviolet robustness in Sec.~\ref{sec:CP-parity-radiative}.

\subsection{Parity as a UV symmetry: left-right and mirror constructions}
\label{sec:parity}

Parity provides another symmetry-based route to the strong $CP$ problem. Since $G^a_{\mu\nu}\widetilde G^{a\mu\nu}$ is odd under spatial parity, $P:\,G^a_{\mu\nu}\widetilde G^{a\mu\nu}\rightarrow-G^a_{\mu\nu}\widetilde G^{a\mu\nu}$, exact parity requires $\theta=-\theta$ modulo $2\pi$, allowing the parity-invariant values $\theta=0$ and $\theta=\pi$. As in the $CP$-based constructions discussed in the previous subsection, we focus on the phenomenologically relevant possibility $\bar\theta=0$, rather than the distinct symmetry-invariant value $\bar\theta=\pi$. We therefore consider constructions in which parity fixes $\theta=0$ and the product of the relevant coloured-fermion mass determinants carries no phase. In the Standard Model this latter condition reduces to $\arg\det(\M_u\M_d)=0$, with the corresponding enlarged mass matrices understood when additional coloured fermions are present.

Unlike the $CP$ symmetry invoked in the Nelson--Barr mechanism, parity is a unitary symmetry and does not require all parameters of the Lagrangian to be real. Instead, it relates left- and right-handed fermions and typically constrains the corresponding flavour structures by Hermiticity, allowing complex flavour couplings and hence weak $CP$ violation. Parity must nevertheless be broken at low energies because the observed weak interactions are chiral. The central model-building challenge is therefore to break parity while preserving the relations that enforce $\bar\theta=0$. These considerations underlie the parity-based solutions proposed in the early literature~\cite{Mohapatra:1978fy,Babu:1988mw,Babu:1989rb,Barr:1991qx}.

\paragraph{Left-right constructions.}

A natural framework is the left-right gauge group
\begin{align}
    SU(3)_c \times SU(2)_L \times SU(2)_R \times U(1)_{B-L},
    \label{eq:LR-gauge-group}
\end{align}
with quarks arranged into $Q_L=(u_L,d_L)^T$ and $Q_R=(u_R,d_R)^T$. Up to an unphysical intrinsic parity phase that can be absorbed into the fermion fields, we choose
\begin{align}
    P:\ Q_L(t,\bm{x}) \leftrightarrow \gamma^0Q_R(t,-\bm{x}),
    \qquad
    SU(2)_L \leftrightarrow SU(2)_R.
    \label{eq:LR-parity}
\end{align}
The gauge and scalar sectors transform correspondingly.

Consider first the conventional scalar bidoublet $\Phi\sim(\bm{1},\bm{2},\bm{2},0)$, $\widetilde\Phi\equiv\tau_2\Phi^*\tau_2$,
where the quantum numbers refer to the factors in Eq.~\eqref{eq:LR-gauge-group}. The bidoublet transforms as $\Phi\to U_L\Phi U_R^\dagger$, and the pseudoreality of the fundamental representation of $SU(2)$ ensures that $\widetilde\Phi$ transforms in the same way. The quark Yukawa interactions therefore contain two independent generation-space matrices,
\begin{align}
    -\mathcal{L}_Y
    = \bar Q_L \left(Y\Phi + \widetilde Y\widetilde\Phi
    \right) Q_R +\hc .
    \label{eq:LR-bidoublet-Yukawa}
\end{align}
Parity acts as $\Phi\leftrightarrow\Phi^\dagger$ and consequently requires $Y=Y^\dagger$, $\widetilde Y=\widetilde Y^\dagger$.
Thus, parity does not require $Y$ and $\widetilde Y$ to be real; they may contain complex off-diagonal entries while remaining Hermitian.\footnote{We display only the bidoublet sector relevant for the quark masses. In realistic left-right models, additional scalar fields are required to break $SU(2)_R\times U(1)_{B-L}$ above the electroweak scale. The same bidoublet also generates charged-lepton and neutrino Dirac masses through analogous Yukawa interactions.}

Preservation of $U(1)_{\rm em}$ requires the charged components of $\Phi$ to have vanishing VEVs. The two neutral expectation values may in general carry two phases, one of which can be removed by an $SU(2)_L\times SU(2)_R$ gauge transformation. We may therefore write
\begin{align}
    \langle\Phi\rangle
    =
    \frac{1}{\sqrt{2}}
    \begin{pmatrix}
        \kappa & 0\\
        0 & \kappa'\e^{\i\alpha}
    \end{pmatrix},
    \qquad
    \kappa,\kappa'\geq0,
    \label{eq:LR-bidoublet-vev}
\end{align}
where the remaining phase $\alpha=\arg\det\langle\Phi\rangle$ is gauge invariant. The quark mass matrices are then
\begin{align}
    \M_u = \frac{1}{\sqrt{2}} \left(\kappa Y + \kappa'\e^{-\i\alpha}\widetilde Y \right),
    \qquad
    \M_d = \frac{1}{\sqrt{2}} \left( \kappa'\e^{\i\alpha}Y + \kappa\widetilde Y \right).
    \label{eq:LR-quark-mass-bidoublet}
\end{align}
For $\alpha=0$, both $\M_u$ and $\M_d$ are Hermitian and hence have real determinants. If $\det(\M_u\M_d)>0$, we then obtain $\bar\theta_{\rm tree}=0$. Hermiticity nevertheless allows a nontrivial complex flavour structure: $\M_u$ and $\M_d$ need not be diagonalised by the same unitary transformation, and their relative left-handed rotation can contain the observed CKM phase.

The residual vacuum phase $\alpha$ exposes an important limitation of the simplest bidoublet construction. For $\alpha\neq0$, the mass matrices in Eq.~\eqref{eq:LR-quark-mass-bidoublet} are no longer Hermitian in general, even though $Y$ and $\widetilde Y$ themselves remain Hermitian. The scalar vacuum can then generate a nonzero phase in $\det(\M_u\M_d)$ already at tree level. Thus, imposing parity in the ultraviolet is not by itself sufficient in a generic left-right scalar sector; the structure of parity breaking, and in particular the phases of the scalar VEVs, must also be controlled.

A particularly transparent alternative avoids the bidoublet altogether and generates ordinary fermion masses through a universal seesaw~\cite{Babu:1988mw,Babu:1989rb,Barr:1991qx}. One introduces parity-related scalar doublets $H_L$ and $H_R$, with
\begin{align}
    \langle H_L\rangle = \frac{1}{\sqrt{2}}
    \begin{pmatrix}
        0\\ v_L
    \end{pmatrix},
    \qquad
    \langle H_R\rangle = \frac{1}{\sqrt{2}}
    \begin{pmatrix}
        0\\ v_R
    \end{pmatrix},
    \qquad
    v_R\gg v_L.
    \label{eq:universal-seesaw-Higgs-vevs}
\end{align}
Their phases can be removed independently by gauge transformations. Since no bidoublet directly couples $Q_L$ to $Q_R$, vector-like $SU(2)_{L,R}$-singlet quarks are introduced: three up-type states $U_{L,R}$ and three down-type states $D_{L,R}$, each transforming as a colour triplet. The relevant interactions may be written as
\begin{align}
    -\mathcal{L}_{Y}\supset \bar Q_L y_dH_LD_R + \bar Q_R y_dH_RD_L +\bar Q_L y_u\widetilde H_L U_R + \bar Q_R y_u\widetilde H_R U_L + \bar U_LM_UU_R + \bar D_LM_DD_R +\hc ,
\end{align}
where $\widetilde H_{L,R}=\i\tau_2H_{L,R}^*$, while parity requires $M_U=M_U^\dagger$ and $M_D=M_D^\dagger$.

After symmetry breaking, the enlarged quark mass matrices take the form
\begin{align}
    \mathbb M_u
    &=
    \begin{pmatrix}
        0 & y_uv_L/\sqrt{2}\\
        y_u^\dagger v_R/\sqrt{2} & M_U
    \end{pmatrix},
    \qquad
    \mathbb M_d
    =
    \begin{pmatrix}
        0 & y_dv_L/\sqrt{2}\\
        y_d^\dagger v_R/\sqrt{2} & M_D
    \end{pmatrix}.
    \label{eq:parity-universal-seesaw-masses}
\end{align}
For three generations their determinants are
\begin{align}
    \det\mathbb M_u = -\left(\frac{v_Lv_R}{2}\right)^3
    |\det y_u|^2,
    \qquad
    \det\mathbb M_d = -\left(\frac{v_Lv_R}{2}\right)^3
    |\det y_d|^2.
    \label{eq:parity-universal-seesaw-determinant}
\end{align}
Thus the complex phases of $y_u$ and $y_d$ cancel from the determinants, and together with $\theta=0$ one obtains $\bar\theta_{\rm tree}=0$. When the vector-like states are heavy, integrating them out gives
\begin{align}
    \M_u^{\rm eff} \approx -\frac{v_Lv_R}{2}\, y_uM_U^{-1}y_u^\dagger, 
    \qquad
    \M_d^{\rm eff}\approx -\frac{v_Lv_R}{2}\, y_dM_D^{-1}y_d^\dagger. 
    \label{eq:parity-universal-seesaw-effective}
\end{align}
These matrices are Hermitian but generically complex, so their relative left-handed diagonalisation can contain a CKM phase. The universal-seesaw construction therefore illustrates particularly clearly how parity can eliminate the tree-level strong phase without eliminating weak $CP$ violation.

\paragraph{Mirror constructions.}

A conceptually distinct implementation introduces a mirror copy of some or all of the Standard Model and lets parity exchange the ordinary and mirror sectors~\cite{Barr:1991qx,Hall:2018let,Dunsky:2019api,Craig:2020bnv,Bonnefoy:2023afx}. Rather than relating left- and right-handed fields within a single enlarged electroweak theory, parity now relates the Standard-Model fermions to a second set with the opposite chirality assignment under the parity-related gauge interactions.

Consider first models in which ordinary and mirror quarks are charged under the same $SU(3)_c$, while the electroweak sector is replicated~\cite{Hall:2018let,Dunsky:2019api}. Denoting mirror-sector quantities by a prime, parity may be represented by $Y'_u=Y_u^*$, $Y'_d=Y_d^*$. If the ordinary and mirror Higgs VEVs are chosen real, the corresponding mass matrices satisfy $\M'_{u,d}\propto\M_{u,d}^*$. Since both sectors carry the same colour charge, their determinant phases contribute to the same strong $CP$ phase and cancel,
\begin{align}
    \arg\det \left( \M_u\M_d\M'_u\M'_d \right) 
    = \arg\det(\M_u\M_d) + \arg\det(\M_u^*\M_d^*)
    = 0.
    \label{eq:mirror-determinant-cancellation}
\end{align}
Parity also sets $\theta=0$ on the branch considered here. Thus, unlike the Hermiticity mechanism of the conventional left-right construction, the ordinary and mirror quarks carry opposite determinant phases whose sum vanishes, while the ordinary Yukawa matrices themselves can retain the complex flavour structure required for the CKM phase.

One may instead mirror the colour group as well~\cite{Bonnefoy:2023afx}. Above the scale at which the two colour sectors are connected, parity relates their invariant strong phases by $\bar\theta'=-\bar\theta$. If subsequent symmetry breaking identifies the two colour groups with their diagonal subgroup, $SU(3)\times SU(3)' \rightarrow SU(3)_c$,
the effective strong $CP$ phase of the diagonal theory receives the sum of the two contributions and therefore vanishes at tree level, provided the colour-breaking sector introduces no additional $CP$-violating phase.

Parity must eventually be broken in mirror theories as well. A common possibility is
\begin{align}
    v' \equiv |\langle H'\rangle|
    \gg v \equiv |\langle H\rangle|,
    \label{eq:mirror-parity-breaking}
\end{align}
so that the mirror fermions and gauge bosons are much heavier than their Standard-Model counterparts. In Higgs-parity constructions this hierarchy arises through spontaneous parity breaking~\cite{Hall:2018let,Dunsky:2019api}. A hierarchy between the magnitudes $v'$ and $v$ does not itself spoil the determinant cancellation, since these VEVs can be chosen real. The relevant question is instead whether parity breaking generates new complex parameters or operators that disturb the relations enforced by parity.

The left-right and mirror constructions thus realise the same underlying principle in complementary ways. In the former, parity constrains the flavour structure through Hermiticity or through the paired structure of a universal seesaw; in the latter, it pairs ordinary and mirror determinant phases with opposite signs. Both mechanisms can accommodate an order-one CKM phase while giving $\bar\theta_{\rm tree}=0$. Their ultimate success depends on whether this protection survives parity breaking, radiative corrections, heavy-particle thresholds, and higher-dimensional operators, which we discuss in Sec.~\ref{sec:CP-parity-radiative}.

\subsection{Radiative stability and UV robustness}
\label{sec:CP-parity-radiative}

The constructions discussed in Secs.~\ref{sec:Nelson-Barr} and~\ref{sec:parity} achieve $\bar\theta=0$ at tree level even after $CP$ or parity is broken, while allowing an order-one CKM phase. The question addressed here is whether the relations responsible for this tree-level result remain stable against radiative corrections and higher-dimensional operators. The general diagnostic was introduced in Eq.~\eqref{eq:radiative-theta}. In theories containing additional coloured fermions, it is useful to organise the fermions at a given scale into mass matrices $\mathbb M_a$, where $a$ labels sectors of states that can mix. To first order in small corrections,
\begin{align}
    \delta\bar\theta = \delta\theta
    + \sum_a {\rm Im}\,\tr
    \left( \mathbb M_a^{-1}\delta\mathbb M_a
    \right) + \mathcal{O}(\delta\mathbb M^2).
    \label{eq:radiative-theta-full}
\end{align}
Since the experimental requirement is $|\bar\theta|\lesssim10^{-10}$, ordinary loop suppression is generally insufficient by itself. The leading corrections must therefore either vanish or be strongly suppressed by the symmetry and flavour structure of the theory.

Renormalisation-group evolution can itself spoil the relations imposed at the high scale~\cite{Peccei:1995fg,Xing:2015sva}. When heavy coloured fermions are integrated out, the matching must also be performed consistently. One may evaluate $\bar\theta$ in the theory in which the heavy states are still dynamical, or match both the fermion mass matrices and the coefficient of $G\widetilde G$ across the heavy threshold. Computing only the phase of the light-quark determinant after removing the heavy states can miss threshold contributions to the topological operator~\cite{Hisano:2023izx}.

\paragraph{Nelson--Barr models.}

The origin of the radiative sensitivity is particularly transparent in the BBP texture of Eq.~\eqref{eq:NB-mass-matrix}. Allowing for small corrections,
\begin{align}
    \mathbb M_d+\delta\mathbb M_d =
    \begin{pmatrix}
        m_d+\delta m_d & \delta A\\
        B+\delta B & M_D+\delta M_D
    \end{pmatrix},
    \label{eq:NB-radiative-matrix}
\end{align}
where $\delta A$ denotes a correction to the block that vanishes at tree level. Neglecting a direct correction to $\theta$ for this illustrative example, Eq.~\eqref{eq:radiative-theta-full} gives
\begin{align}
    \delta\bar\theta = {\rm Im}\,\tr
    \left(m_d^{-1}\delta m_d \right) 
    + {\rm Im} \left( \frac{\delta M_D}{M_D} \right) 
    - {\rm Im}\,\tr \left( m_d^{-1}\delta A\,M_D^{-1}B \right)
    +\cdots .
    \label{eq:NB-radiative-theta}
\end{align}
Corrections confined to the complex mixing block $B$ do not change the determinant as long as the block-triangular structure is maintained. The dangerous corrections are instead those that introduce $CP$-violating phases into $m_d$ or $M_D$, or generate a nonzero upper-right block, since these are precisely the structures that enter Eq.~\eqref{eq:NB-radiative-theta}. A radiatively stable Nelson--Barr construction must therefore ensure that such corrections are either forbidden or sufficiently suppressed.

Whether the dangerous corrections first arise at one, two, or more loops depends on the full particle content and interactions of the model. In generic renormalisable completions, Higgs-portal interactions in the scalar potential can combine with the Nelson--Barr Yukawa couplings to transmit the spontaneous $CP$-violating phase to the ordinary quark mass matrices. In the minimal BBP model, this mechanism generates a complex correction to the down-type Yukawa matrix already at one loop~\cite{Dine:2015jga}. Such one-loop effects depend on interactions involving the scalar $CP$-breaking sector and are therefore model dependent. Even if these effects are suppressed, radiative corrections from the Standard-Model fields and the vector-like quarks required to transmit $CP$ violation remain; Ref.~\cite{Valenti:2021xjp} refers to these as irreducible contributions. For standard constructions with vector-like up- or down-type quarks, analytic contributions to $\bar\theta$ first arise at three loops, while certain nonanalytic two-loop threshold effects decouple as the vector-like quarks become heavy. 

The same dangerous structures can also be generated directly by higher-dimensional operators, providing an additional constraint on the ultraviolet completion. The $\mathbb Z_2$ symmetry of the BBP example enforces the desired mass texture at the renormalisable level, but allows operators such as
\begin{align}
    \Delta\mathcal{L}_{\rm UV}
    \supset 
    \frac{c_S}{\Lambda_{\rm UV}}\, S^2\bar D_LD_R
    + \frac{c_H}{\Lambda_{\rm UV}}\, S\,\bar Q_{iL}H D_R
    +\hc .
    \label{eq:NB-higher-dimensional}
\end{align}
Both operators are even under the $\mathbb Z_2$ symmetry. Upon inserting the $CP$-violating VEV of $S$, the first generates a complex correction to the vector-like mass $M_D$, while the second generates a nonzero upper-right block in Eq.~\eqref{eq:NB-mass-matrix}. They therefore induce precisely the types of corrections appearing in Eq.~\eqref{eq:NB-radiative-theta}. For the first operator, for example,
\begin{align}
    \delta M_D \sim c_S\frac{v_{CP}^2}{\Lambda_{\rm UV}},
    \qquad
    \delta\bar\theta \sim c_S\frac{v_{CP}^2} {M_D\Lambda_{\rm UV}}.
    \label{eq:NB-quality-estimate-general}
\end{align}
If $M_D\sim v_{CP}$, this reduces parametrically to $\delta\bar\theta\sim c_S\, v_{CP}/\Lambda_{\rm UV}$. For $c_S\sim\mathcal{O}(1)$ and $\Lambda_{\rm UV}$ near the reduced Planck scale, the requirement $|\delta\bar\theta|\lesssim10^{-10}$ gives the illustrative upper range $v_{CP}\lesssim10^8$--$10^9\,{\rm GeV}$~\cite{Dine:2015jga}. This sensitivity to higher-dimensional operators is often referred to as the Nelson--Barr quality problem. Its precise severity is model dependent, but it motivates additional structures that suppress the leading dangerous operators, such as gauge or discrete symmetries, sequestering of the $CP$-breaking sector, supersymmetry, or strong dynamics~\cite{Dine:2015jga,Vecchi:2014hpa}.

\paragraph{Parity-based models.}

In parity-based constructions, the radiative correction to $\bar\theta$ depends sensitively on how the relations responsible for $\bar\theta_{\rm tree}=0$ are affected by the parity-breaking sector. This model dependence is already apparent in the constructions discussed in Sec.~\ref{sec:parity}. In the universal-seesaw model of Babu and Mohapatra with softly broken parity, the one-loop contribution vanishes and a nonzero $\bar\theta$ first arises at two loops~\cite{Babu:1989rb}. A modern reanalysis confirmed the one-loop cancellation and showed that the two-loop threshold correction need not be suppressed by the ratio of the electroweak and heavy scales, so additional suppression from the flavour structure or mixing parameters may be required~\cite{deVries:2021pzl}. In related constructions where parity is broken spontaneously by unequal VEVs of parity-related Higgs doublets, stronger cancellations can occur: in certain models the complete two-loop contribution vanishes and the first nonzero term appears at three loops~\cite{Hall:2018let,Craig:2020bnv,Hisano:2023izx}. Thus, there is no universal loop order at which parity-based solutions regenerate $\bar\theta$; it depends on the implementation of parity breaking and the structure of the heavy sector.

Exact ultraviolet parity constrains higher-dimensional operators, but does not prevent parity-invariant operators from inducing $\bar\theta$ once parity is broken. For example, in left-right Higgs-parity models one may have~\cite{Hall:2018let,Craig:2020bnv}
\begin{align}
    \Delta\mathcal{L}_{\rm UV}
    =
    \frac{c_P}{\Lambda_{\rm UV}^2}
    \left(H_R^\dagger H_R - H_L^\dagger H_L\right)
    \frac{g_s^2}{32\pi^2} G^a_{\mu\nu}\widetilde G^{a\mu\nu}.
    \label{eq:parity-higher-dimensional}
\end{align}
Under parity, $H_L\leftrightarrow H_R$, so the combination $H_R^\dagger H_R-H_L^\dagger H_L$ is parity odd. Since $G\widetilde G$ is also parity odd, the full operator is parity invariant. Once parity is spontaneously broken with $v_R\neq v_L$, however, it induces
\begin{align}
    \delta\bar\theta \sim c_P\, \frac{v_R^2-v_L^2}{\Lambda_{\rm UV}^2},
    \label{eq:parity-quality-estimate}
\end{align}
up to factors of order unity depending on the convention for the Higgs VEVs. For $\Lambda_{\rm UV}$ near the reduced Planck scale and $c_P\sim\mathcal{O}(1)$, the present strong $CP$ bound becomes relevant for parity-breaking scales of order $10^{13}\,{\rm GeV}$. Additional higher-dimensional operators can lead to comparable or stronger constraints in concrete models~\cite{Carrasco-Martinez:2023nit}. The quadratic dependence in Eq.~\eqref{eq:parity-quality-estimate} can make the ultraviolet sensitivity milder than in simple Nelson--Barr models, where corrections linear in $v_{CP}/\Lambda_{\rm UV}$ may arise. This comparison is not universal, however, since the leading dangerous operator depends on the field content and symmetries of the particular construction.

Thus, achieving $\bar\theta_{\rm tree}=0$ is only the first requirement. In both Nelson--Barr and parity-based constructions, radiative corrections, heavy-particle thresholds, and higher-dimensional operators must not regenerate an unacceptably large $\bar\theta$. The resulting constraints play an important role in determining the viable models and their phenomenology, to which we turn next.

\subsection{Phenomenology and modern variants}
\label{sec:CP&P-Pheno}

The phenomenology of $CP$- and parity-based solutions is intrinsically model dependent, since these mechanisms do not predict a unique low-energy particle content. Nelson--Barr constructions commonly contain additional coloured fermions and a sector responsible for spontaneous $CP$ breaking, while parity solutions typically introduce a right-handed or mirror sector. Their phenomenology therefore spans flavour and collider observables, EDMs, neutrino physics, dark matter, baryogenesis, and early-Universe cosmology.

\paragraph{Vector-like quarks in Nelson--Barr models.}

Vector-like quarks are among the most characteristic ingredients of Nelson--Barr models. In the minimal construction discussed in Sec.~\ref{sec:Nelson-Barr}, they transmit spontaneous $CP$ violation to the Standard-Model quarks while preserving the determinant structure required for $\bar\theta_{\rm tree}=0$. Their mixing with the ordinary quarks modifies charged-current interactions and the apparent unitarity of the CKM matrix, and can induce flavour-changing neutral couplings of the $Z$ and Higgs bosons, meson mixing, and rare flavour-changing processes~\cite{Bento:1991ez,Dine:2015jga,Cherchiglia:2020kut,Cherchiglia:2021vhe,Valenti:2021rdu}. Dedicated analyses show that a single down-type Nelson--Barr vector-like quark can be comparatively weakly constrained when its couplings to the light generations are strongly hierarchical, whereas other representations or multiple mediators can lead to stronger flavour constraints~\cite{Cherchiglia:2020kut,Cherchiglia:2021vhe,Valenti:2021rdu}.

More general constructions may contain several up- or down-type vector-like quarks. Systematic parametrisations have been developed that retain the measured light-quark masses and CKM parameters as inputs~\cite{Alves:2023cmw}, while a ``seesaw'' regime can generate the observed hierarchy of light-quark masses together with the Nelson--Barr structure~\cite{Cherchiglia:2024ssz}. Models in which spontaneous $CP$ violation is transmitted through a vector-like partner of the Standard-Model quark doublet have also recently been constructed~\cite{Alves:2026hog}. Depending on their electroweak representation and mixing, these states can be probed through QCD pair production, single production, flavour observables, and decays involving $W$, $Z$, or Higgs bosons~\cite{Cherchiglia:2020kut,Cherchiglia:2021vhe,Alves:2023cmw,Alves:2023ufm,Alves:2026hog}.

\paragraph{Modern $CP$-based constructions.}

A substantial body of recent work has developed the Nelson--Barr mechanism and its ultraviolet realisations. Additional gauge and flavour symmetries can improve the ultraviolet quality of the characteristic Nelson--Barr mass structure. Chiral constructions can use an additional continuous gauge symmetry to forbid dangerous higher-dimensional operators~\cite{Asadi:2022vys}, while $U(1)_R$ gauge extensions can realise a high-quality Nelson--Barr structure automatically~\cite{Perez:2023zin}. Horizontal gauge symmetries can combine a Nelson--Barr-type determinant structure with a dynamical explanation of the hierarchy of the Standard-Model Yukawa couplings~\cite{Choi:2019omm}, while minimal-flavour-violation constructions use flavour symmetry to suppress the strong phase and higher-order corrections~\cite{Bai:2022nat}. The minimal BBP setup has also been revisited using an approximate global symmetry~\cite{Murai:2024alz}, and systematic EFT matching through the $CP$-breaking and vector-like-quark thresholds has recently been developed~\cite{Alves:2025owr}.

Other spontaneous-$CP$ solutions control the strong phase more directly through flavour symmetries and special structures of the ordinary quark mass matrices, without relying on the characteristic Nelson--Barr mixing with vector-like quarks. Discrete family symmetries can constrain quark-mass-matrix entries to be real or purely imaginary so that $\arg\det(\M_u\M_d)=0$, while spontaneous $CP$ violation still generates a physical CKM phase~\cite{Antusch:2013rla}. More recently, three-zero quark-mass textures have been proposed with ultraviolet realisations based on extra-dimensional orbifolds~\cite{Liang:2024wbb}; non-invertible symmetries can instead enforce the required texture directly in four dimensions~\cite{Liang:2025dkm}, including GUT-inspired constructions in which spontaneous $CP$ violation generates the CKM phase while preserving a real quark-mass determinant~\cite{Kobayashi:2025thd}. Modular symmetry provides a closely related route, constraining the flavour structure so that the strong phase vanishes while the VEV of the modulus spontaneously breaks $CP$ and generates a large CKM phase~\cite{Feruglio:2023uof,Petcov:2024vph,Penedo:2024gtb,Feruglio:2025ajb}. Finite modular symmetries and more general local flavour symmetries have further extended this framework and connected it to quark-mass hierarchies~\cite{Feruglio:2024ytl,Liu:2026ati}.

A variety of broader ultraviolet embeddings have also been explored. Supersymmetric constructions can protect the strong phase through non-renormalisation properties and additional symmetry structure~\cite{Evans:2020vil,Liu:2025ycm}, while unified models embed the mechanism into larger gauge structures~\cite{Schwichtenberg:2018aqc,Murgui:2026wim}. Composite and strongly coupled realisations instead generate spontaneous $CP$ breaking or the relevant flavour structure through new nonperturbative dynamics~\cite{Perez:2020dbw,Girmohanta:2022giy,Nakagawa:2024ddd,Csaki:2025ikr}. Despite their different ultraviolet implementations, these constructions share the central requirement of generating an order-one weak $CP$ phase while keeping the strong $CP$ phase sufficiently small.

Cosmology provides an important additional constraint on spontaneous-$CP$ solutions. If an exact $CP$ symmetry is spontaneously broken after inflation, the resulting degenerate $CP$-conjugate vacua generically produce domain walls. A viable cosmology therefore requires that the walls be inflated away without subsequent restoration of $CP$, or that the degeneracy be lifted by a small bias. The latter possibility must be implemented carefully, since explicit $CP$ breaking introduced to destabilise the walls can also regenerate an unacceptable contribution to $\bar\theta$~\cite{Asadi:2022vys,Murai:2024alz}. In a recent Nelson--Barr construction, the scalar responsible for spontaneous $CP$ breaking is identified with the inflaton, so that the $CP$-breaking vacuum is selected during inflation and the associated domain walls are not regenerated afterwards~\cite{Takahashi:2026pmf}.

The $CP$-breaking sector can also contain additional light degrees of freedom, particularly in the presence of approximate symmetries or sufficiently flat directions~\cite{Liu:2025ycm}. Examples include ultralight fields whose coherent oscillations modulate CKM elements and can be probed by precision quantum sensors~\cite{Dine:2024bxv}, light scalar dark matter associated with spontaneous-$CP$ solutions~\cite{Feruglio:2024dnc}, and angular modes whose cosmological evolution and domain walls can contribute to the dark-matter abundance or generate a stochastic gravitational-wave background~\cite{Murai:2024bjy}. Connections between spontaneous $CP$ breaking and baryogenesis have likewise been explored in a variety of settings~\cite{Evans:2020vil,Fujikura:2022sot,Suematsu:2023jqa,Jiang:2024frx,Duch:2025abl}. In Nelson--Barr constructions, ultraviolet-quality and domain-wall requirements can constrain the inflationary and reheating scales and create tension with high-scale thermal leptogenesis~\cite{Asadi:2022vys,Murai:2024alz}, while automatic Nelson--Barr models with new chiral fermions can extend spontaneous $CP$ violation to the lepton sector and support high-scale leptogenesis~\cite{Murgui:2025scx}. The Nelson--Barr mechanism can also be combined with other dynamical ideas, as illustrated by the Nelson--Barr relaxion construction~\cite{Davidi:2017gir}.

\paragraph{Low-scale parity and left-right phenomenology.}

If the parity-breaking scale is comparatively low, left-right constructions can lead to experimentally accessible right-handed gauge bosons, additional scalars, and, in universal-seesaw realisations, vector-like fermions. Collider searches, flavour observables, precision measurements, and EDMs therefore provide complementary probes~\cite{Mohapatra:1978fy,Craig:2020bnv}. Mixing between Standard-Model fermions and their heavy vector-like partners can generate flavour-changing neutral interactions and deviations from charged-current unitarity~\cite{Dcruz:2023mvf}. Universal-seesaw parity models can also accommodate light Dirac neutrino masses and baryogenesis through Affleck--Dine or Dirac leptogenesis, with particular constructions simultaneously incorporating inflation and nonthermal warm dark matter~\cite{Babu:2023srr,Babu:2024glr}. Thus, the phenomenology can extend well beyond direct searches for the right-handed gauge sector.

\paragraph{Higgs parity and mirror sectors.}

A particularly active class of parity solutions is based on Higgs parity, in which parity relates the Standard Model to a right-handed or mirror electroweak sector and is broken by a large VEV $v'$ of the parity partner of the Standard-Model Higgs~\cite{Hall:2018let,Dunsky:2019api}. Matching at the parity-breaking scale can yield a small Standard-Model Higgs quartic coupling, relating $v'$ to its renormalisation-group evolution. The inferred parity scale is therefore sensitive to the Higgs and top-quark masses and to the strong coupling, allowing precision measurements of Standard-Model parameters to probe very high scales of parity restoration~\cite{Hall:2018let,Dunsky:2019api,Carrasco-Martinez:2023nit}.

Neutrino physics, dark matter, and cosmology provide further probes. In left-right Higgs-parity models, neutrino masses can arise radiatively, with heavy neutral leptons potentially accessible to intensity-frontier and collider experiments~\cite{Hall:2023vjb}. Right-handed neutrinos can also provide dark matter while a heavier state generates the baryon asymmetry through leptogenesis~\cite{Dunsky:2020dhn}. In mirror constructions with unbroken mirror electromagnetism, the mirror electron is another dark-matter candidate~\cite{Dunsky:2019api}, while the thermal history of the mirror sector can produce dark radiation and, through a first-order mirror-QCD transition, a stochastic gravitational-wave background~\cite{Dunsky:2019upk}. Variants with modified mirror colour structure have distinct collider and dark-matter phenomenology~\cite{Bonnefoy:2023afx}, and the cosmological production and observational constraints on mirror dark matter have been studied in detail~\cite{Bonnefoy:2023yoj,Redi:2023kgu}. Other parity constructions can contain electroweak-charged dark matter whose stability follows accidentally from the gauge structure~\cite{Baldwin:2026mwv}.

The parity-breaking sector can also participate directly in baryogenesis. A first-order $SU(2)_R$-breaking phase transition may generate the baryon asymmetry through $CP$-violating interactions in the right-handed sector~\cite{Harigaya:2022wzt}, while thermal leptogenesis can proceed through heavy right-handed neutrinos~\cite{Carrasco-Martinez:2023nit}. In Higgs-parity constructions, successful nonresonant thermal leptogenesis generally favours a relatively high parity-breaking scale, correlating baryogenesis with precision Standard-Model parameters and with the sensitivity of EDM experiments to higher-dimensional parity-breaking effects~\cite{Carrasco-Martinez:2023nit}. Parity solutions have also been embedded into unified theories, including supersymmetric $SO(10)$ constructions~\cite{Mimura:2019yfi}, Higgs-parity grand unification~\cite{Hall:2019qwx}, $SU(5)_L\times SU(5)_R$ models~\cite{Babu:2023dzz}, and recent $SO(10)$ constructions with spinorial Higgs fields~\cite{Carrasco-Martinez:2025zus}.

\paragraph{EDMs as a common probe.}

Despite the diversity of these constructions, EDMs provide an important common test. Although Nelson--Barr and parity mechanisms give $\bar\theta_{\rm tree}=0$, radiative corrections, threshold effects, and higher-dimensional operators can regenerate a small strong phase whose magnitude depends on the masses, flavour couplings, and symmetry-breaking parameters of the ultraviolet theory. The neutron EDM can therefore directly probe the residual strong phase predicted by these constructions~\cite{Dine:2015jga,Valenti:2021rdu,Mimura:2019yfi,Craig:2020bnv,Carrasco-Martinez:2023nit,Murgui:2025scx,Alves:2026hog,Bonnefoy:2025rvo}.

An EDM signal would be most informative when combined with other observations. A vector-like quark discovered at a collider would not by itself identify a Nelson--Barr mechanism, but its flavour couplings together with EDM measurements could test such an interpretation. Likewise, evidence for a right-handed or mirror sector could be combined with EDM, neutrino, dark-matter, and cosmological information to probe a parity-based explanation. More generally, the common challenge of $CP$- and parity-based solutions is to accommodate the observed weak $CP$ violation while leaving an extraordinarily small strong phase, so testing them typically requires complementary information from EDMs, flavour physics, collider searches, neutrino experiments, and cosmology.

\section{The Peccei--Quinn mechanism} \label{sec:pq-mechanism}

\subsection{Dynamical relaxation of the strong \texorpdfstring{$CP$}{CP} phase}
\label{sec:PQ-dynamical-relaxation}

The PQ mechanism provides a conceptually different approach to the strong $CP$ problem from the $CP$- and parity-based solutions discussed in the previous section. Rather than requiring $\bar\theta$ to vanish as a consequence of a microscopic $CP$ or parity symmetry, the PQ mechanism promotes the effective strong $CP$ phase to a field-dependent quantity by introducing a dynamical degree of freedom $a(x)$. QCD dynamics then determines the VEV of this field such that the effective strong $CP$ phase relaxes to the $CP$-conserving point~\cite{Peccei:1977hh,Peccei:1977ur}. The corresponding light pseudoscalar particle was subsequently identified independently by Weinberg and Wilczek and is now known as the axion~\cite{Weinberg:1977ma,Wilczek:1977pj}; see also Refs.~\cite{Peccei:2006as,Kim:2008hd,Hook:2018dlk,DiLuzio:2020wdo} for reviews.

The essential ingredients can be described without specifying the ultraviolet realisation of the PQ symmetry. Consider a global chiral symmetry $U(1)_{\rm PQ}$ that is spontaneously broken and whose current has a nonvanishing QCD anomaly, so that its divergence contains a term proportional to the gluonic topological density $G^a_{\mu\nu}\widetilde G^{a\mu\nu}$. Spontaneous symmetry breaking produces a Nambu--Goldstone degree of freedom $a(x)$ associated with the phase of the PQ-breaking order parameter. At energies below the PQ-breaking scale, it is convenient to choose a chiral basis in which the quark mass matrix carries no chiral phase, so that the axion dependence relevant for strong $CP$ is represented explicitly by the gluonic topological term. The relevant part of the effective Lagrangian can then be written as
\begin{align}
    \mathcal{L}_{a}
    \supset 
    \frac{1}{2}\partial_\mu a\,\partial^\mu a
    + \frac{g_s^2}{32\pi^2} \left(\bar\theta+\frac{a}{f_a}
    \right) G^a_{\mu\nu}\widetilde G^{a\mu\nu},
    \label{eq:axion-effective-theta}
\end{align}
where $f_a$ denotes the axion decay constant in a normalisation in which the QCD anomaly coefficient has been absorbed into its definition. Other axion interactions, including derivative couplings to fermions and possible anomalous couplings to electroweak gauge fields, are not relevant for the relaxation argument and will be discussed in Part~\ref{part:QCD-axions}. In another chiral basis, part or all of the axion dependence in Eq.~\eqref{eq:axion-effective-theta} may instead reside in chiral phases of the quark mass matrix. The basis-independent statement is that QCD depends on the axion background through the effective strong $CP$ phase
\begin{align}
    \bar\theta_{\rm eff}(x) = \bar\theta+\frac{a(x)}{f_a}.
    \label{eq:axion-effective-angle}
\end{align}
Unlike $\bar\theta$, $\bar\theta_{\rm eff}$ is therefore a dynamical, field-dependent quantity.

This distinction is essential to the relaxation mechanism. As emphasised in Sec.~\ref{sec:diga}, the dependence of the QCD vacuum energy on a fixed parameter $\bar\theta$ does not imply that $\bar\theta$ should take the value that minimises the vacuum energy: different values of a Lagrangian parameter define different theories, and no equation of motion allows such a parameter to relax. With the axion, by contrast, different values of $\bar\theta_{\rm eff}$ correspond to different field configurations within the same theory. The axion can therefore evolve dynamically in response to the $\bar\theta$ dependence of the QCD vacuum energy and settle at its minimum.

More explicitly, let $E(\bar\theta)$ denote the QCD vacuum energy density discussed in Sec.~\ref{sec:chiral-perturbation-theory}. For an axion background that varies slowly on QCD length scales, the QCD degrees of freedom may be integrated out, giving the axion potential
\begin{align}
    V_{\rm QCD}(a) = E\left(\bar\theta+\frac{a}{f_a}
    \right) - E(0).
    \label{eq:QCD-axion-potential-general}
\end{align}
This relation is considerably more general than a semiclassical instanton calculation. Although the axion potential is often loosely described as being generated by instantons, Eq.~\eqref{eq:QCD-axion-potential-general} requires only a nontrivial dependence of the QCD vacuum energy on $\bar\theta$ and does not rely on the DIGA. At zero temperature, this dependence can be determined systematically at low energies using ChPT and nonperturbatively using lattice QCD~\cite{DiVecchia:1980yfw,Leutwyler:1992yt,GrilliDiCortona:2015jxo,DiLuzio:2020wdo}.

Within the conventional treatment, for QCD with positive quark masses, the Euclidean positivity argument underlying the Vafa--Witten result implies
$E(\bar\theta)\geq E(0)$~\cite{Vafa:1983tf,Vafa:1984xg}. This property is also manifest in the leading-order chiral result derived in Eq.~\eqref{eq:chiral-vacuum-energy-two-flavour}. Although $\bar\theta=\pi$ is also a $CP$-symmetric point by virtue of the $2\pi$ periodicity of the theory, it is not the global minimum of the vacuum energy for the physical quark masses. The minimum relevant for the PQ mechanism therefore occurs at $\bar\theta=0$ modulo $2\pi$.

The axion equation of motion following from Eq.~\eqref{eq:QCD-axion-potential-general} is
\begin{align}
    \Box a + \frac{1}{f_a} \left. \frac{\partial E(\vartheta)}{\partial\vartheta} \right|_{\vartheta=\bar\theta+a/f_a}
    = 0.
    \label{eq:axion-relaxation-eom}
\end{align}
For a homogeneous static configuration, this reduces to the stationary condition for the QCD vacuum energy, with the stable vacuum corresponding to its minimum. Choosing the representative minimum at $\bar\theta_{\rm eff}=0$, the axion therefore develops a vacuum expectation value satisfying
\begin{align}
    \left\langle\bar\theta_{\rm eff}\right\rangle = 
    \bar\theta + \frac{\langle a\rangle}{f_a} = 0.
    \label{eq:PQ-relaxation-minimum}
\end{align}
The original value of $\bar\theta$ need not itself be small: the axion VEV dynamically compensates it so that the effective strong $CP$ phase vanishes in the vacuum.

The same QCD dynamics that selects the $CP$-conserving vacuum also lifts the axion direction. Expanding Eq.~\eqref{eq:QCD-axion-potential-general} around the minimum, the axion mass is determined by the curvature of the QCD vacuum energy,
\begin{align}
    m_a^2 = \left. \frac{\partial^2V_{\rm QCD}}{\partial a^2}
    \right|_{\langle a\rangle} = \frac{1}{f_a^2}
    \left. \frac{\partial^2E(\bar\theta)}{\partial\bar\theta^2}
    \right|_{\bar\theta=0} = \frac{\chi}{f_a^2},
    \label{eq:axion-mass-susceptibility}
\end{align}
where $\chi$ is the QCD topological susceptibility introduced in Eq.~\eqref{eq:chiral-topological-susceptibility}. The axion is therefore a pseudo-Nambu--Goldstone boson: in the absence of the nonperturbative QCD effects associated with the anomaly, the axion direction would remain flat, whereas QCD dynamics generates its potential and hence its mass.

The mass can be made explicit using the chiral results derived in Sec.~\ref{sec:chiral-perturbation-theory}. In the two-flavour theory, substituting Eq.~\eqref{eq:two-flavour-susceptibility} into Eq.~\eqref{eq:axion-mass-susceptibility} gives at leading order
\begin{align}
    m_{a,{\rm LO}}^2 = \frac{m_\pi^2F_\pi^2}{f_a^2} \frac{m_um_d}{(m_u+m_d)^2}.
    \label{eq:axion-mass-LO}
\end{align}
Thus, once the normalisation of $f_a$ is fixed, the QCD axion mass is not an independent parameter but is determined by QCD. Higher-order chiral corrections and precise inputs for the light-quark masses allow this relation to be determined at the percent level and better~\cite{GrilliDiCortona:2015jxo,Gorghetto:2018ocs}. Including next-to-next-to-leading-order chiral and electromagnetic corrections, Ref.~\cite{Gorghetto:2018ocs} finds
\begin{align}
    m_a = 5.691(51)\,\mu{\rm eV}
    \left(\frac{10^{12}\,{\rm GeV}}{f_a} \right).
    \label{eq:axion-mass-numerical}
\end{align}
The smallness of the axion mass for $f_a\gg\Lambda_{\rm QCD}$ is therefore a direct consequence of its pseudo-Nambu--Goldstone nature, while the inverse proportionality $m_a\propto f_a^{-1}$ is a characteristic prediction of the QCD axion.

An important assumption in this argument is that QCD provides the dominant explicit breaking of the PQ symmetry. Additional PQ-breaking interactions can generate an independent contribution $\Delta V_{\slashed{\rm PQ}}(a)$ to the axion potential. Unless this contribution is sufficiently small or aligned with the QCD potential, it shifts $\langle\bar\theta_{\rm eff}\rangle$ away from zero. The strong $CP$ bound then requires the residual angle to remain below approximately $10^{-10}$. This sensitivity to additional explicit PQ breaking is the axion quality problem~\cite{Barr:1992qq,Holman:1992us,Kamionkowski:1992mf}, which will be discussed in detail in Sec.~\ref{sec:axion-quality-problem}.

The discussion above is largely independent of how the PQ symmetry is realised microscopically. A complete model must nevertheless provide a field content and interaction structure for which an anomalous global $U(1)_{\rm PQ}$ symmetry exists, is spontaneously broken, and is compatible with the quark Yukawa interactions. The original PQ construction achieved this by extending the electroweak Higgs sector. Its associated axion provides the simplest concrete illustration of the general mechanism and is discussed in the next subsection.

\subsection{The original Peccei--Quinn--Weinberg--Wilczek axion}
\label{sec:PQWW-axion}

The discussion in the previous subsection was independent of the microscopic realisation of the PQ symmetry. Historically, the PQ mechanism was first implemented in an extension of the electroweak theory containing two Higgs doublets~\cite{Peccei:1977hh,Peccei:1977ur}. Weinberg and Wilczek subsequently pointed out that the spontaneous breaking of the anomalous global symmetry necessarily gives rise to a light pseudoscalar particle, now known as the axion~\cite{Weinberg:1977ma,Wilczek:1977pj}. This construction is commonly referred to as the Peccei--Quinn--Weinberg--Wilczek (PQWW) model. It provides a particularly transparent illustration of the mechanism because the axion arises entirely from the electroweak Higgs sector.

The reason for enlarging the Higgs sector can already be understood from the Standard-Model Yukawa interactions. With a single Higgs doublet $H$, the quark Yukawa interactions are
\begin{align}
    -\mathcal{L}_Y = \bar Q_LY_u\widetilde H u_R
    + \bar Q_LY_dH d_R +\hc ,
    \label{eq:SM-one-Higgs-Yukawa}
\end{align}
where $\widetilde H=\i\tau_2 H^*$ and generation indices are suppressed. Suppose that $Q_L$, $u_R$, $d_R$, and $H$ carry charges $X_Q$, $X_u$, $X_d$, and $X_H$, respectively, under an additional global $U(1)$ symmetry. Since $\widetilde H$ carries charge $-X_H$, invariance of the two terms in Eq.~\eqref{eq:SM-one-Higgs-Yukawa} requires
$X_u=X_Q+X_H$ and $X_d=X_Q-X_H$. The corresponding chiral charge differences of the up- and down-type quarks are therefore equal and opposite,
$X_{u_L}-X_{u_R}=X_Q-X_u=-X_H$ and
$X_{d_L}-X_{d_R}=X_Q-X_d=+X_H$. Their contributions to the QCD anomaly of the global $U(1)$ current consequently cancel within each generation. Such a symmetry therefore has no net QCD anomaly and cannot serve as the anomalous PQ symmetry.

The origin of this obstruction is that the same Higgs doublet enters the two Yukawa interactions through $H$ and $\widetilde H$, whose global charges are necessarily opposite. Introducing two Higgs doublets removes this constraint by allowing the up- and down-type quarks to obtain their masses from different scalar fields with different PQ charges. The Yukawa interactions then no longer require the two quark sectors to contribute equally and oppositely to the QCD anomaly, making an anomalous global chiral symmetry compatible with the quark mass terms. 

We take both Higgs doublets $H_u$ and $H_d$ to have hypercharge $+1/2$. The quark Yukawa interactions then take the form
\begin{align}
    -\mathcal{L}_Y = \bar Q_LY_u\widetilde H_u u_R
    + \bar Q_LY_dH_d d_R +\hc .
    \label{eq:PQWW-Yukawa}
\end{align}
A convenient normalisation of the PQ transformation is
\begin{align}
    H_u&\rightarrow\e^{\i\omega}H_u,\qquad
    H_d\rightarrow\e^{-\i\omega}H_d,\qquad
    Q_L\rightarrow Q_L,\qquad
    u_R\rightarrow\e^{\i\omega}u_R,\qquad
    d_R\rightarrow\e^{\i\omega}d_R.
    \label{eq:PQWW-PQ-transformation}
\end{align}
The Yukawa interactions in Eq.~\eqref{eq:PQWW-Yukawa} are invariant under this transformation, while the corresponding PQ current has a nonvanishing QCD anomaly. The charge assignment is not unique, since the PQ generator may be shifted by anomaly-free global symmetries or by a gauge generator without changing the physical mechanism. We leave the lepton-sector assignment unspecified, since it does not affect the QCD anomaly or the relaxation of the strong $CP$ phase.

The scalar potential must respect the same global symmetry. For the PQ charge assignment in Eq.~\eqref{eq:PQWW-PQ-transformation}, the most general renormalisable scalar potential consistent with both the electroweak gauge symmetry and $U(1)_{\rm PQ}$ can be written as
\begin{align}
    V(H_u,H_d) = m_{H_u}^2H_u^\dagger H_u + m_{H_d}^2H_d^\dagger H_d + \frac{\lambda_u}{2} (H_u^\dagger H_u)^2
    + \frac{\lambda_d}{2} (H_d^\dagger H_d)^2
    + \lambda_{ud} (H_u^\dagger H_u)(H_d^\dagger H_d)
    + \lambda_{ud}' (H_u^\dagger H_d)(H_d^\dagger H_u).
    \label{eq:PQWW-scalar-potential}
\end{align}
Compared with the most general renormalisable two-Higgs-doublet potential, the PQ symmetry forbids operators that are sensitive to the relative phase of the two doublets. These include the bilinear term $H_u^\dagger H_d+\hc$ and quartic terms such as $(H_u^\dagger H_d)^2+\hc$, $(H_u^\dagger H_u)(H_u^\dagger H_d)+\hc$, and $(H_d^\dagger H_d)(H_u^\dagger H_d)+\hc$. Such operators would explicitly break $U(1)_{\rm PQ}$ and lift the associated Nambu--Goldstone direction already at the perturbative level. Instead, the explicit breaking relevant for the strong $CP$ mechanism arises from the QCD anomaly, with nonperturbative QCD dynamics generating the axion potential~\cite{Peccei:1977hh,Peccei:1977ur}.

For suitable parameters, the two neutral Higgs fields acquire VEVs,
\begin{align}
    \langle H_u\rangle = \frac{1}{\sqrt{2}}
    \begin{pmatrix}
        0\\ v_u
    \end{pmatrix},
    \qquad
    \langle H_d\rangle = \frac{1}{\sqrt{2}}
    \begin{pmatrix}
        0\\ v_d
    \end{pmatrix},
    \qquad
    v^2=v_u^2+v_d^2,
    \label{eq:PQWW-Higgs-vevs}
\end{align}
where $v\simeq246\,{\rm GeV}$, and we define $\tan\beta=v_u/v_d$. Both the electroweak gauge symmetry and $U(1)_{\rm PQ}$ are then spontaneously broken. To identify the pseudoscalar modes, we parameterise the neutral fields as
\begin{align}
    H_u^0 = \frac{v_u+\rho_u}{\sqrt{2}}\, \e^{\i\eta_u/v_u},
    \qquad
    H_d^0 = \frac{v_d+\rho_d}{\sqrt{2}}\, \e^{\i\eta_d/v_d}.
    \label{eq:PQWW-neutral-Higgs}
\end{align}
Substituting these expressions into the canonical Higgs kinetic terms, $|D_\mu H_u|^2+|D_\mu H_d|^2$, one finds canonical kinetic terms for $\eta_u$ and $\eta_d$, together with a mixing term proportional to
$Z^\mu\partial_\mu(v_u\eta_u+v_d\eta_d)$. The neutral Nambu--Goldstone mode eaten by the $Z$ boson is therefore
$G^0=(v_d\eta_d+v_u\eta_u)/v$. The orthogonal, canonically normalised combination remains as a physical massless pseudoscalar at the classical level, 
\begin{align}
    a_{\rm EW} = \frac{v_u\eta_d-v_d\eta_u}{v}.
    \label{eq:PQWW-axion-mode}
\end{align}
This is the PQWW axion before the QCD-induced potential is taken into account. Its overall sign is conventional.

The relation to the strong $CP$ phase can now be seen directly. In the unitary gauge, $G^0=0$, the definitions above give
$\eta_d=a_{\rm EW}\,v_u/v$ and $\eta_u=-a_{\rm EW}\,v_d/v$.
Substituting these expressions into Eq.~\eqref{eq:PQWW-Yukawa}, the axion-dependent quark mass terms become
\begin{align}
    -\mathcal{L}_{\rm mass} = \bar u_L\M_u \e^{\i(a_{\rm EW}/v)\cot\beta} u_R + \bar d_L\M_d \e^{\i(a_{\rm EW}/v)\tan\beta} d_R +\hc ,
    \qquad
    \M_u = \frac{v_u}{\sqrt{2}}Y_u,
    \quad
    \M_d = \frac{v_d}{\sqrt{2}}Y_d.
    \label{eq:PQWW-axion-mass-phases}
\end{align}
For $N_g$ generations of quarks, with $\M_u$ and $\M_d$ understood as $N_g\times N_g$ matrices in generation space, the invariant strong $CP$ phase defined in Eq.~\eqref{eq:bartheta-full-quark-sector} consequently becomes
\begin{align}
    \bar\theta_{\rm eff}
    = \theta + \arg\det \left[\M_u\e^{\i(a_{\rm EW}/v)\cot\beta}\M_d \e^{\i(a_{\rm EW}/v)\tan\beta}
    \right] = \bar\theta + \frac{N_g\,a_{\rm EW}}{v}
    \left(\cot\beta+\tan\beta \right)
    = \bar\theta + \frac{2N_g\,a_{\rm EW}}{v\sin2\beta}.
    \label{eq:PQWW-effective-theta}
\end{align}
For the PQ charge normalisation in Eq.~\eqref{eq:PQWW-PQ-transformation}, the PQ symmetry-breaking scale associated with the axion is\footnote{Under a PQ transformation, $\eta_u\rightarrow\eta_u+v_u\omega$ and $\eta_d\rightarrow\eta_d-v_d\omega$, so that the physical axion defined in Eq.~\eqref{eq:PQWW-axion-mode} transforms as $a\rightarrow a-v_{\rm PQWW}\,\omega$.}
\begin{align}
    v_{\rm PQWW} = v\sin2\beta = \frac{2v_uv_d}{v}.
    \label{eq:PQWW-fPQ}
\end{align}
The magnitude of the corresponding QCD anomaly coefficient is $2N_g$. Comparing Eq.~\eqref{eq:PQWW-effective-theta} with Eq.~\eqref{eq:axion-effective-angle}, we therefore identify
\begin{align}
    f_a = \frac{v_{\rm PQWW}}{2N_g} = \frac{v\sin2\beta}{2N_g}.
    \label{eq:PQWW-fa}
\end{align}
Different conventions in the literature may instead refer to $f_{\rm PQ}$ as the axion decay constant and keep the QCD anomaly coefficient explicit. Throughout this review, we adopt the anomaly-normalised convention for $f_a$ introduced in Sec.~\ref{sec:PQ-dynamical-relaxation}.

Working in the chiral basis in which the coefficient of the gluonic topological term is $\bar\theta$, while the axion-dependent phases are retained in the quark mass terms rather than transferred to an explicit $aG\widetilde G$ interaction, and expanding Eq.~\eqref{eq:PQWW-axion-mass-phases} to first order in $a_{\rm EW}$, one finds 
\begin{align}
    \mathcal{L}_{aqq} =
    -\frac{\i a_{\rm EW}}{v}
    \left[ \cot\beta \sum_{u_i}m_{u_i}\bar u_i\gamma^5u_i + \tan\beta \sum_{d_i}m_{d_i}\bar d_i\gamma^5d_i \right] +\cdots .
    \label{eq:PQWW-quark-couplings}
\end{align}
The axion couplings are therefore suppressed only by the electroweak scale. Furthermore, since $\sin2\beta\leq1$, Eq.~\eqref{eq:PQWW-fa} gives $f_a\leq v/6\simeq41\,{\rm GeV}$ for the three Standard-Model generations in our anomaly-normalised convention. Equation~\eqref{eq:axion-mass-susceptibility} then implies an axion mass of order $10^2\,{\rm keV}$ or larger, together with comparatively strong couplings to ordinary matter~\cite{Weinberg:1977ma,Wilczek:1977pj}. The PQWW axion is therefore commonly referred to as a \emph{visible axion}.

The visible axion was soon excluded by laboratory searches. Important constraints arose from rare meson and quarkonium decays and from fixed-target experiments; representative early electron beam-dump searches are given in Refs.~\cite{Konaka:1986cb,Riordan:1987aw,Bjorken:1988as}. See Refs.~\cite{Peccei:2006as,Kim:2008hd,DiLuzio:2020wdo} for broader reviews of the experimental history. These exclusions apply to the original electroweak-scale realisation rather than to the PQ mechanism itself, leaving open the possibility that the PQ symmetry is broken at a much higher scale.

\part{QCD axions}
\label{part:QCD-axions}

\section{Invisible axion models and low-energy properties}

\subsection{Benchmark models: KSVZ and DFSZ}
\label{sec:KSVZ-DFSZ}

The key step towards a viable axion model is therefore to separate the PQ-breaking scale from the electroweak scale. This led to the invisible axion models, in which the PQ symmetry is broken predominantly by an electroweak-singlet scalar with a VEV much larger than $v$, so that the axion is correspondingly weakly coupled. The two benchmark realisations are the Kim--Shifman--Vainshtein--Zakharov (KSVZ) model~\cite{Kim:1979if,Shifman:1979if} and the Dine--Fischler--Srednicki--Zhitnitsky (DFSZ) model~\cite{Zhitnitsky:1980tq,Dine:1981rt}; see also Refs.~\cite{Kim:2008hd,DiLuzio:2020wdo,Choi:2020rgn} for reviews. Their main structural difference is the origin of the QCD anomaly and the way the axion couples to the Standard Model: in KSVZ models the anomaly is carried by new heavy coloured fermions, whereas in DFSZ models it is carried by the Standard-Model quarks.

\paragraph{Kim--Shifman--Vainshtein--Zakharov models.}

The KSVZ construction introduces a complex Standard-Model-singlet scalar $\Phi$ together with at least one new heavy coloured fermion~\cite{Kim:1979if,Shifman:1979if}. For illustration, consider a vector-like Dirac fermion $\Psi=\Psi_L+\Psi_R$ transforming in the fundamental representation of $SU(3)_c$. Its electroweak quantum numbers are model dependent. Although $\Psi_L$ and $\Psi_R$ transform identically under the Standard-Model gauge group, they carry different PQ charges, so the fermion is vector-like with respect to the gauge symmetries but chiral under $U(1)_{\rm PQ}$. A minimal set of interactions is
\begin{align}
    -\mathcal{L}_{\rm KSVZ}
    \supset y_\Psi\Phi\,\bar\Psi_L\Psi_R +\hc ,
    \qquad
    V(\Phi) = \lambda_\Phi \left( \Phi^\dagger\Phi-\frac{v_\Phi^2}{2} \right)^2 .
    \label{eq:KSVZ-Lagrangian}
\end{align}
A convenient PQ charge normalisation is
\begin{align}
    \Phi\rightarrow\e^{\i\omega}\Phi,
    \qquad
    \Psi_L\rightarrow\Psi_L,
    \qquad
    \Psi_R\rightarrow\e^{-\i\omega}\Psi_R,
    \label{eq:KSVZ-PQ-transformation}
\end{align}
while the Standard-Model fields may all be taken to be PQ neutral. The Yukawa interaction in Eq.~\eqref{eq:KSVZ-Lagrangian} is then invariant, whereas a bare mass term $\bar\Psi_L\Psi_R$ is forbidden by the PQ symmetry.

When $\Phi$ develops a VEV, we may write
\begin{align}
    \Phi = \frac{v_\Phi+\rho_\Phi}{\sqrt{2}}\, \e^{\i a_\Phi/v_\Phi}, \qquad
    m_\Psi = \frac{y_\Psi v_\Phi}{\sqrt{2}},
    \label{eq:KSVZ-singlet}
\end{align}
where $a_\Phi$ is canonically normalised at leading order. The heavy-fermion mass term consequently contains the phase
\begin{align}
    -\mathcal{L}_{\rm KSVZ} \supset m_\Psi \e^{\i a_\Phi/v_\Phi} \bar\Psi_L\Psi_R +\hc .
    \label{eq:KSVZ-axion-mass-phase}
\end{align}
This makes the connection with the strong $CP$ phase particularly transparent. Above the heavy-fermion threshold, the phase of $m_\Psi$ contributes to the determinant of the full coloured-fermion mass matrix. For one fundamental Dirac fermion and the charge normalisation in Eq.~\eqref{eq:KSVZ-PQ-transformation}, one therefore obtains
\begin{align}
    \bar\theta_{\rm eff} = \bar\theta + \frac{a_\Phi}{v_\Phi},
    \label{eq:KSVZ-effective-theta}
\end{align}
where any constant phase in $y_\Psi$ has been absorbed into $\bar\theta$. Equivalently, after the heavy fermion is integrated out, an anomalous chiral field redefinition transfers the $a_\Phi$ dependence in Eq.~\eqref{eq:KSVZ-axion-mass-phase} to the gluonic topological term. Comparing Eq.~\eqref{eq:KSVZ-effective-theta} with Eq.~\eqref{eq:axion-effective-angle}, we identify $a=a_\Phi$, $f_a=v_\Phi$
for this minimal choice. More generally, additional coloured fermions, different colour representations, or different PQ charge normalisations modify the QCD anomaly coefficient and hence the relation between $f_a$ and the underlying symmetry-breaking scale.

The important point is that $v_\Phi$ is unrelated to electroweak symmetry breaking and can therefore be much larger than $v$. The axion is correspondingly much more weakly coupled than the PQWW axion. In the canonical KSVZ construction, the ordinary quarks and leptons carry no PQ charge, so direct axion couplings to Standard-Model fermions are absent at tree level. Couplings to gluons, photons, and hadrons nevertheless arise through the PQ anomaly and through low-energy QCD effects. Their precise form depends, in particular, on the gauge quantum numbers of the heavy fermions and will be discussed in Sec.~\ref{sec:axion-low-energy-couplings}.

\paragraph{Dine--Fischler--Srednicki--Zhitnitsky models.}

The DFSZ construction achieves the same separation of the PQ and electroweak scales without introducing new coloured fermions~\cite{Zhitnitsky:1980tq,Dine:1981rt}. Instead, it supplements the two-Higgs-doublet structure of the original PQWW model by a complex Standard-Model-singlet scalar $\Phi$. The Standard-Model quarks themselves carry the anomalous PQ charges, while the large singlet VEV makes the physical axion predominantly the phase of an electroweak-singlet field.

To maintain direct continuity with Sec.~\ref{sec:PQWW-axion}, we retain the Higgs fields $H_u$ and $H_d$, their Yukawa interactions in Eq.~\eqref{eq:PQWW-Yukawa}, and the PQ transformations in Eq.~\eqref{eq:PQWW-PQ-transformation}. We further assign the singlet field the PQ transformation
$\Phi\rightarrow\e^{\i\omega}\Phi$. With this charge normalisation, the scalar potential admits the phase-sensitive interaction
\begin{align}
    V_{\rm DFSZ} \supset \lambda_\Phi\, \Phi^2 H_u^\dagger H_d
    +\hc . 
    \label{eq:DFSZ-scalar-interaction}
\end{align}
Indeed, $H_u^\dagger H_d\rightarrow\e^{-2\i\omega}H_u^\dagger H_d$, so Eq.~\eqref{eq:DFSZ-scalar-interaction} is invariant under $U(1)_{\rm PQ}$. This interaction is essential for correlating the singlet phase with the relative phase of the two Higgs doublets. In its absence, the phase of $\Phi$ could be rotated independently, and the renormalisable scalar and Yukawa sectors would possess an additional global $U(1)$ symmetry. After spontaneous symmetry breaking, the QCD-anomalous Goldstone direction would then remain the electroweak-scale PQWW axion, while the independent singlet phase would give rise to an additional massless pseudoscalar. Thus, simply introducing a singlet with a large symmetry-breaking scale would not by itself make the QCD axion invisible. The interaction in Eq.~\eqref{eq:DFSZ-scalar-interaction} locks the singlet and Higgs phases together, breaking the orthogonal global symmetry while preserving the desired $U(1)_{\rm PQ}$. As a result, there is a single PQ Goldstone direction, which can be predominantly singlet when the singlet VEV is much larger than the electroweak scale, while still inheriting the QCD anomaly through the quark sector. The full scalar potential also contains the usual PQ-invariant terms constructed from $\Phi^\dagger\Phi$, $H_u^\dagger H_u$, $H_d^\dagger H_d$, and their products.

We take
\begin{align}
    \langle\Phi\rangle = \frac{v_\Phi}{\sqrt{2}},
    \qquad
    v_\Phi\gg v,
    \qquad
    \Phi = \frac{v_\Phi+\rho_\Phi}{\sqrt{2}}\,\e^{\i\eta_\Phi/v_\Phi}.
    \label{eq:DFSZ-singlet-vev}
\end{align}
The neutral Higgs phases $\eta_u$ and $\eta_d$ are parameterised as in Eq.~\eqref{eq:PQWW-neutral-Higgs}. As discussed there, the electroweak Goldstone mode is
$G^0=(v_u\eta_u+v_d\eta_d)/v$, while the orthogonal electroweak pseudoscalar direction is $a_{\rm EW} \equiv (v_u\eta_d-v_d\eta_u)/v$. Under the PQ transformation, $\eta_\Phi \rightarrow \eta_\Phi+v_\Phi\omega$, $a_{\rm EW}\rightarrow a_{\rm EW}-v\sin2\beta\,\omega $.
The Goldstone direction associated with the broken $U(1)_{\rm PQ}$ is therefore proportional to
$v_\Phi\eta_\Phi-v\sin2\beta\,a_{\rm EW}$.\footnote{More generally, for canonically normalised phase fields $\pi_i$ transforming under a spontaneously broken continuous global symmetry as $\pi_i\rightarrow\pi_i+c_i\omega$, the infinitesimal symmetry transformation displaces the vacuum along the field-space direction $(c_1,c_2,\ldots)$. The corresponding Nambu--Goldstone mode is therefore the canonically normalised combination $a_{\rm NG}=(\sum_i c_i\pi_i)/\sqrt{\sum_i c_i^2}$, which transforms nonlinearly as $a_{\rm NG}\rightarrow a_{\rm NG}+\omega\sqrt{\sum_i c_i^2}$. Combinations orthogonal to this direction do not acquire such a shift and hence are not the Nambu--Goldstone mode associated with this symmetry.} For consistency with the convention
$\bar\theta_{\rm eff}=\bar\theta+a/f_a$ adopted in Eq.~\eqref{eq:axion-effective-angle} and with the definition of the PQWW axion in Eq.~\eqref{eq:PQWW-axion-mode}, we choose the overall sign of the canonically normalised physical axion as
\begin{align}
    a = \frac{-v_\Phi\eta_\Phi+v\sin2\beta\,a_{\rm EW}}{v_{\rm PQ}},
    \qquad
    v_{\rm PQ}^2 = v_\Phi^2 + v^2\sin^22\beta .
    \label{eq:DFSZ-axion-mode}
\end{align}
With this convention, $a\rightarrow a-v_{\rm PQ}\omega$ under the PQ transformation. In the invisible-axion regime $v_\Phi\gg v$, Eq.~\eqref{eq:DFSZ-axion-mode} gives
\begin{align}
    a \approx -\eta_\Phi + \mathcal{O}\left(\frac{v}{v_\Phi}\right),
    \label{eq:DFSZ-singlet-dominated}
\end{align}
so, up to the conventional overall sign, the physical axion is predominantly the phase of the electroweak-singlet field.

The QCD anomaly is nevertheless generated by the ordinary quarks. With the PQ charge assignment inherited from Eq.~\eqref{eq:PQWW-PQ-transformation}, the magnitude of the QCD anomaly coefficient is $2N_g$, as already found in Sec.~\ref{sec:PQWW-axion}. To determine the coupling of the physical axion to QCD, we resolve the two physical pseudoscalar directions into the axion $a$ defined in Eq.~\eqref{eq:DFSZ-axion-mode} and the orthogonal massive pseudoscalar. Setting the latter to zero gives\footnote{The factor in Eq.~\eqref{eq:DFSZ-aEW-axion} follows from the inverse rotation between the original pseudoscalar fields and the mass eigenstate directions. In particular, defining the direction orthogonal to the axion by
$A=[v\sin2\beta\,\eta_\Phi+v_\Phi a_{\rm EW}]/v_{\rm PQ}$, Eq.~\eqref{eq:DFSZ-axion-mode} and this definition imply
$a_{\rm EW}=[v\sin2\beta\,a+v_\Phi A]/v_{\rm PQ}$. Hence, along the axion direction, $A=0$ and Eq.~\eqref{eq:DFSZ-aEW-axion} follows. }
\begin{align}
    a_{\rm EW} = \frac{v\sin2\beta}{v_{\rm PQ}}\,a .
    \label{eq:DFSZ-aEW-axion}
\end{align}
Using the PQWW result in Eq.~\eqref{eq:PQWW-effective-theta}, the effective strong $CP$ phase therefore becomes
\begin{align}
    \bar\theta_{\rm eff} = \bar\theta + \frac{2N_g}{v\sin2\beta}\,a_{\rm EW} = \bar\theta
    + \frac{2N_g}{v_{\rm PQ}}\,a .
    \label{eq:DFSZ-effective-theta}
\end{align}
Comparing with Eq.~\eqref{eq:axion-effective-angle}, we identify
\begin{align}
    f_a = \frac{v_{\rm PQ}}{2N_g} = \frac{\sqrt{v_\Phi^2+v^2\sin^22\beta}}{2N_g} .
    \label{eq:DFSZ-fa}
\end{align}
For the three Standard-Model generations and $v_\Phi\gg v$, this reduces to $f_a\approx \frac{v_\Phi}{6}$. Thus, although the physical axion is predominantly the singlet phase in the invisible-axion regime, the phase-locking interaction in Eq.~\eqref{eq:DFSZ-scalar-interaction} ensures that it retains the QCD anomaly inherited from the quark and Higgs sector. 

Because the Higgs doublets and the Standard-Model fermions carry PQ charge, the DFSZ axion has tree-level derivative couplings to ordinary fermions, although these are suppressed by $f_a$ rather than by the electroweak scale. The charged-lepton sector admits different PQ-compatible Yukawa assignments, leading to the commonly considered DFSZ variants. These choices do not alter the basic strong $CP$ mechanism or the QCD axion mass relation, but they do affect the low-energy axion couplings.

The KSVZ and DFSZ constructions thus provide two complementary benchmark realisations of an invisible QCD axion. Their principal structural difference lies in the origin of the QCD anomaly: in KSVZ models it is carried by new heavy coloured fermions, while the Standard-Model fermions can be PQ neutral, whereas in DFSZ models it is carried by the ordinary quarks through their PQ charges. Neither label denotes a unique model. Different electroweak quantum numbers and multiplicities of the heavy fermions in KSVZ constructions, and different PQ charge and Yukawa assignments in DFSZ constructions, define families of ultraviolet realisations with distinct phenomenology~\cite{Kim:2008hd,DiLuzio:2020wdo}. At low energies, this ultraviolet information is encoded primarily in the anomaly coefficients and in the axion couplings to Standard-Model fermions. We turn next to these low-energy interactions and their model dependence.

\subsection{Low-energy axion couplings and model dependence}
\label{sec:axion-low-energy-couplings}

The QCD axion mass is essentially fixed by QCD once $f_a$ is specified, as discussed in Sec.~\ref{sec:PQ-dynamical-relaxation}. Its low-energy interactions, however, are not determined by $m_a$ or $f_a$ alone. They contain both contributions fixed by QCD and model-dependent contributions inherited from the PQ charges and anomaly coefficients of the ultraviolet theory~\cite{Srednicki:1985xd,Georgi:1986df,GrilliDiCortona:2015jxo,DiLuzio:2020wdo,Choi:2020rgn}. It is therefore useful to organise the axion couplings by separating the QCD-induced contributions common to all QCD axions from the ultraviolet information that distinguishes different realisations of the PQ symmetry.

At energies below the PQ and electroweak scales, but above the scale of QCD confinement, we return to the chiral basis used in Eq.~\eqref{eq:axion-effective-theta}, in which the quark mass matrix carries no chiral phase and the QCD anomaly is represented explicitly by the axion-gluon coupling. After the PQ relaxation, we shift the axion field so that the vacuum lies at $a=0$ and denote the fluctuation again by $a$. Restricting for the moment to flavour-diagonal axion-fermion interactions, the relevant effective Lagrangian can then be written as
\begin{align}
    \mathcal{L}_a
    \supset \frac{1}{2}\partial_\mu a\,\partial^\mu a
    + \frac{g_s^2}{32\pi^2} \frac{a}{f_a}G^a_{\mu\nu}\widetilde G^{a\mu\nu} + \frac{e^2}{32\pi^2} \frac{E}{N} \frac{a}{f_a} F_{\mu\nu}\widetilde F^{\mu\nu}
    + \sum_f C_f(\mu) \frac{\partial_\mu a}{2f_a}
    \bar f\gamma^\mu\gamma^5f +\cdots ,
    \label{eq:axion-general-low-energy-EFT}
\end{align}
where the sum runs over the Standard-Model Dirac fermion mass eigenstates that remain dynamical at the scale $\mu$. 
Here $E$ and $N$ denote the electromagnetic and QCD anomaly coefficients of the PQ current, respectively. We choose the orientation of the PQ symmetry such that $N>0$ and absorb $N$ into the definition of the axion decay constant, $f_a=v_{\rm PQ}/N$, where $v_{\rm PQ}$ denotes the periodicity scale of the canonically normalised axion field, so that $a\sim a+2\pi v_{\rm PQ}$. The gluonic interaction then has unit anomaly coefficient, while the electromagnetic interaction depends on the ratio $E/N$. The coefficients $C_f(\mu)$ describe derivative axion-fermion interactions and in general depend on the renormalisation scale~\cite{Srednicki:1985xd,Georgi:1986df,Bauer:2020jbp,Choi:2021kuy}.

\paragraph{Axion-photon coupling.}

Below the QCD scale, the interaction with photons is conventionally written as
\begin{align}
    \mathcal{L}_{a\gamma\gamma} = \frac{1}{4} g_{a\gamma\gamma}\, aF_{\mu\nu}\widetilde F^{\mu\nu},
    \qquad
    g_{a\gamma\gamma} = \frac{\alpha_{\rm em}}{2\pi f_a}\,
    C_\gamma ,
    \label{eq:axion-photon-coupling}
\end{align}
where, with our convention for the dual field strength,
\begin{align}
    C_\gamma = \frac{E}{N} - \mathcal{C}_{\rm QCD}.
    \label{eq:axion-photon-Cgamma}
\end{align}
The first term is determined by the electromagnetic anomaly of the ultraviolet PQ current and is therefore model dependent. The second is generated by low-energy QCD, in particular through the mixing of the axion with the neutral pseudoscalar mesons, and is common to all QCD axion models~\cite{Srednicki:1985xd,Georgi:1986df,GrilliDiCortona:2015jxo}. Combining NLO two-flavour ChPT with hadronic input gives the widely used result~\cite{GrilliDiCortona:2015jxo}
\begin{align}
    \mathcal{C}_{\rm QCD} = 1.92(4).
    \label{eq:axion-photon-QCD-contribution}
\end{align}
A recent continuum-extrapolated lattice-QCD calculation has provided the first direct nonperturbative determination of this contribution and finds $\mathcal{C}_{\rm QCD}^{\rm lattice}=1.77(8)$~\cite{Brandt:2026wir}.
The precise model-independent contribution to $g_{a\gamma\gamma}$ remains an active precision-QCD problem. In the following, when quoting conventional benchmark values, we use Eq.~\eqref{eq:axion-photon-QCD-contribution}.

The ultraviolet contribution can differ substantially between different axion models. For the simple KSVZ construction introduced in Sec.~\ref{sec:KSVZ-DFSZ}, with a single $SU(3)_c$-fundamental, $SU(2)_L$-singlet heavy fermion of electric charge $Q_\Psi$, the electromagnetic and QCD anomaly coefficients satisfy
\begin{align}
    \frac{E}{N} = 6Q_\Psi^2.
    \label{eq:KSVZ-EN}
\end{align}
The commonly used electrically neutral KSVZ benchmark, $Q_\Psi=0$, therefore has $E/N=0$ and hence $C_\gamma\simeq-1.92$. More general choices of the heavy-fermion electroweak quantum numbers, colour representations, or particle content can lead to different values of $E/N$~\cite{Kim:2008hd,DiLuzio:2020wdo,DiLuzio:2017pfr}. In the minimal DFSZ models, the value of $E/N$ instead depends on the charged-lepton Yukawa assignment~\cite{DiLuzio:2020wdo,Sun:2020iim}. Retaining the Higgs convention of Eq.~\eqref{eq:PQWW-Yukawa}, the two standard assignments may be defined by
$\bar L_LY_eH_de_R$ for DFSZ-I and $\bar L_LY_eH_ue_R$ for DFSZ-II, where generation indices are suppressed and $e_R$ collectively denotes the three right-handed charged leptons. The corresponding anomaly ratios are
\begin{align}
    \left.\frac{E}{N}\right|_{\rm DFSZ-I} = \frac{8}{3},
    \qquad
    \left.\frac{E}{N}\right|_{\rm DFSZ-II} = \frac{2}{3}.
    \label{eq:DFSZ-EN}
\end{align}
Thus, even among the standard KSVZ and DFSZ benchmarks, the axion-photon coupling is not uniquely determined by $m_a$ or $f_a$.

\paragraph{Axion-fermion couplings and renormalisation-group evolution.}

The interaction of the axion with a Dirac fermion is conventionally written in the derivative form
\begin{align}
    \mathcal{L}_{af}
    = C_f(\mu) \frac{\partial_\mu a}{2f_a} \bar f\gamma^\mu\gamma^5f ,
    \label{eq:axion-fermion-derivative}
\end{align}
where $C_f(\mu)$ is a dimensionless, scale-dependent coefficient. For an on-shell fermion, the corresponding interaction can equivalently be expressed in the pseudoscalar form\footnote{This relation follows by integrating Eq.~\eqref{eq:axion-fermion-derivative} by parts and using the axial Ward identity. Neglecting the anomalous terms momentarily, $\partial_\mu(\bar f\gamma^\mu\gamma^5f)=2\i m_f\bar f\gamma^5f$, so that
$C_f(\partial_\mu a)\bar f\gamma^\mu\gamma^5f/(2f_a)\rightarrow-\i C_f(m_f/f_a)a\bar f\gamma^5f$, up to a total derivative. For a gauge-charged fermion, the divergence of the axial current also contains the corresponding gauge anomalies. Hence, the derivative and pseudoscalar operators should not be regarded as isolated operator identities: converting between them simultaneously shifts the coefficients of anomalous interactions such as $aG\widetilde G$ and $aF\widetilde F$. Physical observables are unchanged when all such terms are treated consistently.}
\begin{align}
    \mathcal{L}_{af}
    \supset
    -\i g_{af}\,
    a\bar f\gamma^5f,
    \qquad
    g_{af}
    =
    C_f\frac{m_f}{f_a},
    \label{eq:axion-fermion-pseudoscalar}
\end{align}
provided that the anomalous gauge-field operators are transformed consistently.
The derivative form in Eq.~\eqref{eq:axion-fermion-derivative} is often more convenient because it keeps the approximate axion shift symmetry manifest and provides a natural basis for discussing renormalisation-group evolution.

In minimal KSVZ models, the Standard-Model fermions are PQ neutral, and hence
\begin{align}
    C_f(f_a)=0
    \qquad
    \text{for Standard-Model fermions}
    \label{eq:KSVZ-Cf-tree}
\end{align}
at tree level. This statement does not imply that the corresponding low-energy couplings vanish exactly. They are generated radiatively through the anomalous gauge interactions and renormalisation-group evolution~\cite{Srednicki:1985xd,Choi:2021kuy}. In particular, the axion-electron interaction in KSVZ-like models, although absent at tree level, receives loop-induced contributions.

In the DFSZ models, by contrast, the Standard-Model fermions themselves carry PQ charge and therefore couple to the axion already at tree level. For three generations and the convention $\tan\beta=v_u/v_d$ adopted in Sec.~\ref{sec:PQWW-axion}, the matching-scale coefficients are~\cite{DiLuzio:2020wdo,DiLuzio:2023tqe}
\begin{align}
    C_{u,c,t}(f_a)
    =
    \frac{\cos^2\beta}{3},
    \qquad
    C_{d,s,b}(f_a)
    =
    \frac{\sin^2\beta}{3},
    \label{eq:DFSZ-quark-couplings}
\end{align}
while the charged-lepton couplings are
\begin{align}
    C_{e,\mu,\tau}(f_a)
    =
    \begin{cases}
        \displaystyle
        \frac{\sin^2\beta}{3},
        & \text{DFSZ-I},\\[2mm]
        \displaystyle
        -\frac{\cos^2\beta}{3},
        & \text{DFSZ-II}.
    \end{cases}
    \label{eq:DFSZ-lepton-couplings}
\end{align}
These expressions, together with Eqs.~\eqref{eq:KSVZ-EN} and~\eqref{eq:DFSZ-EN}, summarise the simplest tree-level benchmark patterns. They are collected in Table~\ref{tab:benchmark-axion-couplings}.

\begin{table}[t]
    \centering
    \begin{tabular}{lcccc}
        \hline
        Model
        & $E/N$
        & $C_{u,c,t}(f_a)$
        & $C_{d,s,b}(f_a)$
        & $C_{e,\mu,\tau}(f_a)$
        \\
        \hline\hline
        KSVZ ($Q_\Psi=0$)
        & $0$
        & $0$
        & $0$
        & $0$
        \\
        DFSZ-I
        & $8/3$
        & $\cos^2\beta/3$
        & $\sin^2\beta/3$
        & $\sin^2\beta/3$
        \\
        DFSZ-II
        & $2/3$
        & $\cos^2\beta/3$
        & $\sin^2\beta/3$
        & $-\cos^2\beta/3$
        \\
        \hline
    \end{tabular}
    \caption{Tree-level benchmark axion couplings at the ultraviolet matching scale. The KSVZ entry corresponds to a single electrically neutral heavy colour-triplet fermion. The DFSZ-I and DFSZ-II definitions refer to the charged-lepton Yukawa assignments specified in the text. Renormalisation-group evolution modifies the fermionic coefficients at lower scales.}
    \label{tab:benchmark-axion-couplings}
\end{table}

It is important that the coefficients in Table~\ref{tab:benchmark-axion-couplings} are ultraviolet boundary conditions rather than directly measured low-energy couplings. Axion-fermion interactions evolve under renormalisation-group running and receive threshold corrections when heavy Standard-Model and beyond-Standard-Model states are integrated out~\cite{Bauer:2020jbp,Choi:2021kuy,DiLuzio:2023tqe}. In DFSZ models, the large top-quark Yukawa coupling can induce appreciable corrections to the light-quark and charged-lepton coefficients, with the size of the effect depending on the scale at which the additional Higgs states are integrated out. Corrections at the level of tens of percent can occur in parts of parameter space~\cite{DiLuzio:2023tqe}. The anomalous axion-photon coupling is considerably less sensitive to this running. Thus, precision interpretation of a future axion signal requires evolving the ultraviolet couplings consistently to the scale relevant for the observable.

\paragraph{Axion couplings to hadrons.}

At energies below the QCD confinement scale, the quark- and gluon-level interactions in Eq.~\eqref{eq:axion-general-low-energy-EFT} must be matched onto hadronic degrees of freedom. The leading derivative interaction with nucleons can be parameterised as
\begin{align}
    \mathcal{L}_{aN}
    =
    \sum_{N=p,n}
    C_N
    \frac{\partial_\mu a}{2f_a}
    \bar N\gamma^\mu\gamma^5N,
    \label{eq:axion-nucleon-coupling}
\end{align}
where the dimensionless coefficients $C_p$ and $C_n$ contain both model-dependent contributions inherited from the quark couplings and model-independent contributions originating from the QCD anomaly~\cite{Srednicki:1985xd,Georgi:1986df,GrilliDiCortona:2015jxo,Vonk:2020zfh}.  

To make the origin of the QCD anomaly contribution transparent, it is convenient, for the purpose of matching, to perform a chiral rotation of the light quarks that removes the explicit $aG\widetilde G$ interaction in Eq.~\eqref{eq:axion-general-low-energy-EFT}. A standard choice distributes the rotation among the light flavours in proportion to their inverse masses. The derivative quark couplings are then shifted according to $C_q\rightarrow C_q-m_*/m_q$, and at leading order the nucleon couplings may be written schematically as
\begin{align}
    C_N
    =
    \sum_{q=u,d,s}
    \left[
        C_q(\mu)
        -
        \frac{m_*}{m_q}
    \right]
    \Delta q^{(N)}
    +\cdots ,
    \label{eq:axion-nucleon-matching}
\end{align}
where $m_*^{-1}=m_u^{-1}+m_d^{-1}+m_s^{-1}$ is the reduced light-quark mass introduced in Sec.~\ref{sec:chiral-perturbation-theory}, and
\begin{align}
    \langle N(p,s)|
    \bar q\gamma^\mu\gamma^5q
    |N(p,s)\rangle
    =
    2s^\mu\Delta q^{(N)}
    \label{eq:nucleon-axial-charges}
\end{align}
defines the contribution $\Delta q^{(N)}$ of the quark axial current to the nucleon matrix element. The ellipsis in Eq.~\eqref{eq:axion-nucleon-matching} denotes higher-order chiral corrections, renormalisation effects associated with the singlet axial current, and contributions induced when heavier quarks are integrated out.

The first term in Eq.~\eqref{eq:axion-nucleon-matching} carries the ultraviolet model dependence, whereas the second is generated even when the Standard-Model quarks have no tree-level PQ coupling. Consequently, the KSVZ axion still couples to nucleons. A widely used QCD matching gives~\cite{GrilliDiCortona:2015jxo}
\begin{align}
    C_p^{\rm KSVZ}
    \simeq
    -0.47(3),
    \qquad
    C_n^{\rm KSVZ}
    \simeq
    -0.02(3).
    \label{eq:KSVZ-nucleon-couplings}
\end{align}
The small neutron coefficient results from an accidental numerical cancellation and is not protected by a symmetry. More systematic baryon-ChPT calculations have extended the matching to higher chiral orders and permit axion-nucleon couplings to be determined at the few-percent level~\cite{Vonk:2020zfh}. In DFSZ models, the nonzero quark coefficients in Eq.~\eqref{eq:DFSZ-quark-couplings} modify both $C_p$ and $C_n$, producing a characteristic dependence on $\tan\beta$.

The same chiral matching generates interactions between the axion and the pseudo-Nambu--Goldstone bosons. In a commonly used convention, the leading three-pion interaction is
\begin{align}
    \mathcal{L}_{a\pi}
    \supset
    \frac{C_{a\pi}}{f_aF_\pi}
    \partial_\mu a
    \left[
        2(\partial^\mu\pi^0)\pi^+\pi^-
        -
        \pi^0(\partial^\mu\pi^+)\pi^-
        -
        \pi^0\pi^+(\partial^\mu\pi^-)
    \right],
    \label{eq:axion-pion-coupling}
\end{align}
with the leading-order coefficient~\cite{DiLuzio:2022gsc,DiLuzio:2023tqe}
\begin{align}
    C_{a\pi}
    =
    \frac{1}{3}
    \left[
        \frac{m_d-m_u}{m_u+m_d}
        +
        C_d-C_u
    \right].
    \label{eq:axion-pion-C}
\end{align}
The first term is fixed by QCD, while the second carries the ultraviolet dependence of the quark couplings~\cite{GrilliDiCortona:2015jxo,DiLuzio:2022gsc,DiLuzio:2023tqe}. The axion-pion interaction is particularly relevant for the thermal production of axions around and below the QCD crossover and will reappear in Sec.~\ref{sec:axion-cosmology}.

\paragraph{Domain-wall number and broader model dependence.}

The same QCD anomaly that fixes the relation between $f_a$ and $v_{\rm PQ}$ also determines the number of QCD minima within one PQ period. Before absorbing the anomaly coefficient into $f_a$, the gluonic interaction reads
\begin{align}
    \mathcal{L}
    \supset
    \frac{g_s^2}{32\pi^2}
    N\frac{a}{v_{\rm PQ}}
    G^a_{\mu\nu}\widetilde G^{a\mu\nu}.
    \label{eq:fa-anomaly-normalisation}
\end{align}
Since $a\sim a+2\pi v_{\rm PQ}$, while the QCD potential is periodic under $a\rightarrow a+2\pi f_a$, there are $N$ degenerate QCD minima within one PQ period before any additional gauge or discrete identifications are taken into account. The number of physically inequivalent minima is the domain-wall number $N_{\rm DW}$~\cite{Sikivie:1982qv}. In the simplest case, with no such additional identifications, $N_{\rm DW}=N$.

The minimal KSVZ construction with a single fundamental heavy quark has $N_{\rm DW}=1$, whereas the conventional DFSZ construction with three generations and the quartic phase-sensitive interaction used in Sec.~\ref{sec:KSVZ-DFSZ} has
\begin{align}
    N_{\rm DW}^{\rm DFSZ}
    =
    2N_g
    =
    6.
    \label{eq:DFSZ-domain-wall-number}
\end{align}
This distinction becomes cosmologically important if the PQ symmetry is broken after inflation: a string attached to one wall can collapse, whereas $N_{\rm DW}>1$ leads, in the absence of additional explicit breaking, to a stable string-wall network~\cite{Sikivie:1982qv,Ringwald:2015dsf}. We return to this issue in Sec.~\ref{sec:axion-cosmology}.\footnote{More generally, $N_{\rm DW}$ counts physically inequivalent minima after all gauge and discrete identifications are taken into account, and need not coincide with a naively computed anomaly coefficient before the global structure of the theory is specified. See, for example, Ref.~\cite{Choi:2026oqz} for a recent discussion.}

The benchmark values above should not be interpreted as universal predictions of the QCD axion. The mass relation in Eq.~\eqref{eq:axion-mass-numerical} is determined predominantly by QCD, but $E/N$, the fermionic coefficients $C_f$, and $N_{\rm DW}$ depend on the ultraviolet PQ charge assignments and particle content. Generalised KSVZ constructions can produce a wide range of electromagnetic anomaly coefficients~\cite{DiLuzio:2017pfr,Agrawal:2017cmd}, while generation-dependent PQ charges can suppress particular couplings to electrons or nucleons and can simultaneously generate flavour-changing axion interactions~\cite{Bardeen:1986yb,DiLuzio:2017ogq,MartinCamalich:2020dfe,Sun:2021jpw}. Thus, the familiar KSVZ and DFSZ ``bands'' in axion parameter-space plots are useful benchmark targets rather than strict boundaries on the phenomenology of the QCD axion.

This model dependence is phenomenologically valuable. Measurements of several axion interactions, if eventually available, could distinguish ultraviolet realisations that predict the same axion mass. Conversely, astrophysical, cosmological, and laboratory constraints probe complementary combinations of $g_{a\gamma\gamma}$, $g_{ae}$, $C_{p,n}/f_a$, and the hadronic couplings. We turn to these observational consequences in the next section.

\section{QCD axion cosmology, astrophysics, and searches}
\label{sec:QCD-axion-aspects}

\subsection{Cosmological production and the pre- and post-inflationary histories}
\label{sec:axion-cosmology}

\begin{figure}
    \centering
\includegraphics[width=0.55\linewidth]{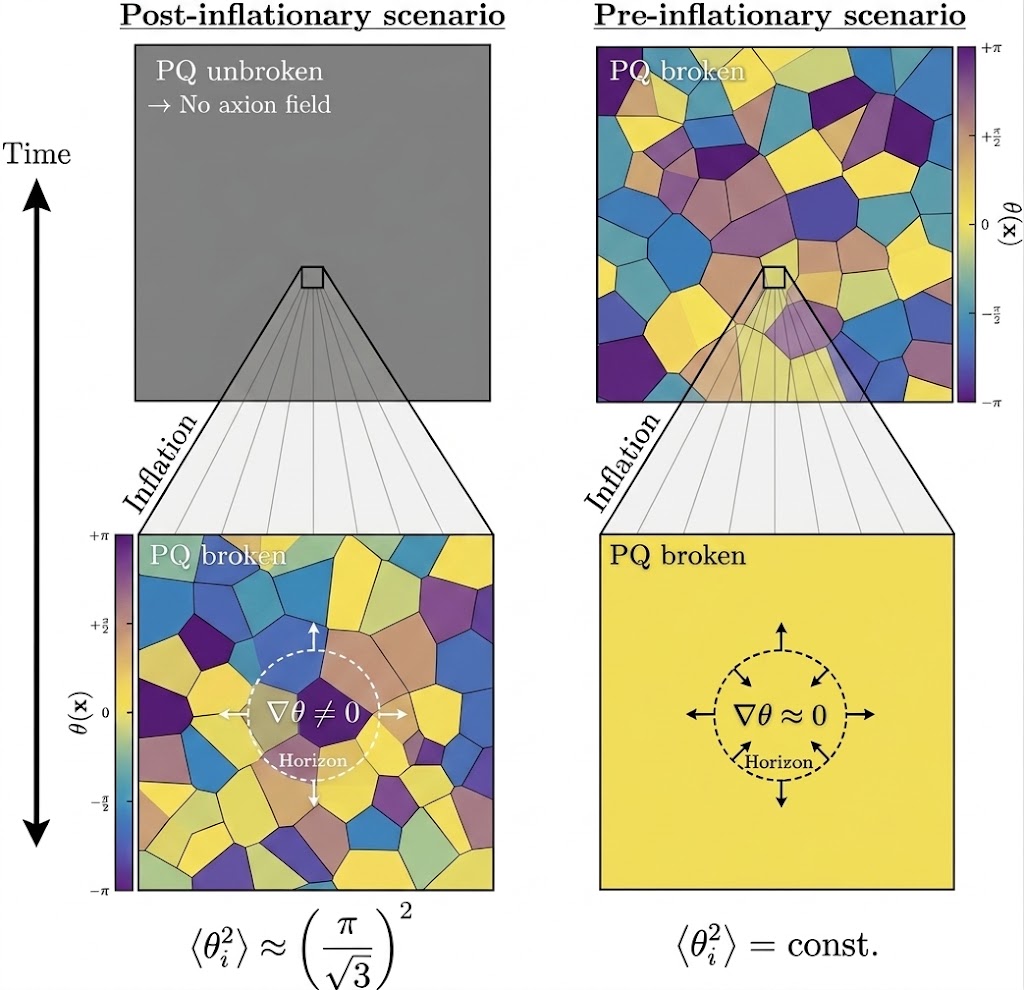}
    \caption{Schematic illustration of the pre- and post-inflationary PQ-breaking scenarios (adapted from Ref.~\cite{OHare:2024nmr} under the permission of the author). In the pre-inflationary scenario, inflation leaves an approximately homogeneous axion field throughout the observable Universe. In the post-inflationary scenario, the PQ symmetry breaks only after inflation, so different causally disconnected regions independently select different PQ phases, leading to the formation of global strings.}
    \label{fig:pre-post}
\end{figure}

The cosmological evolution of the QCD axion depends crucially on when the PQ symmetry is broken relative to inflation. It is therefore useful to distinguish two qualitatively different histories from the outset~\cite{Marsh:2015xka,DiLuzio:2020wdo,OHare:2024nmr}. In the pre-inflationary scenario, the PQ symmetry is already broken during inflation and is not restored afterwards. Inflation then stretches a region in which the PQ order parameter has an approximately uniform phase to encompass our observable Universe, leaving a nearly homogeneous axion field characterised by a single initial misalignment angle, up to inflationary fluctuations. The value of this angle is randomly selected within one period, conventionally $\vartheta_i\in[-\pi,\pi)$, and the relic abundance is consequently controlled primarily by its initial displacement from the QCD minimum. In the post-inflationary scenario, by contrast, the PQ symmetry is unbroken during inflation, or is restored during reheating and subsequently breaks again. Before PQ symmetry breaking, the PQ field remains in the symmetry-preserving phase, and no physical angular direction is defined. At the PQ phase transition, different correlation domains independently select different points on the vacuum manifold, producing a network of global strings. The two cosmological histories are illustrated schematically in Fig.~\ref{fig:pre-post}.

When the QCD potential later becomes dynamically important, domain walls form and become attached to the pre-existing strings. See Fig.~\ref{fig:post-inf} below. The axion abundance then receives contributions from the spatially varying misalignment field as well as from the evolution and decay of these topological defects. These two histories lead to rather different cosmological predictions and experimental target regions.

\paragraph{PQ breaking before inflation.}

In this case, any topological defects produced when the PQ symmetry was initially broken are diluted by inflation, provided that the symmetry is not restored afterwards. The subsequent axion evolution within our observable Universe can therefore be described, to a good approximation, by a spatially homogeneous field with a single initial misalignment angle. Its dynamics is governed by the temperature-dependent QCD potential. Generalising Eq.~\eqref{eq:QCD-axion-potential-general} to finite temperature, one may write
\begin{align}
    V_{\rm QCD}(a,T)
    =
    \mathcal{F}
    \left(
        T,\bar\theta+\frac{a}{f_a}
    \right)
    -
    \mathcal{F}(T,0),
    \label{eq:finite-temperature-axion-potential}
\end{align}
where $\mathcal{F}(T,\bar\theta)$ denotes the QCD free-energy density. The temperature-dependent axion mass is related to the finite-temperature topological susceptibility by
\begin{align}
    m_a^2(T)
    =
    \frac{\chi(T)}{f_a^2},
    \qquad
    \chi(T)
    =
    \left.
    \frac{\partial^2\mathcal{F}(T,\bar\theta)}
         {\partial\bar\theta^2}
    \right|_{\bar\theta=0}.
    \label{eq:finite-temperature-axion-mass}
\end{align}
At temperatures well above the QCD scale, topological fluctuations are strongly suppressed and $m_a(T)$ is much smaller than its zero-temperature value. As the Universe cools through the QCD crossover, $\chi(T)$ grows rapidly and the QCD potential becomes dynamically important. Its temperature dependence, which has been studied extensively using lattice QCD and semiclassical methods, is therefore a central input for quantitative calculations of the axion relic abundance~\cite{Borsanyi:2016ksw}.

For the cosmological discussion, it is convenient to shift the origin of the axion field so that the zero-temperature minimum lies at $a=0$, and we denote the shifted field again by $a$. The homogeneous field then satisfies
\begin{align}
    \ddot a
    +
    3H\dot a
    +
    \frac{\partial V_{\rm QCD}(a,T)}
         {\partial a}
    =
    0,
    \label{eq:axion-cosmological-eom}
\end{align}
where $H$ is the Hubble expansion rate. Its initial displacement from the minimum is conventionally parametrised by the misalignment angle $\vartheta_i\equiv a_i/f_a$. At sufficiently early times, $m_a(T)\ll H$, and Hubble friction keeps the field approximately frozen at $a\simeq a_i$. As the Universe cools, $m_a(T)$ increases while $H$ decreases, and coherent oscillations begin around a temperature $T_{\rm osc}$ conventionally defined by
\begin{align}
    m_a(T_{\rm osc})
    \simeq
    3H(T_{\rm osc}),
    \label{eq:axion-oscillation-condition}
\end{align}
up to an order-one ambiguity in the precise definition of the onset. To obtain an analytic estimate for $T_{\rm osc}$, we approximate the temperature-dependent axion mass by a simple power law,
\begin{align}
    m_a^2(T) =
    \begin{cases}
        m_a^2\left(\dfrac{T_{\rm QCD}}{T}\right)^n, & T>T_{\rm QCD},\\
        m_a^2, & T\leq T_{\rm QCD}.
    \end{cases}
    \label{eq:ma2T}
\end{align}
The high-temperature behaviour of the axion mass can be studied semiclassically in the DIGA~\cite{Gross:1980br}, while phenomenological instanton models and lattice QCD provide complementary information at lower temperatures~\cite{Wantz:2009it,Bonati:2015vqz,Petreczky:2016vrs,Borsanyi:2016ksw,Chen:2022fid,Athenodorou:2022aay}. For the analytic estimates below, we take $T_{\rm QCD}\approx 150\,{\rm MeV}$ and $n\approx 8.16$, motivated by the high-temperature lattice results of Ref.~\cite{Borsanyi:2016ksw}.

We assume that the oscillation begins during radiation domination. If $T_{\rm osc}\leq T_{\rm QCD}$, we obtain
\begin{align}
    T_{\rm osc}
    =\left(m_a M_{\rm Pl} \sqrt{\frac{10}{\pi^2 g_{\star,\rho}(T_{\rm osc})}}\right)^{\frac{1}{2}},
    \qquad
    (T_{\rm osc}\leq T_{\rm QCD}),
\end{align}
where $M_{\rm Pl}$ is the reduced Planck mass and $g_{\star,\rho}$ denotes the effective number of relativistic degrees of freedom. Using the parametric relation $m_a\simeq T_{\rm QCD}^2/f_a$, the condition $T_{\rm osc}\leq T_{\rm QCD}$ requires $f_a$ to approach the Planck scale. For the conventional QCD-axion parameter range, one therefore typically has $T_{\rm osc}>T_{\rm QCD}$. In this case,
\begin{align}
    T_{\rm osc}
    =\left(m_a M_{\rm Pl} \sqrt{\frac{10}{\pi^2 g_{\star,\rho}(T_{\rm osc})}}\right)^{\frac{2}{n+4}} T_{\rm QCD}^{\frac{n}{n+4}}
    \approx \left(\frac{\sqrt{\chi}}{T_{\rm QCD}^2} \frac{M_{\rm Pl}}{f_a} \sqrt{\frac{10}{\pi^2 g_{\star,\rho}(T_{\rm osc})}}\right)^{\frac{2}{n+4}} T_{\rm QCD} ,
\end{align}
where in the second equality we have used $m_a=\sqrt{\chi}/f_a$ with $\chi\approx (75.5\,{\rm MeV})^4$~\cite{Borsanyi:2016ksw,GrilliDiCortona:2015jxo,Gorghetto:2018ocs} being the topological susceptibility at zero temperature.

In the harmonic approximation, the axion energy density at $T_{\rm osc}$ is of order
\begin{align}
    \rho_a(T_{\rm osc})
    \simeq
    \frac{1}{2}
    m_a^2(T_{\rm osc})
    f_a^2\vartheta_i^2.
    \label{eq:misalignment-energy-onset}
\end{align}
After the onset of oscillations, once the evolution becomes adiabatic, the comoving axion number is approximately conserved, $n_a R^3 = (\rho_a/m_a(T)) R^3 \approx {\rm const}$, where $R(t)$ denotes the cosmological scale factor. The coherently oscillating field therefore behaves as cold dark matter~\cite{Preskill:1982cy,Abbott:1982af,Dine:1982ah,Turner:1985si}. If there is no substantial entropy production after the onset of oscillations, conservation of the comoving entropy further implies that $n_a/s$ remains approximately constant, with
$s(T)=2\pi^2g_{\star,s}(T)T^3/45$. Within the harmonic and sudden-onset approximations, the present-day axion energy density is then
\begin{align}
    \rho_a(T_0)
    \simeq
    \frac{1}{2}
    m_a\,m_a(T_{\rm osc})
    f_a^2\vartheta_i^2
    \frac{s(T_0)}{s(T_{\rm osc})} = \frac{1}{2}\chi \left(\frac{T^2_{\rm QCD}}{\sqrt{\chi} M_{\rm Pl}} \sqrt{\frac{\pi^2 g_{\star,\rho}(T_{\rm osc})}{10}}\right)^{\frac{n+6}{n+4}}\left(\frac{g_{\star,s}(T_0) T_0^3 }{g_{\star,s}(T_{\rm osc}) T_{\rm QCD}^3}\right) f_a^{\frac{n+6}{n+4}}\vartheta_i^2.
    \label{eq:axion-misalignment-density-today}
\end{align}
The corresponding relic abundance,
\begin{align}
    \Omega_{a,\rm mis} h^2
    \equiv
    \frac{\rho_a(T_0)}{\rho_c/h^2} \simeq 0.12 ,
    \label{eq:axion-relic-abundance-definition}
\end{align}
can be compared with the observed cold-dark-matter abundance $\Omega_{\rm DM}h^2\simeq0.12$~\cite{Planck:2018vyg}. A more careful numerical calculation that includes an accurate temperature dependence for the axion mass and correctly tracks the temperature dependence of $g_{\star,S}$ etc. yields~\cite{Borsanyi:2016ksw,OHare:2024nmr}
\begin{align}
    \Omega_{a,\rm mis} h^2 \simeq 0.12 \left(\frac{\vartheta_i}{2.155}\right)^2\left(\frac{28\,{\rm \mu eV}}{m_a}\right)^{1.16},
    \label{eq:Omega-a-h2}
\end{align}
where the value of $2.155$ is the the effective misalignment angle in the post-inflationary scenario discussed below.

An important feature of the pre-inflationary history is that $\vartheta_i$ is a genuine cosmological initial condition. The relic abundance is therefore not predicted by $f_a$ alone: a sufficiently small $|\vartheta_i|$ can allow values of $f_a$ well above those that would overproduce dark matter for a generic initial displacement~\cite{Preskill:1982cy,Abbott:1982af,Dine:1982ah,Marsh:2015xka}. Conversely, if the QCD axion is required to account for all of the dark matter, the observed abundance determines a relation between $f_a$ and the particular value of $\vartheta_i$ realised in our observable patch.

Inflation also generates fluctuations of the axion field. If the axion is sufficiently light during inflation, its fluctuations have the characteristic amplitude
\begin{align}
    \delta a
    \simeq
    \frac{H_{\rm inf}}{2\pi},
    \qquad
    \delta\vartheta
    \simeq
    \frac{H_{\rm inf}}{2\pi f_a},
    \label{eq:axion-inflationary-fluctuation}
\end{align}
where $H_{\rm inf}$ is the Hubble scale during inflation. Since these fluctuations perturb the axion density independently of the primordial adiabatic mode, they generate cold-dark-matter isocurvature perturbations. The absence of a sizeable isocurvature component in the cosmic microwave background can therefore impose a stringent constraint on $H_{\rm inf}$ when the axion constitutes an appreciable fraction of the dark matter. The quantitative bound depends on the axion abundance and on the inflationary and PQ-sector dynamics~\cite{Marsh:2015xka,OHare:2024nmr,Graham:2025iwx}; we discuss the observational constraint in Sec.~\ref{sec:axion-astro-cosmo-constraints}. Thus, in the minimal pre-inflationary picture, the freedom to choose a homogeneous initial misalignment angle comes together with the possibility of observable inflationary isocurvature.

\paragraph{PQ breaking after inflation.}

In the post-inflationary scenario, the PQ symmetry is unbroken during inflation, or is restored afterwards and subsequently breaks again. The phase of the PQ order parameter is then selected independently in causally disconnected regions, so the axion field is intrinsically inhomogeneous rather than characterised by a single misalignment angle. The same symmetry-breaking transition generically produces a network of global strings through the Kibble--Zurek mechanism~\cite{Kibble:1976sj,Kibble:1980mv,Zurek:1985qw}.\footnote{See Ref.~\cite{Ai:2026zpy} for a recent discussion of how a bias between vacua that would otherwise be degenerate affects the formation of topological defects.} After formation, the string network evolves towards an attractor commonly described as an approximate scaling regime, although numerical simulations indicate logarithmic deviations from exact scaling, and continuously loses energy through axion radiation~\cite{Gorghetto:2018myk,Buschmann:2019icd,Gorghetto:2020qws}. At this stage, the QCD-induced axion potential is still negligible: the axion direction therefore remains continuously degenerate, and the string network is present while QCD domain walls have not yet formed.

As the Universe approaches the QCD epoch, the temperature-dependent potential in Eq.~\eqref{eq:finite-temperature-axion-potential} becomes dynamically important and lifts the continuous degeneracy of the axion direction. The axion field then begins to evolve locally towards the nearest QCD minimum, with an initial misalignment angle that varies across space. To a first approximation, this angle is uniformly distributed over one period. In the harmonic approximation,
\begin{align}
    \left\langle\vartheta_i^2\right\rangle
    =
    \frac{1}{2\pi}
    \int_{-\pi}^{\pi}\d\vartheta_i\,
    \vartheta_i^2
    =
    \frac{\pi^2}{3}.
    \label{eq:postinflation-misalignment-average}
\end{align}
Thus, unlike in the pre-inflationary scenario, the spatially averaged misalignment is statistically determined rather than set by a freely adjustable homogeneous initial condition. For a quantitative calculation, however, one should average the axion abundance itself rather than simply $\vartheta_i^2$, since regions with $|\vartheta_i|$ close to $\pi$ receive an anharmonic enhancement. In the standard treatment this effect can be represented approximately by an effective misalignment angle $\vartheta_{\rm eff}\simeq2.155$, somewhat larger than the harmonic value $\pi/\sqrt{3}\simeq1.81$~\cite{Borsanyi:2016ksw,OHare:2024nmr}. From Eq.~\eqref{eq:Omega-a-h2}, we see that misalignment alone would reproduce the observed dark-matter abundance for $m_a\simeq28\,\mu{\rm eV}$~\cite{Borsanyi:2016ksw,OHare:2024nmr}. Equation~\eqref{eq:Omega-a-h2} should not be interpreted as the full post-inflationary prediction: spatial gradients, global strings, and the subsequent string-wall evolution provide additional contributions to the axion population. Their quantitative determination requires the nonlinear simulations discussed below and generally shifts the axion mass required to reproduce the observed dark-matter abundance to larger values~\cite{Gorghetto:2018myk,Gorghetto:2020qws,Buschmann:2021sdq,Kim:2024wku,Kaltschmidt:2025nkz}.

The emergence of the QCD potential also leads to the formation of domain walls bounded by the pre-existing strings~\cite{Sikivie:1982qv}. The number of physically inequivalent minima encountered as the axion winds once around its fundamental PQ orbit is the domain-wall number $N_{\rm DW}$ discussed in Sec.~\ref{sec:axion-low-energy-couplings}. Correspondingly, $N_{\rm DW}$ walls are attached to each string. For $N_{\rm DW}=1$, the wall tension destabilises the string-wall network, which eventually collapses and converts a substantial fraction of its energy into axions. For $N_{\rm DW}>1$, by contrast, multiple walls pull on each string in different directions, and the network is stable in the absence of additional PQ-breaking effects. Such a stable network would eventually dominate the energy density of the Universe and is therefore cosmologically unacceptable unless the degeneracy among the axion vacua is lifted~\cite{Sikivie:1982qv,Hiramatsu:2012gg}. The sequence from PQ symmetry breaking and string formation to the appearance of QCD domain walls is illustrated schematically in Fig.~\ref{fig:post-inf}.

\begin{figure}
    \centering
    \includegraphics[width=0.65\linewidth]{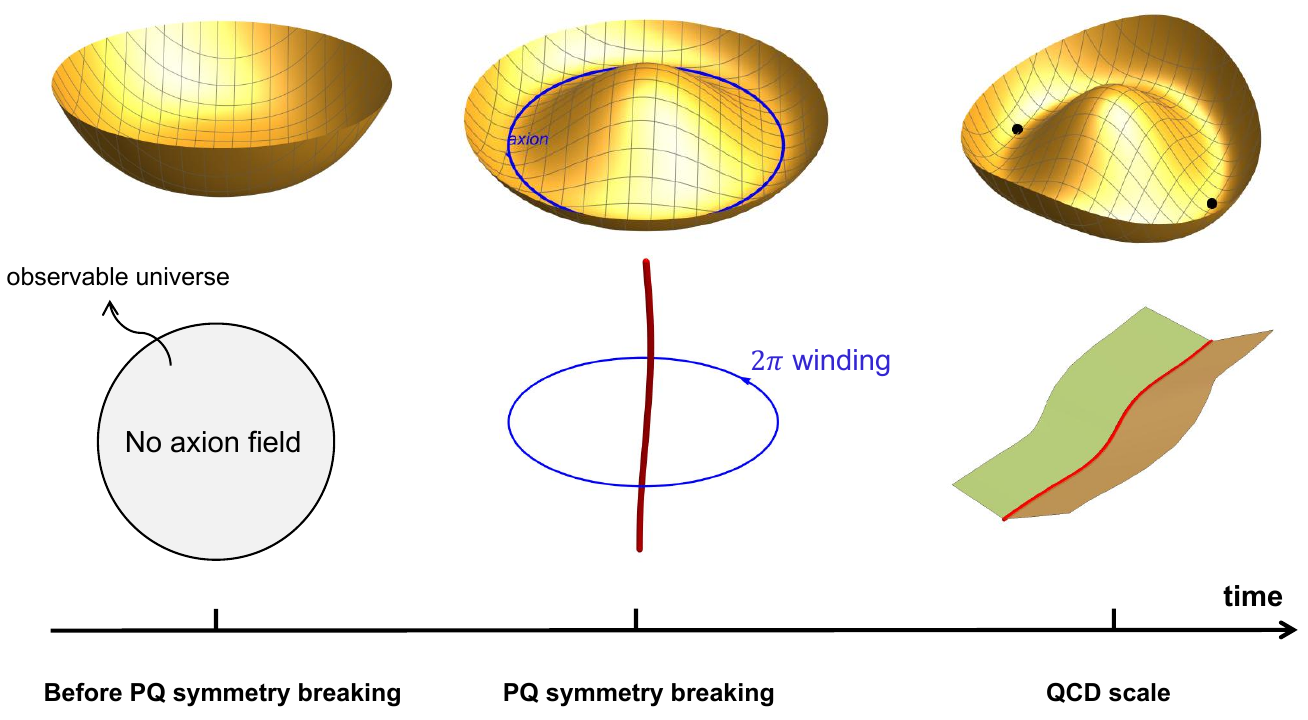}
    \caption{Schematic evolution of the axion field in the post-inflationary scenario. After PQ symmetry breaking, regions separated by more than the correlation length select different points on the vacuum manifold, producing global strings with $2\pi$ winding. At the QCD epoch, the QCD-induced potential lifts the continuous degeneracy and produces $N_{\rm DW}$ domain walls attached to the strings.}
    \label{fig:post-inf}
\end{figure}

For the phenomenologically simplest case $N_{\rm DW}=1$, the final cold-axion abundance therefore receives three closely related contributions: misalignment of the spatially varying axion field, axion radiation from the global-string network before the QCD epoch, and axions produced during the collapse of the string-wall system. Their relative importance has been studied extensively in numerical simulations~\cite{Klaer:2017ond,Gorghetto:2018myk,Buschmann:2019icd,Gorghetto:2020qws,Buschmann:2021sdq,Kim:2024wku,Kaltschmidt:2025nkz}. A major difficulty is the enormous hierarchy between the Hubble scale and the microscopic string-core scale in the physical Universe. Present simulations cannot reproduce this hierarchy directly and must extrapolate over many orders of magnitude. Different treatments of the string density, the emitted axion spectrum, and this extrapolation consequently lead to somewhat different predictions for the final relic abundance. For this reason, we do not adopt a single preferred post-inflationary QCD-axion mass in this review, but instead indicate representative target regions when discussing experimental searches in Sec.~\ref{sec:axion-experimental-searches}.

The spatially varying initial conditions also leave an imprint on small-scale structure. Around the onset of axion oscillations, order-one density contrasts can be present on scales associated with the horizon at that epoch. Their subsequent gravitational evolution can produce dense axion substructures, including axion miniclusters~\cite{Hogan:1988mp,Kolb:1994fi,Fairbairn:2017sil,Vaquero:2018tib}. Gravitational relaxation within sufficiently dense axion clumps may further produce gravitationally bound Bose condensates, commonly referred to as axion stars~\cite{Tkachev:1991ka,Kolb:1993zz,Braaten:2019knj}. The evolution, survival, and internal structure of these objects remain active subjects of numerical study. If a significant fraction of the local axion density resides in such substructures, the phase-space distribution relevant for direct-detection experiments can differ appreciably from that of a smooth Galactic halo, as discussed further in Sec.~\ref{sec:axion-experimental-searches}.

For completeness, the standard misalignment and topological-defect mechanisms described above do not exhaust the possible cosmological production histories of the QCD axion. Thermal interactions with the Standard-Model plasma can produce a relativistic axion population, with the relevant production processes changing across the electroweak and QCD thresholds~\cite{Arias-Aragon:2020shv,DEramo:2021psx,DEramo:2021lgb}. The resulting contributions to dark radiation and hot dark matter are discussed in Sec.~\ref{sec:axion-astro-cosmo-constraints}. The cold axion abundance can also be substantially modified by a nonstandard pre-BBN expansion history, such as an early matter-dominated or kination epoch, or by entropy production associated with late reheating~\cite{Arias:2021rer,Bernal:2021yyb}. Finally, nonstandard initial axion dynamics can lead to qualitatively different relic abundances, as in the stochastic and kinetic-misalignment scenarios~\cite{Graham:2018jyp,Co:2019jts}. These examples illustrate that the relation between the QCD axion mass and its cosmological abundance depends on the early expansion history and the dynamics of the PQ sector. Unless stated otherwise, we use the standard radiation-dominated pre- and post-inflationary histories described above as the reference cosmologies in the remainder of this review.

\subsection{Astrophysical and cosmological constraints}
\label{sec:axion-astro-cosmo-constraints}

Astrophysical and cosmological systems provide exceptionally sensitive probes of the QCD axion, owing to the extreme temperatures and densities, strong gravitational fields, and long timescales that they can realise. Axions can provide additional channels for energy transport in stars and compact objects, modify the cosmological energy density and the growth of structure, generate primordial isocurvature perturbations, and extract angular momentum from rotating black holes through superradiance. These probes are highly complementary. Stellar cooling primarily constrains the axion couplings to photons, electrons, and nucleons; cosmological limits depend additionally on the thermal and inflationary history; while black-hole superradiance is predominantly gravitational and does not require an appreciable cosmological axion abundance. Comprehensive discussions of astrophysical axion constraints can be found in Refs.~\cite{Raffelt:2006cw,DiLuzio:2020wdo,Caputo:2024oqc,Cicoli:2026fqp}.

\paragraph{Stellar cooling.}

The basic stellar-cooling argument is straightforward. If axions can be produced in a stellar interior and escape without being reabsorbed, they carry energy away from the star and provide an additional cooling channel. This modifies the thermal evolution of the star and can affect observable quantities such as stellar lifetimes, luminosities, and cooling rates. Since stellar environments differ substantially in temperature, density, and composition, different systems are sensitive to different axion couplings and production mechanisms~\cite{Raffelt:2006cw,Caputo:2024oqc}.

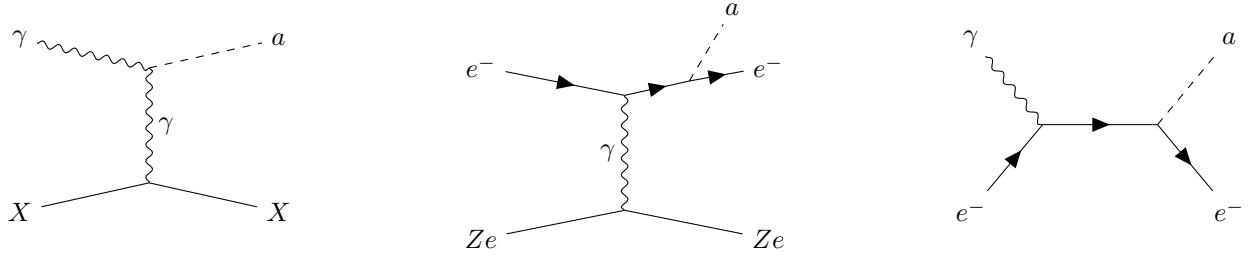
\begin{figure}[t]
    \centering

    \begin{minipage}{0.32\textwidth}
        \centering
        \begin{tikzpicture}[scale=0.95]
            \begin{feynman}
                \vertex (x1) at (-1.8,-1.2) {$X$};
                \vertex (v1) at (0,-0.8);
                \vertex (x2) at (1.8,-1.2) {$X$};
                \vertex (g)  at (-1.8,1.2) {$\gamma$};
                \vertex (v2) at (0,0.8);
                \vertex (a)  at (1.8,1.2) {$a$};

                \diagram*{
                    (x1) -- (v1) -- (x2),
                    (v1) -- [photon, edge label'={$\gamma$}] (v2),
                    (g) -- [photon] (v2),
                    (v2) -- [scalar] (a),
                };
            \end{feynman}
        \end{tikzpicture}
    \end{minipage}
    \hfill
    \begin{minipage}{0.32\textwidth}
        \centering
        \begin{tikzpicture}[scale=0.95]
            \begin{feynman}
                \vertex (z1) at (-2,-1.2) {$Ze$};
                \vertex (vz) at (0,-0.8);
                \vertex (z2) at (2,-1.2) {$Ze$};

                \vertex (e1) at (-2,1.2) {$e^-$};
                \vertex (ve1) at (0,0.8);
                \vertex (ve2) at (0.9,1);
                \vertex (e2) at (2,1.2) {$e^-$};
                \vertex (a) at (1.5,2) {$a$};

                \diagram*{
                    (z1) -- (vz) -- (z2),
                    (e1) -- [fermion] (ve1) -- [fermion] (ve2) -- [fermion] (e2),
                    (ve1) -- [photon, edge label'={$\gamma$}] (vz),
                    (ve2) -- [scalar] (a),
                };
            \end{feynman}
        \end{tikzpicture}
    \end{minipage}
    \hfill
    \begin{minipage}{0.32\textwidth}
        \centering
        \begin{tikzpicture}[scale=0.95]
            \begin{feynman}
                \vertex (e1) at (-1.8,-1.2) {$e^-$};
                \vertex (v1) at (-0.8,0);
                \vertex (v2) at (0.8,0);
                \vertex (e2) at (1.8,-1.2) {$e^-$};

                \vertex (g) at (-1.8,1.2) {$\gamma$};
                \vertex (a) at (1.8,1.2) {$a$};

                \diagram*{
                    (e1) -- [fermion] (v1) -- [fermion] (v2) -- [fermion] (e2),
                    (g) -- [photon] (v1),
                    (v2) -- [scalar] (a),
                };
            \end{feynman}
        \end{tikzpicture}
    \end{minipage}
    \caption{Representative axion-production processes relevant for stellar cooling. The leftmost panel shows Primakoff conversion, $\gamma+X\rightarrow a+X$, in the screened Coulomb field of a charged plasma constituent $X$, which probes the axion-photon coupling. The middle and rightmost panels show electron-ion bremsstrahlung, $e^-+Ze\rightarrow e^-+Ze+a$, and the Compton-like process, $e^-+\gamma\rightarrow e^-+a$, respectively, which probe the axion-electron coupling. The bremsstrahlung and Compton-like channels are shown schematically; in a complete calculation, one should include the full set of relevant diagrams.}
    \label{fig:stellar-axion-processes}
\end{figure}

The axion-photon coupling is particularly well constrained by helium-burning horizontal-branch (HB) stars in globular clusters. Thermal photons in the stellar plasma can convert into axions through the Primakoff process,
$\gamma+X\rightarrow a+X$, where $X$ denotes a charged plasma constituent, such as an electron or ion, as shown in the leftmost panel of Fig.~\ref{fig:stellar-axion-processes}.

Globular clusters are particularly useful because most of their stars formed within a relatively short period, but with a range of initial masses. Since more massive stars evolve more rapidly, stars of slightly different initial masses populate successive late evolutionary stages at a given cluster age. Over the comparatively short red-giant-branch (RGB), HB, and asymptotic-giant-branch (AGB) phases, the rate at which stars pass through this evolutionary sequence varies only slowly. The number of stars observed in each phase is therefore approximately proportional to its duration, allowing ratios of stellar populations to probe relative stellar lifetimes. Additional axion cooling shortens the HB lifetime and hence reduces the so-called $R$ parameter, $R\equiv N_{\rm HB}/N_{\rm RGB}\simeq\tau_{\rm HB}/\tau_{\rm RGB}$,
where $N_{\rm RGB}$ counts RGB stars brighter than the HB luminosity level, and $\tau_{\rm RGB}$ correspondingly denotes the time spent on this upper part of the RGB. Comparing the observed $R$ parameter in 39 Galactic globular clusters with stellar-evolution calculations gives~\cite{Ayala:2014pea}
\begin{align}
    g_{a\gamma\gamma}
    <
    0.66\times10^{-10}\,{\rm GeV}^{-1}
    \qquad
    (95\%~{\rm C.L.}).
    \label{eq:HB-axion-photon-bound}
\end{align}
Since axion cooling affects the HB and AGB lifetimes differently, the ratio
$R_2\equiv N_{\rm AGB}/N_{\rm HB}\simeq\tau_{\rm AGB}/\tau_{\rm HB}$
provides a complementary probe of the axion-photon coupling. A representative analysis obtains~\cite{Dolan:2022kul}
$g_{a\gamma\gamma}<0.47\times10^{-10}\,{\rm GeV}^{-1}$ (95\%~{\rm C.L.}),
although the precise bound remains sensitive to uncertainties in stellar modelling during helium burning. These limits apply in the light-axion regime, where $m_a$ is well below the characteristic thermal scale of the stellar core and the Primakoff production rate is essentially insensitive to the axion mass. At larger masses, thermal production becomes suppressed and the bounds weaken. For the standard QCD-axion benchmarks to which these constraints are usually applied, the light-mass approximation is well satisfied. Solar evolution provides additional constraints on the axion-photon coupling, although they are weaker than those from globular clusters~\cite{Vinyoles:2015aba}.

Stars with degenerate cores provide particularly strong probes of the axion-electron coupling. Axions can be emitted through electron-ion bremsstrahlung and, where relevant, Compton-like processes, as illustrated schematically in Fig.~\ref{fig:stellar-axion-processes}. In red giants, the resulting additional energy loss delays helium ignition and increases the luminosity at the tip of the red-giant branch (TRGB). This effect has been studied from early analytic estimates to modern stellar-evolution calculations~\cite{Raffelt:1994ry,Viaux:2013lha,Straniero:2020iyi,Capozzi:2020cbu}. Current TRGB analyses probe axion-electron couplings at approximately
\begin{align}
    g_{ae}
    \sim
    \mathcal{O}(10^{-13}),
    \label{eq:stellar-gae-scale}
\end{align}
with representative bounds around $g_{ae}\lesssim(1$--$2)\times10^{-13}$, depending on the stellar sample and the treatment of systematic uncertainties~\cite{Capozzi:2020cbu,Dennis:2023aam}. The inferred limits remain sensitive to stellar parameters and their correlations, so no single TRGB bound should be regarded as universally definitive.

White dwarfs provide an independent probe of the axion-electron coupling. Axion emission, dominated by electron-ion bremsstrahlung in the relevant regime, can accelerate their cooling and thereby modify the white-dwarf luminosity function and the evolution of pulsation periods. These observables have long been used to constrain $g_{ae}$ and have occasionally shown a mild preference for additional cooling~\cite{MillerBertolami:2014rka,Giannotti:2017hny}. A recent analysis of the white-dwarf cooling sequence in the globular cluster 47 Tucanae instead finds no preference for axion emission and reports~\cite{Fleury:2025ahw}
\begin{align}
    g_{ae}
    \leq
    0.81\times10^{-13}
    \qquad
    (95\%~{\rm C.L.}),
    \label{eq:WD-axion-electron-bound}
\end{align}
disfavouring the range suggested by some earlier white-dwarf cooling hints. 

Translating these coupling limits into a bound on the QCD axion mass is model dependent. As discussed in Sec.~\ref{sec:axion-low-energy-couplings}, $m_a$ fixes $f_a$ through QCD, but $g_{a\gamma\gamma}$ depends on $E/N$ and the fermionic couplings depend on the PQ charge assignments. Stellar constraints therefore carve out different regions of the KSVZ, DFSZ, and more general QCD-axion parameter spaces rather than providing a completely model-independent upper bound on $m_a$.

\paragraph{Supernovae and neutron stars.}

Core-collapse supernovae provide powerful probes of axion couplings to nucleons. During core collapse, most of the gravitational binding energy of the newly formed proto-neutron star is released through neutrinos. The neutrino burst observed from SN~1987A therefore provides the classic cooling argument: if axions were produced efficiently and escaped freely from the core, they would carry away additional energy and shorten the neutrino signal~\cite{Raffelt:1987yt,Turner:1987by,Mayle:1987as}. At weak coupling, the axion luminosity increases with the coupling, whereas at sufficiently strong coupling axions become trapped. Supernova constraints therefore exhibit distinct free-streaming and trapping regimes rather than depending monotonically on the axion-nucleon coupling. 

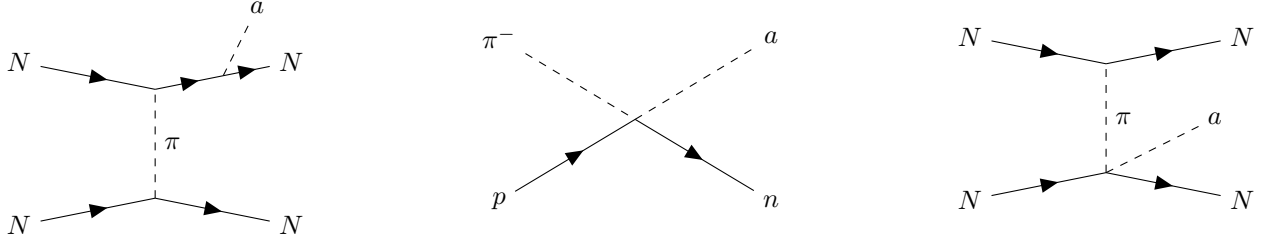
\begin{figure}[t]
    \centering

    \begin{minipage}{0.32\textwidth}
        \centering
        \begin{tikzpicture}[scale=0.9]
            \begin{feynman}
                \vertex (n1i) at (-2,-1.2) {$N$};
                \vertex (v1) at (0,-0.8);
                \vertex (n1f) at (2,-1.2) {$N$};

                \vertex (n2i) at (-2,1.2) {$N$};
                \vertex (v2) at (0,0.8);
                \vertex (v3) at (1,1);
                \vertex (n2f) at (2,1.2) {$N$};

                \vertex (a) at (1.5,2) {$a$};

                \diagram*{
                    (n1i) -- [fermion] (v1) -- [fermion] (n1f),
                    (n2i) -- [fermion] (v2) -- [fermion] (v3) -- [fermion] (n2f),
                    (v1) -- [scalar, edge label'={$\pi$}] (v2),
                    (v3) -- [scalar] (a),
                };
            \end{feynman}
        \end{tikzpicture}
    \end{minipage}
    \hfill
    \begin{minipage}{0.32\textwidth}
        \centering
        \begin{tikzpicture}[scale=0.9]
            \begin{feynman}
                \vertex (p) at (-2,-1.2) {$p$};
                \vertex (v) at (0,0);
                \vertex (n) at (2,-1.2) {$n$};

                \vertex (pi) at (-2,1.2) {$\pi^-$};
                \vertex (a) at (2,1.2) {$a$};

                \diagram*{
                    (p) -- [fermion] (v) -- [fermion] (n),
                    (pi) -- [scalar] (v),
                    (v) -- [scalar] (a),
                };
            \end{feynman}
        \end{tikzpicture}
    \end{minipage}
    \hfill
    \begin{minipage}{0.32\textwidth}
        \centering
        \begin{tikzpicture}[scale=0.9]
            \begin{feynman}
                \vertex (n1i) at (-2,-1.2) {$N$};
                \vertex (v1) at (0,-0.8);
                \vertex (n1f) at (2,-1.2) {$N$};

                \vertex (n2i) at (-2,1.2) {$N$};
                \vertex (v2) at (0,0.8);
                \vertex (n2f) at (2,1.2) {$N$};

                \vertex (a) at (1.6,0) {$a$};

                \diagram*{
                    (n1i) -- [fermion] (v1) -- [fermion] (n1f),
                    (n2i) -- [fermion] (v2) -- [fermion] (n2f),
                    (v1) -- [scalar, edge label'={$\pi$}] (v2),
                    (v1) -- [scalar] (a),
                };
            \end{feynman}
        \end{tikzpicture}
    \end{minipage}

    \caption{Representative axion-production processes in a proto-neutron-star core. The leftmost panel shows nucleon-nucleon bremsstrahlung through one-pion exchange with axion emission from a nucleon line. The middle panel shows the pion-induced process $\pi^-+p\rightarrow n+a$. The rightmost panel shows the tree-level $NN\rightarrow NNa$ contribution generated by the $a\pi NN$ interaction in Eq.~\eqref{eq:universal-axion-pion-nucleon}. Only representative topologies are shown.}
    \label{fig:supernova-axion-processes}
\end{figure}

The conventional axion-production channel in the proto-neutron-star core is nucleon-nucleon bremsstrahlung, $N+N\rightarrow N+N+a$, commonly described at leading order through one-pion exchange~\cite{Carenza:2019pxu}. If the hot and dense medium contains an appreciable population of thermal pions, pion-induced reactions such as $\pi^-+p\rightarrow n+a$ can provide an additional contribution~\cite{Carenza:2020cis}. Representative diagrams are shown in Fig.~\ref{fig:supernova-axion-processes}.

A recent development is the identification of an additional QCD contribution that is independent of the usual model-dependent derivative axion-nucleon couplings. At next-to-leading order in heavy-baryon ChPT, this contribution arises from the isospin-breaking interaction~\cite{Springmann:2024ret}
\begin{align}
    \mathcal{L}^{\rm NLO}_{a\pi N}
    \supset
    -\hat c_5 m_\pi^2
    \frac{4z}{(1+z)^2}
    \frac{a\,\pi^a}{f_aF_\pi}
    \bar N\tau^a N,
    \qquad
    z\equiv\frac{m_u}{m_d},
    \label{eq:universal-axion-pion-nucleon}
\end{align}
where $\hat c_5$ is a standard isospin-breaking pion-nucleon low-energy constant whose value can be fixed from the strong neutron-proton mass splitting~\cite{Bernard:1996gq}. Its coefficient is determined by low-energy QCD and does not depend on the ultraviolet PQ charge assignment. Together with the ordinary pion-nucleon interaction, Eq.~\eqref{eq:universal-axion-pion-nucleon} generates a tree-level contribution to $NN\rightarrow NNa$, illustrated in the rightmost panel of Fig.~\ref{fig:supernova-axion-processes}. Using SN~1987A cooling, Ref.~\cite{Springmann:2024ret} obtains
\begin{align}
    f_a
    >
    1.1^{+0.4}_{-0.6}\times10^8\,{\rm GeV},
    \qquad
    m_a
    <
    5.3^{+7.3}_{-1.3}\times10^{-2}\,{\rm eV}
    \qquad
    (68\%~{\rm C.L.}),
    \label{eq:universal-SN-axion-bound}
\end{align}
where the quoted uncertainties include those from the hadronic inputs and the truncation of the chiral expansion. This bound therefore persists even when the leading derivative nucleon couplings are suppressed, while such model-dependent couplings can strengthen the constraint when present.

Neutron-star cooling provides a related probe of axion-nucleon interactions at lower temperatures and over much longer timescales. Axions can be emitted through nucleon-nucleon bremsstrahlung and, when nucleons become superfluid, through Cooper-pair breaking and formation. Analyses of Cassiopeia~A and other neutron stars have therefore been used to constrain axion-nucleon couplings~\cite{Hamaguchi:2018oqw}. A detailed analysis of five nearby isolated neutron stars finds, within the KSVZ model~\cite{Buschmann:2021juv},
\begin{align}
    m_a
    \lesssim
    16\,{\rm meV}
    \qquad
    (95\%~{\rm C.L.}).
    \label{eq:NS-KSVZ-bound}
\end{align}

The quantitative interpretation of both supernova and neutron-star cooling bounds remains sensitive to dense-matter physics. Nuclear correlations, many-body effects, the equation of state, and the composition of the dense medium can all affect axion production and transport, while neutron-star cooling additionally depends on the envelope composition and nucleon superfluidity. A systematic heavy-baryon ChPT analysis at finite density has shown that the effective axion-nucleon couplings can be substantially modified near nuclear saturation density and, within the regime where the chiral expansion remains under perturbative control, can weaken the inferred neutron-star cooling bounds by a factor of several~\cite{Springmann:2024mjp}. Possible changes in the core composition introduce a further source of uncertainty~\cite{Arias-Aragon:2026mwz}. Supernova and neutron-star cooling therefore provide powerful but actively evolving probes of QCD-axion interactions in dense nuclear matter. Strongly magnetised neutron-star environments can also probe the axion-photon coupling through resonant photon-axion conversion in the magnetosphere~\cite{Bondarenko:2022ngb}.

\paragraph{Cold and thermal axion cosmology.}

The cold axion abundance discussed in Sec.~\ref{sec:axion-cosmology} provides a basic cosmological constraint independent of visible-sector couplings: it must not exceed the observed dark-matter density. In the pre-inflationary scenario, this requirement constrains a combination of $f_a$ and the initial misalignment angle $\vartheta_i$ and, considering the abundance alone, can be relaxed by choosing a sufficiently small initial displacement. In the post-inflationary scenario, the axion field samples different initial angles across causally disconnected regions, so the spatially averaged misalignment contribution is statistically determined. The total abundance also receives contributions from strings and domain walls whose precise magnitude remains subject to numerical uncertainties~\cite{Borsanyi:2016ksw,Gorghetto:2018myk,Gorghetto:2020qws,Buschmann:2021sdq}. We therefore do not associate the dark-matter abundance with a single model-independent QCD-axion mass bound.

A qualitatively different axion population can arise from thermal interactions with the Standard-Model plasma. Depending on the temperature, axions can be produced through quarks and gluons, electroweak and Higgs-sector particles, or hadrons. A consistent calculation must therefore follow the production rate across several mass thresholds and, in particular, through the QCD crossover~\cite{DEramo:2021lgb}. If the resulting relic axions remain relativistic around recombination, they contribute to the radiation energy density, conventionally parametrised by $\Delta N_{\rm eff}$. After becoming nonrelativistic, they behave as hot dark matter and suppress the growth of structure through free streaming, analogously to massive neutrinos.

Cosmological hot-dark-matter limits on axions have a long history~\cite{Hannestad:2005df,Hannestad:2008js}. Modern analyses incorporate improved thermal production rates, the momentum dependence of the axion phase-space distribution, and updated cosmic microwave background (CMB), baryon acoustic oscillation (BAO), large-scale-structure, and primordial light-element-abundance data~\cite{DEramo:2022nvb,Notari:2022ffe,Bianchini:2023ubu}. For example, for the KSVZ axion Ref.~\cite{DEramo:2022nvb} obtains bounds ranging from $m_a<0.16\,{\rm eV}$ to $m_a<0.28\,{\rm eV}$ at $95\%$ confidence, depending on the cosmological dataset, with similar results for the DFSZ axion. Focusing on the minimally coupled QCD-axion case, for which production below the QCD crossover is dominated by axion-pion scattering $(\pi\pi\leftrightarrow a\pi)$, a subsequent analysis solving the momentum-dependent Boltzmann equations finds~\cite{Notari:2022ffe}
\begin{align}
    m_a
    \leq
    0.24\,{\rm eV}
    \qquad
    (95\%~{\rm C.L.}).
\end{align}
A more recent analysis using an improved treatment of the axion-pion production rate and combining CMB, BAO, and primordial-abundance information reports~\cite{Bianchini:2023ubu}
\begin{align}
    m_a
    \leq
    0.16\,{\rm eV}
    \qquad
    (95\%~{\rm credible~interval}).
    \label{eq:thermal-axion-cosmology-bound}
\end{align}
These numerical limits depend on the thermal production history as well as on the cosmological data and assumptions employed. Additional model-dependent axion couplings can modify the production rate and hence the inferred mass bound. Thermal cosmology therefore provides an important constraint on the QCD axion, but not a completely universal upper limit on $m_a$.

\paragraph{Inflationary isocurvature.}

In the pre-inflationary PQ-breaking scenario, the inflationary axion fluctuations discussed in Sec.~\ref{sec:axion-cosmology} generate a cold-dark-matter isocurvature component. In the harmonic regime and for an axion fraction $r_a\equiv\Omega_a/\Omega_{\rm DM}$, the primordial isocurvature power is approximately
\begin{align}
    \mathcal{P}_S
    \simeq
    r_a^2
    \left(
        \frac{H_{\rm inf}}
             {\pi f_a\vartheta_i}
    \right)^2 .
    \label{eq:axion-isocurvature-power}
\end{align}
Planck finds no evidence for primordial cold-dark-matter isocurvature. For an uncorrelated scale-invariant mode, the 2018 analysis gives~\cite{Planck:2018jri}
\begin{align}
    \beta_{\rm iso}
    \equiv
    \frac{\mathcal{P}_S}
         {\mathcal{P}_{\mathcal R}+\mathcal{P}_S}
    <
    0.038
    \qquad
    (95\%~{\rm C.L.}),
    \label{eq:Planck-isocurvature-bound}
\end{align}
where $\mathcal{P}_{\mathcal R}$ is the curvature-perturbation power spectrum. Combining Eqs.~\eqref{eq:axion-isocurvature-power} and~\eqref{eq:Planck-isocurvature-bound} can therefore impose a stringent upper bound on $H_{\rm inf}$ if the QCD axion constitutes most of the dark matter.

This constraint is powerful but conditional. It assumes that the axion is light during inflation, that the PQ symmetry remains broken, and that the axion fluctuations during inflation are normalised by the same $f_a$ that characterises the low-energy theory. Nonminimal inflationary couplings, a large value of the radial PQ field during inflation, or subsequent evolution of the PQ sector can suppress the observable isocurvature fluctuations~\cite{Marsh:2015xka,OHare:2024nmr,Graham:2025iwx}. In particular, a recent analysis has emphasised that even in comparatively minimal PQ sectors the familiar tension between simultaneously large $f_a$ and large $H_{\rm inf}$ can depend sensitively on the ultraviolet dynamics of the PQ-breaking field~\cite{Graham:2025iwx}. Isocurvature bounds should therefore be understood as constraints on a cosmological realisation of the QCD axion rather than on the zero-temperature axion EFT alone.

\paragraph{Black-hole superradiance.}

Rotating black holes provide a conceptually distinct probe of ultralight bosons. A massive scalar whose Compton wavelength is comparable to the gravitational radius of a Kerr black hole can form hydrogen-like bound states. Defining the gravitational fine-structure parameter $\alpha_G\equiv G M_{\rm BH}m_a$, the superradiant instability is efficient for suitable values of $\alpha_G$ when the bound-state frequency satisfies
$\omega<m_{\rm az}\Omega_H$,
where $m_{\rm az}$ is the azimuthal quantum number and $\Omega_H$ is the angular velocity of the black-hole horizon. The occupation number of such a state can then grow exponentially by extracting energy and angular momentum from the black hole, forming a macroscopic axion cloud and spinning the black hole down~\cite{Arvanitaki:2010sy,Brito:2015oca,Arvanitaki:2014wva}.

The observation of a rapidly rotating black hole whose age exceeds the predicted superradiance timescale can therefore exclude the corresponding axion mass range. More generally, an ultralight boson can produce depleted regions in the black-hole mass-spin plane. This probe is complementary to the stellar-cooling constraints discussed above because the instability is predominantly gravitational and does not require the axion to constitute dark matter or to possess appreciable couplings to photons or ordinary fermions~\cite{Arvanitaki:2010sy,Arvanitaki:2014wva,Caputo:2024oqc}.

Quantitative bounds nevertheless depend on the black-hole mass, spin, and age, as well as on astrophysical effects such as accretion and binary evolution. For the QCD axion, self-interactions can further modify the growth and evolution of the cloud. Recent work has incorporated higher superradiant levels and reanalysed X-ray black-hole spin measurements in a Bayesian framework, highlighting the systematic uncertainties associated with spin inference~\cite{Witte:2024drg}.

Gravitational-wave observations provide an increasingly important complementary route. The rapidly spinning black holes in GW231123 have recently been used to constrain axions with masses roughly in the range
\begin{align}
    m_a
    \sim
    (0.6\text{--}5)\times10^{-13}\,{\rm eV},
\end{align}
for sufficiently weak self-interactions, extending the sensitivity of superradiance below the mass range traditionally probed by X-ray binaries~\cite{Caputo:2025oap}. More recently, a hierarchical analysis of 257 binary-black-hole mergers in the Gravitational-Wave Transient Catalog 5 (GWTC-5) reports no evidence for superradiant axions and excludes~\cite{Ning:2026ebu}
\begin{align}
    1.7\times10^{-14}\,{\rm eV}
    \lesssim
    m_a
    \lesssim
    3.3\times10^{-12}\,{\rm eV}
    \qquad
    (95\%~{\rm C.L.}),
    \label{eq:GWTC5-superradiance-bound}
\end{align}
within the adopted black-hole spin-population model. Given the dependence on black-hole formation and spin modelling, this result should not be interpreted as an assumption-independent exclusion.

Superradiant axion clouds can also emit nearly monochromatic gravitational waves through axion annihilations and transitions between bound levels, providing a potential discovery channel in addition to indirect spin constraints~\cite{Arvanitaki:2010sy,Arvanitaki:2014wva,Arvanitaki:2016qwi}. 

\subsection{Experimental searches}
\label{sec:axion-experimental-searches}

The experimental programme for the QCD axion is unusually broad, ranging from resonant microwave detectors and low-frequency quantum sensors to solar helioscopes and spin-precession experiments; see Refs.~\cite{Graham:2015ouw,Irastorza:2018dyq,Baryakhtar:2025jwh} for reviews. Unlike a generic axion-like particle, the QCD axion does not occupy an arbitrary parameter space: its mass and decay constant are related by QCD, as in Eq.~\eqref{eq:axion-mass-numerical}, while its couplings to photons, electrons, and nucleons retain some dependence on the ultraviolet realisation of the PQ symmetry.

The axion-photon coupling provides the most familiar example. Combining Eqs.~\eqref{eq:axion-mass-numerical} and~\eqref{eq:axion-photon-coupling}, one obtains approximately
\begin{align}
    |g_{a\gamma\gamma}|
    \simeq
    2.04\times 10^{-16}\ {\rm GeV}^{-1}
    \left|
        \frac{E}{N}-\C_{\rm QCD}
    \right|
    \left(
        \frac{m_a}{\mu{\rm eV}}
    \right),
    \qquad
    \C_{\rm QCD}\simeq 1.92.
    \label{eq:qcd-axion-photon-trajectory}
\end{align}
For a fixed value of $E/N$, the QCD axion therefore traces an approximately straight line in the $(m_a,|g_{a\gamma\gamma}|)$ plane. The commonly displayed KSVZ and DFSZ lines provide useful benchmarks rather than strict boundaries: different electromagnetic anomaly coefficients can enhance or suppress the photon coupling, and an accidental cancellation between $E/N$ and the QCD contribution can make $g_{a\gamma\gamma}$ particularly small~\cite{Agrawal:2017cmd}.

The interpretation of an experimental limit also depends on the axion source. Dark-matter haloscopes search for an ambient Galactic axion field and hence depend on the local axion density, whereas helioscopes search for axions produced in the Sun and do not require axions to constitute dark matter. Other approaches produce axions in the laboratory or probe their couplings to nuclear and atomic spins. These different assumptions should be kept in mind when comparing experimental exclusions.

\paragraph{Microwave cavity haloscopes.}

The classic haloscope proposed by Sikivie converts nonrelativistic Galactic axions into electromagnetic radiation in a strong static magnetic field~\cite{Sikivie:1983ip,Sikivie:1985yu}. For virialised axion dark matter, the field oscillates with frequency $\omega_a\simeq m_a(1+v^2/2)$, where the Galactic velocity dispersion $v\sim10^{-3}$ gives a fractional linewidth of order $10^{-6}$ and an effective axion quality factor $Q_a\sim10^6$. If a cavity mode is tuned to the axion frequency, the signal power scales schematically as
\begin{align}
    P_{\rm sig}
    \sim
    g_{a\gamma\gamma}^2
    \frac{\rho_a}{m_a}
    B_0^2 V C Q_{\rm eff},
    \qquad
    Q_{\rm eff}\simeq \min(Q_L,Q_a),
    \label{eq:axion-haloscope-power}
\end{align}
where $B_0$ is the magnetic field, $V$ the effective cavity volume, $C$ the mode-overlap form factor, and $Q_L$ the loaded cavity quality factor. Large magnetic volumes, high-quality resonators, cryogenic operation, and low-noise readout are therefore central to this approach.

The Axion Dark Matter eXperiment (ADMX) has reached DFSZ sensitivity over portions of the few-$\mu{\rm eV}$ region~\cite{ADMX:2021nhd} and more recently searched $4.55$--$5.42\,\mu{\rm eV}$ with sensitivity extending beyond the standard KSVZ benchmark~\cite{ADMX:2025vom}. The Center for Axion and Precision Physics Research (CAPP) programme has achieved comparable sensitivity in neighbouring regions, including searches around a few $\mu{\rm eV}$ and near $m_a\simeq24\,\mu{\rm eV}$~\cite{CAPP:2024dtx,Ahn:2025via}. Haloscope At Yale Sensitive To Axion Cold Dark Matter (HAYSTAC), Taiwan Axion Search Experiment with Haloscope (TASEH), QUaerere AXion (QUAX), and Relic Axion Dark-Matter Exploratory Setup (RADES) extend cavity and resonator searches towards higher frequencies~\cite{HAYSTAC:2024jch,TASEH:2022vvu,QUAX:2024fut,Ahyoune:2024klt,QUAX:2025wtb}. Resonant enhancement can reach canonical QCD-axion couplings, but only over a narrow bandwidth at a given tuning, making scan rate a central experimental challenge.

This limitation has motivated quantum-enhanced readout~\cite{Lamoreaux:2013koa,Zheng:2016qjv,Dixit:2020ymh} as well as resonant mode-conversion and multi-mode schemes~\cite{Sikivie:2010fa,Goryachev:2018vjt,Berlin:2019ahk,Berlin:2020vrk}. HAYSTAC has demonstrated the use of squeezed microwave states to increase the scan rate beyond that of a conventional quantum-limited receiver~\cite{HAYSTAC:2020kwv,HAYSTAC:2024jch}, while photon counting, superconducting resonators, entanglement, and correlated sensor arrays are being actively developed. Reference~\cite{Chen:2021bgy} proposed a haloscope array based on parity-time ($\mathcal{PT}$)-symmetric coupled resonators to broaden the signal-response bandwidth, and Ref.~\cite{Chen:2023ryb} developed a multi-mode framework using auxiliary modes and parametric interactions to maintain a strong response over a wider frequency interval. Such techniques may help overcome the intrinsically narrow-band character of conventional resonant searches.

Dark-matter haloscope limits necessarily depend on the assumed Galactic axion distribution. Since $P_{\rm sig}\propto\rho_a g_{a\gamma\gamma}^2$, the coupling reach scales as $\rho_a^{-1/2}$ if axions constitute only a fraction of the local dark matter. The velocity distribution determines the signal lineshape, while miniclusters or other substructure in post-inflationary cosmologies may further modify the local phase-space distribution.

\paragraph{Low-mass searches with lumped-element circuits.}

For $m_a$ well below the $\mu{\rm eV}$ scale, the electromagnetic wavelength is much larger than practical detector dimensions and a conventional microwave cavity becomes inefficient. In this quasistatic regime, the axion-induced effective current can instead generate an oscillating magnetic flux in an inductive pickup~\cite{Kahn:2016aff}, with an inductance-capacitance circuit providing resonant enhancement.

ABRACADABRA has demonstrated sensitivity to the axion-photon interaction in this regime, with ABRACADABRA-10\,cm probing approximately $0.41$--$8.27\,{\rm neV}$~\cite{Salemi:2021gck}. The DMRadio programme aims to extend this approach using larger magnetic volumes and lower-noise readout. DMRadio-$m^3$ targets approximately $20$--$800\,{\rm neV}$~\cite{DMRadio:2022pkf}, while the longer-term DMRadio-GUT and the more recent DMRadio-Core concepts explore complementary portions of the sub-$\mu{\rm eV}$ region~\cite{DMRadio:2022jfv,Ankel:2026zrv}.

\paragraph{Higher masses: multi-cell, dielectric, plasma, and broadband haloscopes.}

For $m_a$ above tens of $\mu{\rm eV}$, conventional microwave cavities face the opposite difficulty: increasing the resonant frequency requires smaller cavity dimensions and therefore reduces the available conversion volume. Mode crowding and tuning also become increasingly challenging. Multi-cell resonators, such as those employed by CAPP and RADES, mitigate this problem by coherently combining several coupled volumes.

A different strategy is the dielectric haloscope, in which the axion field in an external magnetic field induces electromagnetic radiation at interfaces between materials with different dielectric properties. By arranging multiple dielectric disks with suitable separations, the waves emitted from different interfaces can interfere constructively and substantially enhance the signal~\cite{Caldwell:2016dcw}. MADMAX is designed to exploit this principle in the tens-to-hundreds of $\mu{\rm eV}$ region, and a prototype has performed the first corresponding dark-matter search around $77$--$79\,\mu{\rm eV}$~\cite{MADMAX:2024sxs}.

Plasma haloscopes instead exploit resonant axion-photon conversion when the axion mass is matched to the effective plasma frequency of the medium. In a wire-array metamaterial, this plasma frequency is controlled primarily by the wire geometry rather than by the overall detector dimensions, allowing a large conversion volume to be retained at high frequencies. The Axion Longitudinal Plasma HAloscope (ALPHA) programme proposes to realise this concept with arrays of conducting wires, with a Phase-I design targeting approximately $40$--$80\,\mu{\rm eV}$~\cite{ALPHA:2022rxj,ALPHA:2026vwc}.

Broadband reflector or dish-antenna concepts sacrifice resonant enhancement in exchange for a much larger instantaneous bandwidth. In an external magnetic field, the axion field induces an oscillating electromagnetic response; at a conducting surface, the electromagnetic boundary conditions require the emission of a propagating wave that can be collected by a suitably shaped reflector. The Broadband Reflector Experiment for Axion Detection (BREAD) exploits this principle by using a large conducting surface to convert axion dark matter into electromagnetic radiation and focusing the emitted power onto a sensitive receiver~\cite{BREAD:2021tpx}. The GigaBREAD prototype has performed an axion-like-particle search in the $44$--$52\,\mu{\rm eV}$ range~\cite{GigaBREAD:2025lzq}. Although the present sensitivity remains above canonical QCD-axion targets, this broadband approach provides a complementary route at frequencies where conventional resonant volumes become small.

\paragraph{Spin-precession and direct probes of the QCD coupling.}

Photon conversion is not the only route to the QCD axion. Its couplings to QCD and to fermion spins generate oscillatory nuclear and atomic observables that can be searched for with precision magnetometry and nuclear magnetic resonance. The Cosmic Axion Spin Precession Experiment (CASPEr), for example, exploits time-dependent CP-odd nuclear moments induced by an oscillating axion background, which can resonantly drive nuclear-spin precession~\cite{Budker:2013hfa}. Related searches probe derivative axion-nucleon interactions through an effective oscillating ``axion wind.''

Such measurements are particularly valuable for establishing the QCD origin of a possible axion signal. Reference~\cite{Berlin:2022mia}, for example, proposed a polarization haloscope in which QCD-axion-induced CP-odd moments generate an electromagnetic current in a spin-polarised dielectric. Correlations among spatially separated vector sensors can provide additional information on the direction and angular distribution of an ultralight bosonic background and help distinguish axion-fermion signals from vector dark matter~\cite{Chen:2021bdr}. These capabilities would become especially useful after an initial discovery, when the goal shifts from excluding parameter space to identifying the microscopic nature of the new field.

\paragraph{Solar helioscopes.}

Helioscopes search for axions produced inside the Sun rather than for the Galactic dark-matter population. Solar axions produced through the Primakoff process can reconvert into X-ray photons in a laboratory magnetic field. For a homogeneous transverse field and sufficiently small momentum mismatch,
\begin{align}
    P_{a\rightarrow\gamma}
    \simeq
    \left(
        \frac{g_{a\gamma\gamma} B L}{2}
    \right)^2,
    \label{eq:axion-helioscope-conversion}
\end{align}
where $L$ is the magnetic length. At larger masses, coherence between the axion and photon is lost; a buffer gas can give the photon an effective mass and restore conversion over selected mass intervals.

The CERN Axion Solar Telescope (CAST) provides the leading laboratory helioscope constraint over a broad low-mass range,
\begin{align}
    g_{a\gamma\gamma}
    <
    5.8\times 10^{-11}\ {\rm GeV}^{-1}
    \qquad
    (95\%~{\rm C.L.}),
    \qquad
    m_a\lesssim 0.02\ {\rm eV},
    \label{eq:cast-axion-photon-bound}
\end{align}
for the standard Primakoff solar-axion interpretation~\cite{CAST:2024eil}. The International Axion Observatory (IAXO), together with its intermediate stage BabyIAXO, aims to improve substantially on this sensitivity using a larger magnetic aperture, dedicated X-ray optics, and low-background detectors~\cite{IAXO:2019mpb,IAXO:2025ltd}. Unlike dark-matter haloscopes, helioscopes do not depend on the local axion abundance. Their event rate instead depends on the solar production flux; for Primakoff production followed by magnetic reconversion it scales approximately as $g_{a\gamma\gamma}^4$, while axion-electron couplings open additional solar production channels.

\paragraph{Light-shining-through-a-wall searches.}

A complementary, purely laboratory probe of the axion-photon interaction is provided by light-shining-through-a-wall experiments~\cite{Redondo:2010dp,Irastorza:2018dyq}. Laser photons convert into axions in a magnetic field, while an opaque barrier blocks the ordinary photons; axions passing through the barrier can then reconvert into photons in a second magnetic region. Since both production and regeneration involve axion-photon conversion, the regenerated photon rate scales as $g_{a\gamma\gamma}^4$. The Any Light Particle Search II (ALPS II) experiment has recently reported its first science results, obtaining
$g_{a\gamma\gamma}<1.5\times10^{-9}\,{\rm GeV}^{-1}$ for $m_a\lesssim0.1\,{\rm meV}$ at $95\%$ confidence level~\cite{ALPSII:2025eri}. This sensitivity remains well above the standard QCD-axion photon coupling in this mass range, so current light-shining-through-a-wall experiments primarily probe more general axion-like particles. Nevertheless, they provide a particularly clean laboratory test that is independent of astrophysical environments and of any assumption about the axion dark-matter abundance.

\paragraph{} No single experiment or mass interval captures the present status of QCD-axion searches. Although QCD fixes the relation between $m_a$ and $f_a$, the observable couplings remain model dependent, while dark-matter searches depend additionally on the axion abundance and local distribution. The relevant targets therefore depend on both the ultraviolet axion model and its cosmological history. Likewise, a signal observed through a single interaction would not by itself establish that the new particle is the QCD axion. Confirming this interpretation would require testing whether its measured mass and couplings are mutually consistent with the QCD-axion framework, ideally through complementary photon, electron, nucleon, or directly QCD-induced observables. Such complementarity is essential not only for discovery, but also for identifying a future signal with the axion associated with the PQ solution to the strong CP problem.

\section{The axion quality problem}
\label{sec:axion-quality-problem}

\subsection{Quantifying the axion quality problem}
\label{sec:axion-quality-criterion}

As discussed in Secs.~\ref{sec:UV-robustness-viability} and~\ref{sec:PQ-dynamical-relaxation}, the PQ mechanism requires the QCD contribution to the axion potential to dominate sufficiently over any additional explicit PQ breaking whose minimum is not aligned with the $CP$-conserving point~\cite{Barr:1992qq,Holman:1992us,Kamionkowski:1992mf,DiLuzio:2020wdo}. We now quantify this requirement. After shifting the axion field such that the minimum generated by QCD alone lies at $a=0$, the zero-temperature potential may be written as
\begin{align}
    V(a)
    =
    V_{\rm QCD}(a)
    +
    \Delta V_{\slashed{\rm PQ}}(a),
    \qquad
    V_{\rm QCD}(a)
    =
    \frac{1}{2}\chi
    \left(
        \frac{a}{f_a}
    \right)^2
    +\mathcal{O}\left(\frac{a^4}{f_a^4}\right),
    \label{eq:axion-quality-general-potential}
\end{align}
where $\Delta V_{\slashed{\rm PQ}}$ denotes explicit PQ breaking unrelated to the QCD anomaly. If $\Delta V_{\slashed{\rm PQ}}$ is a small perturbation to the QCD potential, minimisation gives
\begin{align}
\left\langle\bar\theta_{\rm eff}\right\rangle
    =
    \frac{\langle a\rangle}{f_a}
    \simeq
    -\frac{f_a}{\chi}
    \left.
\left(\frac{\partial\Delta V_{\slashed{\rm PQ}}}
         {\partial a}\right)
    \right|_{a=0}.
    \label{eq:axion-quality-general-shift}
\end{align}
The axion quality requirement is therefore not simply that all additional contributions to the axion potential be small. Rather, it is the component of the additional potential that exerts a nonvanishing force at the QCD minimum that must satisfy
\begin{align}
    \left|
    \frac{f_a}{\chi}
    \left.
    \left(\frac{\partial\Delta V_{\slashed{\rm PQ}}}
         {\partial a}\right)
    \right|_{a=0}
    \right|
    \lesssim
    10^{-10}.
    \label{eq:axion-quality-general-criterion}
\end{align}
An additional contribution whose minimum is exactly aligned with that of QCD does not regenerate strong $CP$ violation, even if it changes the curvature of the axion potential. This distinction will be important when we discuss modified axion potentials in Sec.~\ref{sec:heavy-QCD-axions}.

For illustration, consider a single PQ-breaking harmonic,
\begin{align}
    \Delta V_{\slashed{\rm PQ}}(a)
    =
    -\Lambda_{\rm br}^4
    \cos\left(
        k\frac{a}{f_a}+\delta
    \right),
    \label{eq:axion-quality-breaking-harmonic}
\end{align}
where $k$ is fixed by the PQ charge carried by the explicit-breaking interaction relative to the QCD anomaly normalisation, and $\delta$ denotes its relative $CP$ phase. Equation~\eqref{eq:axion-quality-general-shift} then gives $\langle\bar\theta_{\rm eff}\rangle\approx -k\Lambda_{\rm br}^4(\sin\delta)/\chi$. For a generic phase, $|\sin\delta|\sim\mathcal{O}(1)$, the neutron-EDM constraint therefore requires $k\Lambda_{\rm br}^4\lesssim 10^{-10}\chi$. 
Since $\chi^{1/4}\approx 75\,{\rm MeV}$~\cite{GrilliDiCortona:2015jxo,Gorghetto:2018ocs}, this corresponds parametrically to $\Lambda_{\rm br}\lesssim{\rm few}\times10^{-1}\,{\rm MeV}$ for $k\sim\mathcal{O}(1)$. 

The severity of this requirement becomes especially transparent for the higher-dimensional operator introduced in Eq.~\eqref{eq:PQ-breaking-UV-operator}. Consider, for illustration, a single PQ-breaking field, $\Phi=[(v_{\rm PQ}+\rho)/\sqrt{2}]\exp(\i a/v_{\rm PQ})$ with  $f_a=v_{\rm PQ}/N$. Writing $c_d=|c_d|\e^{\i\delta_d}$, the operator in Eq.~\eqref{eq:PQ-breaking-UV-operator} induces, up to an irrelevant redefinition of $\delta_d$,
\begin{align}
    \Delta V^{(d)}_{\slashed{\rm PQ}}(a)
    =
    -2|c_d|
    \frac{(v_{\rm PQ}/\sqrt{2})^d}
         {\Lambda_{\rm UV}^{\,d-4}}
    \cos\left(
        d\frac{a}{v_{\rm PQ}}+\delta_d
    \right).
    \label{eq:axion-quality-dim-d-potential}
\end{align}
The induced strong $CP$ phase is then
\begin{align}
    \left|
    \left\langle\bar\theta_{\rm eff}\right\rangle
    \right|
    \approx
    \frac{2d}{N}
    \frac{|c_d\sin\delta_d|}{\chi}
    \frac{(v_{\rm PQ}/\sqrt{2})^d}
         {\Lambda_{\rm UV}^{\,d-4}}.
    \label{eq:axion-quality-dim-d-shift}
\end{align}
Thus, suppression by a large ultraviolet scale does not by itself guarantee sufficient axion quality~\cite{Barr:1992qq,Holman:1992us,Kamionkowski:1992mf,Dobrescu:1996jp}. For example, taking the illustrative single-field normalisation $N=1$, $\Lambda_{\rm UV}= m_{\rm Pl}\approx 2.4\times10^{18}\,{\rm GeV}$, and $|c_d\sin\delta_d|\sim1$, $|\langle \bar{\theta}_{\rm eff}\rangle|<10^{-10}$ requires approximately $d\geq11$ for $f_a=10^{10}\,{\rm GeV}$ and $d\geq14$ for $f_a=10^{12}\,{\rm GeV}$. These numbers are only illustrative: the precise required dimension depends on the anomaly normalisation, the PQ-breaking field content, the ultraviolet scale, and the Wilson coefficients and their phases. A recent statistical analysis, modelling Planck-suppressed PQ-breaking operators with unsuppressed Wilson coefficients and random phases, similarly finds that if the QCD axion accounts for the observed dark-matter abundance at $f_a\simeq10^{11}\,{\rm GeV}$, the leading such operators must be absent up to approximately mass dimension $d\gtrsim12$~\cite{Cheek:2026dvu}. The robust conclusion is that a conventional invisible-axion model generally requires all low-dimensional PQ-violating operators to be absent or highly suppressed.

This observation has traditionally been sharpened by the expectation that exact continuous global symmetries do not survive in a theory of quantum gravity, motivating Planck-suppressed operators and nonperturbative gravitational effects as possible sources of PQ breaking~\cite{Barr:1992qq,Holman:1992us,Kamionkowski:1992mf,Dobrescu:1996jp,Banks:2010zn,Alvey:2020nyh}. The quantitative size of such effects, however, is considerably less universal than the quality criterion itself. A Wilsonian-EFT formulation makes explicit that the relevant question is ultimately how the ultraviolet theory matches onto PQ-breaking operators in the low-energy theory~\cite{Dine:2022mjw}. Heavy thresholds can modify naive estimates based solely on Planck-suppressed operators~\cite{Bonnefoy:2022vop}, while formulations in terms of three-form gauge fields and their dual antisymmetric tensors provide a complementary perspective on the ultraviolet sensitivity of the axion solution~\cite{Dvali:2005an,Burgess:2023ifd}. Moreover, a recent semiclassical reanalysis finds that gravitational-instanton contributions can be extremely small and need not themselves constitute a significant axion quality problem~\cite{Catinari:2024zon}. In the next subsection, we discuss how an ultraviolet theory can naturally enforce a high-quality PQ symmetry.

\subsection{Ultraviolet protection of the Peccei--Quinn symmetry}
\label{sec:axion-quality-UV-protection}

The estimates in Sec.~\ref{sec:axion-quality-criterion} show that suppressing the coefficients of generic PQ-violating operators by hand would merely replace the smallness of $\bar\theta$ by another severe tuning. A more satisfactory possibility is that the PQ symmetry is not a fundamental global symmetry imposed on the ultraviolet theory, but instead emerges accidentally from a more fundamental structure. The familiar baryon and lepton numbers of the renormalisable Standard Model provide simple analogies: they arise because the gauge symmetries and field content forbid operators that violate them at sufficiently low dimension. For the axion, the requirement is considerably stronger, since the first allowed operator that produces a misaligned axion potential must satisfy the quality criterion in Eq.~\eqref{eq:axion-quality-general-criterion}. This observation has motivated a broad class of models in which the approximate PQ symmetry follows from exact gauge symmetries, discrete symmetries, chirality, or strong dynamics~\cite{Randall:1992ut,Redi:2016esr,Lillard:2018fdt,Gavela:2018paw,Ardu:2020qmo,Contino:2021ayn}.

The motivation for such constructions is strengthened by the expectation that exact global symmetries are not fundamental in quantum gravity~\cite{Banks:2010zn,Harlow:2018tng}. As emphasised in the previous subsection, this expectation does not by itself determine the magnitude of PQ violation, but it motivates seeking an ultraviolet origin for the PQ symmetry that is protected by gauge structure rather than postulated as an exact global symmetry. Importantly, the low-energy PQ symmetry itself need not be gauged. Indeed, the PQ current must have a nonvanishing QCD anomaly in order to generate the axion-gluon interaction. The aim is instead to construct an exact ultraviolet gauge or discrete symmetry whose allowed interactions accidentally preserve an anomalous global $U(1)_{\rm PQ}$ to sufficiently high order.

\paragraph{Gauge and discrete protection.}

A simple implementation is to impose an exact discrete gauge symmetry under which the PQ-breaking fields are charged. If, for example, a field $\Phi$ is charged under a sufficiently large $\mathbb Z_n$, operators $\Phi^d$ with $d<n$ can be forbidden, so that the first PQ-violating interaction appears only at high dimension. Unlike a purely global discrete symmetry, a discrete gauge symmetry can arise as the unbroken remnant of a continuous gauge symmetry and is correspondingly expected to be more robust against quantum-gravity effects~\cite{Krauss:1988zc}. Consistency additionally requires the corresponding discrete gauge anomalies to vanish or to be cancelled by an appropriate ultraviolet mechanism. This idea has been implemented in supersymmetric and nonsupersymmetric axion models, sometimes using discrete $R$ symmetries or other symmetries that are independently motivated by the structure of the ultraviolet theory.

Continuous gauge symmetries can play a similar role. The gauge charges may forbid all low-dimensional interactions that violate a particular global combination of field phases, leaving an accidental $U(1)_{\rm PQ}$ after the gauge symmetry is broken~\cite{Barr:1992qq,Fukuda:2017ylt,Qiu:2023los}. Product non-Abelian gauge groups provide another realisation: appropriate representations can make the lowest gauge-invariant operator carrying PQ charge parametrically high dimensional. For example, models based on bifundamental fields of product $SU(N)$ groups can postpone the first breaking of an accidental global $U(1)$ to an operator whose dimension grows with $N$, while other gauge structures can protect the accidental symmetry perturbatively to arbitrarily high operator dimension~\cite{Gavela:2018paw,Ardu:2020qmo}. Gauge and flavour symmetries can also be combined with grand unification, so that the same structure responsible for organising Standard-Model quantum numbers or flavour simultaneously protects the PQ symmetry~\cite{Georgi:1981pu,DiLuzio:2020qio,DiLuzio:2020xgc,DiLuzio:2025jhv}. Recent constructions based on gauged axial symmetries and unified gauge groups provide further examples in which a high-quality PQ symmetry emerges accidentally rather than being postulated as a fundamental global symmetry.

The main lesson is independent of the detailed gauge group. Gauge invariance changes the problem from explaining why a Wilson coefficient is extraordinarily small to explaining why the corresponding operator is forbidden. Once all PQ-violating operators below some dimension $d_{\rm min}$ are absent, the suppression by powers of the ultraviolet scale can become sufficient to satisfy Eq.~\eqref{eq:axion-quality-general-criterion}. The required value of $d_{\rm min}$ remains model dependent, since it depends on $f_a$, the ultraviolet cutoff, anomaly normalisation, and the field content, as illustrated in Sec.~\ref{sec:axion-quality-criterion}.

\paragraph{Composite and dynamical PQ symmetries.}

Strong dynamics provides a particularly natural setting for an accidental PQ symmetry~\cite{Kim:1984pt,Choi:1985cb,Randall:1992ut,Redi:2016esr}. Instead of identifying the PQ order parameter with an elementary scalar $\Phi$, one may identify it with a fermion condensate or another composite operator of a new confining gauge theory. The axion is then a Nambu--Goldstone boson associated with an accidental chiral symmetry of the strong sector. Since any ultraviolet interaction that explicitly breaks this symmetry must itself be constructed from gauge-invariant combinations of the microscopic constituents, the lowest PQ-violating operator can naturally have a large canonical dimension.

This possibility has been developed in a variety of increasingly economical constructions. Product-group theories can protect composite axions to very high order~\cite{Lillard:2018fdt}, while chiral gauge theories can generate the PQ symmetry automatically as a consequence of their gauge representations and fermion content~\cite{Gavela:2018paw,Ardu:2020qmo}. Systematic studies have constructed models in which the first PQ-violating operators occur at dimension twelve or higher, with still higher protection possible for suitable gauge structures~\cite{Contino:2021ayn}. Related constructions embed the strong sector into grand-unified theories or connect it to flavour symmetries~\cite{DiLuzio:2020qio,DiLuzio:2020xgc}. More recently, QCD-like strong dynamics combined with a Pati--Salam embedding has provided a composite axion model in which the first operators capable of modifying the axion potential are eight-fermion operators of dimension twelve~\cite{Gherghetta:2025fip}. Exact supersymmetric chiral dynamics provides another route to a high-quality composite axion, with discrete gauge symmetries suppressing the leading PQ-violating operators~\cite{Gherghetta:2025kff}. A different recent construction identifies the axion with the Nambu--Goldstone boson of baryon number in a new confining sector, for which the high dimension of the gauge-invariant baryon operator automatically suppresses PQ breaking~\cite{Agrawal:2025mke}.

Strong dynamics can also suppress PQ violation through anomalous dimensions rather than solely through the canonical dimension of the leading PQ-breaking operator. If the theory passes through an approximately conformal regime over a sufficiently large range of scales, a PQ-violating operator can acquire a sizeable positive anomalous dimension, increasing its scaling dimension and causing its coefficient to decrease more rapidly under renormalisation-group evolution towards the infrared. The resulting power-law suppression between the ultraviolet and PQ-breaking scales can substantially reduce the effect of otherwise dangerous operators. Superconformal realisations of this mechanism have been constructed explicitly~\cite{Nakai:2021nyf}. Thus, a high-quality PQ symmetry need not arise solely from forbidding all low-dimensional PQ-breaking operators; strong renormalisation-group evolution can also suppress their low-energy effects.

\paragraph{Extra-dimensional realisations.}

A related approach identifies the axion with a component or Wilson line of a gauge field in an extra dimension~\cite{Choi:2003wr,Cox:2019rro}. From the higher-dimensional viewpoint, the axion shift symmetry descends from gauge invariance, which forbids a local potential for the Wilson line. Explicit breaking must therefore arise from effects that are nonlocal in the extra dimension or from suitably charged bulk dynamics, allowing locality and gauge invariance to suppress dangerous contributions to the axion potential exponentially or parametrically. Holographic descriptions connect such constructions to four-dimensional composite axion models, making the relation between extra-dimensional locality, compositeness, and axion quality particularly transparent~\cite{Cox:2019rro,Cox:2023dou}. A recent holographic QCD construction likewise finds that high axion quality is associated with a large degree of compositeness, with the physical axion becoming predominantly a higher-dimensional gauge-field mode in the high-quality limit~\cite{Csaki:2026qjl}.

These examples demonstrate that there is no unique ultraviolet solution to the axion quality problem. Discrete and continuous gauge symmetries can eliminate low-dimensional PQ-breaking operators, strong dynamics can make the PQ order parameter composite or render the dangerous operators highly irrelevant, and higher-dimensional gauge invariance can protect the axion through locality. Each possibility introduces additional structure and its own consistency requirements, including gauge-anomaly cancellation, perturbativity, vacuum structure, and cosmological constraints. The common principle is nevertheless simple: the small quantity required by Eq.~\eqref{eq:axion-quality-general-criterion} should follow from symmetry, dynamics, or locality rather than from tuning unrelated ultraviolet coefficients. A different strategy is to increase the restoring axion potential itself while preserving its alignment with the QCD minimum. This possibility leads to heavy QCD axion constructions, which we discuss next.

\subsection{Heavy QCD axions and the quality problem}
\label{sec:heavy-QCD-axions}

The discussion above suggests a complementary strategy for alleviating the axion quality problem. Rather than suppressing every unwanted PQ-breaking contribution to an extremely high accuracy, one may increase the restoring potential that drives the axion towards the $CP$-conserving point. If this additional contribution is aligned with the ordinary QCD potential, the axion can remain a solution to the strong $CP$ problem while acquiring a mass parametrically larger than the standard QCD prediction~\cite{Fukuda:2015ana,Agrawal:2017eqm,Agrawal:2017ksf,Gaillard:2018xgk,Hook:2019qoh,Gherghetta:2020ofz,Valenti:2022tsc}. We refer broadly to such constructions as heavy QCD axion models. 

To see quantitatively how this can improve axion quality, suppose that in addition to the QCD potential there is an extra contribution $V_{\rm extra}(a)$ whose minimum is aligned with the QCD minimum at $a=0$. We write
\begin{align}
    V(a)
    =
    V_{\rm QCD}(a)
    +
    V_{\rm extra}(a)
    +
    \Delta V_{\slashed{\rm PQ}}(a).
    \label{eq:heavy-axion-total-potential}
\end{align}
Expanding the first two terms around their common minimum, we define
\begin{align}
    V_{\rm QCD}(a)
    +
    V_{\rm extra}(a)
    =
    \frac{1}{2}
    \left(
        \chi+\chi_{\rm extra}
    \right)
    \left(
        \frac{a}{f_a}
    \right)^2
    +\cdots ,
    \label{eq:heavy-axion-aligned-potential}
\end{align}
where $\chi_{\rm extra}\equiv f_a^2V_{\rm extra}''(0)>0$. The physical axion mass is then $m_a^2=(\chi+\chi_{\rm extra})/f_a^2$,
which can be much larger than the standard value $m_{a,{\rm QCD}}^2=\chi/f_a^2$ when $\chi_{\rm extra}\gg\chi$. The same enhancement suppresses the displacement caused by a fixed small misaligned contribution. Generalising Eq.~\eqref{eq:axion-quality-general-shift}, one finds
\begin{align}
    \left\langle\bar\theta_{\rm eff}\right\rangle
    \approx
    -\frac{f_a}{\chi+\chi_{\rm extra}}
    \left.
    \frac{\partial\Delta V_{\slashed{\rm PQ}}}
         {\partial a}
    \right|_{a=0}.
    \label{eq:heavy-axion-quality-shift}
\end{align}
For the same $f_a$ and the same $\Delta V_{\slashed{\rm PQ}}$, the residual strong $CP$ phase is therefore reduced relative to that of the standard QCD axion by
\begin{align}
    \frac{\left|\langle\bar\theta_{\rm eff}\rangle\right|_{\rm heavy}
    }{\left|\langle\bar\theta_{\rm eff}\rangle\right|_{\rm QCD}
    }
    \approx
    \frac{\chi}{\chi+\chi_{\rm extra}}
    =
    \frac{m_{a,{\rm QCD}}^2}{m_a^2}.
    \label{eq:heavy-axion-quality-improvement}
\end{align}
A parametrically heavier axion can therefore tolerate correspondingly larger unrelated PQ-breaking effects. This is the sense in which a heavy QCD axion can alleviate the quality problem.

The alignment assumption is, however, essential. To illustrate this, suppose that the additional contribution is slightly displaced from the QCD minimum,
\begin{align}
    V_{\rm extra}(a)
    \approx
    \frac{1}{2}\chi_{\rm extra}
    \left(\frac{a}{f_a}+\delta_{\rm extra}\right)^2 .
    \label{eq:heavy-axion-misaligned-extra-potential}
\end{align}
Neglecting other PQ-breaking effects, minimisation gives
\begin{align}
    \left\langle\bar\theta_{\rm eff}\right\rangle
    \approx
    -\frac{\chi_{\rm extra}\,\delta_{\rm extra}}{\chi+\chi_{\rm extra}}.
    \label{eq:heavy-axion-extra-phase}
\end{align}
In the regime $\chi_{\rm extra}\gg\chi$ needed for a large mass enhancement, the axion is driven predominantly to the minimum of the new potential, so that
$\langle\bar\theta_{\rm eff}\rangle\approx -\delta_{\rm extra}$. A large additional potential with a generic phase therefore does not solve the quality problem; it simply replaces the original misaligned PQ-breaking contribution by an even more important one. Heavy-axion constructions must consequently explain why the additional nonperturbative dynamics is aligned with QCD to the accuracy required by the strong $CP$ bound~\cite{Hook:2019qoh,Valenti:2022tsc}.

One way to achieve this alignment is through a mirror strong sector coupled to the same axion. If a discrete symmetry relates QCD to a mirror QCD sector, it can correlate their topological angles so that the two axion potentials possess the same $CP$-conserving minimum. If the mirror confinement scale is substantially larger than $\Lambda_{\rm QCD}$, its contribution to the axion potential can dominate the ordinary QCD susceptibility and produce a heavy axion~\cite{Berezhiani:2000gh,Fukuda:2015ana,Hook:2019qoh}. Modern mirror constructions have further developed this idea together with the associated dark-matter and early-Universe cosmology~\cite{Dunsky:2023ucb}. More recently, the mechanism has been embedded into a grand-unified mirror sector that remains unbroken and confines, dynamically generating the additional strong scale responsible for the heavy axion potential~\cite{Cacciapaglia:2026yvm}.

Another broad class of models modifies the ultraviolet structure of the colour interaction itself. QCD may emerge as the diagonal subgroup of a product gauge group, or as the low-energy remnant of a larger colour group. Nonperturbative effects associated with the additional gauge factors, including small-size instantons, can then generate an axion potential at scales above ordinary QCD~\cite{Agrawal:2017ksf,Gaillard:2018xgk,Gherghetta:2020ofz,Gherghetta:2020keg,Gupta:2020vxb,Valenti:2022tsc,Sesma:2024tcd}. In suitable constructions, the gauge and fermion structure correlates the relevant topological phases and preserves the $CP$-conserving vacuum while allowing $\chi_{\rm extra}\gg\chi$. These mechanisms can raise the axion mass by many orders of magnitude and allow values of $f_a$ far below the conventional invisible-axion range.

Small-instanton enhancement must nevertheless be treated with some care. The same ultraviolet dynamics that enhances the aligned axion potential can also enhance contributions carrying additional $CP$-violating phases. A systematic instanton analysis has shown that such misaligned contributions are generically enhanced together with the axion mass, so whether the quality problem is improved depends sensitively on the ultraviolet flavour, gauge, and symmetry structure~\cite{Csaki:2023ziz}. Conversely, in some composite accidental-axion constructions, additional chiral symmetries can suppress the small-instanton contribution so efficiently that the axion mass remains close to its ordinary QCD value~\cite{Aoki:2024usv}. Thus, neither a large nor a negligible ultraviolet instanton contribution is a universal consequence of embedding QCD into a larger gauge theory.

Heavy QCD axion models therefore alleviate the quality problem only if the additional contribution to the axion potential is sufficiently well aligned with the $CP$-conserving QCD minimum; the origin of this alignment must itself be explained by the ultraviolet theory. Since the additional contribution modifies the axion mass, the standard relation $m_a^2f_a^2=\chi$ derived in Sec.~\ref{sec:PQ-dynamical-relaxation} no longer applies, and the resulting cosmology and phenomenology can differ substantially from those of the conventional QCD axion.

\part{Reassessing the strong \texorpdfstring{$CP$}{CP} problem}

\label{part:reassessment}

\section{Is there really a strong \texorpdfstring{$CP$}{CP} problem?}
\label{sec:reassessment}

\subsection{What is being questioned?}
\label{sec:reassessment-what-questioned}

In the preceding parts of this review, we have largely followed the conventional formulation of the strong $CP$ problem. Within this framework, the invariant parameter
$\bar\theta=\theta+\arg\det\M$ is taken to distinguish physically inequivalent QCD theories, and a generic nonzero value of $\bar\theta$ induces $CP$-odd hadronic interactions and observables. Experimental limits on EDMs then imply $|\bar\theta|\lesssim10^{-10}$, giving rise to the strong $CP$ problem. The connection between the parameter $\bar\theta$ in the microscopic Lagrangian and observable $CP$ violation, however, involves several conceptual steps. Recent work has questioned whether all of these steps, as taken in the conventional approach, are really founded in the underlying gauge theory~\cite{Ai:2020ptm,Ai:2024vfa,Ai:2024cnp}.

Schematically, the conventional reasoning follows the logical chain: topological structure of the gauge-field configuration space $\Rightarrow$ $\theta$-vacuum $\Rightarrow$ $\theta$-weighted sum over topological sectors $\Rightarrow$ $\bar\theta$-dependent correlation functions $\Rightarrow$ $CP$-odd observables. The last implication is not controversial once the preceding $\bar\theta$ dependence has been established. For example, as reviewed in Secs.~\ref{sec:chiral-perturbation-theory} and~\ref{sec:theta-CP-effects}, $\bar\theta$-dependent QCD correlation functions can be matched onto a $\bar\theta$-dependent chiral effective theory, which systematically generates $CP$-odd pion-nucleon interactions, nucleon electric dipole moments, and corresponding nuclear and atomic observables. The recent discussion instead concerns the earlier steps in this logical chain: how the physical Hilbert space of the gauge theory is defined, how different topological sectors enter the path integral, and whether the resulting QCD correlation functions necessarily depend on $\bar\theta$.

The first issue was encountered in Sec.~\ref{sec:pre-vacuum} when constructing the conventional $\theta$-vacuum. There, the classification of the pre-vacua $|n_{\rm CS}\rangle$ relies on the assumption that admissible gauge transformations approach a direction-independent constant at spatial infinity; see Eq.~\eqref{eq:constant-assumption}. This restriction separates the space of gauge transformations into distinct homotopy classes: transformations connected to the identity are treated as gauge redundancies, whereas those in disconnected components can act nontrivially on the pre-vacua and lead to the family of $\theta$-vacua. Reference~\cite{Ai:2024vfa} asks what changes if this restriction, which is not built into the canonical Hamiltonian formalism, is not imposed. Once more general gauge transformations are admitted, transformations that would belong to distinct homotopy classes under the conventional restriction can be continuously connected. Together with the requirements of complete gauge fixing and Hermiticity of the Hamiltonian, this leads to a further restriction of the Hilbert space to physical states for which the resulting observables preserve $CP$. We revisit this first caveat in the next subsection.

A logically independent issue appears in the Euclidean path integral. In the conventional treatment reviewed in Sec.~\ref{sec:diga}, the contributions from all integer topological sectors are first summed at finite spacetime volume, and the infinite-volume limit is taken afterwards. This prescription produces the familiar $\bar\theta$ dependence of the partition function and of fermionic correlation functions. References~\cite{Ai:2020ptm,Ai:2024cnp} instead emphasise that, for the formulation on $\mathbb R^4$ considered there, the semiclassical classification into integer topological sectors is tied to the finite-action condition imposed in the infinite-volume limit (cf.~Eqs.~\eqref{eq:finite-action} and~\eqref{eq:pure-gauge}). They therefore argue that the infinite-volume limit should first be taken within each fixed topological sector before the unrestricted sum over sectors is completed. The two limiting procedures do not in general commute. With the latter prescription, the chiral phases induced by the quark masses and by nonperturbative effects become aligned in the fermionic correlation functions, and the dependence on $\bar{\theta}$ drops out of local $CP$-odd observables. This second caveat is examined in Sec.~\ref{sec:reassessment-order-of-limits}.

The issues raised above concern the definition of the physical Hilbert space and the construction of QCD correlation functions. As emphasised in Sec.~\ref{sec:chiral-perturbation-theory}, the low-energy consequences depend on how the chiral theory is matched to these underlying correlation functions. We therefore first revisit the two caveats in detail, then discuss their implications for strong $CP$ violation, and finally review the recent criticisms and related perspectives.

\subsection{Caveat 1 revisited: Gauge redundancy and the physical Hilbert space}
\label{sec:reassessment-physical-Hilbert-space}

We now return to the gauge-redundancy caveat raised in Sec.~\ref{sec:pre-vacuum}. It concerns the canonical quantisation of Yang--Mills theory and, in particular, the consequences of relaxing the condition~\eqref{eq:constant-assumption} imposed on admissible gauge transformations. Reference~\cite{Ai:2024vfa} develops the analysis first on a spatial torus, where no condition at spatial infinity is required, and subsequently extends it to $\mathbb R^3$. For pedagogical clarity, we present the argument directly in Minkowski spacetime.

We work in temporal gauge, $A_0=0$, and denote the CS functional introduced in Eq.~\eqref{eq:winding-number-large-gauge} by $W[\bm A]$. For a gauge transformation $\mathcal U_n$ belonging to homotopy class $n$ under the conventional restriction, its normalisation is such that $W[\bm A^{\mathcal U_n}]=W[\bm A]+n$. It is useful to distinguish three angular parameters in the canonical description. We denote the angle entering the Lagrangian by $\theta_L$. A second angle arises from the freedom in representing the canonical momentum operator,\footnote{Reference~\cite{Ai:2024vfa} uses a gauge field rescaled relative to the convention adopted here: $D_\mu=\partial_\mu-\i A_\mu^{\rm Ref.\,\cite{Ai:2024vfa}}$ there, whereas $D_\mu=\partial_\mu-\i g_sA_\mu$ here, so that $A_\mu^{\rm Ref.\,\cite{Ai:2024vfa}}=g_sA_\mu$.}
\begin{align}
    \bm\Pi^a
    =
    -\i\frac{\delta}{\delta\bm A^a}
    +
    \theta_\Pi
    \frac{\delta W[\bm A]}{\delta\bm A^a},
    \qquad
    \frac{\delta W[\bm A]}{\delta\bm A^a}
    =
    \frac{g_s^2}{8\pi^2}\bm B^a,
    \label{eq:canonical-momentum-thetaPi}
\end{align}
where $\theta_\Pi$ is an arbitrary real constant. The corresponding Hamiltonian is
\begin{align}
    H
    =
    \frac{1}{2}
    \int\d^3x
    \left[
        \left(
            -\i\frac{\delta}{\delta\bm A^a}
            -
            (\theta_L-\theta_\Pi)
            \frac{\delta W[\bm A]}{\delta\bm A^a}
        \right)^2
        +
        (\bm B^a)^2
    \right].
    \label{eq:canonical-Hamiltonian-theta}
\end{align}
A third angle characterises the transformation of the wave functional. Since $\mathcal U_n$ commutes with the Hamiltonian, the states may be chosen to transform as $\Psi[\bm A^{\mathcal U_n}]=\exp(\i n\theta_{\rm wf})\Psi[\bm A]$.

The distribution of the dependence on CP-odd
phases between $\theta_\Pi$ and $\theta_{\rm wf}$ depends on the basis chosen for the wave functionals. Under $\Psi'[\bm A]=\exp[-\i\varphi W[\bm A]]\Psi[\bm A]$, one has $\theta_\Pi\to\theta_\Pi+\varphi$ and $\theta_{\rm wf}\to\theta_{\rm wf}-\varphi$. The physical quantity is the invariant combination $\theta_{\rm inv}\equiv\theta_L-\theta_\Pi-\theta_{\rm wf}$. A particularly convenient choice is the diagonal basis, $\Psi_{\rm diag}[\bm A]=\exp[-\i(\theta_L-\theta_\Pi)W[\bm A]]\Psi[\bm A]$, for which
\begin{align}
H_{\rm diag}=\frac{1}{2}\int\d^3x\left[-\frac{\delta}{\delta A_i^a(\bm x)}\frac{\delta}{\delta A_i^a(\bm x)}+B_i^a(\bm x)B_i^a(\bm x)\right],
\label{eq:canonical-Hamiltonian-diagonal}
\end{align}
where repeated spatial and colour indices are summed. All  the dependence on CP-odd phases is then carried by the transformation of the wave functional, 
\begin{align}
\label{eq:tranf-invariant-angle}
  \Psi_{\rm diag}[\bm A^{\mathcal U_n}]=\e^{-\i n\theta_{\rm inv}}\,\Psi_{\rm diag}[\bm A]. 
\end{align}
This is essentially the basis implicitly adopted in conventional discussions in which only a single $\theta$ angle appears explicitly. In the conventional formulation, $\theta_{\rm inv}$ is regarded as an independent input parameter and may take any value in $[0,2\pi)$. Reference~\cite{Ai:2024vfa} instead argues that, once the restriction in Eq.~\eqref{eq:constant-assumption} is relaxed, gauge invariance of the physical Hilbert space fixes $\theta_{\rm inv}=0$.

To formulate the physical-state condition, Ref.~\cite{Ai:2024vfa} removes the gauge redundancy by defining the inner product on a completely gauge-fixed hypersurface $\mathcal A$ in field-configuration space, chosen such that each gauge orbit is represented once,
\begin{align}
    \langle\Psi|\Phi\rangle_{\rm phys}
    =
    \int_{\mathcal A}
    \mathcal D\bm A\,
    f_{\mathcal A}[\bm A]\,
    \Psi^*[\bm A]\Phi[\bm A],
    \label{eq:physical-gauge-fixed-inner-product}
\end{align}
where the factor $f_{\mathcal A}[\bm A]$ is included to ensure the appropriate covariance of the gauge-fixed measure.\footnote{The measure $\mathcal D\bm A\,f_{\mathcal A}[\bm A]$ is invariant with respect to the translation in $\mathcal A$.} Physical probabilities must be independent of the choice of $\mathcal A$, while the canonical operators must remain Hermitian with respect to this inner product. Gauge independence implies that a wave functional may change along a gauge orbit only by a phase, and Hermiticity of the canonical momentum for arbitrary pairs of states further requires this phase to be state independent~\cite{Ai:2024vfa}. The wave functionals therefore take the form
\begin{align}
    \Psi^{(a)}[\bm A]
    =
    \Psi_{\rm g.i.}^{(a)}[\bm A]\,
    \e^{\i\varphi[\bm A]},
    \label{product:function:phase}
\end{align}
where $(a)$ lable the states, $\Psi_{\rm g.i.}^{(a)}$ is gauge invariant and, the same real phase functional $\varphi[\bm A]$ applies to all states.

To make the distinction between gauge and physical variations explicit in the functional Schr\"odinger equation, let $\{{\bm G}^a(\bm x;\sigma)\}$ denote an orthonormal basis of directions in field-configuration space and define
\begin{align}
    \frac{\delta}{\delta A(\sigma)}
    \Psi[\bm A]
    \equiv
    \int\d^3y\,
    G_i^a(\bm y;\sigma)
    \frac{\delta}{\delta A_i^a(\bm y)}
    \Psi[\bm A].
    \label{eq:configuration-direction-derivative}
\end{align}
At a given configuration, these directions can be separated into those tangent to the gauge orbit, labelled by $\sigma_{\rm gauge}$, and those orthogonal to it, labelled by $\sigma_\parallel$. Correspondingly,
\begin{align}
    \int\limits_\sigma\hskip-.5cm\scalebox{1.}{$\sum$}
    \frac{\delta^2}{\delta A^2(\sigma)}
    =
    \int\limits_{\sigma_{\rm gauge}}\hskip-.7cm\scalebox{1.}{$\sum$}
    \frac{\delta^2}{\delta A^2(\sigma_{\rm gauge})}
    +
    \int\limits_{\sigma_\parallel}\hskip-.5cm\scalebox{1.}{$\sum$}
    \frac{\delta^2}{\delta A^2(\sigma_\parallel)}.
    \label{eq:configuration-direction-decomposition}
\end{align}
The directions $\sigma_\parallel$ need not themselves be tangent to the gauge-fixed hypersurface $\mathcal A$. Their projections onto $\mathcal A$, however, differ from them only by gauge directions and therefore act equivalently on the gauge-invariant part of the wave functional. This decomposition is local in field-configuration space and generally depends on the configuration $\bm A$.

The advantage of the diagonal basis is now apparent. With the orthogonal decomposition of field-configuration space introduced above, the functional Laplacian separates into gauge and physical directions. Moreover, the magnetic energy is gauge invariant and therefore does not vary along a gauge orbit. The Hamiltonian in Eq.~\eqref{eq:canonical-Hamiltonian-diagonal} may thus be written schematically as
\begin{align}
    H_{\rm diag}
    =
    -\frac{1}{2}
    \int\limits_{\sigma_{\rm gauge}}\hskip-.7cm\scalebox{1.}{$\sum$}
    \frac{\delta^2}{\delta A^2(\sigma_{\rm gauge})}
    -\frac{1}{2}
    \int\limits_{\sigma_\parallel}\hskip-.5cm\scalebox{1.}{$\sum$}
    \frac{\delta^2}{\delta A^2(\sigma_\parallel)}
    +\frac{1}{2}
    \int\d^3x\,
    B_i^a(\bm x)B_i^a(\bm x),
    \label{eq:diagonal-Hamiltonian-decomposition}
\end{align}
where the last term depends only on the physical gauge-equivalence class. The functional Schr\"odinger equation can therefore be separated into the two sets of configuration-space directions.

Accordingly, we may use the product ansatz
\begin{align}
    \Psi_{\rm diag}^{(a)}[\bm A]
    =
    \Psi_{{\rm g.i.}, {\rm diag}}^{(a)}[\bm A]\,
    p_{\rm diag}[\bm A],
    \label{product:function}
\end{align}
where $\Psi_{{\rm g.i.},{\rm diag}}^{(a)}$ depends only on the physical directions, while the state-independent factor $p_{\rm diag}$ contains the possible dependence along the gauge orbit. The two factors may therefore be chosen to satisfy
\begin{align}
    \frac{\delta}{\delta A(\sigma_{\rm gauge})}
    \Psi_{{\rm g.i.},{\rm diag}}^{(a)}[\bm A]
    =
    0,
    \qquad
    \frac{\delta}{\delta A(\sigma_\parallel)}
    p_{\rm diag}[\bm A]
    =
    0.
    \label{product:ansatz:indpendences}
\end{align}
Substituting this ansatz into the Schr\"odinger equation separates the dependence along the gauge orbit from the physical dynamics. The gauge-dependent factor satisfies
\begin{align}
    \int\limits_{\sigma_{\rm gauge}}\hskip-.7cm\scalebox{1.}{$\sum$}
    \frac{\delta^2}{\delta A^2(\sigma_{\rm gauge})}
    p_{\rm diag}[\bm A]
    =
    \mu\,p_{\rm diag}[\bm A],
    \label{eq:delta-p-prime}
\end{align}
where $\mu$ is real by Hermiticity. The Schr\"odinger equation  corresponding to a state with energy $E^{(a)}$ then gives
\begin{align}
\left(-\frac{1}{2}\int\limits_{\sigma_{\parallel}}\hskip-.5cm\scalebox{1.}{$\sum$}  \frac{\delta^2}{\delta A^2(\sigma_{\parallel})}
+\frac{1}{2}\,\int {\rm d}^3x\, \mathbf B^2\right){\Psi}_\text{g.i.,diag}^{(a)}
=\left(E^{(a)}+\frac{\mu}{2}\right){\Psi}_\text{g.i.,diag}^{(a)}.
\end{align}
where we have used Eqs.~(\ref{product:ansatz:indpendences}) and~(\ref{eq:delta-p-prime}).

The physical-state condition in Eq.~\eqref{product:function:phase} requires the dependence of $p_{\rm diag}$ along a gauge orbit to be a pure phase. Let $\mathcal{G}_0$ denote the component connected to the identity of the chosen group of admissible gauge transformations. Writing the phase acquired under $U\in\mathcal{G}_0$ as a character $\chi(U)$, group composition requires $\chi(U_2U_1)=\chi(U_2)\chi(U_1)$. Its infinitesimal form defines a one-dimensional representation of the Lie algebra of gauge transformations. For a simple non-Abelian structure group such as $SU(N)$, this Lie algebra is generated by commutators, so any such representation is trivial. Continuity then implies $\chi(U)=1$ throughout $\mathcal{G}_0$. Hence $p_{\rm diag}$ is constant along the corresponding gauge directions, implying $\mu=0$, and the physical wave functionals are invariant under $\mathcal{G}_0$.

With the conventional restriction on gauge transformations at spatial infinity, infinitesimal transformations $\omega(\bm x)=\exp(\i\alpha\Lambda(\bm x))$ with sufficiently rapidly vanishing $\Lambda(\bm x)$ belong to $\mathcal{G}_0$, and the resulting invariance reproduces the familiar Gau{\ss}-law constraint. The key proposal of Ref.~\cite{Ai:2024vfa} is to impose the physical-state condition after dropping the additional restriction in Eq.~\eqref{eq:constant-assumption}. In the enlarged space of admissible gauge transformations, transformations belonging to different homotopy classes under the conventional restriction can now be continuously connected through intermediate transformations with more general behaviour at spatial infinity. A transformation $\mathcal U_n$ conventionally classified as large therefore lies in the identity component of the enlarged gauge-transformation group and must also act trivially,
\begin{align}
    \Psi_{\rm diag}[\bm A^{\mathcal U_n}]
    =
    \Psi_{\rm diag}[\bm A].
\end{align}

Comparing this with $\Psi_{\rm diag}[\bm A^{\mathcal U_n}]=\exp(-\i n\theta_{\rm inv})\Psi_{\rm diag}[\bm A]$ gives $\theta_{\rm inv}=0$ modulo $2\pi$. Within this construction, the phase conventionally associated with large gauge transformations is therefore not an independent superselection parameter. Instead, the phase of the states is fixed consistently with that appearing in the Hamiltonian, leaving no basis-invariant $\theta$ parameter. The argument relies on the simple non-Abelian gauge structure and does not carry over directly to an Abelian theory. Once quarks are included, the chiral phases of the fermion mass matrix enter and the conventional invariant parameter becomes $\bar\theta$, to which we now turn from the complementary path-integral perspective.

\subsection{Caveat 2 revisited: Topological sectors and the order of limits}
\label{sec:reassessment-order-of-limits}

We now return to the second caveat raised in Sec.~\ref{sec:diga}. In the conventional treatment, the path integral at finite spacetime volume $\Omega$ is first decomposed into sectors of integer topological charge and summed over all sectors,
\begin{align}
    Z(\theta,\Omega)
    =
    \sum_{\Delta n=-\infty}^{\infty}
    \e^{\i\theta\Delta n}
    Z_{\Delta n}(\Omega),
    \label{eq:conventional-sector-sum-reassessment}
\end{align}
after which the infinite-volume limit $\Omega\to\infty$ is taken. References~\cite{Ai:2020ptm,Ai:2024cnp} question this order of limits for the path integral on $\mathbb R^4$. The premise is the one already encountered in Sec.~\ref{sec:diga}: for a finite subregion of $\mathbb R^4$ without additional global boundary conditions, finite action does not require the gauge field to approach a pure gauge at the boundary. The familiar classification by an integer winding number emerges only as the boundary is taken to infinity, where the finite-action condition enforces the asymptotic pure-gauge form. Within this formulation, Refs.~\cite{Ai:2020ptm,Ai:2024cnp} therefore argue that the infinite-volume limit should first be taken at fixed topological charge, with the unrestricted sum over integer sectors performed only afterwards.\footnote{This statement concerns the particular construction of the path integral on $\mathbb R^4$ considered here. It should not be confused with the mathematical statement that integer topological charge cannot occur at finite volume. For example, gauge fields on a compact manifold such as $T^4$ may belong to nontrivial bundles and carry integer topological charge already at finite volume~\cite{Leutwyler:1992yt,Ai:2024vfa}.} This ordering has been referred to as the Ai--Cruz--Garbrecht--Tamarit (ACGT) order of limits in some recent studies~\cite{Khoze:2025auv,Aghaie:2026pkf}, and we adopt this terminology below.

More explicitly, introducing a cutoff $N$ on the magnitude of the global winding number, the alternative prescription for a normalised correlation function takes the form
\begin{align}
    \langle\mathcal O\rangle_{\rm ACGT}
    =
    \lim_{N\to\infty}
    \lim_{\Omega\to\infty}
    \frac{
        \displaystyle
        \sum_{\Delta n=-N}^{N}
        \int_{\Delta n}\mathcal D\phi\,
        \mathcal O\,
        \e^{-S_{\rm E}[\phi]+\i\theta\Delta n}
    }{
        \displaystyle
        \sum_{\Delta n=-N}^{N}
        \int_{\Delta n}\mathcal D\phi\,
        \e^{-S_{\rm E}[\phi]+\i\theta\Delta n}
    }.
    \label{eq:alternative-order-general}
\end{align}
Here $\mathcal D\phi$ collectively denotes the functional measure for all fields. The cutoff $N$ restricts only the \emph{net} topological charge; it does not restrict the number of local topological fluctuations within a fixed sector. The conventional calculation instead performs the infinite sector sum at finite $\Omega$ before taking $\Omega\to\infty$. Since these operations involve two distinct infinite limits, their interchange is not guaranteed a priori. Reference~\cite{Ai:2024cnp} further argues, using a construction of the functional integral from steepest-descent contours, that the contours associated with a fixed integer winding number are naturally defined in the infinite-volume configuration space and that the ordering in Eq.~\eqref{eq:alternative-order-general} is compatible with this contour decomposition. Whether this prescription is the appropriate definition of the physical QCD vacuum is precisely one of the central points of the recent debate.

\paragraph{Illustration in the dilute instanton gas.}

The effect of the two orders of limits can be seen particularly transparently in the DIGA already introduced in Sec.~\ref{sec:diga}. We again consider one quark with $M=m\e^{\i\alpha\gamma^5}$ with $m>0$ and denote the number of instantons and anti-instantons by $n$ and $\bar n$, with $\Delta n =n-\bar n$. All ingredients required below, including the approximate Green's function $S_{n,\bar n}$, the instanton density $\kappa$, and the overlap function $\bar h(x,x')$, were defined in Sec.~\ref{sec:diga}. It is therefore sufficient here to retain only the dependence relevant for the topological-sector sum.

Let
\begin{align}
    G_{\Delta n}(x,x')
    \equiv
    \sum_{\substack{n,\bar n\geq0\\ n-\bar n=\Delta n}}
    \frac{1}{n!\bar n!}
    \int
    \mathcal D\hat A_{n,\bar n}\,
    \mathcal D\hat{\bar\psi}\,
    \mathcal D\hat\psi\,
    \hat\psi(x)\hat{\bar\psi}(x')\,
    \e^{-S_{\rm E}}
    \e^{\i\theta\Delta n}
    \label{eq:fixed-sector-unnormalised-correlator}
\end{align}
denote the unnormalised two-point function in a sector of fixed $\Delta n$. Performing the collective-coordinate integrations as in Sec.~\ref{sec:diga} gives
\begin{align}
    G_{\Delta n}(x,x')
    =
    \Bigg\{
        I_{\Delta n}(2\kappa\Omega)
        S_0(x,x')
        +
        \frac{\kappa}{m}
        \bar h(x,x')
        \left[
            \e^{\i\alpha}
            I_{\Delta n+1}(2\kappa\Omega)P_{\rm L}
            +
            \e^{-\i\alpha}
            I_{\Delta n-1}(2\kappa\Omega)P_{\rm R}
        \right]
    \Bigg\}
    \e^{\i\Delta n(\alpha+\theta)}.
    \label{eq:fixed-sector-fermion-correlator}
\end{align}
The corresponding fixed-sector partition function is
\begin{align}
    Z_{\Delta n}
    =
    \sum_{\substack{n,\bar n\geq0\\n-\bar n=\Delta n}}
    \frac{
        (\kappa\Omega)^{n+\bar n}
    }{
        n!\bar n!
    }
    \e^{\i\Delta n(\alpha+\theta)}
    =
    I_{\Delta n}(2\kappa\Omega)
    \e^{\i\Delta n(\alpha+\theta)}.
    \label{eq:Z-Deltan}
\end{align}
The phase $\e^{\i\theta\Delta n}$ is therefore an overall phase within a fixed topological sector. It cancels from the normalised fixed-sector correlator,
\begin{align}
    S_{\Delta n}(x,x')
    \equiv
    \frac{
        G_{\Delta n}(x,x')
    }{
        Z_{\Delta n}
    }
    =
    S_0(x,x')
    +
    \frac{\kappa}{m}
    \bar h(x,x')
    \left[
        \e^{\i\alpha}
        \frac{
            I_{\Delta n+1}(2\kappa\Omega)
        }{
            I_{\Delta n}(2\kappa\Omega)
        }
        P_{\rm L}
        +
        \e^{-\i\alpha}
        \frac{
            I_{\Delta n-1}(2\kappa\Omega)
        }{
            I_{\Delta n}(2\kappa\Omega)
        }
        P_{\rm R}
    \right].
    \label{eq:normalised-fixed-sector-correlator}
\end{align}
Thus, already before taking the thermodynamic limit, the topological angle does not appear in a normalised correlator restricted to a single sector. The remaining dependence on $\Delta n$ resides in the real ratios of Bessel functions.

For fixed integer $\Delta n$ and $\Omega\to\infty$, $I_{\Delta n\pm1}(2\kappa\Omega)/I_{\Delta n}(2\kappa\Omega)\rightarrow1$, and the limiting correlator becomes independent of the global topological sector,
\begin{align}
    \lim_{\Omega\to\infty}
    S_{\Delta n}(x,x')
    =
    S_0(x,x')
    +
    \frac{\kappa}{m}
    \bar h(x,x')
    \left(
        \e^{\i\alpha}P_{\rm L}
        +
        \e^{-\i\alpha}P_{\rm R}
    \right)
    =
    S_0(x,x')
    +
    \frac{\kappa}{m}
    \bar h(x,x')
    \e^{-\i\alpha\gamma^5}.
    \label{eq:fixed-sector-infinite-volume-correlator}
\end{align}
The nonperturbative contribution therefore carries the same chiral phase as the mass dependence already present in $S_0(x,x')$, cf.~Eq.~\eqref{eq:S0}.

Equation~\eqref{eq:fixed-sector-infinite-volume-correlator} also makes the subsequent sector sum particularly simple. For every finite value of $N$, all sectors with $|\Delta n|\leq N$ approach the same normalised correlator as $\Omega\to\infty$. Taking the limits in the order of Eq.~\eqref{eq:alternative-order-general} therefore gives
\begin{align}
    \langle
        \hat\psi(x)\hat{\bar\psi}(x')
    \rangle_{\rm ACGT}
    =
    \lim_{N\to\infty}
    \lim_{\Omega\to\infty}
    \frac{
        \displaystyle
        \sum_{\Delta n=-N}^{N}
        G_{\Delta n}(x,x')
    }{
        \displaystyle
        \sum_{\Delta n=-N}^{N}
        Z_{\Delta n}
    }
    =
    S_0(x,x')
    +
    \frac{\kappa}{m}
    \bar h(x,x')
    \e^{-\i\alpha\gamma^5}.
    \label{correlation:function}
\end{align}
As the chiral phases in the first and the second terms are aligned, they can be removed by a chiral rotation, leaving no $CP$ violation in low-energy observables. By comparison, summing over $n$ and $\bar n$ independently at finite $\Omega$, as in the conventional calculation of Sec.~\ref{sec:diga}, gives Eq.~\eqref{eq:correlation-incorrect-order}. 

It is worth emphasising what is held fixed in this construction. Fixing $\Delta n$ does \emph{not} mean fixing the number of instantons and anti-instantons individually. 
Actually, arbitrarily many instanton--anti-instanton pairs are included even when the net winding number is fixed. Indeed, from Eq.~\eqref{eq:Z-Deltan} we have
\begin{align}
    \langle n+\bar n\rangle_{\Delta n}
    =
    \kappa
    \frac{\partial}{\partial\kappa}
    \log I_{\Delta n}(2\kappa\Omega)
    =
    2\kappa\Omega
    +
    \mathcal O(1),
    \qquad
    (\Omega\to\infty),
    \label{eq:fixed-sector-instanton-density}
\end{align}
so the total number of local topological objects remains extensive. The prescription in Eq.~\eqref{eq:alternative-order-general} therefore constrains the global difference $n-\bar n$ while retaining a finite density of local topological fluctuations. This distinction will be important when discussing the topological susceptibility and the $\eta'$ in Sec.~\ref{sec:reassessment-consequences}.

The role of normalisation is equally important. At fixed $\Delta n$,
\begin{align}
    I_{\Delta n}(2\kappa\Omega)
    =
    \frac{
        \e^{2\kappa\Omega}
    }{
        \sqrt{4\pi\kappa\Omega}
    }
    \left[
        1+\mathcal O\left(\frac{1}{\kappa\Omega}\right)
    \right],
    \qquad
    \Omega\to\infty,
    \label{eq:Bessel-large-volume}
\end{align}
whose leading large-volume factor is independent of $\Delta n$. Thus both the fixed-sector partition functions and unnormalised correlators contain the same divergent vacuum-normalisation factor, whose logarithm is extensive in $\Omega$. In Eq.~\eqref{eq:alternative-order-general}, the numerator and denominator are kept together at finite regulator $N$, so that this common factor cancels before the unrestricted sector sum is taken. The resulting normalised correlation functions may therefore have a well-defined limit even though the unnormalised quantity $\sum_{\Delta n}\lim_{\Omega\to\infty}Z_{\Delta n}$ does not by itself define an ordinary convergent partition function. Whether one should assign a separate thermodynamic interpretation to the latter is an additional question and is not required for the definition of the normalised observables.

\paragraph{Beyond the dilute instanton gas.}

The DIGA provides a useful illustration because the two limits can be evaluated analytically, but the claim of Refs.~\cite{Ai:2020ptm,Ai:2024cnp} is not intended to rely on the dilute gas as a quantitative description of zero-temperature QCD. A more general argument can be formulated using the composition of fixed-topology path integrals together with the index theorem.

Suppose that a spacetime region is decomposed into two sufficiently separated subregions, $\Omega=\Omega_1\cup\Omega_2$, and denote their topological charges schematically by $\Delta n_1$ and $\Delta n_2$, with $\Delta n=\Delta n_1+\Delta n_2$. Cluster decomposition motivates the convolution relation
\begin{align}
    Z_{\Delta n}(\Omega)
    =
    \sum_{\Delta n_1}
    Z_{\Delta n_1}(\Omega_1)
    Z_{\Delta n-\Delta n_1}(\Omega_2).
    \label{eq:fixed-sector-convolution}
\end{align}
For a quark mass $m\e^{\i\alpha\gamma^5}$, the index theorem fixes the phase of the fermionic determinant in a background of integer topological charge, so that one may write
\begin{align}
    Z_{\Delta n}(\Omega)
    =
    \e^{\i(\theta+\alpha)\Delta n}
    \widetilde Z_{\Delta n}(\Omega),
    \qquad
    \widetilde Z_{\Delta n}(\Omega)\in\mathbb R.
    \label{eq:fixed-sector-general-phase}
\end{align}
Reference~\cite{Ai:2020ptm} shows that, with parity considerations and analyticity in the spacetime volume, the infinite set of composition relations in Eq.~\eqref{eq:fixed-sector-convolution} admits the form
\begin{align}
    Z_{\Delta n}(\Omega)
    =
    \e^{\i(\theta+\alpha)\Delta n}
    I_{\Delta n}(2\beta\Omega),
    \label{eq:fixed-sector-general-Bessel}
\end{align}
where $\beta$ is a real function of the microscopic parameters. The parameter $\beta$ reduces to the instanton density $\kappa$ in the DIGA, but Eq.~\eqref{eq:fixed-sector-general-Bessel} is obtained without assuming that the relevant gauge configurations form a dilute gas. 

The phase of the corresponding fermionic two-point functions can then be extracted without an instanton expansion. Introducing the complex mass $\mathfrak m\equiv
m\e^{\i\alpha}$ one may regard $\mathfrak m$ and $\mathfrak m^*$ as sources for the chiral bilinears,
\begin{align}
    \frac{\partial Z_{\Delta n}}
         {\partial\mathfrak m}
    &=
    -\int\d^4x\,
    \left\langle
        \hat{\bar\psi}P_{\rm R}\hat\psi
    \right\rangle_{\Delta n},
    \qquad
    \frac{\partial Z_{\Delta n}}
         {\partial\mathfrak m^*}
    =
    -\int\d^4x\,
    \left\langle
        \hat{\bar\psi}P_{\rm L}\hat\psi
    \right\rangle_{\Delta n}.
    \label{eq:mass-source-chiral-condensates}
\end{align}
Since the real function $\beta$ can depend on the mass through the invariant combination $\mathfrak m\mathfrak m^*=m^2$, applying the order of limits in Eq.~\eqref{eq:alternative-order-general} gives~\cite{Ai:2020ptm}
\begin{align}
    \frac{1}{\Omega}
    \int\d^4x\,
    \left\langle
        \hat{\bar\psi}P_{\rm R}\hat\psi
    \right\rangle_{\rm alt}
    &=
    -2\mathfrak m^*
    \frac{\partial\beta}
         {\partial(\mathfrak m\mathfrak m^*)},
    \notag\\
    \frac{1}{\Omega}
    \int\d^4x\,
    \left\langle
        \hat{\bar\psi}P_{\rm L}\hat\psi
    \right\rangle_{\rm alt}
    &=
    -2\mathfrak m
    \frac{\partial\beta}
         {\partial(\mathfrak m\mathfrak m^*)}.
    \label{eq:general-chiral-condensate-alignment}
\end{align}
The two correlation functions therefore carry phases $\e^{-\i\alpha}$ and $\e^{\i\alpha}$, respectively, exactly as required for alignment with the chiral phase of the quark mass. The same qualitative conclusion obtained in Eq.~\eqref{correlation:function} is thus recovered without identifying $\beta$ with a semiclassical instanton density.

The central claim of this path-integral viewpoint can therefore be stated independently of the quantitative validity of the DIGA. Within the prescription of Eq.~\eqref{eq:alternative-order-general}, local correlation functions are first defined in the thermodynamic limit at fixed global topological charge. At fixed $\Delta n$, the overall $\bar\theta$-dependent factor $\exp(\i\bar\theta\Delta n)$ is common to the numerator and denominator and therefore cancels from the normalised fermionic correlators. The nonperturbative chiral phase then remains aligned with the quark mass phase, and the subsequent sum over $\Delta n$ does not generate a relative chiral phase between the nonperturbative contribution and the quark mass term. By contrast, performing the unrestricted sector sum before taking the thermodynamic limit gives the conventional $\bar\theta$-dependent result. Which order of limits describes the physical QCD vacuum is not determined by the algebra above alone and is the subject of the debate discussed in Sec.~\ref{sec:recent-strong-CP-debate}. In the next subsection, we examine the physical consequences that follow if the ACGT prescription is adopted.

\subsection{Consequences for strong \texorpdfstring{$CP$}{CP} violation}
\label{sec:reassessment-consequences}

Both caveats discussed above imply that the conventional strong $CP$ phase does not survive as an independent $CP$-violating parameter in physical observables~\cite{Ai:2020ptm,Ai:2024vfa,Ai:2024cnp}. We now summarise the consequences of this conclusion and clarify which nonperturbative features of QCD remain unchanged.

\paragraph{The 't Hooft vertex and $CP$ violation.}

For the 't Hooft vertex, the conventional order of limits gives Eq.~\eqref{eq:Delta-LM1}, whereas the ACGT prescription gives Eq.~\eqref{eq:Delta-LM2}. A chiral field redefinition that renders the quark masses real simultaneously removes the phase in Eq.~\eqref{eq:Delta-LM2}, so that both the mass terms and the anomalous interaction can be made real. The ACGT prescription does not, however, remove the 't Hooft interaction or restore the anomalous $U(1)_A$ symmetry. Instanton-induced axial-symmetry violation and the associated selection rule remain unchanged; only the chiral phase of the effective interaction relative to the quark masses is altered.

As discussed in Sec.~\ref{sec:chiral-perturbation-theory}, chiral symmetry alone does not fix the phase of the anomalous singlet interaction in ChPT; it must be determined by matching to the underlying QCD correlation functions~\cite{Ai:2020ptm,Ai:2024cnp,Ai:2025quf}. The conventional and ACGT prescriptions correspond to the two different matchings discussed there. In the latter case, no residual $CP$-violating phase remains in the chiral effective theory.

Consequently, within the ACGT prescription, hadronic $CP$ violation sourced solely by $\bar\theta$ is absent, including the $\bar\theta$-induced $CP$-odd meson and pion-nucleon interactions and the corresponding contributions to nucleon, nuclear, atomic, and molecular electric dipole moments. The experimental EDM bounds reviewed in Sec.~\ref{sec:theta-CP-effects} would therefore no longer imply $|\bar\theta|\lesssim10^{-10}$. This conclusion concerns only the strong $CP$ phase: the CKM phase and possible higher-dimensional $CP$-violating operators remain physical sources of $CP$ violation.

\paragraph{Topological susceptibility and the \texorpdfstring{$\eta'$}{eta-prime}.}

A potentially confusing point concerns the topological susceptibility. In a sector of fixed global topological charge $\Delta n$, the quantity $(\Delta n)^2/\Omega$ vanishes as $\Omega\to\infty$. Accordingly, taking the infinite-volume limit at fixed $\Delta n$ before summing over topological sectors gives a vanishing $\langle\Delta n^2\rangle/\Omega$. This does not, however, imply that the susceptibility associated with local vacuum fluctuations must vanish. It is therefore useful to distinguish the local vacuum susceptibility~\cite{Ai:2020ptm,Ai:2025quf}
\begin{align}
    \chi_{\rm loc}
    \equiv
    \lim_{p\to0}
    \int\d^4x\,
    \e^{\i p\cdot x}
    \langle0|
        T q(x)q(0)
    |0\rangle ,
    \label{eq:local-topological-susceptibility-reassessment}
\end{align}
from the global quantity
\begin{align}
    \chi_{\rm glob}
    \equiv
    \frac{\langle\Delta n^2\rangle}{\Omega}
    =
    \frac{1}{\Omega}
    \int_\Omega\d^4x
    \int_\Omega\d^4y\,
    \langle q(x)q(y)\rangle .
\end{align}
In $\chi_{\rm loc}$, the correlator is understood as the infinite-volume vacuum correlator, so that the thermodynamic limit is taken before the spacetime integral and the limit $p\to0$. By contrast, $\chi_{\rm glob}$ probes fluctuations of the total topological charge over the full spacetime volume and is therefore sensitive to the order of limits: it vanishes in the ACGT prescription, whereas the conventional order gives a nonzero result. A nonvanishing $\chi_{\rm loc}$ is nevertheless compatible with fixed global $\Delta n$, because fixing $\Delta n$ constrains only the spacetime integral of $q(x)$ and does not eliminate local topological fluctuations. In particular, the fixed-$\Delta n$ ensemble still contains an extensive number of instanton--anti-instanton pairs, as illustrated by Eq.~\eqref{eq:fixed-sector-instanton-density}. Equivalently, for a finite subvolume $\Omega_1\subset\Omega$, the enclosed topological charge $\Delta n|_{\Omega_1}$ can fluctuate even when the total $\Delta n$ is fixed, so that $\langle(\Delta n|_{\Omega_1})^2\rangle/\Omega_1$ can remain nonzero. Thus the ACGT order of limits can yield $\chi_{\rm glob}=0$ while retaining a nonvanishing $\chi_{\rm loc}$.

The same distinction applies to the $U(1)_A$ problem. The anomalous Ward identity in Eq.~\eqref{eq:singlet-anomaly} is a local operator relation and is unaffected by the order in which global topological sectors are summed. The anomalous determinant interaction therefore remains present and gives a nonzero contribution to the $\eta'$ mass also for the $CP$-conserving matching $\xi=-\bar\alpha$ in Eq.~\eqref{eq:options}. The ACGT prescription thus neither restores the continuous $U(1)_A$ symmetry nor predicts an additional light Nambu--Goldstone boson.

Reference~\cite{Ai:2025quf} further argues that a vanishing global susceptibility need not conflict with the Witten--Veneziano relation. In this interpretation, the susceptibility relevant for the anomalous singlet dynamics is the zero-momentum limit of the vacuum correlator of the local topological charge density, as in Eq.~\eqref{eq:local-topological-susceptibility-reassessment}, rather than the variance of the total topological charge evaluated with the ACGT order of limits. The former can remain nonzero even when the latter vanishes. In a different construction, Kaplan and Sen formulate the four-dimensional gauge theory of interest as the boundary theory of a five-dimensional regulator and find that the resulting effective boundary theory is restricted to the sector with vanishing global topological charge, $Q=0$, while retaining local topological fluctuations and a  massive $\eta'$~\cite{Kaplan:2024ezz}. This provides an independent example in which anomalous singlet dynamics survives without summing over sectors of nonzero global topology.\footnote{Very recently, Ref.~\cite{DiVecchia:2026mrz} discussed a tension between the absence of strong $CP$ violation and a nonvanishing $\eta'$ mass if $\bar\theta$ is not fine-tuned. This may not constitute a contradiction as their 1PI construction starts from the matching $\xi=\theta$ in Eq.~\eqref{eq:large-Nc-chiral-Lagrangian}. On the other hand, the ACGT order of limits gives correlation functions mathematically equivalent to those in a fixed topological sector (discussed further in the next section). Kaplan and Sen, and others, have shown that in a fixed topological sector in the large-volume limit, there is no $U(1)_A$ problem~\cite{Brower:2003yx,Aoki:2007ka,Kaplan:2024ezz}.} Whether these observations are sufficient to reproduce the full nonperturbative content conventionally associated with the QCD topological susceptibility, including the Witten--Veneziano relation, remains part of the recent debate discussed in Sec.~\ref{sec:recent-strong-CP-debate}.

\paragraph{What becomes of the strong \texorpdfstring{$CP$}{CP} problem?}

If either construction reviewed above correctly describes the physical QCD vacuum, the usual strong $CP$ problem would no longer arise: $\bar\theta$ would not control strong-interaction $CP$ violation, and there would be no physical observable requiring $|\bar\theta|\lesssim10^{-10}$. The anomaly, fermionic zero modes, 't Hooft interactions, chiral symmetry breaking, and local topological fluctuations would remain; what changes is whether these nonperturbative effects render $\bar\theta$ observable. These conclusions remain under active debate, particularly concerning the physical Hilbert space, the order of limits, and chiral matching. We now turn to these criticisms and related perspectives.

\section{Recent debate and related perspectives}
\label{sec:recent-strong-CP-debate}

The proposals reviewed in Sec.~\ref{sec:reassessment-physical-Hilbert-space}--\ref{sec:reassessment-consequences} have generated a substantial recent debate~\cite{Albandea:2024fui,Benabou:2025viy,Ai:2025quf,Khoze:2025auv,Bhattacharya:2025qsk,Ringwald:2026apz,Aghaie:2026pkf,Kobakhidze:2026ymc}. The criticisms address different steps of the argument, including the construction of the physical Hilbert space, the order of the thermodynamic and topological-sector limits, and the compatibility of the proposed prescription with topological susceptibility, instanton calculus, and chiral effective theory. We summarise these issues below.

\paragraph{The physical Hilbert space and the status of the \texorpdfstring{$\theta$}{theta} parameter.}

Reference~\cite{Kaplan:2025bgy} emphasises a Hamiltonian viewpoint in which $\theta$ labels quantum states rather than the Yang--Mills Hamiltonian itself. A $CP$-symmetric Hamiltonian can therefore admit distinct $\theta$-labelled superselection sectors, much as a periodic Hamiltonian admits Bloch states with different crystal momenta. Imposing parity as a symmetry of the Hamiltonian does not, by itself, eliminate the $\theta$ angle. The interpretation of $\theta$ as a label of quantum states closely parallels the diagonal basis discussed in Sec.~\ref{sec:reassessment-physical-Hilbert-space}, in which the only dependence on CP-odd
phases resides in the transformation of the wave functional; see Eq.~\eqref{eq:tranf-invariant-angle}. At this level, the two viewpoints in Refs.~\cite{Kaplan:2025bgy} and~\cite{Ai:2024vfa} are not in direct contradiction. Reference~\cite{Ai:2024vfa} further argues that, once the restriction in Eq.~\eqref{eq:constant-assumption} is removed and the enlarged class of transformations is treated as gauge redundancy, gauge invariance fixes $\theta_{\rm inv}=0$.

\paragraph{The order of limits, fixed topology, and boundary conditions.}

The ACGT order of limits has also been applied to quantum-mechanical analogues. Studies of the quantum rotor, and more recently of the rotor and pendulum, find that taking the long-time limit at fixed winding number before completing the winding sum does not reproduce the known $\theta$-dependent spectrum~\cite{Albandea:2024fui,Aghaie:2026pkf}. These examples show that the alternative order of limits is not a universal prescription for systems with winding sectors. Refs.~\cite{Ai:2020ptm,Ai:2024cnp} argue, however, that Yang--Mills theory on $\mathbb R^4$ is qualitatively different, because its topological classification is tied to gauge redundancy and the asymptotic finite-action condition, rather than to distinct homotopy classes of physical trajectories satisfying the same boundary conditions, which already exist at finite spacetime volume, or, in the quantum-mechanical analogues, at finite time extent.

A related criticism concerns finite-volume topology and fixed-topology ensembles~\cite{Bhattacharya:2025qsk,Ringwald:2026apz}. Integer topological charge can occur at finite volume on compact manifolds such as $T^4$; the issue in Sec.~\ref{sec:reassessment-order-of-limits} is instead the construction of the path integral on $\mathbb R^4$ without imposing additional global boundary data on a finite subregion. Standard fixed-topology results also show that local observables at fixed topological charge $Q$ approach those of the $\bar\theta=0$ vacuum as the volume becomes large~\cite{Brower:2003yx,Aoki:2007ka}. From the conventional viewpoint, the ACGT prescription therefore selects the $\bar\theta=0$ vacuum, whereas a fixed nonzero $\bar\theta$ requires contributions from topological charges that grow with the volume. The alternative viewpoint instead regards the fixed-$Q$ limit as the appropriate intermediate definition and questions whether the unrestricted sector sum can be interchanged with the thermodynamic limit. Reference~\cite{Ringwald:2026apz} further argues for the conventional result by invoking analyticity in the complex quark mass near the chiral limit, after rotating the full $\bar\theta$ dependence into its phase. In response, Ref.~\cite{Ai:2024cnp}\footnote{A response to the recent debate was added to the updated arXiv version of Ref.~\cite{Ai:2024cnp}; see also \href{http://users.ph.nat.tum.de/t70/CPconservation/reply_to_recent.pdf}{this online copy}.} argues that the relevant thermodynamic function is already nonanalytic in $\bar\theta$ under the ACGT prescription, so that imposing the analyticity assumed in Ref.~\cite{Ringwald:2026apz} effectively presupposes the conventional order of limits. The disagreement therefore again reduces to the question of which thermodynamic limiting prescription defines the physical QCD vacuum.

\paragraph{Instantons, susceptibility, and chiral matching.}

Reference~\cite{Khoze:2025auv} argues that, in the DIGA with the conventional order of limits, the partition function takes a simple exponential form whose exponent is given by the connected vacuum diagrams, as expected from the standard field-theoretic organisation, while correlation functions factorise into connected contributions multiplied by the same vacuum exponential. In the ACGT order, this simple exponential form of the partition function is lost. In response, Ref.~\cite{Ai:2024cnp} shows explicitly that, at fixed $\Delta n$, the correlator can still be factorised into the same connected one-instanton and one-anti-instanton contributions multiplied by a common sum of disconnected vacuum diagrams, even though the partition function itself is no longer a simple exponential. The additional large-volume prefactor identified in Ref.~\cite{Khoze:2025auv} is then an overall normalisation that cancels from normalised correlators. The disagreement therefore concerns whether a simple exponential form of the partition function is itself a necessary condition for obtaining consistent physical observables.

Reference~\cite{Benabou:2025viy} raised related objections based on the topological susceptibility, the $\eta'$ mass, and chiral effective theory. Reference~\cite{Ai:2025quf} responds that the vanishing global quantity $\langle\Delta n^2\rangle/\Omega$ in the ACGT prescription should not be identified with the local vacuum susceptibility entering anomalous singlet dynamics; the latter can remain nonzero, as discussed in Sec.~\ref{sec:reassessment-consequences}. Likewise, the conventional $\bar\theta$ dependence of ChPT follows once the anomalous singlet phase is matched conventionally to QCD. Spurion symmetry alone does not determine this matching. ChPT therefore propagates the assumed QCD matching to low-energy observables but does not independently decide between the two proposed definitions of the underlying generating functional. Whether the alternative prescription reproduces all standard nonperturbative properties of QCD remains one of the central points of the debate.

\paragraph{Boundary degrees of freedom and related perspectives.}

Reference~\cite{Kobakhidze:2026ymc} has argued that a finite region with open boundaries requires additional edge degrees of freedom that retain the topological information associated with gauge transformations acting at the boundary. If such degrees of freedom are essential to the finite-volume construction relevant to Sec.~\ref{sec:reassessment-order-of-limits}, they may restore the conventional $\theta$ structure even before the boundary is taken to infinity. Reference~\cite{Ai:2024cnp} questions whether the proposed boundary contribution constitutes an independent gauge-invariant quantity once the bulk and exterior fields are treated consistently. This issue is closely related to the broader problem of factorising the Hilbert space of a gauge theory across spatial boundaries.

Several other proposals reach related conclusions about strong $CP$ conservation through different mechanisms. Nakamura and Schierholz argue that the renormalised vacuum angle flows towards zero in the infrared~\cite{Nakamura:2021meh}, while Schierholz has separately proposed that local hadronic observables decouple from the global topological charge in the thermodynamic limit~\cite{Schierholz:2024var}. Kaplan and Sen formulate the four-dimensional gauge theory as the boundary theory of a five-dimensional regulator, in which bulk fermionic zero modes suppress sectors with nonzero global winding while local topological fluctuations and anomalous singlet dynamics remain~\cite{Kaplan:2024ezz}. Williams has likewise emphasised that local anomalous QCD dynamics need not require the full conventional global construction of integer sectors and $\theta$-vacua~\cite{Williams:2026cec,Williams:2026tuw}. These approaches are not equivalent to Refs.~\cite{Ai:2020ptm,Ai:2024vfa,Ai:2024cnp}, but share the broader question of how global topological information enters local observables.

\paragraph{Current status.}

The conventional $\bar\theta$-dependent formulation remains the mainstream framework for QCD phenomenology and underlies the results reviewed in Parts~\ref{part:origin}--\ref{part:QCD-axions}. The present debate concerns a question that takes logical precedence: whether the conventional global gauge structure, Hilbert-space sectors, and thermodynamic limiting procedure are uniquely required by the underlying gauge theory. Existing criticisms argue for the conventional formulation from the perspectives of finite-volume topology, fixed-topology ensembles, boundary data, chiral matching, and quantum-mechanical analogues, but they do not yet provide a generally accepted resolution of the gauge-theory-specific questions raised in Sec.~\ref{sec:reassessment}. Although the order-of-limits issue has attracted substantial attention and has become the main focus of the recent debate, the separate caveat concerning gauge redundancy and the construction of the physical Hilbert space has received comparatively less scrutiny, apart from the discussions in Refs.~\cite{Ai:2024vfa,Williams:2026cec}.

\section{Summary and outlook}
\label{sec:summary-outlook}

The strong $CP$ problem lies at the intersection of several characteristic features of non-Abelian gauge theory: the axial anomaly, topologically nontrivial gauge configurations, the structure of the Yang--Mills vacuum, and chiral symmetry breaking. In the conventional formulation, these ingredients lead to the invariant phase $\bar\theta=\theta+\arg\det\M$, which induces $CP$-odd hadronic interactions. Current EDM bounds then require $|\bar\theta|\lesssim10^{-10}$. This conclusion is supported by a well-developed chain of calculations connecting QCD to ChPT, hadronic and nuclear effective theories, and ultimately atomic and molecular observables. Improving this chain remains important independently of how the strong $CP$ problem is ultimately resolved. In particular, more precise lattice-QCD determinations of nucleon EDMs and other $CP$-odd hadronic matrix elements, together with improved nuclear and atomic calculations, will be essential for interpreting the next generation of EDM experiments.

Within the conventional framework, the proposed solutions differ in how they eliminate or suppress the strong $CP$ phase. An exactly massless light quark would render $\bar\theta$ unphysical, but modern lattice calculations exclude the minimal massless-up-quark possibility within the Standard Model. $CP$- and parity-based solutions instead impose a symmetry in the ultraviolet and arrange its breaking so that an order-one CKM phase can coexist with an extremely small strong $CP$ phase. Nelson--Barr, left-right, mirror, modular, and related constructions demonstrate possible realisations of this idea, but their viability depends sensitively on radiative corrections, heavy thresholds, higher-dimensional operators, and the detailed structure of symmetry breaking. Progress in this direction is therefore closely tied to ultraviolet model building and to complementary probes from EDMs, colliders, and cosmology.

The PQ mechanism remains the most extensively developed solution and has the distinctive advantage of being experimentally testable through the QCD axion. Invisible-axion constructions separate the PQ-breaking scale from the electroweak scale, while the low-energy axion mass is fixed by QCD once $f_a$ is specified. The observable couplings, however, retain substantial dependence on the ultraviolet PQ charges and particle content. Likewise, the cosmological abundance is not determined by the axion mass alone: it depends on whether PQ breaking occurs before or after inflation, on the initial conditions, and, in the post-inflationary scenario, on the evolution of strings and domain walls. There is therefore no single model- and cosmology-independent ``axion window.'' The expanding experimental programme, ranging from microwave and lumped-element haloscopes to dielectric and plasma haloscopes, helioscopes, spin-precession experiments, and purely laboratory searches, is correspondingly important because different techniques probe complementary masses, couplings, and cosmological assumptions.

A central theoretical issue for the PQ mechanism is its ultraviolet quality. Generic explicit PQ-breaking interactions can displace the axion from the $CP$-conserving QCD minimum by an unacceptable amount even when they are suppressed by very high scales. This motivates constructions in which PQ symmetry emerges accidentally from gauge or discrete symmetries, strong dynamics, compositeness, or extra-dimensional gauge invariance. Heavy QCD axions provide a complementary strategy by increasing the restoring potential, provided that the additional contribution is sufficiently well aligned with the ordinary QCD minimum. Understanding whether high-quality PQ symmetry and such alignment can arise naturally in ultraviolet-complete theories remains an important model-building problem.

The final part of this review has considered a logically distinct possibility: that the conventional interpretation of a constant strong $CP$ phase may itself require reassessment. Two complementary arguments were reviewed. In canonical quantisation, relaxing the conventional restriction on gauge transformations at spatial infinity has been argued to enlarge the gauge redundancy and remove an independent invariant $\theta$ parameter from the physical Hilbert space. In the path integral, taking the infinite-volume limit at fixed topological charge before completing the sum over sectors has been argued to yield fermionic correlation functions in which the nonperturbative chiral phase aligns with the quark mass phase. If either proposal are established as the appropriate formulation of physical QCD, the usual strong $CP$ fine-tuning problem would have to be reconsidered. These constructions nevertheless retain the axial anomaly, fermionic zero modes, the 't Hooft interaction, local topological fluctuations, and the anomalous dynamics responsible for the $\eta'$ mass.

These proposals remain under active debate. The conventional $\bar\theta$-dependent formulation remains the standard framework for QCD phenomenology, while several recent analyses have defended it or raised objections to the alternative constructions from complementary perspectives. At present, there is no generally accepted resolution of the underlying questions concerning the physical Hilbert space, the treatment of topological sectors, the order of limits, and the matching to low-energy chiral dynamics. Clarifying these issues would sharpen the theoretical foundations of the strong $CP$ problem independently of which viewpoint ultimately proves correct.

\section{Acknowledgements}

First, I am deeply grateful to my closest collaborators, Björn Garbrecht and Carlos Tamarit, who led our earlier work on this subject and from whom I learned much of what I know about its theoretical foundations. I would like to thank Björn Garbrecht for reading the manuscript carefully and for the very helpful comments. I am grateful to Hai-Yang Cheng, Juan S. Cruz, Sang Hui Im, Maxim Pospelov and Josef Pradler for helpful discussions. I would also like to thank Andrea Caputo and Tim Cohen for organising \href{https://indico.cern.ch/event/1617813/}{the Rapid Response Workshop on the Strong CP Problem}, the speakers Joshua N. Benabou, Björn Garbrecht and David E. Kaplan, the panel members Raffaele Tito D'Agnolo, Massimo D'Elia, Nick Dorey and Edward Witten, and all the participants for the stimulating discussions. I also benefited from online discussions with Jordy de Vries, Andrea Shindler and Giorgio Torrieri, together with Björn Garbrecht and Carlos Tamarit. Although my own contribution to these exchanges was limited, I found them highly illuminating. At last, I would like to thank Ciaran A. J. O’Hare for permission to adapt a figure from Ref.~\cite{OHare:2024nmr}, which is used in Fig.~\ref{fig:pre-post}.

\appendix
\renewcommand*{\thesection}{\Alph{section}}

\section{Fujikawa derivation of the chiral anomaly}
\label{app:Fujikawa-anomaly}

The spectral argument in Sec.~\ref{sec:chiral-charge-violation} makes transparent the connection between the integrated axial anomaly and the index of the Euclidean Dirac operator. The corresponding local Ward identity requires ultraviolet regularisation because the coincident-point trace over fermionic eigenmodes is formally divergent. In this appendix, we review the Fujikawa derivation~\cite{Fujikawa:1979ay,Fujikawa:1980eg}, in which the local anomaly arises from the non-invariance of the fermionic path-integral measure under a chiral rotation.

We begin with one Dirac fermion in the fundamental representation of the colour gauge group and take its mass to be real. The fermionic Euclidean action is
\begin{align}
    S_{\rm E}^{\rm fermion}
    =
    \int \d^4x\,
    \hat{\bar\psi}
    \left(
        \hat{\slashed{D}}+m
    \right)
    \hat\psi .
    \label{eq:Fujikawa-Euclidean-action}
\end{align}
Consider an infinitesimal local axial transformation,
\begin{align}
    \hat\psi(x)
    &\rightarrow
    \hat\psi'(x)
    =
    \left[1+\i\beta(x)\gamma^5\right]\hat\psi(x),
    \qquad
    \hat{\bar\psi}(x)
    \rightarrow
    \hat{\bar\psi}'(x)
    =
    \hat{\bar\psi}(x)
    \left[1+\i\beta(x)\gamma^5\right].
    \label{eq:Fujikawa-local-axial-transformation}
\end{align}
Using $\{\hat{\slashed{D}},\gamma^5\}=0$, the corresponding variation of the fermionic action is
\begin{align}
    \delta S_{\rm E}^{\rm fermion}
    =
    \i\int \d^4x\,\left[(\partial_\mu\beta)\hat{\bar\psi}\hat\gamma_\mu\gamma^5\hat\psi + 2m\beta\, \hat{\bar\psi}\gamma^5\hat\psi\right]
    =
    \i\int \d^4x\,
    \beta(x)
    \left[-\partial_\mu\hat J_\mu^5 + 2m\hat{\bar\psi}\gamma^5\hat\psi\right],
    \label{eq:Fujikawa-action-variation}
\end{align}
where $\hat J_\mu^5=\hat{\bar\psi}\hat\gamma_\mu\gamma^5\hat\psi$, and in the second line we have integrated by parts. The significance of Eq.~\eqref{eq:Fujikawa-action-variation} is most transparent in the path integral. A change of integration variables cannot alter its value. If the fermionic measure were invariant under Eq.~\eqref{eq:Fujikawa-local-axial-transformation}, this would imply $\langle\delta S_{\rm E}^{\rm fermion}\rangle=0$. Since $\beta(x)$ may be chosen as an arbitrary smooth function of compact support, one would obtain the classical Ward identity
\begin{align}
    \partial_\mu\hat J_\mu^5 = 2m\hat{\bar\psi}\gamma^5\hat\psi,
\end{align}
understood inside correlation functions, with the appropriate contact terms when additional operators are inserted. The quantum anomaly arises because the fermionic measure is not invariant under the chiral change of variables.

To see this explicitly, expand the fermion fields in a complete orthonormal basis of eigenfunctions of the massless Euclidean Dirac operator, as in Eq.~\eqref{eq:thooft_mode_expansion}, now with $N_f=1$. The fermionic measure is
\begin{align}
    \mathcal D\hat{\bar\psi}\,\mathcal D\hat\psi
    = \prod_n \d\bar b_n\,\d a_n.
\end{align}
Under Eq.~\eqref{eq:Fujikawa-local-axial-transformation}, the expansion coefficients transform as $a'_n=C_{nm}a_m$ and $\bar b'_n=\bar b_m C_{mn}$, where
\begin{align}
    C_{nm} = \delta_{nm} + \i \int\d^4x\, \beta(x) \psi_n^\dagger(x) \gamma^5 \psi_m(x).
    \label{eq:Fujikawa-transformation-matrix}
\end{align}
Equivalently, if $\bar b_n$ are regarded as a column vector, they transform with $C^T$. Since Grassmann integration measures transform with the inverse determinant and $\det C^T=\det C$, the $\hat\psi$ and $\hat{\bar\psi}$ measures each contribute a factor $(\det C)^{-1}$. Hence,
\begin{align}
    \mathcal D\hat{\bar\psi}'\, \mathcal D\hat\psi'
    = \mathcal J[\beta]\, \mathcal D\hat{\bar\psi}\, \mathcal D\hat\psi,
    \qquad
    \mathcal J[\beta] = \e^{-2\i\int\d^4x\,\beta(x)\,\mathcal A(x)},
    \label{eq:Fujikawa-Jacobian}
\end{align}
where formally $\mathcal A(x)=\sum_n\psi_n^\dagger(x)\gamma^5\psi_n(x)$. This expression is ultraviolet divergent and must be regulated. Fujikawa's prescription is to introduce a gauge-covariant regulator constructed from the Dirac operator,
\begin{align}
    \mathcal A(x) 
    = \lim_{\Lambda\rightarrow\infty}
    \sum_n \psi_n^\dagger(x) \gamma^5 \e^{\lambda_n^{2}/\Lambda^2} \psi_n(x)
    = 
    \lim_{\Lambda\rightarrow\infty}{\rm tr}_{D,c}
    \left\langle x\left| \gamma^5 \e^{\hat{\slashed{D}}^{2}/\Lambda^2}
    \right|x\right\rangle .
    \label{eq:Fujikawa-regulated-trace}
\end{align}
Here ${\rm tr}_{D,c}$ denotes the trace over Dirac and colour indices. Since $\hat{\slashed{D}}$ is anti-Hermitian, $\lambda_n^2\leq0$, and the regulator exponentially suppresses high-frequency modes.

With $\hat D_\mu=\partial_\mu-\i g_s\hat A_\mu^aT^a$ and $[\hat D_\mu,\hat D_\nu]=-\i g_s\hat G_{\mu\nu}^aT^a$, the square of the Dirac operator is
\begin{align}
    \hat{\slashed{D}}^{2} = \hat D_\mu\hat D_\mu -
    \frac{\i g_s}{4} [\hat\gamma_\mu,\hat\gamma_\nu] \hat G_{\mu\nu}^a T^a.
    \label{eq:Fujikawa-Dirac-square}
\end{align}
Only terms containing four Euclidean gamma matrices survive the trace with $\gamma^5$. With our conventions $\gamma^5=-\hat\gamma_1\hat\gamma_2\hat\gamma_3\hat\gamma_4$ and $\epsilon_{1234}=+1$,
\begin{align}
    {\rm tr}_{D}
    \left(\gamma^5 \hat\gamma_\mu \hat\gamma_\nu \hat\gamma_\rho \hat\gamma_\sigma \right)
    = -4\epsilon_{\mu\nu\rho\sigma},
    \qquad
    {\rm tr}_{D} \left[ \gamma^5 [\hat\gamma_\mu,\hat\gamma_\nu] [\hat\gamma_\rho \hat\gamma_\sigma] \right]
    = -16\epsilon_{\mu\nu\rho\sigma}.
    \label{eq:Fujikawa-gamma-trace}
\end{align}
Expanding the regulated trace to the first nonvanishing order in the field strength, and using ${\rm Tr}(T^aT^b)=\delta^{ab}/2$ for the fundamental representation and
\begin{align}
    \int\frac{\d^4p}{(2\pi)^4}\, \e^{-p^2/\Lambda^2} = \frac{\Lambda^4}{16\pi^2},
\end{align}
gives
\begin{align}
    \mathcal A(x)
    = \frac{1}{16\pi^2} \frac{1}{2} \left(-\frac{\i g_s}{4}\right)^2 {\rm tr}_{D}
    \left[\gamma^5[\hat\gamma_\mu,\hat\gamma_\nu][\hat\gamma_\rho,\hat\gamma_\sigma]\right]
    {\rm Tr}(T^aT^b) \hat G_{\mu\nu}^a \hat G_{\rho\sigma}^b
    =
    \frac{g_s^2}{32\pi^2} \hat G_{\mu\nu}^a \widetilde{\hat G}^{a\mu\nu}.
    \label{eq:Fujikawa-regulated-anomaly}
\end{align}
Terms with additional derivatives or higher powers of the background field are suppressed by inverse powers of $\Lambda$ and vanish when the regulator is removed. The regulated fermionic Jacobian is therefore
\begin{align}
    \mathcal J[\beta]
    = \e^{ -\i \frac{g_s^2}{16\pi^2} \int\d^4x\, \beta(x) \hat G_{\mu\nu}^a \widetilde{\hat G}^{a\mu\nu}}
    \label{eq:Fujikawa-final-Jacobian}
\end{align}
for one Dirac fermion in the fundamental representation.

We can now derive the local Ward identity. Performing Eq.~\eqref{eq:Fujikawa-local-axial-transformation} as a change of integration variables leaves the path integral unchanged, so to first order in $\beta(x)$ one has $\langle\delta\ln\mathcal J-\delta S_{\rm E}^{\rm fermion}\rangle=0$. The common factor of $\i$ in the two variations cancels, and since $\beta(x)$ is arbitrary, Eqs.~\eqref{eq:Fujikawa-action-variation} and~\eqref{eq:Fujikawa-final-Jacobian} give
\begin{align}
    \partial_\mu\hat J_\mu^5 = 2m\hat{\bar\psi}\gamma^5\hat\psi + \frac{g_s^2}{16\pi^2} \hat G_{\mu\nu}^a \widetilde{\hat G}^{a\mu\nu}.
    \label{eq:Fujikawa-local-Ward-one-flavour}
\end{align}
Additional operator insertions produce the corresponding contact terms from their axial transformations. For the flavour-singlet current of $N_f$ fundamental quarks, the Jacobian receives the same contribution from each flavour, giving
\begin{align}
    \partial_\mu\hat J_\mu^5 = 2\hat{\bar\psi}M\gamma^5\hat\psi + \frac{N_f g_s^2}{16\pi^2}
    \hat G_{\mu\nu}^a \widetilde{\hat G}^{a\mu\nu},
    \label{eq:Fujikawa-local-Ward-Nf}
\end{align}
which reproduces Eq.~\eqref{eq:anomalous-Ward-Euclidean}. The derivation extends directly to a chiral mass phase. For $M=m\e^{\i\alpha\gamma^5}$, one has $[M,\gamma^5]=0$, so the mass contribution in Eq.~\eqref{eq:Fujikawa-action-variation} becomes $2\hat{\bar\psi}M\gamma^5\hat\psi$, while the Jacobian is unchanged.

\bibliography{Refs}

\end{document}